\documentclass[a4paper,11pt]{article}
\usepackage[dvipsnames]{xcolor}
\usepackage[utf8]{inputenc}
\usepackage[bbgreekl]{mathbbol}
\usepackage{geometry}

\usepackage{xcolor}
\usepackage{amsmath, array, amssymb, amsfonts,amsthm}
\usepackage{upgreek}
\usepackage{xfrac}
\usepackage[inline]{enumitem}
\setlist{nolistsep}
\usepackage[french, english]{babel}
\usepackage[all]{xy}
\usepackage{txfonts}  
\usepackage{sectsty}
\usepackage{booktabs}
\usepackage{caption}
\usepackage{dsfont}
\usepackage{mathtools}
\usepackage{slashed}
\usepackage[makeroom]{cancel}
\usepackage[hidelinks]{hyperref}
\usepackage{textcomp}
\usepackage{multicol}
\usepackage[title]{appendix}
\usepackage[square,numbers,sort,compress,semicolon,merge]{natbib}
\let\cite\citep 
\usepackage{hyperref}
\hypersetup{colorlinks=true, urlcolor=blue, citecolor=blue, linktoc=page}
\allowdisplaybreaks

\usepackage{tikz}
\usepackage{tikz-cd}
\usetikzlibrary{cd}
\tikzcdset{
arrow style=tikz,
diagrams={>={Straight Barb[scale=0.8]}}
}

\DeclareMathAlphabet{\mathpzc}{OT1}{pzc}{m}{it}

\usepackage{mathrsfs}
\usepackage{footmisc}

\makeatletter
\renewcommand*\env@matrix[1][\arraystretch]{%
  \edef\arraystretch{#1}%
  \hskip -\arraycolsep
  \let\@ifnextchar\new@ifnextchar
  \array{*\c@MaxMatrixCols c}}
\makeatother

\newcommand{\defeq}{\vcentcolon=}
\newcommand{\rdefeq}{=\vcentcolon}
\renewcommand\P{\mathcal{P}}
\newcommand\M{\mathcal{M}}

\newcommand\RR{\mathbb{R}}
\newcommand\CC{\mathbb{C}}

\newcommand\id{\textit{id}}
\newcommand\T{\mathcal{T}}
\newcommand\G{\mathcal{G}}
\renewcommand\H{\mathcal{H}}
\newcommand\A{\mathcal{A}}
\newcommand\B{\mathcal{B}}
\newcommand\E{\mathcal{E}}

\renewcommand\S{\mathcal{S}}

\newcommand\SU{\mathcal{SU}}
\newcommand\U{\mathcal{U}}
\newcommand\SO{\mathcal{SO}}

\newcommand\Q{\mathcal{Q}}
\newcommand\K{\mathcal{K}}
\newcommand\J{\mathcal{J}}
\renewcommand\O{\mathcal{O}}
\newcommand\GL{\mathcal{GL}}
\newcommand\D{\mathcal{D}}

\newcommand\vphi{\varphi}

\renewcommand\epsilon{\varepsilon}

\newcommand\rarrow{\rightarrow}

\newcommand\su{\mathfrak{su}}
\newcommand\so{\mathfrak{so}}

\newcommand\co{\mathfrak{co}}

\renewcommand\t{\tilde}

\renewcommand\b{\bar }
\newcommand\w{\wedge}
\renewcommand\d{\partial}
\newcommand\s{\sigma}
\newcommand\bs{\boldsymbol}

\renewcommand\-{^{-1}}
\newcommand\Ad{\text{Ad}}
\newcommand\ad{\text{ad}}
\renewcommand\id{\text{id}}

\def\munderline#1#2{\color{#1}\underline{{\color{black}#2}}\color{black}}

\makeatletter

\newcommand{\Rmnum}[1]{\expandafter\@slowromancap\romannumeral #1@}
\makeatother

\makeatletter
\newcommand{\leqnomode}{\tagsleft@true\let\veqno\@@leqno}
\newcommand{\reqnomode}{\tagsleft@false\let\veqno\@@eqno}
\makeatother

\DeclareMathOperator{\Diff}{Diff}
\DeclareMathOperator{\Aut}{Aut}
\DeclareMathOperator{\Tr}{Tr}
\DeclareMathOperator{\vol}{vol}

\newtheorem{thm}{Theorem}

\newtheorem{prop}[thm]{Proposition}
\theoremstyle{definition}
\newtheorem{definition}[thm]{Definition}

\makeatletter
\newcounter {subsubsubsection}[subsubsection]
\renewcommand\thesubsubsubsection{\thesubsubsection .\@alph\c@subsubsubsection}
\newcommand\subsubsubsection{\@startsection{subsubsubsection}{4}{\z@}%
                                     {-3.25ex\@plus -1ex \@minus -.2ex}%
                                     {1.5ex \@plus .2ex}%
                                     {\normalfont\normalsize\bfseries}}
\newcommand*\l@subsubsubsection{\@dottedtocline{3}{10.0em}{4.1em}}
\newcommand*{\subsubsubsectionmark}[1]{}
\makeatother

\begin{document}


\title{Bundle geometry of the connection space, covariant Hamiltonian formalism, the problem of boundaries in gauge theories, and the dressing field method}
\author{J. François${\,}^{a}$}
\date{}

\maketitle
\begin{center}
\vskip -0.8cm
\noindent
${}^a$ Service de Physique de l'Univers, Champs et Gravitation, Universit\'e de Mons -- UMONS\\
20 Place du Parc, B-7000 Mons, Belgique.
\end{center}
\vspace{-3mm}

\begin{center}
\parbox{11cm}{ \begin{center}{\small 
\emph{In loving memory of my grandparents, Monique and Albert ANDR{\'E}, \\
in whose home my intellectual life began
, and has thrived ever since.}}
\end{center}
}
\end{center}
\vspace{-3mm}

\begin{abstract}
We  take advantage of the principal bundle geometry of the space of \mbox{connections}  to obtain general results on the presymplectic structure of two classes of (pure) gauge theories: invariant theories, and non-invariant theories satisfying two restricting hypothesis. In particular, we  derive the general field-dependent gauge transformations of the presymplectic potential and presymplectic 2-form in both cases.  
We point-out that a generalisation of the standard bundle geometry, called  \emph{twisted} geometry, arises naturally in the study of non-invariant gauge theories (e.g. non-Abelian Chern-Simons theory).  
These  results prove that the well-known problem of associating a symplectic structure to a gauge theory over bounded regions is a generic feature of both classes. 
The \emph{edge modes} strategy, recently introduced to address this issue, has been actively  developed in various contexts by several authors. 
We draw attention to the \emph{dressing field method} as the geometric framework underpinning, or rather encompassing, this strategy. The geometric insight afforded by the method both clarifies it and clearly delineates its potential shortcomings as well as its conditions of success. 
 Applying our general framework to various examples allows to straightforwardly recover several results of the recent literature on edge modes and on the presymplectic structure of general relativity.  \end{abstract}

\textbf{Keywords} : Differential geometry, covariant Hamiltonian formalism, boundaries in gauge theory, edge modes.

\vspace{-3mm}

\tableofcontents

\bigskip


\section{Introduction}  

The covariant Hamiltonian formalism, whose inception is due to \cite{Zuckerman1986, Witten1986, CrnkovicWitten1986, Crnkovic1987}), aims at providing an analogue of the canonical formalism for field theory,  especially gauge field theory, that preserves relativistic covariance. In this framework, the physical - or reduced - phase space associated to a gauge theory with Lagrangian $L$ defined over a region $\Sigma$ is the space $\S/\H$ of orbits of solutions $\S$,  determined by $L$,  under the action of the gauge group $\H$ of the theory (we neglect the question of constraints).  From $L$ still, one derives the so-called presymplectic potential $\bs\theta_\Sigma$ and its associated presymplectic 2-form $\bs\Theta_\Sigma$.  Under adequate boundary conditions (either the region is s.t. $\d\Sigma=\emptyset$ or the gauge fields fall-off quickly enough), the latter are gauge-invariant and thus induce a well-behaved symplectic structure  on the physical phase space. One leading motivation was that from such a covariant symplectic phase space, a covariant canonical quantization \cite{Witten1986, Crnkovic1987} -  or a geometric quantization \cite{Zuckerman1986} - of  the theory may be within~reach. 


Yet, a ``boundary problem"  arises in this framework: 
due to the lack of invariance of $\bs\theta_\Sigma$ and $\bs\Theta_\Sigma$  in the presence of boundaries, one cannot assign
a symplectic structure to a gauge field theory over a bounded  region, or 
decompose
the symplectic structure of a boundaryless region $\Sigma$ into symplectic sub-structures associated to an arbitrary partition of $\Sigma$ into  subregions $\cup_i \Sigma_i$  sharing  fictitious boundaries $\d\Sigma_i$. 
The latter circumstance is the classical analogue of the problem of factorisation of a Hilbert space associated to a region into a tensor product of Hilbert subspaces associated to subregions, which is relevant e.g. to the problem of entanglement entropy, or to the problem of how a (semi)-classical world emerges from quantum theory (see e.g. \cite{Carroll-Singh2020}).

One attempt to address this boundary problem is the quite recent proposal by \cite{DonnellyFreidel2016} to introduce so-called \emph{edge modes}, i.e. degrees of freedom (d.o.f.) living on the boundary and whose role is to restore the gauge invariance of the bare presymplectic structure. It is argued that the introduction of edge modes not only gives an \emph{extended} presymplectic structure that circumvents the boundary problem, but also reveals a new kind of physical symmetries, often called \emph{boundary} or \emph{surface symmetries}, whose associated charges are new observables. The proposal have been followed by many and applied to various contexts, see e.g. \cite{Geiller2017, Geiller2018, Speranza2018, Geiller2019} and references therein. Nonetheless, both the origin and the physical meaning of these edge modes remain obscure and a topic of active discussion, even among  philosophers of physics interested in foundational problems in gauge theories \cite{Teh2020}. 
\medskip

Our aim here is twofold. First, we want to give general results about the presymplectic structure of large classes of gauge theories by tacking advantage of the natural principal bundle geometry of field space. Then, using these results, we want to show that a gauge symmetry reduction scheme known as the \emph{dressing field method} (DFM) provides a systematic strategy to deal with the boundary problem, and that it is the geometric  underpinning of the edge mode proposal (as first noted in \cite{Teh2020}). Actually the DFM framework encompasses and clarifies it. 

We try to provide an account that is as synthetic and self-contained  as possible. We believe this to be necessary  because we describe notions that are not well-known to a wider audience (especially in section \ref{Twisted bundle geometry} and \ref{The dressing field method}), so that their precise articulation with better known concepts need to be spelled out in some details. To this end, the main body of the paper is completed by appendices collecting standard material whose knowledge is often tacitly assumed, or  results that are either hard to find or scattered among several sources. It has been useful to the author to collect this material, it may be useful to  some readers too. 
The plan of the paper is thus as follows. 

In section \ref{Connections on principal bundles} we synthesise relevant facts about the framework of principal bundles,  Ehresmann and Cartan connections underlying classical gauge theories. This gives a template to describe a  generalisation of this framework called ``twisted geometry", that will turn out to be relevant to our purpose. 
In section \ref{The dressing field method} we review  the dressing field method. 
In section \ref{The space of connections as a principal bundle}
 we describe the bundle geometry of the space of connections $\A$ and show how the twisted geometry naturally arises in gauge theories. Most of what is done there is indifferent as to whether we consider $\A$ to be the space of connections on a principal bundle $\P$ or to be the space of local representatives of such connections - i.e. YM potentials - on the base manifold $\M$ of $\P$. We here give geometric substance to several heuristic computations found in the recent literature. 
  
 In section \ref{Covariant Hamiltonian formalism (and the puzzle of boundaries)} we reach the main goal of the paper. We begin by briefly reviewing the basics of the covariant Hamiltonian formalism. Then, thanks to the  material articulated in previous sections, we are able to give general results about the presymplectic structure of two classes of gauge theories. 
 
Considering first the class of invariant theories, we define integrable Noether currents and charges and we prove that the Poisson algebra of Noether charges, equipped with the Poisson bracket induced by the presymplectic \mbox{2-form}, is isomorphic to the Lie algebra of (field-independent) gauge transformations. We then derive from first principles the general field-dependent gauge transformations of the presymplectic potential $\bs\theta_\Sigma$ and 2-form $\bs\Theta_\Sigma$. 

Considering next non-invariant theories, we work under two restricting assumptions that allow us to define \mbox{integrable} Noether currents and charges conserved on-shell. We then prove that the Poisson algebra of such charges, equipped again with the Poisson bracket induced by the presymplectic 2-form, is a central extension of the Lie algebra of (field-independent) gauge transformations. We again derive from first principles the field-dependent gauge transformations of the presymplectic structure associated to such theories. 

The general results on the field dependent-gauge transformations of $\bs\theta_\Sigma$ and $\bs\Theta_\Sigma$ highlight the generic character of the boundary problem. Relying on these results, we show that via the DFM one can  define  \emph{dressed} presymplectic structures that may circumvent this  problem, and that it is the geometric underpinning of the edge modes strategy as introduced in \cite{DonnellyFreidel2016}.
But we offer some caveats as to why this strategy might fail: namely, the problem may re-emerge with respect to  a new (non-physical) symmetry encoding an ambiguity in the choice of \emph{dressing field} (identified with the so-called \emph{surface symmetries}). In particular, when the latter is introduced by fiat, this ambiguity is essentially a replica of the initial gauge symmetry. We conclude that the only way the DFM can help with the boundary problem is to 1) build a dressing field from the original field content of the theory in such a way that 2) the ambiguity in the constructive procedure is minimal. 
We also argue that this might be achieved only by 
sacrificing the locality of the theory if its gauge symmetry is \emph{substantial}. Only for theories with \emph{artificial} gauge symmetries can the boundary problem be solved via the DFM without losing  locality (but then the problem was  fictitious  to begin with). 
%

Results of a recent literature on the symplectic structure of General Relativity, unrelated to edge modes, are also recovered as special cases of the DFM. 
  We review our results in the conclusion and hint at forthcoming extensions.


\section{Connections on principal bundles}
\label{Connections on principal bundles}

In sections \ref{Principal bundles and their smooth structure} and \ref{Ehresmann and Cartan connections} we review in a synthetic and self-contained  manner many elementary definitions and facts about fiber bundles and connections so as to fix some notations, and to lay the ground for both the description of a generalisation of these well-known notions in section \ref{Twisted bundle geometry}, and to their extension to the less familiar context of infinite dimensional manifolds in section \ref{The space of connections as a principal bundle}.

\subsection{Principal bundles and their smooth structure}
\label{Principal bundles and their smooth structure}

The primary object to consider is a principal bundle $\P$ over a base $n$-dimensional manifold $\M$ supporting a smooth free right action by a Lie group $H$, called its structure group. Given $p\in \P$ and $h\in H$ this action $\P \times H \rarrow \P$ is noted $(p, h) \mapsto R_h p \defeq ph$. It defines an equivalence relation on $\P$: any $p,p'\in\P$ such that (s.t.) $p'=R_hp=ph$ belong to the same fiber. Alternatively, the fiber to which $p$ belongs  is its $H$-orbit. The base manifold parametrizes the set of fibers, $\P/H \simeq \M$, and one has the projection $\pi : \P \rarrow \M$, $p\mapsto \pi(p)=x$, such that $\pi \circ R_h=\pi$. 

A bundle is locally trivial in that given $U \subset \M$, $\P_{|U}\simeq U \times H$. Locally it is always possible to find a section of $\pi$, called a trivialising (or local) section, $\s:U \rarrow \P$, s.t. $\pi \circ \s=\id_U$.  If $\exists$ a global section $\s:\M \rarrow \P$, then the bundle is trivial, $\P \simeq \M \times H$.  
Given $\s_i$ and $\s_j$ sections over  $U_i,U_j\subset \M$ s.t. $U_i \cap U_j \neq \emptyset$, on the overlap $\s_j=\s_i\, g_{ij}$ where $g_{ij}:  U_i \cap U_j \rarrow H$ is a \emph{transition function}. The set $\{g_{ij}\}$ of transition functions  subordinated to a covering $\{U_i\}_{i\in I \subset\mathbb{N}}$ of $\M$ are local data from which it is possible to reconstruct the bundle $\P$. 

The automorphism group of $\P$ is the subgroup of its diffeomorphisms that commute with the right $H$-action, thereby preserving the fibration structure: $\Aut(\P)\defeq \left\{\, \Psi \in \Diff(\P)\,|\, \Psi \circ R_h = R_h \circ \Psi\, \right\}$. Thus any $\Psi \in \Aut(\P)$ projects to a well-defined $\psi=\pi\circ\Psi \circ \pi\-\rdefeq \t\pi(\Psi)\in \Diff(\M)$.  As the name suggest, $\Aut(\P)$ is the natural transformation group of $\P$. The subgroup of \emph{vertical automorphisms} further preserves the fibers, $\Aut_v(\P)\defeq\left\{\, \Psi \in \Aut(\P)\, |\, \pi \circ \Psi = \pi\, \right\}$, its elements project to the identity transformations $\id_x$ on $\M$. It is easy to see that it is isomorphic to the \emph{gauge group} $\H\defeq\left\{\, \gamma : \P \rarrow H\, |\, \gamma(ph)=h\- \gamma(p) h\, \right\}$, with the identification given by $\Psi(p)=p\gamma(p)$. We have the following short exact sequence (SES) of groups: 
\begin{align}\label{SESgroups}
\makebox[\displaywidth]{
\hspace{-18mm}\begin{tikzcd}[column sep=large, ampersand replacement=\&]
\&0     \arrow [r]         \& \Aut_v(\P) \simeq \H     \arrow[r, "\iota"  ]          \& \Aut(\P)       \arrow[r, "\t\pi"]      \&  \Diff(\M) \simeq \Aut(\P)/\H         \arrow[r]      \& 0.  
\end{tikzcd}}  \raisetag{3.4ex}
\end{align}
Without a splitting of this SES, one cannot lift a $\psi \in \Diff(\M)$ into an element of $\Aut(\P) $, or decompose uniquely an element of $\Aut(\P)$ into a vertical automorphism and a diffeomorphism on  $\M$. 
\medskip

As a smooth manifold, $\P$ has a tangent bundle $T\P\defeq \bigcup_{p\in \P} T_p\P$ whose sections $\Gamma(T\P)$ are vector field given in terms of their flow $\phi_\tau:\P \rarrow \P$ by $X_p=\tfrac{d}{d\tau} \phi_\tau(p)\big|_{\tau=0}$. A $H$-right invariant vector field $X\in \Gamma_H(T\P)$ is s.t. $R_{h*} X_p=X_{ph}$, so that $\pi_*X_{ph}=\pi_*R_{h*} X_p=\pi_*X_p$ is a well defined vector fields $\t X \in \Gamma(T\M)$ on $\M$. Under the bracket $[\,,\,]$ of vector fields, $\Gamma_H(T\P)$ is stable and therefore a Lie subalgebra of $\Gamma(T\P)$. Then $\pi_* :  \Gamma_H(T\P) \rarrow  \Gamma(T\M)$ is a morphism of Lie algebras. Furthermore, it is easy to see that the flow of $X\in \Gamma_H(T\P)$ is s.t. $\phi_\tau \circ R_h =R_h \circ \phi_\tau$, i.e. $\phi_\tau \in \Aut(\P)$. In other words $\Gamma_H(T\P)=$ Lie$\Aut(\P)$.

The right action of $H$ on $\P$ induces a canonical subbundle $V\P \subset T\P$. An element $X \in$ Lie$H$ gives a vertical vector $X^v_p\defeq  \tfrac{d}{d\tau} p e^{\tau X}\big|_{\tau=0}$  tangent to a curve $pe^{\tau X}$ contained in the fiber through $p$. The vector space of vertical vectors at $p$ is $V_p\P \subset T_p\P$, and the vertical subbundle is $V\P=\cup_{p\in\P}V_p\P$.  Clearly $\pi_*X^v=0$, and  one easily proves $R_{h*}X^v_p=(h\-Xh)^v_{ph}$. A vertical vector field is $X^v \in \Gamma(V\P)$, and the map Lie$H \rarrow \Gamma(V\P)$ is an injective morphism of Lie algebras. Any element of the Lie algebra of the gauge group Lie$\H\defeq \left\{\, \chi : \P \rarrow \text{Lie}H\, |\, \chi(ph)=h\-\chi(p)h\, \right\}$ induces a $H$-right invariant vertical vector field via $\chi^v\defeq  \tfrac{d}{d\tau} p e^{\tau \chi}\big|_{\tau=0}$. The map Lie$\H\rarrow \Gamma_H(V\P)$ is a Lie algebra \emph{anti}-isomorphism.\footnote{One shows that $[\chi^v, \eta^v]_p=\left(-[\chi, \eta] \right)^v_p$. We have a Lie algebra isomorphism with the definition $\chi^v\defeq  \tfrac{d}{d\tau} p e^{-\tau \chi}\big|_{\tau=0}$. See appendix \ref{Lie algebra (anti)-isomorphisms} for a technical proof.} Thus, corresponding to \eqref{SESgroups}, we have the SES of Lie algebras
\begin{align}\label{SESLieAlg}
\makebox[\displaywidth]{
\hspace{-18mm}\begin{tikzcd}[column sep=large, ampersand replacement=\&]
\&0     \arrow [r]         \& \Gamma_H(V\P) \simeq \text{Lie}\H     \arrow[r, "\iota"  ]          \&  \Gamma_H(T\P)     \arrow[r, "\pi_*"]      \&  \Gamma(T\M)  \simeq \Gamma_H(T\P)/\Gamma_H(V\P)         \arrow[r]      \& 0
\end{tikzcd}}  \raisetag{3.4ex}
\end{align}
(also kown as the \emph{Atiyah Lie algebroid} associated to $\P$). Here again, a splitting is needed to either lift a $X\in \Gamma(T\M)$, or decompose uniquely $X\in\Gamma_H(T\P)$ into a vertical vector field and an `horizontal' part. 
\medskip

As a smooth manifold still, $\P$ has a de Rham complex $\left(\Omega^\bullet(\P), d \right)$, where $d$ is the exterior de Rham (degree 1) derivative defined the usual way via Kozsul formula. The interior product $\iota : \Gamma(T\P) \times \Omega^\bullet(\P) \rarrow \Omega^{\bullet -1}(\P)$, $(X, \omega) \mapsto \iota_X \omega$, is a degree $-1$ derivative $\forall X$. Then, one can define the Lie derivative as the degree 0 derivation $L_X\defeq [\iota_x, d]=\iota_X d + d\iota_X$.\footnote{The vector space of derivation Der$(\sf A)$ of an algebra $\sf A$ is a Lie algebra under the graded bracket $[d_1, d_2]\defeq  d_1 \circ d_2 - (-)^{|d_1|\cdot|d_2|}d_2 \circ d_1$.}
It satisfies $[L_X, \iota_Y]=\iota_{[X, Y]}$, and it is a Lie algebra morphism $[L_X, L_Y]=L_{[X, Y]}$.\footnote{Thus formulated, the action of $\Gamma(T\P)$ - as a Lie algebra - on $\Omega^\bullet(\P)$ via $\iota_X$ and $L_X$ is the motivating example for the abstract notion of \emph{Cartan operation} (after Henri Cartan, son of \'{E}lie Cartan after who are named e.g. Cartan's structure equation and Cartan connections - see section \ref{Ehresmann and Cartan connections}).}

 An exterior product $\w$ is defined the usual way on the space $\Omega^\bullet(\P, \sf A)$ of differential forms with values in an algebra $(\sf A, \cdot)$ - and kept tacit throughout the paper to lighten the notation - so that  $\big( \Omega^\bullet(\P, \sf A), \w, d \big)$ is a differential graded algebra. Such a product is not defined on $\Omega^\bullet(\P, V)$ where $V$ is merely a vector space. But if $(V, \rho)$ is a representation for $H$, 
 then one can define the vector space of \emph{equivariant} forms on $\P$ as those for which the pullback by $R_h$ is well defined: $\Omega_\text{eq}^\bullet(\P, V)=\left\{ \, \omega \in \Omega^\bullet(\P, V)\,|\, R^*_h\omega_{|ph}=\rho(h)\-\omega_{|p},\, \text{ or }\ L_{X^v}\omega=-\rho_*(X) \omega\, \right\}$. Forms with trivial equivariance constitute the subspace of invariant differential forms: $\Omega_\text{inv}^\bullet(\P, V)=\left\{ \, \omega \in \Omega^\bullet(\P, V)\,|\, R^*_h\omega_{|ph}=\omega_{|p}, \, \text{ or }\ L_{X^v}\omega=0 \,\right\}$. 

The space $\Omega_\text{hor}^\bullet(\P)=\left\{ \, \omega \in \Omega^\bullet(\P)\,|\, \iota_{X^v}\omega=0 \right\}$ of \emph{horizontal} forms could have been defined earlier. A form which is both horizontal and equivariant is said \emph{tensorial}: $\Omega_\text{tens}^\bullet(\P, V)=\left\{ \, \omega \in \Omega^\bullet(\P, V)\,|\, R^*_h\omega=\rho(h)\-\omega,\, \text{ and }\ \iota_{X^v}\omega=0\, \right\}$. 
Notice that obviously $\Omega_\text{tens}^0(\P, V)=\Omega_\text{eq}^0(\P, V)$. 
Finally a form which is both horizontal and invariant is said \emph{basic}: $\Omega_\text{basic}^\bullet(\P, V)=\left\{ \, \omega \in \Omega^\bullet(\P, V)\,|\, R^*_h\omega=\omega,\, \text{ and }\ \iota_{X^v}\omega=0\, \right\}$. Basic forms are remarkable since, as their name suggests, they descend as well-defined forms on $\M$. Actually, an alternative definition is that basic forms are those that belong to Im$(\pi^*)$, that is: $\Omega_\text{basic}^\bullet(\P, V)=\left\{ \, \omega \in \Omega^\bullet(\P, V)\,|\, \exists\, \beta \in \Omega^\bullet(\M, V) \text{ s.t. } \omega=\pi^*\beta \,  \right\}$. 

\medskip
The action of $\Aut_v(\P)\simeq\H$ on a differential form $\omega$ defines its \emph{active gauge transformation}, $\omega^\gamma\defeq \Psi^*\omega$. Since we have on the one hand that $\Psi^*\omega_{|\Psi(p)} (X_p)=\omega_{|\Psi(p)}(\Psi_*X_p)$ for a generic $X\in \Gamma(T\P)$, and on the other hand the well known result  that for $\Psi \in \Aut_v(\P)$,
\begin{align}
\label{Pushforward-X}
\Psi_*X_p&= R_{\gamma(p)*} X_p + \left\{ \gamma\- d\gamma_{|p}(X_p)\right\}^v_{p\gamma(p)}
               = R_{\gamma(p)*} \left( X_p + \left\{ d\gamma {\gamma\- }_{|p}(X_p)\right\}^v_p \right),
\end{align}
it is clear that the gauge transformation of a differential form is given by its equivariance and verticality properties. In particular the gauge transformations of tensorial forms are immediate: if $\omega \in \Omega_\text{tens}^\bullet(\P, V) \Rightarrow \omega^\gamma= \rho(\gamma)\-\omega$. As a special case, basic forms are gauge-invariant: $\omega \in \Omega_\text{basic}^\bullet(\P, V) \Rightarrow \omega^\gamma= \omega$. In more than one way, the latter are of paramount importance in gauge theories (as they encompass e.g. Lagrangians, action functionals, or candidate observables). 

Given a trivialising section $\s$ over $U\subset \M$, a local representative of a form $\alpha$ on $\P$ is $a\defeq\s^*\alpha \in \Omega^\bullet(U)$. Local representatives $a'=\s'^{*}\alpha$ and $a=\s^*\alpha$ on open sets $U' \cap U \neq \emptyset$ must, on the overlap where $\s'=\s\,g$, satisfy gluing relations often called \emph{passive gauge transformations}. Since on the one hand 
$a_{|x}(X_x)\defeq\s^*\alpha_{|\s(x)}(X_x)=\alpha_{|\s(x)}(\s_{*}X_x)$, for a generic $X\in \Gamma(T\M)$, and on the other hand we have the well known result
\begin{align}
\label{Push-X-local}
\s'_*X_x = R_{g(x)*} \left( \s_*X_x \right) +  \left\{  g\-dg_{|x}(X_x)  \right\}^v_{\s'(x)}= R_{g(x)*} \left( \s_*X_x +  \left\{  dgg\-_{|x}(X_x)  \right\}^v_{\s(x)} \right) ,
\end{align}
it is again clear that the passive gauge transformation of the a local representative $a$ is determined by the equivariance and verticality properties of $\alpha$. It is in particular immediate that $\alpha \in \Omega_\text{tens}^\bullet(\P, V) \Rightarrow a'= \rho(g)\-a$, and that $\alpha \in \Omega_\text{basic}^\bullet(\P, V) \Rightarrow a'= a$. Again, for obvious reasons the latter are of special significance to gauge theories. 

Passive gauge transformations are akin to coordinate changes on $\M$ (in GR), while active gauge transformations are akin to action by diffeomorphisms. Passive and active gauge transformations are formally alike, and locally indistinguishable.\footnote{Meaning that the pullback on $U$ by $\s$ of $\alpha^\gamma=\rho(\gamma)^{-1}\alpha$ cannot be told apart from $a'=\rho(g)^{-1}a$.} In the rest of this paper, we will  deal only with active gauge transformations. 
\medskip

The notion of associated bundles is of special importance. Given a representation $(V, \rho)$ of $H$, define a right action of $H$ on $\P \times V$  by $\left((p, v), h \right) \mapsto (ph, \rho(h)\-v)$, and consider the equivalence relation under this action $(p, v)\sim (ph, \rho(h)\-v)$. One builds an associated bundle to $\P$ as the space of equivalence classes under this action, noted $E=\P \times_H V \defeq \P\times V/\sim$ with projection $\pi_E(q)=\pi_E([p, v])\defeq \pi(p)=x \in \M$. The space $\Gamma(E)$ of sections of $E$ is isomorphic to the space 
$\Omega^0_\text{eq}(\P, V)=\left\{\, \vphi : \P \rarrow V\,|\, R^*_h\vphi=\rho(h)\-\vphi\, \right\}$ of equivariant functions (0-forms) on $\P$ by the correspondence:  $\vphi \Rightarrow s(x)\defeq [p, \vphi(p)]\sim [ph, \vphi(ph)]=[ph, \rho(h)\-\vphi(p)]$. Sections of $E$/equivariant functions (or rather their local representatives $\phi\defeq \s^*\vphi:U \rarrow V$) represent various kinds of matter fields.

\subsection{Ehresmann and Cartan connections}
\label{Ehresmann and Cartan connections}

All of the above 
come, freely, from the smooth structure of $\P$ (and the representations of its structure group). Now, the latter can be endowed with an additional structure: an Ehresmann - or principal - connection form, i.e. a Lie$H$-valued 1-form $A$ s.t. 
\begin{enumerate}[label=\roman*, leftmargin=*, itemsep=1pt]
\item $R^*_hA_{|ph}=\Ad_{h\-}A_{|p}$, i.e. $A \in \Omega_\text{eq}^1(\P, \text{Lie}H)$,
\item $A_{|p}(X^v_p)=X \in$ Lie$H$, where $X^v_p \in V_p\P$.
\end{enumerate}
The set $\A$  of Ehresmann connections on $\P$ is an affine space modelled on the vector space $\Omega_\text{tens}^{1}(\P, \text{Lie}H)$: given $A', A \in \A$, it is easy to see that  $ A'-A \in \Omega_\text{tens}^{1}(\P, \text{Lie}H)$. Or, given $A\in \A$ and $\beta\in \Omega_\text{tens}^{1}(\P, \text{Lie}H)$, $A'\defeq A+\beta \in \A$. 

Any connection $A\in \A$ splits the SES \eqref{SESLieAlg}, because it is a retraction of the map $\iota$: $A \circ \iota  = \id_{\text{Lie}\H}$ .  At any  $p\in \P$, it allows to define a horizontal complement to $V_p\P$ in $T_p\P$ by $H_p\P\defeq \ker A_{|p}$, so that any $X_p\in T_p\P$ has horizontal component $X_p^h\defeq X_p-\{A_{|p}(X_p)\}_p^v$. $A$ thus defines a non-canonical subbundle $H\P=\cup_{p\in\P}H_p\P \subset T\P$.  The horizontal lift of a curve $c_\tau$ in $\M$ is a curve $c_\tau^h$ in $\P$ whose tangent vector field is horizontal. Correspondingly, the horizontal lift of $\t X\in \Gamma(\M)$ is ${\t X}^h \in \ker A$ s.t $\pi_*\t X^h=\t X$. 

Furthermore, and most importantly to us, an Ehresmann connection allows to define a covariant derivative  $D^A\!:~\!\Omega_\text{eq}^\bullet(\P, V) \rarrow \Omega_\text{tens}^{\bullet+1}(\P, V)$, which on $\Omega^\bullet_\text{tens}(\P, V)$ (thus in particular  on sections $\Gamma(E)\simeq \Omega_\text{eq}^0(\P, V)$) is given algebraically by $D^A\!= d\, +\rho_*(A)$. 
It is easy to show that $D^A\circ D^A=\rho_*(F)$,
 where $F$ is the curvature 2-form of $A$, given algebraically by Cartan's structure equation
 : $F=dA+\sfrac{1}{2}[A, A]$. By definition, $F \in \Omega_\text{tens}^2(\P, \text{Lie}H)$, thus $D^A$ acts on it, trivially so, giving the Bianchi identity $D^AF=0$. From the above general discussion follows that its gauge transformation is $ F^\gamma=\gamma\-F\gamma$. Which can also be found via Cartan's structure equation and the fact that, due to the axioms i-ii and eq.\eqref{Pushforward-X}, the gauge transformations of the connection is $A^\gamma = \gamma\-A\gamma+\gamma\-d\gamma$.

Locally, on $U\subset \M$, the local representative of $A$ via a section $\s$ is a Yang-Mills (YM) potential, while  $\s^*F$ is the YM field strength, and $\s^*(D^A\vphi)$ is the minimal coupling between the YM potential and a matter field $\vphi$. Ehresmann connections are thus the geometric underpinning of (classical) YM gauge theories. 
\medskip

For gravitational gauge theories, another kind of connection is best suited: Cartan connections (see \cite{Sharpe, Cap-Slovak09}). 
Given Lie$G\supset$ Lie$H$,\footnote{It is not necessary to assume that Lie$G$ exponentiates into a Lie group $G$, but  in the following we will nonetheless tacitly admit it does.} a Cartan connection $\b A$ on $\P$ is a Lie$G$-valued 1-form s.t. 
\begin{enumerate}[label=\roman*, leftmargin=*, itemsep=1pt]
\item $R^*_h \b A_{|ph}=\Ad_{h\-}\b A_{|p}$, i.e. $\b A \in \Omega_\text{eq}^1(\P, \text{Lie}G)$,
\item $\b A_{|p}(X^v_p)=X \in$ Lie$H$, where $X^v_p \in V_p\P$,
\item $\forall p\in \P$, $\b A_{|p} : T_p\P \rarrow$ Lie$G$ is a linear isomorphism. 
\end{enumerate}
The set $\B$  of Cartan connections on $\P$ is an affine space modelled on the vector space $\Omega_\text{tens}^{1}(\P, \text{Lie}G)$: given $\b A', \b A \in \B$, it is easy to see that  $ \b A'-\b A \in \Omega_\text{tens}^{1}(\P, \text{Lie}G)$. Or, given $\b A\in \B$ and $\beta\in \Omega_\text{tens}^{1}(\P, \text{Lie}G)$, $\b A'\defeq \b A+\beta \in \B$. 

A pair $(\P, \b A)$ is a Cartan geometry. Contrary to an Ehresmann connection, a Cartan connection is not designed to split the sequence \eqref{SESLieAlg} and to define an horizontal subbundle $H\P$. Rather, the distinguishing axiom iii has several important consequences. First, obviously $\dim\P=\dim G$ and $\dim\M= \dim G/H$.  Then, $\b A$ induces a \emph{soldering} on $\M$, i.e. due to $\ker \b A= \emptyset$ one has the bundle isomorphism: $T\M \simeq \P \times_H \text{Lie}G/\text{Lie}H$.\footnote{If Lie$H$ is only a subalgebra of Lie$G$ instead of an ideal, \text{Lie}G/\text{Lie}H is merely a vector space - not a subalgebra - and  $H$ acts on it via the $\Ad$ representation. As for Lie$G$, it acts via the $\ad$ representation. }  

Relatedly, with $\tau : \text{Lie}G \rarrow \text{Lie}G/\text{Lie}H$ the projection,  $e\defeq \tau(\b A) \in \Omega^1_\text{tens}(\P, \text{Lie}G/\text{Lie}H)$ is a 
 \emph{soldering form}. Given a non-degenerate bilinear form $\eta$ on $\text{Lie}G/\text{Lie}H$, a Cartan connection induces a metric on $U \subset \M$ via $g\defeq \eta(\s^*e, \s^*e)$. 
If  $H$ preserves $\eta$,  $g$ is independent of $\s$ and $\H$-invariant, thus well-defined across $\M$. Otherwise a gauge classe $[g]$ is induced (e.g. in  conformal Cartan geometry $[g]$ is a conformal class) either passively by  changing $\s$, or actively via the action of $\H$ on $\b A$. On  account of axioms i-ii and eq.\eqref{Pushforward-X}, the latter is again given by ${\b A}^\gamma = \gamma\-\b A\gamma+\gamma\-d\gamma$.\footnote{Notice that there are no gauge transformations corresponding to the whole group $G$/algebra Lie$G$!}
From these facts alone one already appreciates how  a Cartan geometry $(\P, \b A)$  reflects and encodes the geometry of $\M$, making it the right fit for (classical) gauge theories of gravity.

Consider a  $(\text{Lie}G, H)$-module $V$ , i.e. $V$ supports  actions of Lie$G$ via $\rho'_*$ and of $H$ via $\rho$ which are s.t.  ${\rho'_*}_{|\text{Lie}H}=\rho_*$. Then, $\b A$   defines a covariant derivative $D^{\b A}\!\defeq d\, +\rho'_*(\b A) : \Omega_\text{tens}^\bullet(\P, V) \rarrow  \Omega_\text{tens}^{\bullet+1}(\P, V)$\footnote{In some other contexts it is known as a \emph{tractor connection} \cite{Cap-Slovak09}}  (acting  in particular  on $\Gamma(E)$). 
One  finds again that $D^{\b A}\circ D^{\b A}\!=\rho'_*(\b F)$, where $\b F$ is the curvature of $\b A$  defined via Cartan's structure equation, $\b F\defeq d\b A +\sfrac{1}{2}[\b A,\b A] \in \Omega^2_\text{tens}(\P, \text{Lie}G)$, and satisfy a Bianchi identity $D^{\b A}\b F=0$. The \emph{torsion} of the connection is $T\defeq \tau(\b F) \in \Omega^2_\text{tens}(\P, \text{Lie}G/\text{Lie}H)$, a notion obviously absent for Ehresmann connections. 

Constraints on $\b F$  can be imposed so that $\b A$'s only degrees of freedom (d.o.f.) are those of its soldering form $e$. These \emph{normality} conditions, which most often comprise at least torsionlessness $T\equiv 0$, single out a unique \emph{normal} Cartan connection. Up to gauge transformations that is: since $\b F \in \Omega^2_\text{tens}(\P, \text{Lie}G)$  one has still $\b F^\gamma=\gamma\- \b F \gamma$, so that normality conditions (torsionlessness in particular) are preserved by the action of $\H$.

It may be noticed that  $\b F=0$ would imply that the base manifold is an homogeneous space $\M\simeq G/H$.\footnote{The Lie group $G$ (if it exists) is a $H$-bundle over the homogeneous space $G/H$, and the Maurer-Cartan form $\theta \in \Omega^1(G, \text{Lie}G)$, satisfying $d\theta +\sfrac{1}{2}[\theta, \theta]=0$, is an instance of flat Cartan connection. The pair $(G, \theta)$ is called a Klein geometry, and it is the homogeneous model that is generalised by a Cartan geometry $(\P, \b A)$.} Flatness in the sense of Cartan therefore generalises flatness in the sense of (pseudo-) Riemannian geometry, for which the homogeneous model is (pseudo-) Euclidean. 

Regarding physics, especially noteworthy is the subclass of  \emph{reductive}  Cartan geometries, where there is a $\Ad(H)$-invariant decomposition Lie$G=\text{Lie}H+V^n$. It implies a clean split of the Cartan connection as $\b A= A+e$, where $A$ is an Ehresmann connection on $\P$. The curvature splits accordingly $\b F=F+T$. In this class we find e.g. pseudo-Riemannian geometry\footnote{As reformulated by Cartan via his ``moving frame'', and Einstein via his ``vierbein/vielbein" or tetrad field - i.e. the soldering $e$.} 
based on $\mathfrak{iso}(r,s)=\so(r,s)+\RR^n$. 

\emph{Parabolic} Cartan geometries are another remarkable subclass where one has a $|k|$-grading of Lie$G$, 
  i.e. Lie$G=\bigoplus_{-k\leq i \leq k} \text{Lie}G_i$ s.t. $[\text{Lie}G_i, \text{Lie}G_i] \subset \text{Lie}G_{i+j}$, and $H$ is s.t. Lie$H=\bigoplus_{0 \leq i \leq k} \text{Lie}G_i$.  Both $\b A$ and $\b F$ split along the $|k|$-grading, and here also the Lie$H$-part of $\b A$ is a Ehresmann connection. An important example is conformal Cartan geometry, which is based on the $|1|$-graded algebra $\so(r+1, s+1)=\RR^n \oplus \co(r, s) \oplus \RR^{n*}$, with $\co(r, s)=\so(r, s) \oplus \RR$, and where $H$ is s.t. Lie$H=\co(r, s) \oplus \RR^{n*}$. The spin version in case $n=4$, based on the $|1|$-grading of $\su(2, 2)$, is relevant to twistor geometry \cite{Penrose-Rindler-vol1, Penrose-Rindler-vol2}. 
  
  It is understood that the local representatives on $\M$ of the Cartan connection $\b A$ and its curvature $\b F$ represent respectively the gravitational gauge potential and its curvature/field strength.  Sections $\Gamma(E)$ of associated bundles $E$ built via spin representations of $H$ represent spinorial matter fields. Then,  in both the above subclasses, the local representative of the covariant derivative induced by the Lie$H$-part of  $\b A$ acting on $\Gamma(E)$ represents the minimal coupling of matter field  to gravity.

\subsection{Twisted bundle geometry}
\label{Twisted bundle geometry}

The main takeaway of the above detailled review is that starting with a $H$-principal bundle, representations $(V, \rho)$ of $H$ allows to define $\Omega_\text{eq}^\bullet(\P, V)/ \Omega_\text{tens}^\bullet(\P, V)$ - and in particular associated bundles $E$ with sections $\Gamma(E)\simeq \Omega_\text{eq}^0(\P, V)$ - and that  connections are needed to obtain  covariant derivatives on these spaces of forms. 

Recently, a conservative extension of this state of affair was proposed and named ``twisted geometry"\cite{Francois2019_II}. It is conservative because it still starts with a $H$-principal bundle $\P$. Yet it extends the previous scheme by considering not representations of $H$, but 1-cocycles $C$ for the action of $H$ on $P$ with values in a Lie group $G$ \cite{Cartan-Eilenberg1956}, i.e.
\begin{align}
\label{1Cocycle}
C: P \times H &\rarrow G, \notag\\
      (p, h) &\mapsto C(p, h)  \qquad \text{s.t.} \quad C(p, hh')=C(p, h)\,C(ph, h'). 
\end{align}
Clearly,  $p$-independent cocycles  are just group morphisms (these are the trivial cocycles). 
In the following we describe the salient features of this generalised geometry, and refer to \cite{Francois2019_II} for a more complete exposition presenting all the technical proofs. 
\medskip

Given  representations $(V, \rho)$ of $G$ (not of the structure group $H$), one defines the space of $C$-equivariant forms on $\P$ as 
$\Omega^\bullet_\text{eq}(\P, V)^C\defeq \left\{ \, \omega \in \Omega^\bullet(\P, V)\, |\, R^*_h\omega_{ph} = \rho\left( C(p, h)\- \right)\omega_p \,\right\}$. Obviously we have $\Omega^\bullet_\text{inv}(\P, V)^C = \Omega^\bullet_\text{inv}(\P, V)$. 
The~spaces $\Omega^\bullet_\text{hor}(\P)$  of horizontal forms is defined as usual, thus so are basic forms $\Omega^\bullet_\text{basic}(\P, V)$. But then we have a space of $C$-tensorial forms  defined as  $\Omega^\bullet_\text{tens}(\P, V)^C\defeq \left\{ \, \omega \in \Omega^\bullet(\P, V)\, |\, R^*_h\omega_{ph} = \rho\left( C(p, h)\- \right)\omega_p\, \text{ and }\,  \iota_{X^v} \omega=0\,\right\}$. 

In particular, \emph{twisted} bundles associated  to $\P$  can be defined following the standard procedure: One considers the action of $H$ on $\P \times V$ twisted by the cocycle $C$, i.e. $\left( \P \times V\right) \times H \rarrow \P \times V$ is $\left( (p, v), h\right) \mapsto \left(ph, \rho\left( C(p, h)\-\right) v \right)$. Thanks to \eqref{1Cocycle} it is a well-defined right action. One then builds the twisted bundle $E^C$ as the space of equivalence classes under this action: $E^C=\P \times_{C(H)} V \defeq \P \times V / \sim$. 
By the usual correspondance, its space of sections $\Gamma\big(E^C\big)$ is isomorphic to the space of $C$-equivariant functions on $\P$, 
$\Omega^0_\text{eq}(\P, V)^C\!=\Omega^0_\text{tens}(\P, V)^C=\left\{\, \vphi :\P \rarrow V \,|\, \vphi(ph)=\rho\left(C(p, h)\- \right)\vphi(p) \,\right\}$.
\medskip

As we've seen, gauge transformations are  defined by the action of $\Aut_v(\P)\simeq \H$ and, by virtue of eq.\eqref{Pushforward-X},  determined by the equivariance and verticality properties of a form on $\P$. Then, the gauge transformations of \mbox{$C$-tensorial} forms are immediate: 
$\omega \in \Omega^\bullet_\text{tens}(\P, V)^C\, \Rightarrow\, \omega^\gamma=\rho\left( C(\gamma)\- \right) \omega$, where we introduce for convenience the notation 
$C(\gamma):\P \rarrow G$, $p\mapsto C\left(\gamma(p)\right)\defeq C\left(p, \gamma(p)\right)$. 

In the same way, the gluing relations of local representatives $a'=\s'^*\alpha$ and $a=\s^*\alpha$ on $U'\cap U \neq \emptyset$ of a form $\alpha$ on $\P$ are, on account of eq.\eqref{Push-X-local}, also determined by its equivariance and verticality properties. Thus, the gluing relations (passive gauge transformations) of the local representatives of a $C$-tensorial form is also immediate: $\alpha \in \Omega^\bullet_\text{tens}(\P, V)^C\, \Rightarrow\, a'=\rho\left( C(g)\- \right) a$, with the convenient notation $C(g):U'\cap U \rarrow G$, $x \mapsto C\left( g(x) \right)\defeq C\left( \s(x), g(x) \right)$.

\subsubsection{Twisted connections}
\label{Twisted connections}

It is clear that a new notion of connection on $\P$ is needed to obtain a covariant derivation on $C$-equivariant forms. One thus defines a \emph{twisted} connection (or $C$-connection) $\t A$ as a Lie$G$-valued 1-form s.t. 
\vspace{1.5mm}
\begin{enumerate}[label=\roman*, leftmargin=*, itemsep=4pt]
\item $R^*_h\t A_{|ph}=C(p, h)\- \t A_{|p} C(p, h) + C(p, h)\-dC(\,,h)_{|p}$
\item $\t A_{|p}(X^v_p)= \tfrac{d}{dt} C\big(p, e^{\tau X}\big)\big|_{\tau=0} \in$ Lie$G$, where $X^v_p \in V_p\P$ and $X\in$ Lie$H$.
\end{enumerate}
\vspace{1.5mm}
The set $\t\A$ of twisted connections is an affine space modelled  on the vector space $\Omega_\text{tens}^{1}(\P, \text{Lie}G)^C$: given $\t A', \t A \in \t\A$, clearly $\t A'-\t A \in \Omega_\text{tens}^{1}(\P, \text{Lie}G)^C$. Or, given $\t A\in \t\A$ and $\beta\in \Omega_\text{tens}^{1}(\P, \text{Lie}G)^C$, $\t A'\defeq \t A+\beta \in \t\A$. 

Like a Cartan connection, a twisted connection does not define an horizontal subbundle by splitting the SES~\eqref{SESLieAlg} as does an Ehresmann connection. Rather, as stated above, it is defined only so as to get a covariant derivative $D^{\t A}\defeq d\, + \rho_*(\t A): \Omega^\bullet_\text{tens}(\P, V)^C \rarrow  \Omega^{\bullet+1}_\text{tens}(\P, V)^C$, which in particular acts on sections $\Gamma\big( E^C \big)$ of twisted  bundles. 

 One shows that as usual $D^{\t A}\circ D^{\t A} =\rho_*\big( \t F \big)$, where $\t F\defeq d\t A+\sfrac{1}{2}[\t A, \t A]$ is the curvature 2-form of  $\t A$. It is a non-trivial task to prove that $\t F \in \Omega^2_\text{tens}(\P, \text{Lie}G)^C$, but it then easily found that we have a Bianchi identity $D^{\t A}\t F=0$, and it follows immediately that $\t F^\gamma= C(\gamma)\- \t F \,C(\gamma)$.  
As can be checked via Cartan's structure equation, the latter result is consistent with the gauge transformation of the twisted connection, 
\begin{align}
\label{GT-TwistedConnection}
\t A^\gamma= C(\gamma)\- \t A \,C(\gamma) + C(\gamma)\-dC(\gamma),
\end{align}
which follows from \eqref{Pushforward-X} and the axioms i-ii above. By analogous reasoning, the local representatives  $\t a\defeq \s^*\t A$ and $\t f\defeq \s^*\t F$ have gluing relations on $U'\cap U \neq \emptyset$ given by $\t a' =C(g)\- \t a \,C(g) + C(g)\-dC(g)$ and $f'=C(g)\- \t f \,C(g)$. From the viewpoint of physics, these represent \emph{twisted gauge fields} pertaining to a new class of gauge theories. 
\medskip

Twisted connections  generalises Ehresmann connections since $\A$ is isomorphic to the subspace of  $\t \A$ for which $C$'s are $H$-valued trivial cocycles. We note $ \t \A \supset \A$.   
As is obvious from section \ref{Ehresmann and Cartan connections}, Cartan connections are a subset of Ehresmann connections since they satisfy an additional axiom; $\A \supset \B$. It so happens, as one would have rightly suspected,  that there is within  $\t\A$ a subspace of \emph{twisted Cartan connections} $\t\B$ that  has a subset isomorphic to $\B$. As in the case of a regular Cartan connection, any element of $\B$ induces a soldering on $T\M$, as well as a twisted soldering form giving rise to a metric on $\M$. See \cite{Francois2019_II} for details. 
\medskip

Finally, consider a principal bundle with structure group $H\times K$. Assume there is a $G$-valued $H$-1-cocycle as before, but that it satisfies  $C(pk, h)=k\-C(p, h)\, k$, for $k\in K$.\footnote{This follows from requiring compatibility with the commutativity of the right actions of $H$ and $K$ on $\P$: $R_{hk}=R_k\circ R_h=R_h\circ R_k=R_{kh}$.} Assume also that we have representations $(V, \rho)$ of the (inner) semidirect product group $G \rtimes K$.  It it then possible to define  $C(H)\rtimes K$-equivariant forms, or \emph{mixed} equivariant for short:
$\Omega^\bullet_\text{eq}(\P, V)^M\defeq \left\{\, \omega \in \Omega^\bullet(\P)\, |\, R^*_{hk}\omega_{|phk} = \rho\left( k\-C(p, h)\-\right)\omega_{|p} \, \right\}$. 
Among these, those that are also horizontal - i.e. vanishing along vectors of $V\P$ generated by  Lie$H\oplus$ Lie$K$ or either factors - are  mixed tensorial forms, 
$\Omega^\bullet_\text{tens}(\P, V)^M$. As usual, one defines mixed associated bundles  $\E$ with space of section $\Gamma\big(\E\big)\simeq\Omega^0_\text{eq}(\P, V)^M$. 

To get a covariant derivative adapted to the above spaces of forms, one defines mixed connections:  these are twisted connection w.r.t. $H$ but Ehresmann w.r.t. $K$. Mixed Cartan connections exist as a special case. We won't enter into further details, save to mention that the bundle of local twistors and its twistor connection provide an example (slightly degenerate) or mixed bundle $\E$ equipped with a mixed Cartan connection. In this case $K$ is the Lorentz group and $H$ is the abelian group of Weyl rescalings: $H\times K= \RR \times S\!O(1,3)=C\hspace{-1pt}O(1,3)$ (see \cite{Francois2019_II})

\medskip

In section \ref{The space of connections as a principal bundle} we will briefly indicate how the twisted geometry based on 1-cocycles appears naturally in the study of anomalies in QFT.  But most importantly for the concern of this paper, it turns out to be  relevant to the analysis of the presymplectic structure of gauge theories, see  section \ref{Covariant Hamiltonian formalism (and the puzzle of boundaries)}. 

Before all this, we next describe the \emph{dressing field method} which will help  clarify a recent proposal regarding the handling of the problem of boundaries in gauge theories via  edge modes \cite{DonnellyFreidel2016, Geiller2017, Speranza2018}.

\section{The dressing field method}
\label{The dressing field method}

From the viewpoint of gauge theory, the dressing field method (DFM from now on) can be seen as a tool of gauge symmetry reduction. It is distinct from other means to achieve similar results, such as gauge fixing or spontaneous symmetry breaking (SSB) mechanisms, and closer in spirit to  the Bundle Reduction Theorem (for various equivalent formulations of which see  \cite{Trautman, Westenholz, Sternberg, Sharpe}). First formulated in \cite{GaugeInvCompFields}, it was more systematically explored in \cite{Francois2014} and a review  appeared in \cite{Attard_et_al2017}, while its philosophical implications have been studied in \cite{Francois2018}. 
In this section we report the main results of the DFM and refer to the above references for detailed proofs.

\subsection{Reduction of gauge symmetries}
\label{Reduction of gauge symmetries}

 To appreciate how the DFM achieves a reduction of a gauge symmetry, let us define its central object.
 
\begin{definition}
\label{Def1}
Suppose $\exists$  subgroups $K \subseteq H$ of the structure group, to which corresponds a subgroup $\K \subset \H$ of the gauge group, and $G$  s.t.  $K\subseteq G \subseteq H$. A \emph{dressing field}  is a map $u:\P \rarrow G$ defined by its $K$-equivariance~$R^*_k u=k\-u$. 
Denote the space of $G$-valued $K$-dressing fields on $\P$ by $\D r[G, K]$  ($\defeq \Omega^0_\text{eq}(\P, G, \ell(K))$, where $\ell$ if the left action). 
From this  follows immediately that the $\K$-gauge transformation of a dressing field is $u^\gamma=\gamma\- u$, for $\gamma \in\K$.
\end{definition} 

A dressing field allows to build a quotient subbundle $\P/K \subset \P$ via the map $f_u:\P \rarrow \P/K$ defined as $p \mapsto f_u(p)\defeq pu(p)$, 
and thus s.t. $f_u \circ R_k =f_u$. It means that the bundle factorises along the subgroup $K$ as $\P\simeq\P/K \times K$. 
Not~only that, we have the following
\begin{prop} \normalfont 
\label{Prop1}
From $A \in \A$ and  $\alpha \in \Omega^\bullet_\text{tens}(\P, V)$, one defines the following \emph{dressed fields}
\begin{align}
\label{dressed-fields}
A^u\defeq f^*_u A = u\- Au +u\-du, \quad \text{ and } \quad \alpha^u \defeq f^*_u \alpha = \rho(u)\-\alpha,  \
\end{align}
which have trivial $K$-equivariance (as is easily seen from $R^*_k \circ  f^*_u=f^*_u$), are $K$-horizontal, thus are $K$-basic on $\P$. It~follows that they are $\K$-invariant: $(A^u)^\gamma=A^u$ and $(\alpha^u)^\gamma=\alpha^u$, for $\gamma \in \K$, as is easily checked. 

As an instance of $\alpha^u$ we have the  curvature of $A^u$, the dressed curvature $F^u =dA^u+\sfrac{1}{2}[A^u, A^u]= u\- F u$, which~appears when squaring the dressed covariant derivative  $ D^{A^u}\defeq d\, + \rho_*(A^u)=\rho(u)\-D^A$, and satisfies the Bianchi identity $D^{A^u}F^u=0$. 
\end{prop}

Remark that in case the equivariance group of $u$ is $K=H$,  first the bundle is trivial $\P\simeq\M \times H$,  second  $\alpha^u \in \Omega^\bullet_\text{basic}(\P, V)$ and  $A^u\in \Omega^1_\text{basic}(\P, \text{Lie}H)$ are $\H$-invariant (and project as, or come from, forms on $\M$). 

We also highlight that the above results can make sense for $G \supset H$: One needs only to assume that $G$ is (a~subgroup~of) the structure group of a bigger principal bundle of which $\P$ is a subbundle/a reduction (as is typically the case for Cartan geometries\footnote{Indeed the $H$-bundle of a Cartan geometry $(\P, \b A)$  always can be embedded as a subbundle of a $G$-principal bundle $\Q\defeq \P \times_H G$ - on which $\b A$ can be lifted to an Ehresmann connection (yet not all Ehresmann connections on $\Q$ restricts to Cartan connections on $\P$, see \cite{Sharpe}, appendix A).  More is true, as the bundle of a Cartan geometry $(\P, \b A)$ is always a reduction of the frame bundle $L\M$ of $\M$ with structure group $GL(n)$, or of the $r^\text{th}$-order frame bundle $L^r\!\M$ (a jet bundle) \cite{Kobayashi1972}. For example, conformal Cartan geometry is a reduction of  $L^2\!\M$ \cite{Ogiue}. }), and that representations $(V, \rho)$ of $H$ extend to representations of $G$.

Finally, let us emphasize an important fact: It should be clear from its  definition that $u \notin \K$, so $f_u \notin \Aut_v(\P, K)$, and therefore that \eqref{dressed-fields} are \emph{not} gauge transformations, despite the formal resemblance. This means, in particular, that the dressed connection is no more a $H$-connection, $A^u \notin \A$, and a fortiori is not a point in the gauge $\K$-orbit $\O_\K[A]$ of $A$, so that $A^u$ must not be confused with a gauge-fixing of $A$. 
\medskip

%
%

Let us indulge in a brief digression that is also a segue to the results of the next section. 
In the BRST framework,  infinitesimal gauge transformations are encoded as $\boldsymbol{s} A=-D^A \boldsymbol v$ and $\boldsymbol{s} \alpha = -\rho_*(\boldsymbol{v})\alpha$,  where  $\boldsymbol s$ is the nilpotent BRST operator and $\boldsymbol v$ the ghost field. The latter has values in Lie$\H$ and satisfies $\boldsymbol{sv}+\sfrac{1}{2}[\boldsymbol{v}, \boldsymbol{v}]=0$. 
For~this reason, $\boldsymbol s$ is best interpreted geometrically as the de Rham derivative on $\H$ and $\boldsymbol v$ as its Maurer-Cartan form \cite{Bonora-Cotta-Ramusino}. 
One shows that, at a purely formal level, the dressed variables satisfy a modified BRST$^u$ algebra:  $\boldsymbol{s} A^u=-D^{A^u} \boldsymbol{v}^u$ and $\boldsymbol{s} \alpha^u = -\rho_*(\boldsymbol{v}^u)\alpha^u$, where one defines the \emph{dressed ghost} $\boldsymbol{v}^u\defeq u\- \boldsymbol{v} u + u\- \boldsymbol{s} u$. 

In the special  case where $u$ is a $H$-dressing, its defining equivariance translates as $\bs{s}u=-\bs{v} u$. Then the dressed ghost is $\bs{v}^u=0$ and  BRST$^u$ is trivial, $\bs{s}A^u=0$ and $\bs{s}\alpha^u=0$, as one would expect.
In the more general case of a $K$-dressing $u$ achieving only partial gauge reduction by Proposition \ref{Prop1}, BRST$^u$ only makes sense if it encodes some residual  transformations that one can speak about meaningfully.
We may then inquire about potential residual transformations of the dressed fields \eqref{dressed-fields}, which actually come in more than one way, as we discuss in the following.

\subsection{Residual gauge transformations (first kind)}
\label{Residual gauge transformations (first kind)}

To  speak meaningfully about residual gauge transformations of the dressed fields, we need some assumptions. First, we must assume that $K$ is a \emph{normal} subgroup, $K \triangleleft H$, so that  $J\defeq H/K$ is indeed a group, to which corresponds the (residual) gauge subgroup $\J \subset \K$. It follows that $\P/K=\P'$ is a $J$-principal bundle with gauge group $\J \simeq \Aut_v(\P')$, and $\A'$ is its space of Ehresmann connections. 

Now, the action of $\J$ on the initial variables $A$ and $\alpha$ is known. Therefore, what will determine the $\J$-residual gauge transformations of the dressed fields is the action of $\J$ on the dressing field. And this in turn is determined by its $J$-equivariance. In that regard two possibilities are especially noteworthy. We consider them in turn, as two propositions.

\begin{prop} \normalfont 
\label{Residual1}
Suppose the dressing field $u$ has $J$-equivariance given by $R^*_j u =j\-uj$. Then $\alpha^u \in \Omega^\bullet_\text{tens}(\P', V)$,  while  $A^u \in \A'$ with curvature $F^u \in \Omega^2_\text{tens}(\P', \text{Lie}H)$,  and $D^{A^u}: \Omega^\bullet_\text{eq}(\P', V) \rarrow \Omega^{\bullet+1}_\text{tens}(\P', V)$. 

As immediate corollary,  the dressing field has $\J$-gauge transformation $u^\eta=\eta\-u\,\eta$ for $\eta \in \J$, and
the residual gauge transformations of the dressed fields are: $(A^u)^\eta=\eta\- A^u \eta + \eta\-d\eta$ and $(\alpha^u)^\eta=\rho(\eta)\-\alpha^u$.  
\end{prop}

In the BRST language, the normality of $K$ in $H$ implies $\bs{v}=\bs{v}_K+\bs{v}_J$, where $\bs{v}_K$ and $\bs{v}_J$ are respectively Lie$\K$- and Lie$\J$-valued, and in accordance $\bs{s}=\bs{s}_K+\bs{s}_J$. The defining $K$-equivariance of the dressing field translates as $\bs{s}_Ku=-\bs{v}_K u$, while its $J$-equivariance assumed in Proposition \ref{Residual1} is encoded as $\bs{s}_Ju=[u, \bs{v}_J]$. The dressed ghost field is thus 
$\bs{v}^u=u\-(\bs{v}_K+\bs{v}_J) u + u\- (\bs{s}_K + \bs{s}_J)u = u\-(\bs{v}_K+\bs{v}_J) u + u\- (-\bs{v}_K u + [u, \bs{v}_J]) = \bs{v}_J$. Therefore, the modified (actually \emph{reduced}) BRST$^u$ algebra is: $\bs{s}_J A^u = -D^{A^u} \bs{v}_J$ and $\bs{s}_J \alpha^u = -\rho_*(\bs{v}_J)\alpha^u$. It encodes the residual gauge transformations of the dressed fields.  
\medskip

Under the assumption of Proposition \ref{Residual1}, the dressed fields are `standard' geometrical objects, usual gauge fields. So, if a \emph{second} dressing field is available, one may apply Proposition \ref{Prop1} again to further reduce the gauge symmetry.  

 As a matter of definition, such a $J$-dressing field would be $u':\P' \rarrow J$ s.t. $R^*_j u' =j\- u'$, so that the dressed fields $(A^u)^{u'}\defeq {u'}\- A^u u' +{u'}\-du'$ and $(\alpha^u)^{u'}\defeq \rho(u')\-\alpha^u$ are $J$-basic, thus $\J$-invariant, by Proposition \ref{Prop1}. 
But in order not to spoil the $K$-basicity (thus the $\K$-invariance) obtained via the first $K$-dressing $u$, the second $J$-dressing field must further satisfy the \emph{compatibility condition} $R^*_k u'=u'$ (implying $u'^\gamma=u'$, for $\gamma \in \K$). 

Indeed, collecting the equivariance properties of the two dressing fields 
\begin{align}
\label{CompCond}
R^*_ku= k\- u, \qquad R^*_j u= j\- u\, j, \qquad R^*_j u' =j\- u', \qquad \text{and} \qquad R^*_k u' =u',
\end{align}
one finds  on the one hand $R^*_k \, uu'= k\- u u'$, and on the other hand $R^*_j \, uu'= j\- u u'$. That is $uu': \P \rarrow H$ is a $H$-dressing field, and by Proposition \ref{Prop1} one has: $A^{uu'}=(A^u)^{u'} \in \Omega^1_\text{basic}(\P, \text{Lie}H)$ and $\alpha^{uu'}=(\alpha^u)^{u'}\in\Omega^\bullet_\text{basic}(\P, V)$. Obviously, the scheme extends  to multiple dressings $u^{(r)}$ satisfying a tower of compatibility conditions, as e.g. in the context of jet bundles and higher-order $G$-structures (see \cite{Francois2014} for not so enlightening details).

\bigskip
We now turn to the second noteworthy possibility alluded to above. Consider a $J$-cocycle  $C:\P' \times J \rarrow G'$, with $G'\supset H$ s.t.  $(V, \rho)$ extends to representations of $G'$.
 
\begin{prop} \normalfont 
\label{Residual2}
Suppose the dressing field $u$ has $J$-equivariance given by $R^*_j u =j\-u\,C(\  , j)$.
Then the dressed fields are \emph{twisted} gauge fields: $\alpha^u \in \Omega^\bullet_\text{tens}(\P', V)^C$,  while  $A^u \in \t\A'$ with curvature $F^u \in \Omega^2_\text{tens}(\P', \text{Lie}G)^C$,  and $D^{A^u}: \Omega^\bullet_\text{tens}(\P', V)^C \rarrow \Omega^{\bullet+1}_\text{tens}(\P', V)^C$. 

As  corollary,  the dressing field has $\J$-gauge transformation $u^\eta=\eta\-u\,C(\eta)$ for $\eta \in \J$, and
the residual gauge transformations of the twisted dressed fields are: $(A^u)^\eta=C(\eta)\- A^u C(\eta) + C(\eta)\-dC(\eta)$ and $(\alpha^u)^\eta=\rho\big(C(\eta)\-\big)\alpha^u$.  
\end{prop}

The BRST version of the $J$-equivariance assumed in Proposition \ref{Residual2} is: $\bs{s}_j=-\bs{v}_J\, u + u\,c(\bs{v}_J)$, where $c(X)\defeq \tfrac{d}{d\tau}C(e^{\tau X})\big|_{\tau=o}$ for $X\in$ Lie$J$. In a manner analogous to the first case, the dressed ghost is then $\bs{v}^u=c(\bs{v}_J)$, and BRST$^u$ encodes the residual gauge  transformations of the  dressed fields: $\bs{s}_J A^u = -D^{A^u} c(\bs{v}_J)$ and $\bs{s}_J \alpha^u = -\rho_*\big(c(\bs{v}_J)\big)\alpha^u$.

We refrain from giving further details, except for noticing that the conformal tractor bundle and connection as well as the bundle of local twistors and the twistor connection both can be obtained, via dressing, from the conformal Cartan geometry  \cite{Attard-Francois2016_I, Attard-Francois2016_II, Attard_et_al2017}. Proposition \ref{Residual2} in particular is brought to bear w.r.t. Weyl rescalings.\footnote{Remark also that the DFM  applies to twisted gauge fields. Indeed, if a second $J$-dressing as above exists then \mbox{$C(u'):\P' \rarrow G'$} is a twisted dressing field, $R^*_jC(u') =C(\ , j)^{-1}C( u' )$, and it preserves the $K$-basicity if $C$ has trivial $K$-equivariance so that $R^*_kC(u')=C(u')$. See \cite{Francois2019} for an application.}

\subsection{Residual transformations (second kind) : ambiguity in choosing a dressing field}
\label{Residual transformations (second kind) : ambiguity in choosing a dressing field}

From the inception of the DFM \cite{GaugeInvCompFields}, the problem was addressed of residual transformations that are not quite of the kind discussed above, as they result from a potential ambiguity in choosing a dressing field. Even in the case of a full gauge symmetry reduction - either through a single dressing $\in \D r[G, H]$ or multiple ones combining to the same effect -  the dressed fields may nonetheless exhibit residual transformations of this other kind. 

A priori two dressings $u, u' \in \D r[G, K]$ may be related by $u'=u\xi$, where $\xi:\P \rarrow G$.\footnote{We have again that $G$ can be either s.t. $K\subseteq G \subseteq H$, or s.t. $G\supseteq H$.} Since by definition $R^*_k u =k\-u$ and $R^*_k u' =k\-u'$, one has $R^*_k \xi = \xi$. Let us denote the group of such maps $\t \G \defeq \left\{\,   \xi:\P \rarrow G\, |\,  R^*_k \xi = \xi  \, \right\}$, and by analogy with the notation for the action of the gauge group $\H$, have its action on $\D r[G, H]$ noted as $u^\xi=u\xi$. 
It is clear that by definition $\t\G$ has no action on the space of connections $\A$ (or $\B$): we may denote this by $A^\xi=A$. On the other hand, it is clear how $\t\G$ acts on  the space of dressed  connections $\A^u$: 
\begin{align}
\label{ResidualGT-2Kind}
(A^u)^\xi \defeq A^{u^\xi}= A^{u\xi}=\xi\- A^u \xi + \xi\-d\xi, 
\end{align}
which implies for the dressed curvature: $(F^u)^\xi=\xi\- F^u \xi$ (analogous formulae hold for $\b A^u \in \B^u$ and $\b F^u$). 
The  new dressed field  $(A^u)^\xi$ is also $K$-basic, and therefore $\K$-invariant. It means that the bijective correspondance between the $K$-dressings $(A^u)^\xi$ of a connection $A\in \A$ and its gauge $\K$-orbit $\O_\K[A]$ holds $\forall \xi \in \t\G$. So there is a $1:1$ correspondance $\O_\K[A] \sim \O_{\t\G}[A^u]$.

What it tells us is that if a dressing is introduced by \emph{fiat}, the reduced $\K$ gauge symmetry is replaced with a local symmetry which is (at least) as big. 
As a matter of fact, it was shown in \cite{FLM2015_I} that by freely introducing a dressing field $u \in \D r[H, H]$ into a theory, one has  tacitly assumed at the onset that the underlying bundle $\P$ is trivial so that $\H\simeq \H_0$, with  $\H_0$  the gauge group of the trivial bundle $\M \times H$.  
It was further shown that  $\t\G$, in this case renamed $\t\H$,\footnote{Notice that in this case the $\xi$'s are basic functions, i.e. they project to (come from) globally defined $H$-valued functions on $\M$, $C^\infty(\M, H).$}  
 is actually isomorphic to $\H_0$, so that $\H\simeq \t\H$. From the viewpoint of gauge theory this is a priori a problem, as it seems that nothing of substance has been achieved by thus introducing a dressing field. 

The situation is not necessarily so bad though. As discussed in \cite{GaugeInvCompFields}, the only  way in which a meaningful constraint on this arbitrariness in choosing a dressing field could arise is if the latter is built from the  gauge variables ($A$, $\b A$, and/or $\alpha$) already at hand, as is the case in most fruitful applications \cite{Attard-Francois2016_I, Attard-Francois2016_II, Attard_et_al2017, Francois2018}. In such cases, even if $\t\G$ is not `small' it  may nevertheless be an interesting symmetry, as we will see.  In this paper we will consider specifically the case of $\A$-dependent dressing fields $\bs{u}: \A \rarrow \D r[G, K]$, i.e. $A \mapsto \bs{u}(A)$, with by definition $\bs{u}(A)^\gamma\defeq\bs{u}(A^\gamma)=\gamma\-\bs{u}(A)$, for $\gamma \in \K$. 

Further developments on the matter drafted above  will have to await section \ref{A-dependent dressing fields and basic variational forms on A}, the time to get first a taste of the principal bundle geometry of the space $\A$ in the next section. 
\bigskip

\medskip

We end this review by stressing the fact that the DFM provides a framework for  an idea that has a long history. The earliest example of (abelian) dressing field is probably the so-called \emph{Stueckelberg field}, introduced in \cite{Stueckelberg1938-I, Stueckelberg1938-II-III}, see \cite{Ruegg-Ruiz} for a review. 
Dirac's gauge-invariant formulation of QED - conceived as better amenable to quantization, first proposed in a 1955 paper \cite{Dirac55} and developed in the 1958 fourth edition of his \emph{Principles of Quantum Mechanics} \cite{Dirac58} (section 80) - is also  seen to be an (abelian) application of the DFM. 

 Subsequently, the core idea behind the DFM  has resurfaced multiple times in many area of gauge theories.\footnote{But seldom with the conceptual clarity about what was indeed achieved, as a dressing was often mistaken for a gauge-fixing, or associated to a SSB mechanism.}
Let us mention e.g.  the study of anomalies in QFT \cite{Garajeu-Grimm-Lazzarini, Manes-Stora-Zumino1985, Stora1984}, some formal explorations in quantum gravity \cite{Maas2020, Lazzarini2008, Polyakov1989}, the construction of Wess-Zumino functionals \cite{Attard-Lazz2016} (more on this in section \ref{Classical gauge anomalies}), the so-called  `proton spin decomposition controversy' \cite{LorceGeomApproach, Leader-Lorce, FLM2015_I}, the question of how constituent quarks arise in QCD \cite{McMullan-Lavelle97}, and - most notably - reformulations of  theories undergoing SSB \cite{Maas2019, Ilderton-Lavelle-McMullan2010, Faddeev2009, Chernodub2008, McMullan-Lavelle95, Frohlich-Morchio-Strocchi81} (and going as far back as the pioneering works of Higgs \cite{Higgs66} and Kibble \cite{Kibble67}). 

In recent years, the fact that such reformulations  cast a new light on these theories, on the electroweak model in particular, has been appreciated by philosophers of physics \cite{Smeenk2006, Lyre2008, Struyve2011, Friederich2013, Friederich2014}. For a  discussion of this issue as situated within the broader philosophical question of distinguishing \emph{substantial} from \emph{artificial} gauge symmetries, see \cite{Francois2018}. 

As philosophers noticed first \cite{Teh-et-al2020, Teh2020}, the last example to date of an  unwitting application of the DFM are the so-called ``edge modes"  invoked as a way to deal with the problem of boundaries in the symplectic structure of gauge theories \cite{DonnellyFreidel2016, Geiller2017, Speranza2018}. 
This is of direct concern to this paper, and  will be addressed explicitly in section \ref{Boundaries and dressed presymplectic structure}. 

\section{The space of connections as a principal bundle}
\label{The space of connections as a principal bundle}

In this section and the next, we will   apply the material described above in the context of infinite dimensional vector spaces  \cite{Frolicher-Kriegl1988} and more generally to infinite dimensional manifolds \cite{Kriegl-Michor1997}. In doing so, we will be guilty of ignoring a host of subtleties, referring to the relevant literature to back the soundness of extending any notion defined in the finite dimensional context to its infinite dimensional counterpart. Our aim is to paint a broadly correct conceptual landscape rather than being perfectly mathematically rigorous. 
So, in several instances the arguments adduced to support our results do not exactly to rise to the level of mathematical proofs, but we are confident that these are sound enough that such proofs could be produced by more expert colleagues. 

Since it is the smooth structure on infinite dimensional manifolds that will be of interest, and in order that most tools of the finite setting pass on to the infinite one, we will admit that we deal essentially with  Banach manifolds (tacit weakening to Fréchet manifolds or specialization to Hilbert manifolds may occasionally be needed). Most $\A$-dependent object are boldfaced, and so are variational object (vector fields, forms) and operators.

\subsection{Bundle geometry of $\A$}
\label{Bundle geometry of  A}

The space of (Ehresmann) connections $\A$  of a $H$-principal bundle of $\P$ is an infinite dimensional Banach manifold, so is its gauge group $\H$ as an infinite dimensional  Lie group. Under proper restrictions (of either $\A$ or $\H$ \cite{Singer1978, Singer1981, Mitter-Viallet1981, Cotta-Ramusino-Reina1984, Abbati-et-al1989, Fuchs1995}), the moduli space $\A/\H$ is  well-behaved as a manifold. Then,  $\A$ is a principal bundle over $\A/\H$  with structure group $\H$, whose right action we denote $(A, \gamma) \mapsto R_\gamma A \defeq A^\gamma$.\footnote{Since $(A^\gamma)^{\gamma'}=(A^{\gamma'})^{\gamma^{\gamma'}}=(A^{\gamma'})^{\gamma'^{-1} \gamma \, \gamma'}=A^{\gamma\gamma'}$, this is indeed a right action: $R_{\gamma'} \circ R_\gamma=R_{\gamma\gamma'}$.} The gauge orbit $\O_\H[A]$ of $A\in \A$ is a fiber over the gauge class $[A]\in \A/\H$. The projection $\pi:\A \rarrow \A/\H$, $A \mapsto \pi(A)=[A]$, is s.t. $\pi \circ R_\gamma = \pi$.

 The natural transformation group of $\A$ is its automorphism group 
 $\bs{\Aut}(\A)\defeq\left\{\,  \bs{\Psi}: \A \rarrow \A\, |\,  \bs{\Psi} \circ R_\gamma =R_\gamma \circ \bs{\Psi} \,\right\}$. Only $\bs{\Psi} \in \bs{\Aut}(\A)$ project to  well-defined $\bs{\psi} \in \bs{\Diff}(\A/\H)$. As usual, the subgroup of vertical automorphisms 
 $\bs{\Aut}_v(\A) \defeq \left\{\,  \bs{\Psi} \in \bs{\Aut}(\A)  \, |\,  \pi \circ \bs{\Psi} = \pi\,\right\}$ is isomorphic to the gauge group $\bs{\H}\defeq \left\{ \bs{\gamma}:\A \rarrow \H \, |\, \bs{\gamma}(A^\gamma)=\gamma\- \bs{\gamma}(A) \gamma \right\}$ by the correspondance $\bs{\Psi}(A)=R_{\bs{\gamma}(A)}A=A^{\bs{\gamma}(A)}$. We have the SES, 
 \begin{align}\label{SESgroups-inf}
\makebox[\displaywidth]{
\hspace{-18mm}\begin{tikzcd}[column sep=large, ampersand replacement=\&]
\&0     \arrow [r]         \& \bs{\Aut}_v(\A) \simeq \bs{\H}     \arrow[r, "\iota"  ]          \& \bs{\Aut}(\A)       \arrow[r, "\t\pi"]      \&  \bs{\Diff}(\A/\H)        \arrow[r]      \& 0.  
\end{tikzcd}}  \raisetag{3.4ex}
\end{align}
The gauge group $\bs{\H}$ gives geometric substance to the so-called \emph{field-dependent gauge transformations} sometimes alluded to in the physics literature.\footnote{Of course not all maps  $\A \rarrow \H$  ($\A$-dependent elements of $\H$) belong to $\bs{\H}$, yet the latter contains all such maps relevant for physics. } Notice that $\bs{\Diff}(\A/\H)$ is the \emph{physical} transformation group sending physical states to physical states, it contains the Hamiltonian flow of the covariant Hamiltonian formalism (see section~\ref{Covariant Hamiltonian formalism (and the puzzle of boundaries)}). 

As already noticed, $\A$ is an affine space modelled on $\Omega^1_\text{tens}(\P, \text{Lie}H)$. Therefore, the tangent space at $A \in \A$ is $T_A\A \simeq \Omega^1_\text{tens}(\P, \text{Lie}H)$, and a generic vector $\bs{X}_A \in T_A\A$ with flow $\phi_\tau : \A \rarrow \A$ is s.t.   $\bs{X}_A =\tfrac{d}{d\tau}\phi_\tau(A)\big|_{\tau=0}$. 
Formally, we can  write a vector field $\bs{X} \in \Gamma(T\A)$  as a variational operator $\bs{X}_A= \bs{X}(A) \tfrac{\delta}{\delta A}$, with $\bs{X}(A)=\tfrac{d}{d\tau}\phi_\tau(A)\big|_{\tau=0} \in \Omega^1_\text{tens}(\P, \text{Lie}H)$ the `component' of $\bs{X}$. 
Only right-invariant vector fields, s.t. $R_{\gamma\star}\bs{X}_A=\bs{X}_{A^\gamma}$, project to well-defined vector fields on the base, and $\pi_\star : \Gamma_\H(T\A) \rarrow \Gamma(T\A/\H)$ is a morphism of Lie algebras. The flow of a right-invariant vector field belongs to $\bs{\Aut}(\A)$, so that $\Gamma_\H(T\A)\simeq \text{Lie}\bs{\Aut}(\A)$. 

Any $\chi \in \text{Lie}\H$ induces a vertical vector $\chi^v_A\defeq \tfrac{d}{d\tau}  A^{\exp(\tau \chi)} \big|_{\tau=0} = D^A\chi$ tangent to the fiber $\O_\H[A]$ at $A$. \mbox{Vertical} vector fields $\chi^v \in \Gamma(V\A)$ are s.t. $\pi_\star \chi^v=0$ and $R_{\gamma\star}\chi_A^v=(\gamma\- \chi \gamma)^v_{A^\gamma}$ (see appendix \ref{Lie algebra (anti)-isomorphisms}). We have the injective morphism of Lie algebras  Lie$\H \rarrow \Gamma(V\A)$. Elements of the Lie algebra of the gauge group Lie$\bs{\H} \defeq \left\{ \, \bs{\chi}:\A \rarrow \text{Lie}\H\, |\,  R^\star_\gamma \bs{\chi}= \gamma\- \bs{\chi} \gamma    \, \right\}$ induce  $\H$-right invariant vertical vector fields $\bs{\chi}^v_A\defeq \tfrac{d}{d\tau}  A^{\exp(\tau \bs{\chi}(A))} \big|_{\tau=0}$, s.t.  $R_{\gamma\star} \bs{\chi}^v_A=\bs{\chi}^v_{A^\gamma}$, so that the map Lie$\bs{\H} \rarrow \Gamma_\H(V\A)$ is a Lie algebra \emph{anti}-isomorphism (appendix \ref{Lie algebra (anti)-isomorphisms}). Corresponding to \eqref{SESgroups-inf} we have the SES of Lie algebras
\begin{align}\label{SESLieAlg-inf}
\makebox[\displaywidth]{
\hspace{-18mm}\begin{tikzcd}[column sep=large, ampersand replacement=\&]
\&0     \arrow [r]         \& \Gamma_\H(V\A) \simeq \text{Lie}\bs{\H}     \arrow[r, "\iota"  ]          \&  \Gamma_\H(T\A)     \arrow[r, "\pi_\star"]      \&  \Gamma(T \A/\H)         \arrow[r]      \& 0.
\end{tikzcd}}  \raisetag{3.4ex}
\end{align}
It is the Atiyah Lie algebroid associated to $\A$. To split  this SES, $\A$ would need to be endowed with an Ehresmann connection $\bs{A} \in \bs{\A}$. One special type known as  Singer (-deWitt) connections \cite{Singer1978, Singer1981} has been used in \cite{Gomes-Riello2018, Gomes-et-al2018} regarding the problem of defining a symplectic structure for gauge theories on bounded regions (as an alternative to the proposal of edge modes, see \cite{Gomes2019, Gomes2019-bis} for a philosophical discussion). We comment further on  this  in section \ref{Covariant Hamiltonian formalism (and the puzzle of boundaries)}.
\medskip

The de Rham complex  is $\left( \Omega^\bullet(\A), \bs{d}\right)$ with $\bs d$ the variational exterior derivative defined via a Kozsul formula. We have an interior product $\iota: \Gamma(T\A) \times \Omega^\bullet(\A) \rarrow \Omega^{\bullet-1}(\A)$,  $(\bs{X}, \bs{\alpha}) \rarrow \iota_{\bs X} \bs{\alpha}$, and the Lie derivative is $\bs{L_X}\defeq [\iota_{\bs X}, \bs{d}]$.\footnote{ \label{Lie-BRST} Remark that $\bs{L}_{\chi^v}$ is a geometric realisation of the BRST operator $\bs s$, where the concrete element $\chi$  replaces the Lie$\H$-valued ghost  $\bs v$. }
An exterior product is defined on algebra-valued variational forms $\Omega^\bullet(\A, \sf{A})$. Given representations $(\bs{V}, \rho)$ of $\H$, the spaces of equivariant $\Omega^\bullet_\text{eq}(\A, \bs{V})$, tensorial  $\Omega^\bullet_\text{tens}(\A, \bs{V})$, and basic forms $\Omega^\bullet_\text{basic}(\A, \bs{V})$,   are defined in complete analogy with the finite dimensional case. A  variational Ehresmann connection $\bs{A}\in \bs{\A}$ on $\A$ induces a variational  covariant derivative  $\bs{D}^{\bs{A}}: \Omega^\bullet_\text{eq}(\A, \bs{V}) \rarrow  \Omega^{\bullet+1}_\text{tens}(\A, \bs{V})$, which obviously reduces to $\bs d$ on $\Omega^\bullet_\text{basic}(\A, \bs{V})$. 

The action of $\bs{\Aut}_v(\A)\simeq \bs{\H}$ on variational forms  defines their gauge transformations. Since, in analogy with~\eqref{Pushforward-X},  the action of $\bs{\Psi}\in \bs{\Aut}_v(\A)$ on a generic $\bs{X} \in \Gamma(T\A)$  is
\begin{align}
\label{Pushforward-X-inf}
\bs{\Psi}_\star \bs{X}_A&= R_{\bs{\gamma}(A)\star} \bs{X}_A + \left\{ \bs{\gamma}\- \bs{d}\bs{\gamma}_{|A}(\bs{X}_A)\right\}^v_{A^{\bs{\gamma}(A)}}
               = R_{\bs{\gamma}(A)\star} \left( \bs{X}_A + \left\{ \bs{d}\bs{\gamma} {\bs{\gamma}\- }_{|A}(\bs{X}_A)\right\}^v_A \right),
\end{align}
(see appendix \ref{Proofs of pushforward formulae for variational vector fields} for a proof) the $\bs\H$-gauge transformations of a variational form is controlled by its verticality\footnote{That a variational form would fail to be horizontal is sometimes loosely expressed as it lacking `gauge invariance'. Thus are characterised the pre-symplectic potential and 2-form (see  section \ref{Covariant Hamiltonian formalism (and the puzzle of boundaries)}) e.g. in \cite{Lee-Wald1990}, \cite{DonnellyFreidel2016} or \cite{De-Paoli-Speziale2018}. }
and $\H$-equivariance properties. From this fact follows immediately that the gauge transformations of tensorial variational forms are given solely by their $\H$-equivariance, and that basic variational forms are $\bs \H$-invariant. A fact that we will use again in section \ref{Covariant Hamiltonian formalism (and the puzzle of boundaries)} to compute the `field-dependent gauge transformations' of the pre-symplectic potential and associated pre-symplectic form.

As an example relevant to our purpose, consider $\bs{d}A \in \Omega^1(\A)$ (seen as a basis for variational forms). On a generic $\bs{X}\in \Gamma(T\A)$ with flow $\phi_\tau$, by definition $\bs{d}A_{|A}(\bs{X}_A)=\tfrac{d}{d\tau} \phi_\tau(A)\big|_{\tau=0}$ ($=\bs{X}(A)$). 
Then,  evaluated at a point $A\in \A$ on a vertical vector field generated by $\chi \in \text{Lie}\H$, it gives the corresponding infinitesimal gauge transformation of~$A$:   
\begin{align}
\label{Vert-dA}
\bs{d}A_{|A}(\chi^v_A)=D^A\chi,      \quad \text{or} \quad       \iota_{\chi^v}\bs{d}A =D\chi.
\end{align}
That's its verticality property. Thus we have, $R^\star_\gamma \bs{d}A_{|A^\gamma}(\chi^v)=  \bs{d}A_{|A^\gamma}(R_{\gamma \star} \chi^v_A) = \bs{d}A_{|A^\gamma} (\gamma\- \chi \gamma)^v_{A^\gamma}= D^{A^\gamma}(\gamma\-\chi \gamma)= \gamma\- (D^A\chi) \gamma=\gamma\- \bs{d}A_{|A}(\chi^v_A) \gamma$, by \eqref{Vert-dA}. 
Then we have the $\H$-equivariance, holding on $\Gamma(V\A)$,
\begin{align}
\label{Equiv-dA}
R^\star_\gamma \bs{d}A_{|A^\gamma} = \gamma\- \bs{d}A_{|A} \, \gamma,  \quad \text{or} \quad   R^\star_\gamma \bs{d}A = \gamma\- \bs{d}A \, \gamma,
\end{align}
and we require that it holds $\forall \bs{X} \in \Gamma(T\A)$. From \eqref{Pushforward-X-inf}, it is then easy to find the $\bs \H$-gauge transformation of $\bs{d}A$~to~be, 
\begin{align}
\label{GT-dA}
\bs{d}A^{\bs \gamma}_{|A}(\bs{X}_A)&\defeq (\bs{\Psi}^\star \bs{d}A)_{|A}(\bs{X}_A)=\bs{d}A_{|A^\gamma} \left( \bs{\Psi}_\star \bs{X}_A \right),   \notag\\
						        & = \bs{d}A_{|A^\gamma} \left(R_{\bs{\gamma}(A)\star} \left[ \bs{X}_A + \left\{ \bs{d}\bs{\gamma} {\bs{\gamma}\- }_{|A}(\bs{X}_A)\right\}^v_A \right] \right)
						         = (R_{\bs{\gamma}(A)}^\star  \bs{d}A_{|A^\gamma})  \left( \bs{X}_A + \left\{ \bs{d}\bs{\gamma} {\bs{\gamma}\- }_{|A}(\bs{X}_A)\right\}^v_A \right), \notag \\
						        & = \bs{\gamma}(A)\-   	\bs{d}A_{|A}  \left( \bs{X}_A + \left\{ \bs{d}\bs{\gamma} {\bs{\gamma}\- }_{|A}(\bs{X}_A)\right\}^v_A \right) 		\bs{\gamma}(A) 
						        =    \bs{\gamma}(A)\-  \left(    \bs{d}A_{|A}(\bs{X}_A) + D^A \left\{ \bs{d}\bs{\gamma} {\bs{\gamma}\- }_{|A}(\bs{X}_A)\right\} \right)  \bs{\gamma}(A),   \notag\\
						        & = \left[     \bs{\gamma}(A)\-   \left( \bs{d}A_{|A}  + D^A\left\{ \bs{d}\bs{\gamma} {\bs{\gamma}\- }_{|A}\right\} \right)    \bs{\gamma}(A)    \right]     (\bs{X}_A ). 
\end{align}
Or in short, $ \bs{d}A^{\bs \gamma} =  \bs{\gamma}\-   \left( \bs{d}A  + D\left\{ \bs{d}\bs{\gamma} {\bs{\gamma}\- }\right\} \right)    \bs{\gamma}$.  
 This results gives a geometric interpretation to the heuristic computation performed e.g. in \cite{DonnellyFreidel2016} and \cite{Geiller2017}. 
 
 Now, consider also the curvature map  $F: \A \rarrow \Omega^2_\text{tens}(\P, \text{Lie}H)$, $A \mapsto F(A)$,  s.t. $R^\star_\gamma F =\gamma\- F \gamma$ by definition. 
 Given a vector field  $\bs{X} \in \Gamma(T\A)$ with flow $\phi_\tau$, we have that:
 \begin{align*}
 \bs{d}F_{|A} (\bs{X}_A)&
 =\bs{X}\big(F\big)(A)=\tfrac{d}{d\tau} F\big( \phi_\tau(A) \big) \big|_{\tau=0} = \tfrac{d}{d\tau} d\phi_\tau(A) + \sfrac{1}{2}[\phi_\tau(A), \phi_\tau(A)] \big|_{\tau=0}= D^A\big( \tfrac{d}{d\tau}\phi_\tau(A)\big|_{\tau=0} \big), \\
        				    &= D^A\big( \bs{d}A_{|A}(\bs{X}_A)\big),
	 \quad \text{or simply} \quad \bs{d} F=D\big( \bs{d}A\big).
 \end{align*}
 From this follows that evaluated on a vertical vector field, $\bs{d}F \in \Omega^1(\A)$ gives the infinitesimal $\H$-transformation~of~$F$:
\begin{align}
\label{Vert-dF}
\bs{d}F_{|A}(\chi^v_A)=D^A\big( \bs{d}A_{|A}(\chi^v_A) \big)= D^A\big( D^A\chi \big)= [F(A), \chi],      \quad \text{or} \quad       \iota_{\chi^v}\bs{d}F =[F, \chi].
\end{align}
We also find its $\H$-equivariance to be 
\begin{align}
\label{Equiv-dF}
R^\star_\gamma \bs{d}F_{|A^\gamma}(\bs{X}_A)&=  \bs{d}F_{|A^\gamma}(R_{\gamma\star}\bs{X}_A) = D^{A^\gamma}\big( \bs{d}A_{|A^\gamma}(R_{\gamma\star}\bs{X}_A) \big) =D^{A^\gamma}\big( R^\star_\gamma \bs{d}A_{|A^\gamma}(\bs{X}_A) \big),  \notag\\
									& = D^{A^\gamma}\big( \gamma\- \bs{d}A_{|A}(\bs{X}_A) \gamma \big) = \gamma\- D^A\big( \bs{d}A_{|A}(\bs{X}_A) \big) \gamma = \gamma\- \bs{d}F_{|A} (\bs{X}_A) \gamma, \qquad \text{or}\quad R^\star \bs{d}F =\gamma \- \bs{d}F \gamma.
\end{align}
This is also recovered simply from the fact that pullbacks and $\bs d$ commute, so that $R^\star_\gamma \bs{d} F= \bs{d} R^\star_\gamma F = \gamma\- \bs{d}F \gamma$. 
From  \eqref{Vert-dF}-\eqref{Equiv-dF} and \eqref{Pushforward-X-inf},  we thus find  the $\bs \H$-gauge transformation of $\bs{d}F$ to be 
\begin{align}
\label{GT-dF}
\bs{d}F^{\bs \gamma}_{|A}(\bs{X}_A)&\defeq (\bs{\Psi}^\star \bs{d}F)_{|A}(\bs{X}_A)=\bs{d}F_{|A^\gamma} \left( \bs{\Psi}_\star \bs{X}_A \right),   \notag\\
						        & = \bs{d}F_{|A^\gamma} \left(R_{\bs{\gamma}(A)\star} \left[ \bs{X}_A + \left\{ \bs{d}\bs{\gamma} {\bs{\gamma}\- }_{|A}(\bs{X}_A)\right\}^v_A \right] \right)
						         = (R_{\bs{\gamma}(A)}^\star  \bs{d}F_{|A^\gamma})  \left( \bs{X}_A + \left\{ \bs{d}\bs{\gamma} {\bs{\gamma}\- }_{|A}(\bs{X}_A)\right\}^v_A \right), \notag \\
						        & = \bs{\gamma}(A)\-   	\bs{d}F_{|A}  \left( \bs{X}_A + \left\{ \bs{d}\bs{\gamma} {\bs{\gamma}\- }_{|A}(\bs{X}_A)\right\}^v_A \right) 		\bs{\gamma}(A) 
						        =    \bs{\gamma}(A)\-  \left(    \bs{d}F_{|A}(\bs{X}_A) + \big[F(A), \bs{d}\bs{\gamma} {\bs{\gamma}\- }_{|A}(\bs{X}_A)\big] \right)  \bs{\gamma}(A),   \notag\\
						        & = \left\{     \bs{\gamma(A)}\-   \left( \bs{d}F_{|A}  + \big[F(A), \bs{d}\bs{\gamma} {\bs{\gamma}\- }_{|A}\big] \right)    \bs{\gamma}(A)    \right\}     (\bs{X}_A ). 
\end{align}
In short, $ \bs{d}F^{\bs \gamma} =  \bs{\gamma}\-   \left( \bs{d}A  + \big[F, \bs{d}\bs{\gamma} {\bs{\gamma}\- }\big] \right)    \bs{\gamma}$.  
 This  is  the geometric counterpart to computations found e.g.~in~\cite{DonnellyFreidel2016}. 
\medskip

 All of the above hold for the space of Cartan connections as well: $\B$ is also a $\H$-principal bundle, whose tangent bundle and vertical subbundle are defined in the same manner as above. For $\bs{d} \b A \in \Omega^1(\B)$  we have the exact analogues of equations \eqref{Vert-dA}-\eqref{GT-dA}, and for the bundle map $\b F:\B \rarrow \Omega^2_\text{tens}(\P, \text{Lie}G)$ we have that $\bs{d}\b F \in \Omega^1(\B)$ satisfies analogues of equations \eqref{Vert-dF}-\eqref{GT-dF}. 
 
 In the especially interesting  case of reductive Cartan geometries recall that the connection splits as $\b A=A+e$, with $e$ the soldering form, so we have $\bs{d}\b A = \bs{d} A+ \bs{d}e$ with $\H$-equivariance and verticality properties
 \begin{align}
  R^\star_\gamma \bs{d}\b A = \gamma\- \bs{d}\b  A \gamma \qquad  \Rightarrow \qquad   R^\star_\gamma \bs{d} A = \gamma\- \bs{d} A \gamma \qquad \text{and} \qquad R^\star_\gamma \bs{d} e = \begin{cases} \ \gamma\- \bs{d}e \,\gamma, \\[-1.5mm] \ \gamma\- \bs{d}e,  \end{cases}  \\
 \iota_{\chi^v}\bs{d}\b A =D^{\b A}\chi \qquad  \Rightarrow \qquad    \iota_{\chi^v}\bs{d}A =D\chi \qquad \text{and} \qquad \iota_{\chi^v}\bs{d}e =-\ad(\chi)e=\begin{cases}\  [e, \chi], \\[-1.5mm] \ -\chi e.  \end{cases}
 \end{align}
The third equalities reflect the possibility that the action of $H$ on Lie$G$/Lie$H$ reduces to a left action, $\Ad(H)=\ell(H)$. 
Then, the formula for $\bs{d}\b A^{\bs \gamma}$ splits as
  \begin{align}
  \label{GT-dA-Cartan}
\bs{d}A^{\bs \gamma} =  \bs{\gamma}\-   \left( \bs{d}A  + D\left\{ \bs{d}\bs{\gamma} {\bs{\gamma}\- }\right\} \right)    \bs{\gamma}
\quad \text{and} \quad
\bs{d}e^{\bs \gamma} = \begin{cases}\  \bs{\gamma}\-   \left( \bs{d}e  + \big[e, \bs{d}\bs{\gamma} {\bs{\gamma}\- }\big] \right)    \bs{\gamma}, \\
						          \  \bs{\gamma}\-   \left( \bs{d}e   -  \bs{d}\bs{\gamma} {\bs{\gamma}\- e}\big] \right).   
				      \end{cases}
 \end{align} 
This gives  geometric substance to computation found e.g. in \cite{Geiller2017}. In the same way we have $\bs{d}\b F=\bs{d}F + \bs{d}T$, whose $\H$-equivariance $R^\star_\gamma \bs{d}\b F = \gamma\- \bs{d}\b F \gamma$ and  verticality $\iota_{\chi^v}\bs{d}\b F = [\b F, \chi]$  decompose as above, so that  $\bs{d}\b F^{\bs \gamma}$ splits as
   \begin{align}
   \label{GT-dF-Cartan}
\bs{d}F^{\bs \gamma} =  \bs{\gamma}\-   \left( \bs{d}F  + \big[F, \bs{d}\bs{\gamma} {\bs{\gamma}\- }\big] \right)    \bs{\gamma}
\quad \text{and} \quad
\bs{d}T^{\bs \gamma} =   \begin{cases}\ \bs{\gamma}\-   \left( \bs{d}T  + \big[T, \bs{d}\bs{\gamma} {\bs{\gamma}\- }\big] \right)    \bs{\gamma},\\ 
						          \  \bs{\gamma}\-   \left( \bs{d}T   -  \bs{d}\bs{\gamma} {\bs{\gamma}\- T}\big] \right).   
				      \end{cases}
 \end{align} 
\bigskip

Bundles associated to $\A$ are $\bs{E}\defeq \A \times_\H \bs{V}$ and $\Gamma(\bs{E})\simeq \Omega^0_\text{eq}(\A, \bs{V})$. Consider in particular the following examples relevant to classical and quantum gauge theories. A  $\H$-invariant Lagrangian functional  $L: \A \rarrow \Omega^n_\text{basic}(\P, \RR)$, $A \mapsto L(A)$, with   $n=\dim\M$, is a section of the associated bundle corresponding to the trivial representation $\bs{V}=\Omega^n_\text{basic}(\P, \RR)$. Then, $L \in \Omega^0_\text{basic}(\A)$ and descends to (comes from) a well-defined $\ell \in \Omega^0(\A/\H)$,  $L=\pi^\star \ell$.\footnote{Not to be confused with $l \in \Omega^n(\M)$, the globally defined local representative of $L$ on $\M$, functional of the YM potential $\s^*A$.} 
The~corresponding action functional $S:\A \rarrow \RR$, $A\mapsto S[A]=\int L(A)$,\footnote{Where it is understood that the integration is over the image $\s(U)\subset \P$ of a (compact) domain $U\subseteq \M$ by a local section $\s$. }
 is a section of the bundle $\bs E$ built from the  trivial representation $\bs{V}=\RR$. So $S \in \Omega^0_\text{basic}(\A)$ and descends to $\A/\H$ also. 

In quantum gauge theories, the generating functional (path integral) $Z: \A \rarrow \CC$, $A \mapsto Z[A]=\int d\mu_{|A}\, e^{\sfrac{i}{\hbar} S[A]}$, if well-defined and if the measure is $\H$-invariant, $R^\star_\gamma d\mu=d\mu$, is formally a section of $\bs E$ for the trivial representation $\bs{V}= \CC$. In that case also $Z\in \Omega^0_\text{basic}(\A)$. Of course it is a divergent quantity, one must quotient out the volume of $\H$ by selecting a single representative in each orbit $\O_\H[A]$: This may be done simply by a gauge-fixing, i.e. via a section $\bs{\s}: \U \subset \A/\H \rarrow \A$, so that $\bs{\s}^\star Z < \infty$. Of course the fact that there exists no such global section of $\A$ (at~least in the case of $S\!U(N)$-gauge theories) is well-known as the \emph{Gribov obstruction} (or ambiguity) \cite{Gribov, Singer1978}. 

On  the above basic 0-forms, valued in trivial representations of $\H$, a functional covariant derivative $\bs{D}^{\bs{A}}$ reduces to $\bs d$. So that actually  a functional Ehresmann connection $\bs{A}$ is not needed. This is obviously not the case for non-basic  forms. Again, all this holds when substituting $\A$ by $\B$, i.e. by working with gravitational gauge theories.

\subsection{Anomalies in gauge theories and twisted geometry on $\A$}
\label{Anomalies in gauge theories and twisted geometry on A}

One would think that in physics, generically, $\H$-non-invariant theories would be sections of `standard' bundles associated to $\A$ via plain representations of $\H$. Actually, several prominent examples turns out to be sections of \emph{twisted} associated bundles of the kind described in section \ref{Twisted bundle geometry}. 

\subsubsection{Quantum gauge anomalies}
\label{Quantum gauge anomalies}

A well-known manifestation of gauge non-invariance is the phenomenon of quantum anomalies: it occurs when the $\H$-invariance of a classical theory $S[A]$ , $R^\star_\gamma S =S$, fails to be upheld by the corresponding quantum theory~$Z[A]$. 
This happens when the measure is not $\H$-invariant, $R^\star_\gamma d\mu_{|A^\gamma}=d\mu_{|A}\, e^{\sfrac{i}{\hbar}\, {\bf c}(A, \gamma)}$ where the phase - sometimes called `Wess-Zumino (WZ) term' -  is
s.t.  $\tfrac{d}{d\tau} {\bf c}(A, e^{\tau \chi})\big|_{\tau=0} \rdefeq a(\chi, A)$ is the (integrated) quantum gauge anomaly, obviously linear in $\chi \in$ Lie$\H$.
 First discovered via perturbative methods, (consistent) anomalies were found to be characterized by BRST cohomological methods through  the Stora-Zumino descent equations. They finally came to be understood as degree $1$ elements in the cohomology of Lie$\H$ \cite{Bonora-Cotta-Ramusino}.

From the non-invariance of the measure follows that $Z[A^\gamma]= C(A, \gamma)\- Z[A]$, with $C(A, \gamma)=e^{-\sfrac{i}{\hbar}\, {\bf c}(A, \gamma)}$  (see e.g. \cite{Bertlmann} chap.5 and 11). It turns out that the map $C:\A \times \H \rarrow U(1)$ is a 1-cocycle for the action of $\H$ on~$\A$, since indeed $Z(A^{\gamma\gamma'})=Z\big((A^\gamma)^{\gamma'}\big)$ implies the cocycle relation  $C(A, \gamma\gamma')=C(A, \gamma)\ C(A^\gamma, \gamma')$ - or, for the phase ${\bf c}(A, \gamma\gamma')={\bf c}(A, \gamma)+{\bf c}(A^\gamma, \gamma')$ -  which is a form of the Wess-Zumino consistency condition for the gauge anomaly. 
The fact that consistent anomalies are cocycles of the gauge group $\H$ as long been known \cite{Faddeev-Shatashvili1984, Reiman-et-al1984, Falqui-Reina1985}.

It then follows that an anomalous quantum functional $Z$  is - in the language of section \ref{Twisted bundle geometry} -  a $C$-equivariant functional on $\A$, $R^\star_\gamma Z= C(\ ,\gamma)\- Z$,  i.e. a section of the \emph{twisted} associated line bundle $\bs{L}^C:=\A \times_{C(\H)} \CC$. In our notation $Z\in \Omega^0_\text{eq}(\A, \CC)^C \simeq \Gamma(\bs{L}^C)$.
A simple functional differentiation $\bs{d}Z$ of such sections would not be geometrically sound, only a twisted functional covariant derivative $\bs{D^A}Z$ is, and for that $\A$ would need to be endowed with a twisted connection $\bs{\t A} \in \bs{\t\A}$. In this case  $\bs{\t A} \in \Omega^1\left(\A, \text{Lie}U(1)\right)$ is by definition s.t. 
\begin{align}
\label{axiom-i}
R^\star_\gamma \bs{\t A}_{|A^\gamma}= C(A, \gamma)\- \bs{\t A}_{|A} C(A, \gamma)+ C(A, \gamma)\-\bs{d} C(\ \, , \gamma)_{|A}  
							  = \bs{\t A}_{|A}  - i \bs{d} {\bf c}(\ \, ,\gamma)_{|A}, 
\end{align}
by axiom i. And by axiom ii, 
\begin{align}
\label{axiom-ii-anomaly}
\bs{\t A}_{|A} \big( \chi^v_A\big)&= \tfrac{d}{d\tau} C(A, e^{\tau \chi}) \big|_{\tau=0} \ \in \text{Lie}U(1),  \notag \\
						 &=\tfrac{d}{d\tau} \exp\left( - \tfrac{i}{\hbar} {\bf c}(A, e^{\tau \chi}) \big|_{\tau=0} \right) = - \tfrac{i}{\hbar}\,\tfrac{d}{d\tau} {\bf c}(A, e^{\tau \chi}) \big|_{\tau=0} \rdefeq -\tfrac{i}{\hbar} \, a(\chi, A). 
\end{align}
This close link between the gauge anomaly and the twisted connection is remarkable. So is the fact that since the curvature two form is $\bs{\t F}=\bs{d\t A} + \sfrac{1}{2}[\bs{\t A}, \bs{\t A} ]= \bs{d\t A}\ \in \Omega^2_\text{tens}\left(\A, \text{Lie}U(1)\right)$, we have in particular (by Koszul formula): 
\begin{align}
\label{curvature-WZ-cc}
\bs{\t F}_{|A} \big(\chi^v_A, \chi'^v_A \big)&=\bs{d\t A}_{|A} \big(\chi^v_A, \chi'^v_A \big) = \chi^v [\bs{\t A}_{|A}(\chi'^v_A)] - \chi'^v [\bs{\t A}_{|A}(\chi^v_A)] - \bs{\t A}_{|A}([\chi^v, \chi'^v]_A) \equiv 0,  \notag\\*[2mm]
								\Rightarrow &\quad  \chi^v [a(\chi', A)] - \chi'^v[a(\chi, A)] - a([\chi, \chi'], A) = 0, \quad \text{by \eqref{axiom-ii-anomaly}.}
\end{align}
This is none other than the WZ consistency condition for the gauge anomaly (see e.g. Eq.(8.62) and Eq.(10.76) in \cite{Bertlmann}, or Eq.(12.25) in \cite{GockSchuck}), which is thus encoded in the tensoriality of the twisted curvature. 
Notice that  from section  \ref{Twisted bundle geometry} we have  the $\bs \H$-gauge transformations: 
$\bs{\t F^\gamma}=\bs{\t F}$
 and 
$\bs{\t A^\gamma}=C(\bs{\gamma})\- \bs{\t A}\,C(\bs{\gamma})+ C(\bs{\gamma})\- \bs{d}C(\bs{\gamma})= \bs{\t A} -i \bs{d}{\bf c}(\, \, ,\bs{\gamma})$.

The fact that anomalous functionals  are sections of `special' line bundle was noted in \cite{Falqui-Reina1985, Catenacci-et-al1986, Catenacci-Pirola1990}, see also  \cite{Mickelsson1986, Mickelsson1987}, and   \cite{Ferreiro-Perez2018} more recently. But as far as we know, particular emphasis on the peculiar geometrical nature of these  bundles is first found in \cite{Blau1988, Blau1989}\footnote{In the introduction of \cite{Blau1989} we read, ``\emph{[...] recently objects (called generalized associated bundles hereafter) have appeared in the physics literature, about whose general structure little seems to be known}", and further in the text  ``\emph{bundles of this kind have recently appeared in the physics literature (mainly in relation with anomalies). Their geometrical structure, however, was not further investigated.}"} where, by the way, an instance of flat twisted connection is  built from a twisted (local) dressing field. See \cite{Francois2019_II} section 10.4 for  a  discussion of the details (and the next section for a classical analogue related to WZ functionals).

It is not the goal of this paper to explore further applications of the twisted geometry to quantum gauge theories and anomalies. This will be done elsewhere. Rather we now focus on its relevance to classical gauge theories.

\subsubsection{Classical gauge anomalies}
\label{Classical gauge anomalies}

Suppose we have a non-invariant classical  theory whose Lagrangian $L: \A \rarrow \Omega^n(\P, \RR)$ has generic equivariance  $R^\star_\gamma L = L + c(\, \, ,\gamma)$, i.e. $L(A^\gamma)=L(A)+ c (A, \gamma)$. The corresponding action is $R^\star_\gamma S =S + {\bf c}(\ , \gamma)$, with ${\bf c}=\int c$. 
 Consistency with the right $\H$-action $R^\star_\gamma R^\star_{\gamma'}  = R^\star_{\gamma\gamma'}$ - i.e.  $L[(A^\gamma)^{\gamma'}]=L[A^{\gamma\gamma'}]$ - implies $c(A, \gamma\gamma')=c(A, \gamma)+c(A^\gamma, \gamma')$. 
 This in turn implies that $C: \A \times \H \rarrow \CC$, $(A, \gamma) \mapsto C(A, \gamma)\defeq e^{i {\bf c}(A, \gamma)}$, satisfies $C(A, \gamma\gamma')=C(A, \gamma)C(A^\gamma, \gamma')$ and is thus a 1-cocycle. Then $Z: \A \rarrow \CC $ defined by  $Z[A]\defeq e^{iS[A]}$ is a $C$-equivariant functional,  $Z\in \Omega^0_\text{eq}(\A, \CC)^C \simeq \Gamma(\bs{L}^C)$.

 A Lagrangian whose non-invariance manifests in this way will be called $c$-equivariant. As a 0-form on $\A$ it is  even $c$-tensorial, thus its $\bs \H$-gauge transformation - or `field dependent' gauge transformation - is controlled by its $\H$-equivariance so that: $L^{\bs \gamma}= L + c(\ \, ,\bs{\gamma})$.
  Notice that it encompasses  \emph{quasi-invariant} Lagrangians, for which the linearized cocycle is $d$-exact, $c=db$, as well as  cases where $c$ is a trivial cocycle (i.e. $C$ is a group morphism).

  By analogy with the quantum case, we call $\tfrac{d}{d\tau}  c(A, e^{\tau \chi})\big|_{\tau=0} \rdefeq \alpha_\text{cl}(\chi, A)$, with $\chi \in \text{Lie}\H$, the \emph{classical} anomaly. While $\tfrac{d}{d\tau} {\bf c}(A, e^{\tau \chi})\big|_{\tau=0} \rdefeq a_\text{cl}(\chi, A)$ is  the integrated classical anomaly, $a_\text{cl}=\int \alpha_\text{cl}$. Since from now on we will deal only with classical anomalies, we drop the subscript. The anomaly features in the infinitesimal equivariance of $L$, which is given by the variational Lie derivative: $\bs{L}_{\chi^v} L= \iota_{\chi^v}\bs{d}L= \chi^v(L)=\tfrac{d}{d\tau} R^\star_{e^{\tau \chi}} L \big|_{\tau=0} =\alpha(\chi, A)$. 
 Consider the following examples, which will be further studied - among others - in section \ref{Covariant Hamiltonian formalism (and the puzzle of boundaries)}. 
 
 \paragraph{Massive Yang-Mills theory}  
 The Lagrangian of the theory is   $L_\text{{\tiny mYM}}(A)=\tfrac{1}{2}\Tr(F *\!F)- \tfrac{1}{2}m^2\Tr(A*\!A)$, where $m$ is the~mass of  $A$, and $*:\Omega^\bullet(\P) \rarrow \Omega^{\text{dim}\P-\bullet}(\P)$ is the Hodge operator.\footnote{This tacitly presumes that we have a (pseudo) Riemannian metric on $\P$. It would be more customary to assume that the metric is on $\M$ only, then $\A$ must be seen as the space of YM potential on $\M$ - i.e. local representatives of connections on $\P$ - and $\H$ is the pullback of the gauge group. Nothing of substance is affected by this.} A quick computation gives the $\H$-equivariance
 \begin{align}
 \label{MYM}
 L_\text{{\tiny mYM}}(A^\gamma) &= L_\text{{\tiny mYM}}(A) + c(A,\gamma),  \notag\\
 						   &= L_\text{{\tiny mYM}}(A) - m^2\Tr \left(  A *\!d\gamma\gamma\- - \tfrac{1}{2} d\gamma\-*\!d\gamma  \right), \quad \gamma \in \H=\SU(n).
 \end{align}
The proof of the cocycle relation $c(A,\gamma\gamma')= c(A,\gamma) + c(A^\gamma, \gamma')$ is straightforward but relegated to appendix \ref{Cocycle relations for c-equivariant theories}. 
The $\bs \H$-gauge transformation is obvious. The classical anomaly is $\bs{L}_{\chi^v} L_\text{{\tiny mYM}}=\alpha(\chi, A)=-m^2\Tr(A*\!d\chi)$.

The~abelian limit  gives massive Maxwell theory $L_\text{{\tiny mM}}(A)$, where $c(A,\gamma)= -m^2\left( A*\!d\chi + \tfrac{1}{2}d\chi *\!d\chi \right)$ with $\gamma=e^\chi$, $\chi \in$ Lie$\H$. Since here $\H=\U(1)$ is abelian, one has $c(A, \gamma\gamma')=c(A, \gamma'\gamma)$. See again appendix  \ref{Cocycle relations for c-equivariant theories}.

 \paragraph{3D non-Abelian Chern-Simons theory} 
 \label{3D-CS-theory}
   The Lagrangian of the theory is $L_\text{{\tiny CS}}(A)=\Tr(AdA + \tfrac{2}{3}A^3)$. A~fastidious but straightforward calculation gives the well-known $\H$-equivariance
  \begin{align}
  \label{CS}
 L_\text{{\tiny CS}}(A^\gamma) &= L_\text{{\tiny CS}}(A) + c(A,\gamma),  \notag\\
 						   &= L_\text{{\tiny CS}}(A) + \Tr \left(  d\big(  \gamma d\gamma\- A  \big) - \tfrac{1}{3}\big(  \gamma\-d\gamma \big)^3  \right), \quad \gamma \in \H=\SU(n).
 \end{align}
 Again, the proof of the cocycle relation for $c(A,\gamma)$ is in the dedicated appendix. The $\bs \H$ gauge transformation is easily read-off, as well as the classical anomaly  
 $\bs{L}_{\chi^v} L_\text{{\tiny CS}}=\alpha(\chi, A)= -d\Tr( d\chi\, A)$.  The fact that the latter is $d$-exact will be relevant when analysing the presymplectic structure of the theory in section \ref{Presymplectic structure for non-invariant Lagrangians}. Remark also that in the BRST language (cf footnote \ref{Lie-BRST}) we have $\bs{s}L_\text{{\tiny CS}} = -dQ(\bs{v}, A)$, one of the Stora-Zumino descent equations, where $Q(\bs{v}, A)=\Tr(\bs{v}dA)$ is the 2D consistent \emph{quantum} non-Abelian anomaly (see e.g. \cite{Bertlmann} p.382-389).
  
 The abelian theory $L_\text{{\tiny AbCS}}=AdA$ is quite degenerate since its cocycle is not only $d$-exact, $c(A, \gamma)=d(\chi dA)$  - the Lagrangian is thus quasi-invariant - but also trivial:  $c(A, \gamma\gamma')=d\big((\chi+\chi')dA\big)=c(A, \gamma) + c(A, \gamma')$ - so $C(A, \gamma)$ is clearly a group morphism.
We should then be wary of generalising results  holding in the Abelian theory to the general non-Abelian case.  Here also we have $c(A, \gamma\gamma')=c(A, \gamma'\gamma)$.

 \paragraph{3D-$\bs \CC$-gravity $\bs{\Lambda\!=\!0}$}
\label{3D-C-grav-noLambda}
 In term of Cartan geometry $(\P, \b A)$, with $\P$ a $H$-principal bundle equipped with a Lie$G$-valued Cartan connection $\b A$, the theory is based on the pair of groups   $(G, H)=\big(S\!U(2)\ltimes \text{Herm}(2, \CC), S\!U(2)\big)$ in Euclidean signature, or $\big(S\!U(1, 1) \ltimes \text{Herm}(2, \CC), S\!U(1,1)\big)$ in Lorentzian signature.\footnote{In the real case we would have $(G, H)=\big(IS\!O(3), S\!O(2)\big)$ in Euclidean signature, or $\big(IS\!O(1, 2), S\!O(1,2)\big)$ in Lorentzian signature.} The homogeneous space (and ground state of the theory) is $\RR^3$ which is mapped into $\text{Herm}(2, \CC)$ via $x=\{x^i\}_{i=1,2,3}  \mapsto \b x=x^i \s_i$ where $\s_i$ are the Pauli matrices, and on which  $S\!U(2)$/$S\!U(1,1)$ acts by conjugation. 
 
 The geometry $(\P, \b A)$ is reductive, so the connection splits as $\b A=A+e$ and its $\H$-gauge transformation is $\b A^\gamma = \gamma \- A \gamma + \gamma\-d\gamma$, which gives $\b A^\gamma = \gamma \- A \gamma + \gamma\-d\gamma$ and $e^\gamma =\gamma\- e \gamma$. The curvature $\b F=R+T$, where $R=dA +\sfrac{1}{2}[A,A]$ and $T=de +[A,e]$,  transforms as $\b F^\gamma =\gamma\- \b F \gamma$. If $T=0$ then $\b A$ is normal, so that $A=A(e)$.
  The~Lagrangian of the theory, $L(\b A)=L(A, e)= \Tr \left(  eR  \right)$, is  clearly $\H$-invariant ($R^\star_\gamma L=L$). 
  
  From a Cartan theoretic point of view, there is no meaning to `translational' gauge symmetry. But if one insist on considering them, one can see $\b A$ as lifted to an Ehresmann connection on a $G$-bundle $\Q \supset \P$ whose gauge group $\G$ contains both $\H$, acting on $\b A$ as above, and the abelian (additive) subgroup $\T\defeq \left\{ \, t: \Q \rarrow \text{Herm}(2, \CC)\, |\, \ldots\, \right\}$ acting as $\b A^t=\b A + D^At$, which implies $A^t=A$ and $e^t =e + D^A t$. The $\T$-equivariance of the Lagrangian is then, 
 \begin{align}
 \label{3D-grav-cocycle}
 L(\b A^t)&= L(\b A) + c(\b A, t), \notag\\[-4mm]
 	      &= L(\b A) + \Tr\big(D^At \, R\big)
	      =L(\b A) + d\Tr(t\, R) - \Tr\big(t \, \cancelto{0 \text{ by Bianchi}}{D^AR\big),}
 \end{align} 
 from which is deduced its $\bs \T$-gauge transformation. Since $\T$ is abelian, $c(\b A, t+t')=c(\b A, t'\!+t)$.
 The cocycle is not only exact, $c=db$,  the Lagrangian thus being quasi-invariant, but also clearly trivial  $c(A, t+t')=c(\b A, t)+c(\b A, t')$. The classical anomaly is $\bs{L}_{\tau^v} L = \alpha(\tau, \b A)= d\Tr(\tau\, R)$, with $\tau \in$ Lie$\T$ an infinitesimal gauge translation.  Again, the fact that it is $d$-exact is relevant to the analysis of the presymplectic structure of the theory (section \ref{Presymplectic structure for non-invariant Lagrangians}).
\bigskip

\medskip
  
  For  $Z=e^{iS}\in \Omega^0_\text{eq}(\A, \CC)^C \simeq \Gamma(\bs{L}^C)$,  the covariant derivative w.r.t. a twisted connection $\bs{\t A} \in \bs{\t\A}$ on $\A$ is $\bs{D^{\t A}} Z = \big(\bs{d}S + \bs{\t A} \big) Z$. Since $\bs{D^{\t A}} Z$ and $Z$ have the same equivariance, it follows that $i \bs{d}S + \bs{\t A}$ is a basic 1-form.  Indeed  by the axiomatic properties of $\bs{\t A}$, we have on the one hand
  \begin{align*}
  R^\star_\gamma \big( i\bs{d}S +\bs{\t A} \big) = i\bs{d}R^\star_\gamma S + R^\star_\gamma \bs{\t A}= i\bs{d} \big(S + {\bf c}(\ \, ,\gamma) \big) +  \bs{\t A} - i\bs{d} {\bf c}(\ \, ,\gamma) = i\bs{d} S + \bs{\t A},
  \end{align*}
 which checks the trivial equivariance. On the other hand we get
  \begin{align*}
  \big( i\bs{d} S + \bs{\t A} \big)_{|A}(\chi^v_A) = i \chi^v\big( S \big)(A) + \bs{\t A} (\chi^v_A) =  i \tfrac{d}{d\tau} \big( R^\star_{e^{\tau \chi}} S \big) (A) \big|_{\tau=0} - i a(\chi, A)
   								     = i \tfrac{d}{d\tau} {\bf c}(A, e^{\tau \chi}) \big|_{\tau=0} - i a(\chi, A) \equiv 0, 
   \end{align*}
 which checks the horizontality. Thus $i \bs{d}S + \bs{\t A}$ descends to the physical space $\A/\H$. 
 
 The special case of a flat twisted connection built via a  dressing field happens to be connected to the Wess-Zumino functional (as defined by Stora) \cite{Manes-Stora-Zumino1985, Attard-Lazz2016} and alluded to in section \ref{Residual transformations (second kind) : ambiguity in choosing a dressing field}. Consider a $\A$-dependent dressing field $\bs{u}: \A \rarrow \D r[H, H]$, and have the corresponding twisted dressing field\footnote{Notice that the cocycle property of $c$, ${\bf c}$ or $C$, owes nothing to the fact that $\gamma \in \H$. It holds for any map $\P \rarrow H$, then for $\D r[H, K]$. } $C\big(A, \bs{u}(A)\big)=e^{i\,{\bf c}(A,\, \bs{u}(A))}$. It satisfies 
 \begin{align}
 \label{Twisted-DF}
 \big[R^\star_\gamma C\big(\ \, ,\bs{u}\big)\big](A)&= C\big(A^\gamma, \bs{u}(A^\gamma)\big)=C\big(A^\gamma, \gamma\-\bs{u}(A)\big)= C\big(A^\gamma, \gamma \big)\, C\big(  A, \bs{u}(A) \big)=C\big( A, \gamma \big)\- C\big( A, \bs{u}(A)\big) \\
 							      &= \big[ C\big(\ \, , \gamma \big)\- C\big( \ \, , \bs{u}\big)\big](A),
 \end{align}
 as it should. Correspondingly the linearized cocycle satisfies, 
  \begin{align}
  \label{Twisted-DF-linear}
 \big[R^\star_\gamma {\bf c}\big(\ \, ,\bs{u}\big)\big](A)&= {\bf c}\big(A^\gamma, \bs{u}(A^\gamma)\big)={\bf c}\big(A^\gamma, \gamma\-\bs{u}(A)\big)= {\bf c}\big(A^\gamma, \gamma \big)+ {\bf c}\big(  A, \bs{u}(A) \big)=-{\bf c}\big( A, \gamma \big) + {\bf c}\big( A, \bs{u}(A)\big) \\
 							      &= \big[ -{\bf c}\big(\ \, , \gamma \big) + {\bf c}\big( \ \, , \bs{u}\big)\big](A).
 \end{align}
 Thus, it is s.t. 
 \begin{align}
 \label{lin-c-WZ-term}
 \big[\bs{L}_{\chi^v} {\bf c}(\ \,, \bs{u})\big](A) = \iota_{\chi^v}\bs{d} {\bf c}(\ \,, \bs{u})_{|A}= \big[\chi^v\big( {\bf c}(\ \,, \bs{u}) \big)\big](A) = \tfrac{d}{d'\tau} \big( R^\star_{e^{\tau \chi}} {\bf c}(\ \,, \bs{u}) \big)(A) \big|_{\tau=0}= - \tfrac{d}{d\tau} {\bf c}(A, e^{\tau \chi}) \big|_{\tau=0} \rdefeq - a(\chi, A). 
 \end{align}
 The linearized twisted dressing field ${\bf c}\big(A, \bs{u}(A)\big)$ is then none other than the WZ functional, defined usually as satisfying $\bs{s} \Gamma_{WZ}(u, A)= -a(\bs{v}, A)$   in the BRST language (see e.g. Eq.(42) in \cite{Manes-Stora-Zumino1985}  or   Eq.(11) and (A.5) in \cite{Attard-Lazz2016}). 
 
 It is easy to see that $\bs{\t A}_0 \defeq C(\ \,, \bs{u})\, \bs{d} C (\ \, ,\bs{u} )\-=i\bs{d} {\bf c}(\ \, ,\bs{u})$ satisfies \eqref{axiom-i}-\eqref{axiom-ii-anomaly}, and is s.t. $\bs{d \t A}_0 + \sfrac{1}{2}[ \bs{\t A}_0, \bs{\t A}_0]\equiv0$. It is therefore a \emph{flat} twisted connection. So, we have that  $i\bs{d}S +\bs{\t A}_0=i \bs{d} \big( S + {\bf c}(\ \, ,\bs{u}) \big)$ is a basic 1-form. And since $\bs d$ preserve basicity, this means that the 0-form $S \!+ {\bf c}(\ \, ,\bs{u})$ is also basic, thus  $\bs \H$-invariant, and descends to $\A/\H$. This~is an action  `improved' by a WZ functional. We see how it nicely arises from a twisted covariant derivative. 
  \medskip
  
  The improvement of an action by a WZ term can also be seen as a special case of the construction of a basic form via dressing. 
  The general scheme of this procedure on $\A$ is detailed in the next section. As~we will see next 
  in section \ref{Boundaries and dressed presymplectic structure}, the construction of and `improved' pre-symplectic structure via edge modes  comes precisely under such an application of the DFM.

\subsection{$\A$-dependent dressing fields and basic variational forms on $\A$}
\label{A-dependent dressing fields and basic variational forms on A}

We've seen that basic forms on $\P$ or on $\A$ are of special interest in gauge theories. Often, one looks for tricks, or specifics in the situation at hands, to find one. In the following, we complete section \ref{The dressing field method} and consider how \mbox{$\A$-dependent} dressing fields allow to systematically build  basic forms on the $\H$-bundle $\A$.

\subsubsection{The dressing field method, complement}
\label{The dressing field method, complement}

 Let $\Omega^\bullet_{K\text{-basic}}\big(\P, \text{Lie}G\big)$ denote the vector space of $K$-basic Lie$G$-valued forms on $\P$,
 with $G\supset H$ for generality. 
 %
 Consider a $\A$-dependent dressing field $\bs u : \A \rarrow \D r[G, K]$, so s.t. $R^\star_\gamma \bs{u}=\gamma\- \bs{u}$ for $\gamma\in \K$, and define the map 
 \begin{align}
 \label{Dressing-map}
 \text{F}_{\bs{u}} : \A &  \rarrow  \A/\K  \subset \Omega^1_{K\text{-basic}}\big(\P, \text{Lie}G\big)  \notag\\
 			     A & \mapsto \text{F}_{\bs{u}}(A) \defeq A^{\bs u}=\bs{u}\-A\bs{u} + \bs{u}\- d\bs{u},
 \end{align}
  so that on $\A$ by construction $\text{F}_{\bs u}\circ R_\gamma=\text{F}_{\bs u}$  for $\gamma \in \K \subset \H$. Remark that F$_{\bs u}$ formally resembles $\bs{\Psi} \in \Aut_v(\A)$, but since 
  $\bs{u} \notin \bs\H$, $\text{F}_{\bs u} \notin \Aut_v(\A)$.  Yet, since it is clear from the proof in appendix \ref{Proofs of pushforward formulae for variational vector fields} that the formula of the pushforward of $\bs{X} \in \Gamma(T\A)$ by $\bs{\Psi}$ does not depend on the equivariance of $\bs\gamma$,  we have an analogue of  \eqref{Pushforward-X-inf}:
  \begin{alignat}{2}
  \label{Dressed-vector}
   \text{F}_{\bs{u}\star} : T_A\A & \rarrow T_{A^{\bs{u}}} \A/\K   \notag\\
 			         \bs{X}_A & \mapsto  \text{F}_{\bs{u}\star} \bs{X}_A 
					       									    = \bs{u}\-\!\left( \bs{X}(A) + D^A\left\{    \bs{du}\bs{u}_{|A}\-(\bs{X}_A  )  \right\}  \right)  \bs{u} \ \tfrac{\delta}{\delta [A]}.
  \end{alignat}
In particular for $\chi^v  \in \Gamma(V\A)$ generated by $\chi \in$ Lie$\K$ we have $\text{F}_{\bs{u}\star} \, \chi^v_A = \tfrac{d}{d\tau} \big( \text{F} \circ R_{e^{\tau \chi}}\big)(A) \big|_{\tau =0} = \tfrac{d}{d\tau} \text{F}(A)\big|_{\tau=0}=0$. 
\medskip
Denote $[A] \sim A^{\bs u}$ a point in $\A/\K$, we have the pullback application
 \begin{align}
 \label{Dressed-basic-forms}
 \text{F}^\star_{\bs{u}} : \Omega^\bullet \big(\A/\K\big) &  \rarrow   \Omega^\bullet\big(\A \big)  \notag\\*
 			    \b{\bs{\alpha}}_{|[A]} & \mapsto \text{F}_{\bs{u}}^\star  \b{\bs{\alpha}}_{\, |A}  \rdefeq    {\bs{\alpha}^{\bs u}}_{|A}.
 \end{align}
The form $\bs{\alpha}^{\bs u}$ is $\K$-basic on $\A$. Indeed,  we easily prove the trivial $\K$-equivariance $R^\star_\gamma \bs{\alpha}^{\bs u} = R^\star_\gamma \text{F}_{\bs u}^\star \b{\bs{\alpha}} = \text{F}_{\bs u}^\star \b{\bs{\alpha}} = \bs{\alpha}^{\bs u}$, with $\gamma \in \K$, and the $\K$-horizontality as well $\bs{\alpha}^{\bs u}\big( \chi^v \big) = \big(  \text{F}_{\bs u}^\star  \b{\bs{\alpha}} \big) (\chi^v)= \b{\bs{\alpha}}\big(  \text{F}_{\bs{u} \star}\, \chi^v\big) =0$. It follows that it is $\bs \K$-gauge invariant, $(\bs{\alpha}^{\bs u})^{\bs \gamma} = \bs{\alpha}^{\bs u}$ for $\bs{\gamma} \in \bs{\K}$. 

At this stage $\bs{\alpha}^{\bs u}$ is just a convenient notation for the basic form $\text{F}_{\bs u}^\star \b{\bs{\alpha}}$. We do not have a functional expression for it in terms of the dressing field $\bs u$. By virtue of the formal similarity between $\text{F}_{\bs u}$ and $\bs \Psi$, their actions turn out to be similar, so that the  rule of thumb to obtain the basic/dressed version of a given form on $\A$ is to replace $\bs\gamma$ in its $\bs \H$-gauge transformation by $\bs u$.  

To see this, let us first consider 
 the case of the basis of $\K$-basic 1-forms $\bs{d}A^{\bs{u}} \defeq \text{F}_{\bs u}^\star \bs{d}[A]$ where $\bs{d}[A] \in \Omega^1(\A/\K)$. 
%
By definition, using   \eqref{Dressed-vector} we have,
\begin{align}
\label{Dressed-dA}
{\bs{d}A^{\bs{u}}}_{|A} (\bs{X}_A)  \defeq &\ \big( \text{F}_{\bs u}^\star \bs{d}[A]_{|[A]} \big) (\bs{X}_A)= \bs{d}[A]_{|[A]} \big(\text{F}_{\bs u\star} \bs{X}_A \big)=\bs{u}\-\!\left( \bs{X}(A) + D^A\left\{    \bs{du}\bs{u}_{|A}\-(\bs{X}_A  )  \right\}  \right)  \bs{u}, \notag\\
						 = &\ \bs{u}\-\!\left( \bs{d}A_{|A}(\bs{X}_A) + D^A\left\{    \bs{du}\bs{u}_{|A}\-(\bs{X}_A  )  \right\}  \right)  \bs{u} , \notag \\
						  =&  \left[ \bs{u}\-\!\left( \bs{d}A_{|A}+ D^A\left\{    \bs{du}\bs{u}_{|A}\- \right\}  \right) \bs{u} \right](\bs{X}_A) . 
\end{align}
or simply $\bs{d}A^{\bs{u}}= \bs{u}\-\!\left( \bs{d}A+ D\left\{    \bs{du}\bs{u}\- \right\}  \right) \bs{u}$. Comparing with \eqref{GT-dA}, we see our rule of thumb  justified in this case. Separate from the above general argument, it is easy to explicitly check the $\K$-basicity of this expression for $\bs{d}A^{\bs{u}}$. Indeed 
$R^\star_\gamma \ D^A\big\{\bs{du u}_{|A}\- \big\}=  D^{A^\gamma} \big\{ \bs{d}(\gamma\- \bs{u} \bs{u}\- \gamma) \big\} = \gamma\- D^A\big\{\bs{du u}_{|A}\- \big\} \, \gamma$, and since $R^\star_\gamma \bs{d}A = \gamma\- \bs{d}A \gamma$ it is clear that the expression \eqref{Dressed-dA} has trivial $\H$-equivariance. 
Also $\bs{du u}\-_{|A} (\chi^v_A)= \big( \chi^v \bs{u} \big)(A)\,\bs{u}(A)\-= -\chi \bs{u}(A) \bs{u}(A)\-=-\chi$, so that $\left(   \bs{d}A_{|A}+ D^A\left\{ \bs{du u}\-_{|A} \right\} \right) (\chi^v_A)= D^A\chi + D^A(-\chi)=0$, which proves the horizontality of \eqref{Dressed-dA}. Therefore  $\bs{d}A^{\bs{u}}$ is $\K$-basic, and $\bs\K$-gauge invariant, $(\bs{d}A^{\bs{u}})^{\bs \gamma}= \bs{d}A^{\bs{u}}$.
An identical reasoning for Cartan connections, $\b \A$, would give the same result for $\bs d \b A^{\bs u}$, which in the case of reductive geometries would split as
  \begin{align}
  \label{Dressed-dA-Cartan}
\bs{d}A^{\bs u} =  \bs{u}\-   \left( \bs{d}A  + D\left\{ \bs{d}\bs{u} {\bs{u}\- }\right\} \right)    \bs{u}
\quad \text{and} \quad
\bs{d}e^{\bs u} = \begin{cases}\  \bs{u}\-   \left( \bs{d}e  + \big[e, \bs{d}\bs{u} {\bs{u}\- }\big] \right)    \bs{u}, \\
						          \  \bs{u}\-   \left( \bs{d}e   -  \bs{d}\bs{u} {\bs{u}\- e}\big] \right),
				      \end{cases}
 \end{align} 
according to the  action of $H$ on Lie$G$/Lie$H$. Compare to   \eqref{GT-dA-Cartan}. By the same process one finds $\bs d F^{\bs u}$ and $\bs d \b F^{\bs u}$ whose expressions are obtained by replacing $\bs\gamma$ by $\bs u$ in  \eqref{GT-dF}  and \eqref{GT-dF-Cartan} respectively.

To treat the general case, denote a  form $\bs{\alpha} \in \Omega^\bullet(\A)$ 
as $\bs{\alpha}_{|A} = \alpha\big(  \Lambda^\bullet \bs{d}A ; A \big)$ where $\alpha$ is generically a functional on $\Omega^\bullet(\P, \text{Lie}G)$ 
 (e.g. a $\Ad(G)$-invariant polynomial~$P$, such as a trace).
We have seen that the $\bs\H$-gauge transformation of $\bs \alpha$ can be obtained  in a geometric way via \eqref{Pushforward-X-inf} using its $\H$-equivariance and verticality properties, but it can also be seen as resulting algebraically from the functional properties of $\alpha$ and from \eqref{GT-dA}  since we have,
 \begin{align}
 \label{GT-var-form}
  {\bs{\alpha}^{\bs \gamma}}_{|A} \defeq \bs{\Psi}^\star {\bs{\alpha}}_{\, |A} = \alpha\big(  \bs{\Psi}^\star \Lambda^\bullet \bs{d}A; \bs{\Psi}(A) \big)=\alpha \big(  \Lambda^\bullet\bs{d}A^{\bs \gamma} ; A^{\bs \gamma}   \big).
 \end{align}

Now, since both $\A$ and $\A/\K$ can be domains for $\alpha$, corresponding to $\bs{\alpha}$ we can define a form $\b{\bs{\alpha}} \in \Omega^\bullet(\A/\K)$ as $\b{\bs{\alpha}}_{|[A]} = \alpha \big(  \Lambda^\bullet \bs{d}[A] ; [A] \big)$. Then, 
 \begin{align}
 \label{Dressed-variational-form}
  {\bs{\alpha}^{\bs u}}_{|A} \defeq \text{F}^\star_{\bs u} \b{\bs{\alpha}}_{\, |A} = \alpha\big(  \text{F}^\star_{\bs u}\Lambda^\bullet \bs{d}[A]; \text{F}_{\bs u}(A) \big)=\alpha \big(  \Lambda^\bullet\bs{d}A^{\bs u} ; A^{\bs u}   \big)
  \ \  \in \Omega^\bullet_{\K\text{-basic}}(\A),
 \end{align} 
   is  well-defined, and is referred to  as the \emph{dressing}, or the dressed version, of $\bs\alpha$. Comparison with  \eqref{GT-var-form} justifies our rule of thumb in the general case: once one has the $\bs\H$-gauge transformation $\bs{\alpha^\gamma}$ of a form $\bs\alpha$, one obtains its dressing $\bs{\alpha^u}$ simply by replacing $\bs\gamma$ by $\bs u$.  
 \medskip
   
   The first relevant examples for gauge theory are the Lagrangian and action 0-forms on $\A$. We have defined the Lagrangian as a functional $L:\A \rarrow \Omega^n(\P, \mathbb{K})$, with $\mathbb{K}= \RR, \CC$. Given that it actually extends as a functional on $\Omega^1(\P, \text{Lie}G)$, we define the dressed Lagrangian as $L^{\bs u} \defeq \text{F}_{\bs u}^\star L=L \circ \text{F}_{\bs u}$, i.e. $L^{\bs u}(A)= L(A^{\bs u})$. The corresponding action $S:\A \rarrow \RR$, is dressed as  $S^{\bs u}\defeq S \circ \text{F}_{\bs u}$, $S^{\bs u}[A]=S[A^{\bs u}]$.  If $L$ was already $\K$-invariant/basic due to its  functional properties, $L^{\bs u}$ is by virtue of taking as variable the $\K$-invariant dressed field $A^{\bs u}$. 
   
   If on the other hand we have the general $c$-equivariance $R^\star_\gamma L = L + c(\ \, ,\gamma)$, implying the $\bs \H$-gauge transformation $L^{\bs \gamma}= L +  c(\ \, ,\bs{\gamma})$, then the  dressed Lagrangian is $L^{\bs u}=L + c(\ \, ,{\bs u})$. Correspondingly, the action is s.t. $S^{\bs\gamma}= S + {\bf c}(\ \, , \bs{\gamma})$, so dresses up as $S^{\bs u}= S + {\bf c}(\ \, , \bs{u})$.
    The (integrated) anomaly being $\bs{L}_{\chi^v} S(A) = a(\chi, A)$, if $\bs{u} \in \D r[H, H]$ then $\bs{L}_{\chi^v} S^{\bs u}=0$, and the linearised cocycle is s.t. $\bs{L}_{\chi^v} {\bf c}(A, \bs{u}) = -a(\chi, A)$. The dressed action $S^{\bs u}$ is thus  in effect an action improved by a WZ functional. This  viewpoint on the matter is complementary to the one appealing to a flat twisted connection given at the end of section \ref{Classical gauge anomalies}. 
 \medskip
   
 The question of residual transformations of dressed forms $\bs{\alpha}^{\bs u}$ is addressed in the next section, where the bundle geometry of the space of dressed connections $A^{\bs u}$ plays an important role.

\subsubsection{Residual transformations and the bundle structure of the space of dressed connections}
\label{Residual transformations and the bundle structure of the space of dressed connections}

Suppose $\bs u$ satisfies Proposition \ref{Residual1} of section \ref{Residual gauge transformations (first kind)}, 
so that $R^\star_\eta \bs{u} = \eta\- \bs{u}\eta$, with $\eta \in \J$, and its $\bs \J$-gauge transformation is $\bs{u^\eta}=\bs{\eta}\- \bs{u \eta}$, with $\bs{\eta} \in \bs{\J}$. 
As we have seen, in this case $A^{\bs u}\in  \A'$ the space of Ehresmann connections of the $J$-bundle~$\P'$, so that $(A^{\bs u})^{\bs \eta}= \bs{\eta}\- A^{\bs u} \bs{\eta} + \bs{\eta}\- d \bs{\eta}$.
Then, $\A/\K \simeq \A' \subset \A$  is a $\J$-principal subbundle of $\A$ with gauge group $\bs \J$, and
 the residual $\bs \J$-gauge transformation of the dressed form $\bs{\alpha^u}=\alpha\big( \Lambda^\bullet \bs{d}A^{\bs u}; A^{\bs u}\big)$ is obtained  geometrically as usual  from its $\J$-equivariance and  verticality properties.
 In particular, since $R^\star_\eta \bs{d}A^{\bs u}= \eta\- \bs{d}A^{\bs u} \eta$ and $\bs{d}A^{\bs u}(\lambda^v)= D^{A^{\bs u}}\!\lambda$, for $\lambda \in \text{Lie}\J$, we have the analogue of~\eqref{GT-dA},
\begin{align}
\label{J-GT-dressed-dA}
\big( \bs{d}A^{\bs{u}} \big)^{\bs \eta}= \bs{\eta}\-\!\left( \bs{d}A^{\bs u}+ D^{A^{\bs u}}\!\left\{    \bs{d\eta}\bs{\eta}\- \right\}  \right) \bs{\eta}.
\end{align}
Then,  the residual $\bs \J$-gauge transformation of $\bs{\alpha}^{\bs u}$ is also found algebraically from the functional properties of $\alpha$ via 
\begin{align}
\label{ResidualGT-dressed-alpha}
 \big( \bs{\alpha}^{\bs u} \big)^{\bs \eta} = \alpha \left( \Lambda^\bullet \big(\bs{d}A^{\bs u}\big)^{\bs \eta}; (A^{\bs u})^{\bs \eta} \right).
\end{align}
 Anyway, the result is that $(\bs{\alpha^u})^{\bs \eta}$ with $\bs{\eta} \in \bs{\J}$ is formally identical to $\bs{\alpha^\gamma}$ with $\bs{\gamma} \in \bs{\H}$. 
 
Our most elementary examples are Lagrangians and actions. If the $\bs \H$-gauge transformation of a $c$-equivariant Lagrangian is $L^{\bs \gamma} = L + c(\ \, ,\bs{\gamma})$, then the residual $\bs \J$-gauge transformation of its $\bs \K$-invariant dressed counterpart is 
$(L^{\bs u})^{\bs \eta}(A)= L((A^{\bs u})^{\bs \eta})=L(A^{\bs u})+ c(A^{\bs u}, \bs{\eta})$, 
that is $\big(L^{\bs u}\big)^{\bs \eta } = L^{\bs u} + c(\ \, ,\bs{\eta})$.
 Correspondingly, if we have $S^{\bs \gamma} = S + {\bf c}(\ \, ,\bs{\gamma})$ for the action, we have $\big( S^{\bs u}\big)^{\bs \eta} = S^{\bs u} + {\bf c}(\ \, ,\bs{\eta})$ for its dressed counterpart. 
 
 In case $\bs u$ satisfies Proposition \ref{Residual2} then $A^{\bs u} \in \t\A$, the space of twisted connection on the $J$-bundle~$\P'$. The latter is also a $\J$-principal bundle, and we should be able to generalise everything that as been done in section \ref{Bundle geometry of A}  to such a space. This will be addressed in still another follow-up paper, where we will show that it applies immediately to the study of the pre-symplectic structure of conformal gauge gravity - in either its real or twistorial guise. 
 \medskip
 
 As the discussion in section \ref{Residual gauge transformations (first kind)} has prepared us to appreciate, a further dressing operation with a second dressing field $\bs{u}':\A' \rarrow \D r[J,J]$ s.t. $R^\star_\eta \bs{u}'= \eta\- \bs{u}'$ and $R^\star_\gamma \bs{u}'=\bs{u}'$ for $\gamma \in \K$, would give us $\bs{\alpha}^{\bs{uu}'}_{|A}=\big( \bs{\alpha^u} \big)^{\bs{u}'}_{|A}=\alpha\big(\Lambda^\bullet \bs{d}A^{\bs{uu}'} ; A^{\bs{uu}'}  \big) \in \Omega^\bullet_\text{basic}(\A)$. For a $c$-equivariant Lagrangian this is 
 \begin{align*}
  L^{\bs{uu}'}\!(A)\defeq L(A^{\bs{uu}'})=L( (A^{\bs u})^{\bs{u}'} )= L(A^{\bs u}) + c(A^{\bs u}, \bs{u}') = L(A) + c(A, \bs{u}) + c(A^{\bs u}, \bs{u}')
  			       = L(A) + c(A, \bs{uu}').
 \end{align*}
 Similarly for the action,  $S^{\bs{uu}'}\!(A)=S(A) + {\bf c}(A, \bs{uu}')$. 
The same result would have been obtained with a single dressing $\bs{u} \in \D r[G, H]$, so that $\bs{\alpha^u} \in \Omega^\bullet_\text{basic}(\A)$ descends to, or comes from, a form on the moduli space 
$\A/\H$. 
\bigskip

Let us now complete the discussion on the residual transformation of the second kind described in section~\ref{Residual transformations (second kind) : ambiguity in choosing a dressing field}. 
As we have acknowledged there, the ambiguity in choosing a dressing field $\bs u$ manifests as a right action of the group $\t\G\defeq \left\{ \xi :\P \rarrow G\, |\, R^*_k\xi = \xi \right\}$ on the space of dressed connection $\A^{\bs u}$ given by $\A^{\bs u}  \times \t\G \rarrow \A^{\bs u}$, $(A^{\bs u}, \xi) \mapsto R_\xi A^{\bs u}\defeq (A^{\bs u})^\xi = \xi\- A^{\bs u} \xi + \xi\-d \xi$.
 The manifold $\A^{\bs u}$ is fibered by this action, whose orbits are the fibers, so that it is $\t\G$-principal bundle over $\A^{\bs u}/\t\G$.  
As such it gives rise to a SES,
 \begin{align}
 \label{SESgroups-dressed}
\makebox[\displaywidth]{
\hspace{-18mm}\begin{tikzcd}[column sep=large, ampersand replacement=\&]
\&0     \arrow [r]         \& \bs{\Aut}_v(\A^{\bs u}) \simeq \bs{\t\G}     \arrow[r, "\iota"  ]          \& \bs{\Aut}(\A^{\bs u})       \arrow[r, "\t\pi"]      \&  \bs{\Diff}(\A^{\bs u}/\t\G)        \arrow[r]      \& 0,
\end{tikzcd}}  \raisetag{3.4ex}
\end{align}
 where $\bs{\Aut}(\A^{\bs u})$ is the automorphism group defined as usual, and the subgroup of vertical automorphisms is isomorphic to the gauge group $\bs{\t\G}\defeq \left\{ \bs{\xi}: \A^{\bs u} \rarrow \t\G\, | \ R^\star_\xi\bs{\xi}=\xi\-\bs{\xi} \xi \right\}$. Its infinitesimal version is,
\begin{align}
\label{SESLieAlg-dressed}
\makebox[\displaywidth]{
\hspace{-18mm}\begin{tikzcd}[column sep=large, ampersand replacement=\&]
\&0     \arrow [r]         \& \Gamma_{\!\t\G}(V\A^{\bs u}) \simeq \text{Lie}\bs{\t\G}     \arrow[r, "\iota"  ]          \&  \Gamma_{\!\t\G}(T\A^{\bs u})     \arrow[r, "\pi_\star"]      \&  \Gamma(T\A^{\bs u}/\t\G)         \arrow[r]      \& 0,
\end{tikzcd}}  \raisetag{3.4ex}
\end{align}
with $\t\G$-invariant vector fields defined as usual.
Since there is a 1:1 correspondence between $\K$-orbits $\O_\K[A]$ and $\t\G$-orbits $\O_{\!\t\G}[A^{\bs u}]$, we have that $\A^{\bs u}/\t\G \simeq \A/\K$. 
In particular, in case $\bs{u}:\A \rarrow \D r[G, H]$, we have $\A^{\bs u}/\t\G \simeq \A/\H$, which means that 
 $\bs{\Diff}(\A^{\bs u}/\t\G) \simeq \bs{\Diff}(\A/\H)$ is the physical transformation group with infinitesimal counterpart $\Gamma(\A^{\bs u}/\t\G) \simeq \Gamma(\A/\H)$. 
 This remark is relevant to the interpretation of what is achieved by the introduction of edge modes in gauge theory, as discussed in section \ref{Boundaries and dressed presymplectic structure}.

As usual, the action by pullback of $\bs{\Aut}_v(\A^{\bs u})$ on elements of $\Omega^\bullet(\A^{\bs u})$ defines their $\bs{\t\G}$-gauge transformations which is thus determined by their $\t\G$-equivariance and verticality properties. In particular, for $\bs{d}A^{\bs u} \in \Omega^1(\A^{\bs u})$ we have  $R^\star_\xi \bs{d}A^{\bs u} =\xi\-   \bs{d}A^{\bs u} \xi$, and  $\bs{d}A^{\bs u}_{|A^{\bs u}}\big( \zeta^v_{A^{\bs u}} \big)=D^{A^{\bs u}}\!\zeta$, where $\zeta^v \in \Gamma(V\A^{\bs u})$ and $\zeta \in \text{Lie}\t\G$,%
\footnote{These relations can be taken as axiomatic, or be motivated by the fact that since by assumption $A^\xi=A$ then we can formally admit $\bs{d}A(\zeta^v)=0$, and since by definition $\bs{u}^\xi=\bs{u}\xi$, infinitesimally we can formally admit $\bs{du}(\zeta^v)=\zeta^v(\bs{u})=\bs{u}\zeta$. Using then the definition \eqref{Dressed-dA} of $\bs{d}A^{\bs u}$, its $\t\G$-equivariance is clear, while  $\bs{d}A^{\bs u}_{|A^{\bs u}}\big( \zeta^v_{A^{\bs u}} \big)=\bs{u}^{-1} D^A\left\{\bs{u} \zeta \bs{u}^{-1} \right\} \bs{u}=D^{A^{\bs u}}\!\zeta$.} 
 in analogy with \eqref{J-GT-dressed-dA} we get,
\begin{align}
\label{Dressed-dA-residual-2nd-kind}
\big(\bs{d}A^{\bs u}\big)^{\bs \xi} = \bs{\xi}\- \left( \bs{d}A^{\bs u} + D^{A^{\bs u}}\!\left\{\bs{d\xi\xi}\- \right\}  \right) \bs{\xi}.
\end{align}
By this  geometric method we obtain the $\bs{\t\G}$-gauge transformation of a dressed form $\bs{\alpha^u}$, that we can also find via
\begin{align}
\label{Dressed-alpha-residual-2nd-kind}
 \big( \bs{\alpha}^{\bs u} \big)^{\bs \xi} = \alpha \left( \Lambda^\bullet \big(\bs{d}A^{\bs u}\big)^{\bs \xi}; (A^{\bs u})^{\bs \xi} \right).
\end{align}
Again, the result is that $(\bs{\alpha^u})^{\bs \xi}$  is formally identical to $\bs{\alpha^\gamma}$, with $\bs{\gamma} \in \bs{\H}$, due to the functional properties of $\alpha$. 

Taking again the elementary example of a $c$-equivariant Lagrangian whose $\bs \H$-gauge transformation is thus $L^{\bs \gamma}=L+c(\ \, ,\bs{\gamma})$ and dressed as $L^{\bs u}= L + c(\ \, ,\bs{u})$, we have $(L^{\bs u})^{\bs \xi} (A)=L^{\bs{u\xi}}(A)=L(A^{\bs{u\xi}}) =L(A)+ c(A ,\bs{u\xi})=L(A)+c(A, \bs{u})+c(A^{\bs u}, \bs{\xi})=L(A^{\bs{u}}) + c(A^{\bs u} ,\bs{\xi})$, that is $(L^{\bs u})^{\bs \xi} = L^{\bs u}+ c(\ \, ,{\bs \xi})$. Correspondingly, for the action we have $(S^{\bs u})^{\bs \xi} = S^{\bs u}+ {\bf c}(\ \, ,{\bs \xi})$. We will encounter more elaborate examples in section
 \ref{Boundaries and dressed presymplectic structure}.

We should remind again that since the dressing field $\bs u$ is $\A$-dependent, it may be that $\t\G$ is reduced to a small group (perhaps even a discrete one) or that it represents another interesting gauge symmetry. In any event, it is not a permutation group of physical states.
\bigskip

To give substance to this formal framework, one should mention that there are several interesting examples of field-dependent dressing fields. In gauge formulations of gravity, the tetrad field is actually a $\B$-dependent Lorentz dressing field, with $\B$ the space of Cartan connections. 
We have
$\bs u : \B \rarrow \D r(G, H)$, $\b A \mapsto \bs u(\b A)$,  for $G=GL(4)$ and  $H=S\!O(1,3)$  \cite{GaugeInvCompFields, Francois2014, Attard_et_al2017}. 
This fact will be exploited  in section \ref{Boundaries and dressed presymplectic structure}, subsection \ref{4D gravity} in particular, see also \ref{G}.

Considering the conformal Cartan geometry $(\P, \b A)$ and its associated $\RR$- and $\CC$-vector bundles, $E$ and $\sf E$, one can build a $\B$-dependent dressing field  $\bs u : \B \rarrow \D r(K, K)$, $\b A \mapsto \bs u(\b A)$, for $K$ the (abelian) subgroup of special conformal transformations. Using it to dress the conformal Cartan connection $\b A$ itself one gets $\b A^{\bs u}$, which turns out to be a \emph{twisted} (actually \emph{mixed}) connection as described in  section \ref{Twisted connections}. It is otherwise known as the conformal tractor connection, or the twistor connection in its spin representation. 
Conformal tractors and twistors are obtained by dressing sections of $E$ and $\sf E$ respectively. This was alluded to at the end of section \ref{Residual gauge transformations (first kind)} and described in details in \cite{Attard-Francois2016_I} and \cite{Attard-Francois2016_II}, see also \cite{Attard_et_al2017}.

Finally, another example is provided by the electroweak model, where one can build a SU(2)-dressing field from the $\CC^2$-scalar (Higgs) field \cite{Attard_et_al2017, Francois2018}. To formalise this case, one would need to extend the above analysis from  $\A$ to $\Gamma(E)$, or even $\A \times \Gamma(E)$, with $E$ an associated bundle to a principal bundle $\P$. The dressing field of the electroweak model is then $\bs u : \Gamma(E) \rarrow \D r(K, K)$, $\phi \mapsto \bs u(\phi)$,  for $K=S\!U(2)$. We further comment on this in the conclusion. 

Notice that all the field-dependent dressing fields mentioned above are local, which is interesting to bear in mind when comes
  the discussion of section \ref{Dressed presymplectic structure for invariant Lagrangians}, just before \ref{Yang-Mills theory}. 

\bigskip

This whole section \ref{The space of connections as a principal bundle}  has laid the ground for our main purpose, 
 that is to show that by paying attention to the bundle structure of $\A$ one obtains quite general results about the pre-symplectic structure of gauge-invariant theories, as well as of a reasonably broad class of non-invariant theories. We tackle this in the next section, 
 in which we also assess how the DFM fares when confronted to the problem of boundaries in gauge theory. In this regard, what we just did in this last subsection \ref{A-dependent dressing fields and basic variational forms on A} will prove illuminating. 

\clearpage

\section{Covariant Hamiltonian formalism, and the boundary problem}
\label{Covariant Hamiltonian formalism (and the puzzle of boundaries)}

\medskip

A gauge theory is specified by a choice of Lagrangian $L: \A \rarrow \Omega^n(\P, \mathbb{K})$\footnote{In this paper we focus on pure gauge theories, i.e. without coupling to matter fields, so we present a minimal version of the formalism. Of course it extends to coupled gauge theories, see e.g. \cite{Compere-Fiorucci2018, Harlow-Wu2020} for  modern complete presentations (\cite{Farajollahi-Luckock2002} gives a lucid short summary). As discussed in our conclusion, it is one of the aims of follow-up papers to extend the results presented in this section to coupled theories.} with associated action $S=\int_U L$.\footnote{Since we took the viewpoint that $L$ is a n-form on $\P$, we  have actually  $S=\int_{\s(U)} L=\int_U \s^*L$ with $\s:U\rarrow \P$ a local section and $\s^*L$ written in terms of the gauge potential $\s^*A$. For convenience, and as it is suggestive enough, we shall omit $\s$ in writing integration domains.}
For $S$ to be finite, one usually assumes that $U$ is compact or closed, or that  the fields are either compactly supported or satisfy sufficiently fast fall-off conditions at infinity (which amounts to an effective compactification of $\M$). 

The variational principle stipulates that the field equations are found from requiring $S$ to be stationary, that is $\delta S=0$ $\forall \delta A$, under well-defined boundary conditions. In the formulation adopted here this translates as $\bs{d}S(\bs{X})=0$ $\forall \bs{X} \in \Gamma(T\A)$ , i.e the functional $S$ is closed, $\bs{d}S=0$. Admitting that $\bs d$ and $\int$ commute, this gives 
\begin{align}
\label{variational_principle}
\bs{d}S=\int_U \bs{d}L= \int_U \bs{E} + d\bs{\theta}=\int_U \bs{E}\ + \int_{\d U} \bs{\theta} =0
\end{align}
where $\bs{E}_{|A}=E(\bs{d}A; A)$ is the field equations 1-form and $\bs{\theta}_{|A}=\theta(\bs{d}A; A)$ is the presymplectic potential current \mbox{1-form}. Here $E$ and $\theta$ are different functionals of $A$,  both linear in $\bs{d}A$ and based on the same functional as $L$. 
In field theory, the latter is often simply an $\Ad(H)$-invariant symmetric multilinear map $P:  \otimes ^k \text{Lie}H \rarrow \mathbb{K}$, $(X_1, \ldots, X_k) \mapsto P(X_1, \ldots, X_k)$. Its invariance $P(\Ad_h \, A_1, \ldots, \Ad_h \, A_k)= P(A_1,\ldots, A_K)$ for $h\in H$ and $A_i \in \Omega^{p_i}(\P, \text{Lie}H)$, gives the useful identity 
\begin{align}
\label{usefull-identity}
\Sigma_i(-)^{p(p_1+\ldots+p_i)} P(A_1, \ldots, [A_i, \eta], \ldots, A_k)=0,
\end{align}
with $\eta \in \Omega^{p}(\P, \text{Lie}H)$. Most often $P=\Tr$, in gravity $P$ is derived from the Pfaffian (the square root of the determinant), as we will see.
\medskip

The point of the covariant Hamiltonian approach - whose inception is due to  \cite{Zuckerman1986,CrnkovicWitten1986, Crnkovic1987} (see also \cite{Lee-Wald1990, Ashtekar-et-al1990}) 
is to associate a phase space equipped with a symplectic structure 
to a field theory over a region $U \subseteq \M$, and doing so while keeping all spacetimes symmetries manifest.  The configuration space is the field space, here the $\H$-bundle $\A$. 
The covariant phase space is the solution space $\S$ - the \emph{shell} - defined by $\boldsymbol E=0$. 
The physical, or reduced, phase space is  $\S/\H$, or rather it is if it can be endowed with a well-defined symplectic 2-form. Notice then that the Hamiltonian flow belongs to the physical transformation group $\bs\Diff(\A/\H)$ in the SES \eqref{SESgroups-inf}, and the corresponding Hamiltonian vector field thus belongs to $\Gamma(T\A/\H)$ in the SES \eqref{SESLieAlg-inf}. 

The preseymplectic potential $\bs\theta$ allows to define the Noether currents and charges associated to the action of $\H$, and a 
 natural candidate symplectic form is derived  from it. Since $[\bs d, d]=0$,  we have $0\equiv \bs{d}^2 L = \bs{dE}+d(\bs{d\theta})$. So, the 2-form $\bs \Theta \defeq \bs{d\theta}$ is $d$-closed on-shell, $d \bs\Theta =0_{\, |\S}$.
 Given  a $(n-1)$-dimensional Cauchy surface $\Sigma \subset U$,  we have  $\bs\Theta_\Sigma \defeq \int_\Sigma \bs\Theta \in \Omega^2(\A, \mathbb{K})$.\footnote{A more conventional choice of notations for the 2-forms $\bs\Theta$ and $\bs\Theta_\Sigma$ is  $\omega$ and  $\Omega$, respectively.  } 
 We can also define the presymplectic potential $\bs\theta_\Sigma\defeq \int_\Sigma \bs\theta$, so that $\bs{\Theta}_\Sigma=\bs{d} \bs{\theta}_\Sigma$. Since $\bs{d} \bs{\Theta}_\Sigma=0$, $\bs{\Theta}_\Sigma$ is a presymplectic 2-form (hence the name given to $\bs\theta$ and $\bs{\theta}_\Sigma$). It allows to define a Poisson bracket between charges.
 
 Now, in order for $\bs\theta_\Sigma$ and $\bs\Theta_\Sigma$ to induce a symplectic structure on  $\S/\H$, it must be on the one hand that on-shell the $\H$-equivariance and verticality properties of $\bs \theta$ and $\bs \Theta$ are right, and on the other hand that adequate boundary conditions are specified, so that in the end $\bs\theta_\Sigma$ and $\bs\Theta_\Sigma$ are basic on $\A$. This last requirement is jeopardised when considering bounded regions, or when one considers the mereological problem of  partitioning  a region into subregions sharing a fictitious boundary. 
 
 We say more about this problem, and consider one of its proposed solution in light of the DFM, in section
   \ref{Boundaries and dressed presymplectic structure}. But first, we study the presymplectic structure on $\A$ for invariant theories in section \ref{Presymplectic structure for invariant Lagrangians}, and for $c$-equivariant theories in section \ref{Presymplectic structure for non-invariant Lagrangians}. There we are interested in identifying the Noether currents and charges, the Poisson bracket of charges, and finally the $\bs\H$-gauge transformations of $\bs \theta_\Sigma$ and $\bs \Theta_\Sigma$ for which will  need their respective $\H$-equivariance and verticality properties. In the spirit of the Hamiltonian formalism, whenever possible we will give these results as functions of the field equations $\bs E$ so that on-shell restriction is read-off immediately. 
 
\vspace{2cm}

\subsection{Presymplectic structure for invariant Lagrangians}
\label{Presymplectic structure for invariant Lagrangians}

We start here from an invariant Lagrangian $L: \A \rarrow \Omega^n(\P, \mathbb{K})$, which is  a basic 0-form on $\A$: its $\H$-equivariance is trivial $R^\star_\gamma L=L$ for  $\gamma \in \H$, infinitesimally this is $\bs L_{\chi^v} L = 0$ for $\chi^v \in \Gamma(V\A)$ and $\chi \in \text{Lie}\H$. Since the de Rham derivative preserves basicity, $\bs d : \Omega^\bullet_\text{basic}(\A) \rarrow \Omega^{\bullet +1}_\text{basic}(\A)$, we have that $\bs d L=\bs E + d \bs \theta$ is  a basic 1-form. Both its equivariance and  horizontality  give useful informations. 


First, from its equivariance $R^\star_\gamma \bs d L=\bs d L$ we obtain a priori $R^\star_\gamma \bs E= \bs E + t(\gamma)$ and $R^\star_\gamma d \bs \theta= d \bs \theta - t(\gamma)$. But on the basis the geometric Utiyama theorem (see \cite{Utiyama1956, Kolar-Michor-Slovak} also \cite{Castrillon-Lopez-et-al_2019}) the stronger result $R^\star_\gamma \bs E= \bs E$ and $R^\star_\gamma \bs \theta =\bs \theta$ can be argued. 
Indeed, the theorem states that a $\H$-gauge invariant Lagrangian factors as $L=\t L \circ F$, where $F: \A \rarrow \Omega^2_\text{tens}(\P, \text{Lie}H)$, $A \mapsto F(A)$, is the curvature map and $\t L : \Omega^2_\text{tens}(\P, \text{Lie}H) \rarrow \Omega^n_\text{inv}(\P, \mathbb{K})$ is an $\Ad(H)$-invariant functional.\footnote{In the case of gravity, an invariant $L:\b \A \rarrow \Omega^n(\P, \mathbb{K})$ factors in a similar way so as to depend only on $\b F$ and the soldering form $e$ \cite{Bruzzo1987}.} 
As mentioned above $\t L$ is virtually always an invariant multilinear map, so we have $\bs d L_{|A}= \bs d \t L \circ \bs d F_{|A}= \t L \circ D^A(\bs d A)_{|A} = \t L\big(D^A(\bs d A); [F]\big)$ which is linear in the first argument and $[F]$ denote $F$-dependant terms. Using  $\Ad$-invariance and Leibniz identity, we can thus write $\bs d L_{|A}= d \t L(\bs d A; [F]) + \t L(\bs d A ; D^A[F]) \rdefeq d \theta(\bs d A; A) + E(\bs d A ; A)$, where now the  $\H$-invariance of both $\bs E$ and $\bs\theta$ is clear.  If the  argument sketched here fails to convince, just admit the result as a working hypothesis (to be checked when specific examples are considered) and have $\bs E = E(\bs d A; A) = \t E (\bs d A ; D^A[F])$ and $\bs \theta = \theta(\bs d A; A) = \t \theta (\bs d A ; [F])$ as a notational book keeping of their respective invariance.

We are interested first in the Noether currents and associated charges, then in the Poisson bracket of charges, and finally in the field-dependent gauge transformations of the presymplectic potential and 2-form. 
\medskip

\noindent {\bf Noether current and charge: } 
Given  $\chi^v \in \Gamma(V\A)$, $\chi \in \text{Lie}\H$, the quantity $J(\chi; A)\defeq \iota_{\chi^v} \bs \theta$ is the Noether current associated to $\H$-gauge transformations. 
By using an argument similar as above we have that,
\begin{align}
\label{Noether-current-invariant}
J(\chi;A)\defeq \bs\theta_{|A}(\chi^v_A)= \theta\big(D^A\chi; A\big)=\t\theta\big(D^A\chi; [F]\big)= d\t\theta(\chi; [F]) - \t E\big(\chi; D^A[F]\big)= d\theta(\chi; A) - E(\chi; A). 
\end{align}
Clearly, on-shell $J(\chi;A)$ is $d$-exact, thus $d$-closed. This is crossed-checked by another argument: From the horizontality of $\bs d L$, $\iota_{\chi^v}\bs dL=0$, we get $\iota_{\chi^v}\bs E = - d(\iota_{\chi^v} \bs \theta)=-dJ( \chi; A)$, from which follows that the Noether current is $d$-closed on-shell, $dJ(\chi; A)=0_{\, |\S}$. %
The  Noether charge is $Q_\Sigma( \chi;A)\defeq \int_\Sigma J( \chi;A)$, also written  $Q_\Sigma(\chi;A)=\iota_{\chi^v}\bs \theta_\Sigma$, so 
\begin{align}
\label{Noether-charge-invariant}
Q_\Sigma(\chi;A)= \int_{\d\Sigma} \theta(\chi; A) - \int_\Sigma E(\chi; A). 
\end{align}
On-shell, the Noether charge reduces to a boundary charge $Q_\Sigma(\chi;A)= \int_{\d\Sigma} \theta(\chi; A)_{\, |\S}$. 
\medskip

\noindent {\bf Poisson bracket: } 
We now turn to the Poisson bracket of charges defined by  the presymplectic 2-form.
 The trivial $\H$-equivariance of $\bs \theta$ is infinitesimally $\bs L_{\chi^v} \bs\theta= \iota_{\chi^v} \bs{d \theta} + \bs d \iota_{\chi^v} \bs \theta \rdefeq  \iota_{\chi^v} \bs \Theta + \bs d \iota_{\chi^v} \bs \theta=0$. This gives us a relation between the Noether charge and the presymplectic 2-form,
\begin{align}
\label{presympl-2form-Noether-charge}
\iota_{\chi^v} \bs \Theta = - \bs d J(\chi; A), \quad \text{ so that } \quad \iota_{\chi^v} \bs \Theta_\Sigma = - \bs d Q_\Sigma(\chi; A) 
																			= - \int_{\d\Sigma} \bs d \theta(\chi; A) + \int_\Sigma \bs d E(\chi; A). 
\end{align}
From this, and using $[\bs L_{\bs Y}, \iota_{\bs X}]= \iota_{[\bs Y, \bs X]}$ as well as $\bs L_{\chi^v} \bs \theta=0$ (since $R^\star_\gamma \bs \theta=\bs \theta$), we further obtain that  for $\chi^v, \eta^v \in \Gamma(V\A)$
\begin{align}
\label{verticality-Theta}
\bs\Theta(\chi^v, \eta^v)=\iota_{\eta^v} \big(\iota_{\chi^v} \bs \Theta \big) = - \iota_{\eta^v} \bs d \iota_{\chi^v}\bs \theta = -\bs L_{\eta^v} \iota_{\chi^v} \bs \theta= -\iota_{\chi^v} \bs L_{\eta^v}\bs\theta - \iota_{[\eta^v, \chi^v]} \bs \theta 
					 = \iota_{[\chi, \eta]^v} \bs \theta . 
\end{align}
where in the last step we use the fact that the map Lie$\H \rarrow \Gamma(V\A)$ is a isomorphism (section \ref{Bundle geometry of A} and appendix \ref{Lie algebra (anti)-isomorphisms}). This shows that the antisymmetric Poisson bracket of charges defined by the presymplectic 2-form is~s.t.
\begin{align}
\label{Poisson-bracket}
\big\{ Q_\Sigma(\chi; A) ,  Q_\Sigma(\eta; A)\big\}\defeq\, \bs\Theta_\Sigma(\chi^v, \eta^v) = \int_\Sigma \iota_{[\chi, \eta]^v} \bs \theta= \int_\Sigma J([\chi. \eta]; A) 
										    = Q_\Sigma([\chi, \eta]; A). 
\end{align}
That is, the map Lie$\H \rarrow \big( Q_\Sigma(\ ;A), \big\{\, ,\, \big\})$ is a Lie algebra morphism. The Jacobi identity is satisfied for the Poisson bracket because it holds in Lie$\H$. 
Written in a functional way it reproduces the Peierls-DeWitt bracket (see \cite{Forger-VieraRomero2005} Theorem 4, also \cite{Harlow-Wu2020}).
Noether charges are also called generators of $\H$-gauge transformations since we have indeed that the action of $\big\{Q_\Sigma(\chi; A),  \ \ \big\}$ on  functionals on $\A$ generates the infinitesimal action by $\chi^v$, see appendix \ref{Noether charges as generators of gauge transformations}. 
\bigskip

\noindent {\bf Field-dependent gauge transformations: } 
We now seek to obtain the $\bs \H$-gauge transformations of both $\bs \theta$ and $\bs\Theta$, so as to see under which circumstances they induce a symplectic structure on $\S/\H$. As we have now often stressed, all that is needed is their $\H$-equivariance and verticality properties. Considering $\bs \theta$ first, we have already argued that under the hypothesis that $R^\star_\gamma L=L$, we have $R^\star_\gamma \bs \theta=\bs \theta$. Furthermore, its verticality property is none other than the definition of the Noether current \eqref{Noether-current-invariant}. So we have all we need: given $\bs\gamma \in \bs\H$ corresponding to $\bs\Psi \in \bs\Aut_v(\A)$ we get,
\begin{align}
\label{Field-depGT-presymp-pot-current}
{\bs\theta^{\bs\gamma}}_{|A}(\bs X_A) &\defeq \big( \bs\Psi^\star \bs \theta \big)_{|A}(\bs X_A) = \bs \theta_{A^{\bs\gamma}}\big(\bs\Psi_\star \bs X_A\big)
					=   \bs \theta_{|A^{\bs\gamma}} \left(R_{\bs{\gamma}(A)\star} \left( \bs{X}_A + \left\{ \bs{d}\bs{\gamma} {\bs{\gamma}\- }_{|A}(\bs{X}_A)\right\}^v_A \right) \right), \quad \text{ using \eqref{Pushforward-X-inf}} \notag \\[1mm]
					&= R^\star_{\bs{\gamma}(A)} \bs \theta_{|A^{\bs\gamma}}  \left( \bs{X}_A + \left\{ \bs{d}\bs{\gamma} {\bs{\gamma}\- }_{|A}(\bs{X}_A)\right\}^v_A \right)
					= \bs \theta_{|A} \left( \bs{X}_A + \left\{ \bs{d}\bs{\gamma} {\bs{\gamma}\- }_{|A}(\bs{X}_A)\right\}^v_A \right),  \notag \\[1mm]
					&=  \bs \theta_{|A} \big( \bs{X}_A \big) + d \theta\big( \{ \bs{d\gamma\gamma}_{|A}\-(\bs X_A) \}; A \big) - E \big( \{ \bs{d\gamma\gamma}_{|A}\-(\bs X_A) \}; A \big), \quad \text{ using \eqref{Noether-current-invariant}} \notag \\[1mm]
		\text{that is \quad} \bs\theta^{\bs \gamma}&= \bs \theta + d\theta \big(  \bs{d\gamma\gamma}\- ;  A \big) - E \big( \bs{d\gamma\gamma}\- ; A \big). 
\end{align}
We have  on-shell $\bs\theta^{\bs\gamma}=\bs\theta + d\theta \big( \bs{d\gamma\gamma}\-; A \big)_{\, |\S}$. This implies for the presymplectic potential 
\begin{align}
\label{Field-depGT-presymp-pot}
\bs\theta_\Sigma^{\bs \gamma} = \bs \theta_\Sigma + Q_\Sigma(\bs{d\gamma\gamma}\-; A) =  \bs \theta_\Sigma + \int_{\d\Sigma} \theta( \bs{d\gamma\gamma}\- ; A) - \int_\Sigma E(\bs{d\gamma\gamma}\-; A),
\end{align}
so that on-shell $\bs\theta_\Sigma^{\bs\gamma}=\bs\theta_\Sigma + \int_{\d\Sigma} \theta \big( \bs{d\gamma\gamma}\-; A \big)_{\, |\S}$. It is clear from \eqref{Field-depGT-presymp-pot}, but could already have been inferred from the trivial $\H$-equivariance of $\theta$ and \eqref{Noether-current-invariant}, that the presymplectic potential is basic, $\bs\H$-invariant, if we are on-shell and if either $\d\Sigma=\emptyset$ or $A$ and/or the gauge parameter $\chi$/$\gamma$
 are required to vanish at $\d\Sigma$ or at infinity.\footnote{Since $\bs\gamma=\bs\gamma(A)$, this requirement on $A$ implies the same constraint for $\bs\gamma$.}
 
 A similar computation holds for $\bs E$. Using $R^\star_\gamma \bs E=\bs E$ and $\iota_{\chi^v}\bs E= -d (\iota_{\chi^v} \bs \theta)$ as well as \eqref{Noether-current-invariant}, we get 
 \begin{align}
\label{Field-depGT-FieldEq}
{\bs E^{\bs\gamma}}_{|A}(\bs X_A) &\defeq \big( \bs\Psi^\star \bs E \big)_{|A}(\bs X_A) = \bs E_{A^{\bs\gamma}}\big(\bs\Psi_\star \bs X_A\big)
					=   \bs E_{|A^{\bs\gamma}} \left(R_{\bs{\gamma}(A)\star} \left( \bs{X}_A + \left\{ \bs{d}\bs{\gamma} {\bs{\gamma}\- }_{|A}(\bs{X}_A)\right\}^v_A \right) \right), \quad \text{ using \eqref{Pushforward-X-inf}} \notag \\[1mm]
					&= R^\star_{\bs{\gamma}(A)} \bs E_{|A^{\bs\gamma}}  \left( \bs{X}_A + \left\{ \bs{d}\bs{\gamma} {\bs{\gamma}\- }_{|A}(\bs{X}_A)\right\}^v_A \right)
					= \bs E_{|A} \left( \bs{X}_A + \left\{ \bs{d}\bs{\gamma} {\bs{\gamma}\- }_{|A}(\bs{X}_A)\right\}^v_A \right),  \notag \\[1mm]
					&=  \bs E_{|A} \big( \bs{X}_A \big) + d E \big( \{ \bs{d\gamma\gamma}_{|A}\-(\bs X_A) \}; A \big),  \notag \\[1mm]
		\text{that is \quad} \bs E^{\bs \gamma}&= \bs E + dE \big(  \bs{d\gamma\gamma}\- ;  A \big).
\end{align}
Fortunately, this shows that the action of $\bs\H$ does not take us off-shell. 

Turning our attention to the $\bs\H$-gauge transformation of $\bs\Theta$, since we already have its verticality given by \eqref{verticality-Theta}, we need only its $\H$-equivariance to proceed. It is simply $R^\star_\gamma \bs\Theta= R^\star_\gamma \bs{d\theta}=\bs{d}R^\star_\gamma\bs\theta=\bs{d\theta}=\bs\Theta$, since pullbacks commute with the exterior derivative. So, 
\begin{align}
\label{1}
{\bs\Theta^{\bs\gamma}}_{|A}\big(\bs X_A, \bs Y_A \big)\defeq&\, \big(  \bs\Psi^\star \bs\Theta \big)_{|A}(\bs X_A, \bs Y_A) = \bs\Theta_{|A^{\bs\gamma}} \left(\bs\Psi_\star \bs X_A, \bs\Psi_\star\bs Y_A  \right), \notag\\
			=&\, \bs\Theta_{|A^{\bs\gamma}} \left( R_{\bs{\gamma}(A)\star} \left( \bs{X}_A + \left\{ \bs{d}\bs{\gamma} {\bs{\gamma}\- }_{|A}(\bs{X}_A)\right\}^v_A   \right) ,  R_{\bs{\gamma}(A)\star} \left( \bs Y_A + \left\{ \bs{d}\bs{\gamma} {\bs{\gamma}\- }_{|A}(\bs Y_A)\right\}^v_A   \right)  \right), \notag \\[1mm]
			=&\, R^\star_{\bs\gamma(A)} \bs \Theta_{|A^{\bs\gamma}} \left( \bs{X}_A + \left\{ \bs{d}\bs{\gamma} {\bs{\gamma}\- }_{|A}(\bs{X}_A)\right\}^v_A,     \bs{Y}_A + \left\{ \bs{d}\bs{\gamma} {\bs{\gamma}\- }_{|A}(\bs{Y}_A)\right\}^v_A \right) ,\notag\\[1mm]
			=&\, \bs\Theta_{|A} \big(\bs X_A, \bs Y_A \big) +  \bs\Theta_{|A} \left(  \left\{ \bs{d}\bs{\gamma} {\bs{\gamma}\- }_{|A}(\bs{X}_A)\right\}^v_A, \bs Y_A  \right) + \bs\Theta_{|A}\left( \bs X_A,  \left\{ \bs{d}\bs{\gamma} {\bs{\gamma}\- }_{|A}(\bs{Y}_A)\right\}^v_A\right)  \notag\\
			&\hspace{5cm} +  \bs\Theta_{|A} \left(  \left\{ \bs{d}\bs{\gamma} {\bs{\gamma}\- }_{|A}(\bs{X}_A)\right\}^v_A,  \left\{ \bs{d}\bs{\gamma} {\bs{\gamma}\- }_{|A}(\bs{Y}_A)\right\}^v_A \right), \notag\\[1mm]
			=&\,  \bs\Theta_{|A} \big(\bs X_A, \bs Y_A \big)  - \iota_{\bs Y} \bs d\, \theta\left( D^A\big\{ \bs{d\gamma\gamma}\-_{|A}(\bs X_A) \big\}; A \right)  +  \iota_{\bs X} \bs d\, \theta\left( D^A\big\{ \bs{d\gamma\gamma}\-_{|A}(\bs Y_A) \big\} ; A \right) \notag\\ 
			&\hspace{5cm}  +  \bs\theta_{|A} \left(\big[\bs{d\gamma\gamma}\-_{|A}(\bs X_A), \bs{d\gamma\gamma}\-_{|A}(\bs Y_A)\big]^v_A\right), \notag \\[1mm]
			=&\, \bs\Theta_{|A} \big(\bs X_A, \bs Y_A \big) - \bs Y \cdot \theta\left( D^A\big\{ \bs{d\gamma\gamma}\-_{|A}(\bs X_A) \big\}; A \right)  + \bs X \cdot \theta\left( D^A\big\{ \bs{d\gamma\gamma}\-_{|A}(\bs Y_A) \big\}; A \right) \notag \\
			&\hspace{5cm} +  \theta\left( D^A\big\{ [\bs{d\gamma\gamma}\-_{|A}(\bs X_A), \bs{d\gamma\gamma}\-_{|A}(\bs Y_A)\big] \big\}; A \right). 
\end{align}
by \eqref{presympl-2form-Noether-charge}  and \eqref{verticality-Theta} in step before last.
Considering the second and third terms of the above expression,  only  the underlined $A$'s in $\theta\left( D^{\munderline{blue}{A}}\big\{ \bs{d\gamma\gamma}\-_{|A}(\bs Z_A) \big\}; \munderline{blue}{A} \right)$ are acted upon. 
 Observe that the quantity $\bs d\, \theta\left( D^A\big\{ \bs{d\gamma\gamma}\-_{|A} \big\}; A \right)$ is a 2-form on $\A$ so that, by the Kozsul formula, evaluated on two vectors it gives
 
 \begin{align}
 \label{2}
 \bs d\, \theta\left( D^A\big\{ \bs{d\gamma\gamma}\-_{|A} \big\}; A \right) \big( \bs X_A, \bs Y_A \big)&= \bs X \cdot \theta\left( D^A\big\{ \bs{d\gamma\gamma}\-_{|A}(\bs Y_A) \big\}; A \right)
 																			- \bs Y \cdot \theta\left( D^A\big\{ \bs{d\gamma\gamma}\-_{|A}(\bs X_A) \big\}; A \right) \notag\\*
															  & \hspace{5cm} - \theta\left(    D^A\big\{ \bs{d\gamma\gamma}\-_{|A}([\bs X, \bs Y]_A) \big\}; A  \right),
 \end{align}
where all the $A$'s in the terms $\theta\left( D^A\big\{ \bs{d\gamma\gamma}\-_{|A}(\bs Z_A) \big\}; A \right)$ are acted upon. Observe also that 
\begin{align*}
\big[ \bs{d\gamma\gamma}\-(\bs X), \bs{d\gamma\gamma}\-(\bs Y) \big] &= \bs{d\gamma\gamma}\-(\bs X) \bs{d\gamma\gamma}\-(\bs Y) - \bs{d\gamma\gamma}\-(\bs Y) \bs{d\gamma\gamma}\-(\bs X) 
														= -\bs{d\gamma}(\bs X)\bs{d\gamma}\-(\bs Y) + \bs{d\gamma}(\bs Y)\bs{d\gamma}\-(\bs X), \\
														&= \big(\! -\bs{d\gamma}\bs{d\gamma}\- \big)(\bs X, \bs Y) 
														= \bs d \big( \bs{d\gamma\gamma}\-\big) (\bs X, \bs Y)
\end{align*}
which is indeed simply the ``flatness" - or Maurer-Cartan like - condition  $\bs d \big(\bs{d\gamma\gamma}\-\big) -\sfrac{1}{2}\big[\bs{d\gamma\gamma}\-, \bs{d\gamma\gamma}\- \big]=0$ for the 1-form $\bs{d\gamma\gamma}\-$. 
But then, again by Kozsul we have, 
\begin{align*}
\big[ \bs{d\gamma\gamma}\-_{|A}(\bs X_A), \bs{d\gamma\gamma}\-_{|A}(\bs Y_A) \big] = \bs d \big( \bs{d\gamma\gamma}\-\big)_{|A} (\bs X_A, \bs Y_A) 
																  = \bs X \cdot \big\{  \bs{d\gamma\gamma}\-_{|\munderline{blue}{A}} (\bs Y_{\munderline{blue}{A}})\big\} 
																  - \bs Y \cdot \big\{  \bs{d\gamma\gamma}\-_{|\munderline{blue}{A}} (\bs X_{\munderline{blue}{A}})\big\}   
																  -  \bs{d\gamma\gamma}\-_A\big( \big[\bs X, \bs Y\big]_A \big),
\end{align*}
where we stressed that the underlined $A$'s are acted upon.
Inserting this in the last term of  \eqref{1},  remembering that $\theta$ is linear in the first argument and using \eqref{2}, we have 
\begin{align*}
{\bs\Theta^{\bs\gamma}}_{|A}\big(\bs X_A, \bs Y_A \big)   =&\,   \bs\Theta_{|A} \big(\bs X_A, \bs Y_A \big)   + \bs X \cdot \theta\left( D^{\munderline{blue}{A}}\big\{ \bs{d\gamma\gamma}\-_{|A}(\bs Y_A) \big\}; \munderline{blue}{A} \right) - \bs Y \cdot \theta\left( D^{\munderline{blue}{A}}\big\{ \bs{d\gamma\gamma}\-_{|A}(\bs X_A) \big\}; \munderline{blue}{A} \right)  \notag \\
			&\hspace{3cm}  + \theta\left( D^A\big\{ \bs X \cdot \big\{  \bs{d\gamma\gamma}\-_{|\munderline{blue}{A}} (\bs Y_{\munderline{blue}{A}})\big\} 
																  - \bs Y \cdot \big\{  \bs{d\gamma\gamma}\-_{|\munderline{blue}{A}} (\bs X_{\munderline{blue}{A}})\big\}   
																  -  \bs{d\gamma\gamma}\-_A\big( \big[\bs X, \bs Y\big]_A \big) \big\}; A \right), \notag\\
										=&\,  \bs\Theta_{|A} \big(\bs X_A, \bs Y_A \big)  + \bs X \cdot \theta\left( D^{\munderline{blue}{A}}\big\{ \bs{d\gamma\gamma}\-_{|\munderline{blue}{A}}(\bs Y_{\munderline{blue}{A}}) \big\}; \munderline{blue}{A} \right)
 																			- \bs Y \cdot \theta\left( D^{\munderline{blue}{A}}\big\{ \bs{d\gamma\gamma}\-_{|\munderline{blue}{A}}(\bs X_{\munderline{blue}{A}}) \big\}; \munderline{blue}{A} \right) 
															  - \theta\left(    D^A\big\{ \bs{d\gamma\gamma}\-_{|A}([\bs X, \bs Y]_A) \big\}; A  \right), \notag\\
										=&\,  \bs\Theta_{|A} \big(\bs X_A, \bs Y_A \big)  +  \bs d\, \theta\left( D^A\big\{ \bs{d\gamma\gamma}\-_{|A} \big\}; A \right) \big( \bs X_A, \bs Y_A \big).
\end{align*}
Which is finally, using \eqref{Noether-current-invariant},
\begin{align}
\label{Field-depGT-Theta}
\bs\Theta^{\bs\gamma} = \bs\Theta +  \bs d\, \theta\left( D^A\big\{ \bs{d\gamma\gamma}\- \big\}; A \right) = \bs\Theta + \bs d \left( d\theta\big(\bs{d\gamma\gamma}\-; A\big) - E\big(\bs{d\gamma\gamma}\-; A\big) \right), 
\end{align}
consistent with \eqref{Field-depGT-presymp-pot-current} - given again that $\bs d$ commute with pullbacks (here $[\bs\Psi^\star, \bs d]=0$).
This gives us the $\bs \H$-gauge transformation of the presymplectic 2-form, 
\begin{align}
\label{Field-depGT-presymp-form}
\bs\Theta_\Sigma^{\bs\gamma} = \bs\Theta_\Sigma + \int_{\d\Sigma}   \bs d \theta\big(\bs{d\gamma\gamma}\-; A\big) - \int_\Sigma \bs d E\big(\bs{d\gamma\gamma}\-; A\big). 
\end{align}
This is indeed consistent with  \eqref{Field-depGT-presymp-pot}. As for  $\bs\theta_\Sigma$, the presymplectic 2-form is then basic, $\bs\H$-invariant, if we are on-shell and if either $\d\Sigma=\emptyset$ or $A$ and/or the gauge parameter $\chi$/$\gamma$
 are required to vanish at $\d\Sigma$ or at infinity. In which case it induces a symplectic 2-form on $\S/\H$. 
 \medskip
 
 We now apply the above general results to obtain the Noether charges as well as the presymplectic potential and 2-forms (and their field-dependent gauge transformations) for Yang-Mills theory, for  3D-$\CC$-gravity with cosmological constant, and for 4D  gravity with or without cosmological constant. May the reader excuse the repetitive nature of the exposition, which is design so that each example can be read independently of the others.

\subsubsection{Yang-Mills theory}
\label{The case of Yang-Mills theory}

The Lagrangian of  the theory $L_\text{\tiny YM}(A)=\tfrac{1}{2}\Tr(F *F)$ is invariant under $\H=\SU(n)$, and $\bs d L_\text{\tiny YM}$ gives  the field equations $\bs E_\text{\tiny YM} = E_\text{\tiny YM}(\bs d A; A) = \Tr\big( \bs d A \, D^A\!*\!F\big)$ and the presymplectic potential current $\bs \theta_\text{\tiny YM} =   \theta_\text{\tiny YM}(\bs d A; A) = \Tr\big( \bs d A *\!F\big)$. 
 By~\eqref{Noether-charge-invariant},~the Noether  charge is thus
\begin{align}
\label{Noether-current-YM}
Q^\text{\tiny YM}_\Sigma(\chi;A)&= \int_{\d\Sigma} \theta_\text{\tiny YM}(\chi; A) - \int_\Sigma E_\text{\tiny YM}(\chi; A) , \notag \\
			  &= \int_{\d\Sigma} \Tr\big(\chi *\!F \big) - \int_\Sigma \Tr\big( \chi\, D^A\!*\!F \big).
\end{align}
The presymplectic 2-form current  is  $\bs\Theta^\text{\tiny YM}_\Sigma=\int_\Sigma \bs{d\theta}_\text{\tiny YM}=-\int_\Sigma \Tr\big( \bs d A *\!\bs d F  \big)$ and  by \eqref{presympl-2form-Noether-charge} relates to the charge as
\begin{align}
 \iota_{\chi^v} \bs \Theta^\text{\tiny YM}_\Sigma = - \bs d Q^\text{\tiny YM}_\Sigma(\chi; A) 
 										           = -\int_{\d\Sigma} \Tr\big(\chi *\!\bs dF \big) + \int_\Sigma \Tr\big( \chi\, \bs d D^A\!*\!F \big).
\end{align} 
 By \eqref{Poisson-bracket} it induces the Poisson bracket of charges $\big\{ Q^\text{\tiny YM}_\Sigma(\chi; A) ,  Q^\text{\tiny YM}_\Sigma(\eta; A)\big\}=Q^\text{\tiny YM}_\Sigma([\chi, \eta]; A)$, as can be checked explicitly. 
From \eqref{Field-depGT-presymp-pot} and \eqref{Field-depGT-presymp-form}
we get the  $\bs\H=\bs \SU(n)$ gauge transformations of the presymplectic potential and~2-form
\begin{align}
(\bs\theta^\text{\tiny YM}_\Sigma)^{\bs \gamma}&= \bs \theta^\text{\tiny YM}_\Sigma + \int_{\d_\Sigma}\theta_\text{\tiny YM} \big(  \bs{d\gamma\gamma}\- ;  A \big) - \int_\Sigma E_\text{\tiny YM} \big( \bs{d\gamma\gamma}\- ; A \big), \notag\\
				                  &= \bs \theta^\text{\tiny YM}_\Sigma + \int_{\d _\Sigma }\Tr\big(\bs{d\gamma\gamma}\- *\!F \big) - \int_\Sigma\Tr\big( \bs{d\gamma\gamma}\-\, D^A\!*\!F\big),     \label{GT-thetaYM} \\[1mm]
(\bs\Theta^\text{\tiny YM}_\Sigma)^{\bs\gamma} &= \bs\Theta^\text{\tiny YM}_\Sigma +  \int_{\d \Sigma}\bs d \theta_\text{\tiny YM}\big(\bs{d\gamma\gamma}\-; A\big) - \int_\Sigma \bs d E_\text{\tiny YM}\big(\bs{d\gamma\gamma}\-; A\big), \notag\\
				                    &= \bs\Theta^\text{\tiny YM}_\Sigma + \int_{\d \Sigma} \bs d \Tr\big( \bs{d\gamma\gamma}\- *\!F \big)  - \int_\Sigma \bs d  \Tr\big(  \bs{d\gamma\gamma}\-\, D^A\!*\!F \big).   \label{GT-ThetaYM}
\end{align}
This is  verified algebraically, by using \eqref{GT-dA}  and \eqref{GT-dF} in  $(\bs\theta^\text{\tiny YM}_\Sigma)^{\bs \gamma}=\int_\Sigma \Tr\big( \bs dA^{\bs \gamma} *\!F^{\bs \gamma} \big)$ and $(\bs \Theta^\text{\tiny YM}_\Sigma)^{\bs \gamma}= -\int_\Sigma\Tr\big( \bs d A^{\bs \gamma}*\!\bs d F^{\bs \gamma}\big)$.  
Clearly, only on-shell and under proper boundary conditions are these  basic forms on $\A$, and thus induce a symplectic structure on  $\S/\H$.

Finally, we can illustrate \eqref{Field-depGT-FieldEq} by giving the $\bs\H$-gauge transformation of the field equations  
\begin{align}
 \bs E_\text{\tiny YM}^{\bs \gamma}= \bs E_\text{\tiny YM} + dE_\text{\tiny YM} \big(  \bs{d\gamma\gamma}\- ;  A \big) = \bs E + d\Tr\big(  \bs{d\gamma\gamma}\- D^A*\!F  \big),
\end{align}
 which is verified algebraically by $\bs E_\text{\tiny YM}^{\bs \gamma}=E_\text{\tiny YM}\big( \bs d A^{\bs \gamma}; A^{\bs \gamma}\big)$.


\subsubsection{3D-$\mathbb{C}$-gravity $\Lambda \neq 0$}
\label{3D gravity}

We describe the theory in terms the underlying Cartan geometry $(\P, \b A)$ with $\P$ a $H$-principal bundle equipped with a Lie$G$-valued Cartan connection $\b A$, specifying the pair of groups $(G, H)$ on which it is based. In Euclidean signature, for a positive cosmological constant $\Lambda > 0$ we have $\big(S\!pin(4), S\!U(2)\big)$, with $S\!pin(4) \simeq S\!U(2) \times S\!U(2)$, while for $\Lambda < 0$ we have $\big(S\!L(2, \CC), S\!U(2)\big)$. In Lorentzian signature,  for $\Lambda >0$ we have $\big( S\!L(2, \CC), S\!U(1,1) \big)$, and for  $\Lambda <0$ we have  $\big( S\!pin(2,2), S\!U(1,1)  \big)$ with $S\!pin(2,2)\simeq S\!L(2, \RR) \times S\!L(2,\RR)$ and $S\!U(1,1) \simeq S\!L(2,\RR)$. 
 In all cases the Cartan connection splits as $\b A=A +\tfrac{1}{\ell}e$, with $\tfrac{1}{\ell^2}= \tfrac{2|\Lambda|}{(n-1)(n-2)}=|\Lambda|$ for $n=3=$ dim$\M$, and correspongingly the curvature splits as $\b F=F+\tfrac{1}{\ell}T$. The gauge group $\H$ acts as $R_\gamma A = A^\gamma=\gamma\- A \gamma +\gamma\- d\gamma $ and $R_\gamma e=e^\gamma =\gamma\- e\gamma$, so that $F^\gamma=\gamma\- F\gamma$ and $T^\gamma=\gamma\- T \gamma$. 
 
  The Lagrangian is $L(\b A)=L(A, e)= \Tr\big( eF\big) = \Tr\left\{ e\big(R-\tfrac{\epsilon}{3\ell^2}ee\big) \right\}$, with $\epsilon=\pm 1$  the sign of $\Lambda$.  It is $\H$-invariant, and from the viewpoint of Cartan geometry, there is no other gauge symmetry. 
  From $\bs d L$, since $\bs d \b A=\bs d A + \bs d e$, one finds the field equations $\bs E =E(\bs d \b A; \b A)= \Tr\big(\bs d e\, F+ \bs d A\, T \big)= \Tr\left( \bs d e  \big(R- \tfrac{\epsilon}{\ell^2} ee \big) + \bs d A\, D^Ae  \right)$, and the presymplectic potential current $\bs\theta=\theta(\bs d \b A; \b A)= \Tr\big(  \bs d A \, e  \big)$. 
  By  \eqref{Noether-charge-invariant} the Noether   charge is 
\begin{align}
\label{Noether-charge-3D-grav}
Q_\Sigma(\chi; \b A)= \int_{\d\Sigma} \theta(\chi; \b A) - \int_\Sigma E(\chi; \b A) 
			      = \int_{\d\Sigma} \Tr\big(\chi \, e \big) - \int_\Sigma \Tr\big( \chi\, D^Ae \big).
\end{align}
Remark that since $\chi \in$ Lie$\H$, only the piece of the field equations linear in $\bs dA$ contributes to the transformation formula. 
By  \eqref{presympl-2form-Noether-charge}, the presymplectic 2-form    $\bs\Theta_\Sigma=\int_\Sigma \bs{d\theta}= -\int_\Sigma \Tr\big( \bs d A \,\bs d e  \big)$  relates to the charge as
\begin{align}
 \iota_{\chi^v} \bs \Theta_\Sigma = - \bs d Q_\Sigma(\chi; \b A) 
 										           = -\int_{\d\Sigma} \Tr\big(\chi \bs de \big) + \int_\Sigma \Tr\big( \chi\, \bs d D^Ae \big).
\end{align} 
and by  \eqref{Poisson-bracket} it  induces the Poisson bracket of charges $\big\{ Q_\Sigma(\chi; \b A) ,  Q_\Sigma(\eta; \b A)\big\}=Q_\Sigma([\chi, \eta]; \b A)$, as is easily verified. 
By  \eqref{Field-depGT-presymp-pot} and \eqref{Field-depGT-presymp-form}
we have the field-dependent $\bs\H$-gauge transformations
\begin{align}
\bs\theta_\Sigma^{\bs \gamma}&= \bs \theta_\Sigma + \int_{\d_\Sigma}\theta \big(  \bs{d\gamma\gamma}\- ;  \b A \big) - \int_\Sigma E \big( \bs{d\gamma\gamma}\- ; \b A \big)  
				  = \bs \theta_\Sigma +  \int_{\d \Sigma}\Tr\big(\bs{d\gamma\gamma}\- e \big) -  \int_{\Sigma}\Tr\big( \bs{d\gamma\gamma}\-\, D^Ae\big).  \label{Field-depGT-theta-3D-grav}\\[1mm]
\bs\Theta_\Sigma^{\bs\gamma} &= \bs\Theta_\Sigma +  \int_{\d \Sigma}\bs d \theta\big(\bs{d\gamma\gamma}\-; \b A\big) - \int_\Sigma \bs d E\big(\bs{d\gamma\gamma}\-; \b A\big), 
				    = \bs\Theta_\Sigma +  \int_{\d \Sigma} \bs d \Tr\big( \bs{d\gamma\gamma}\- e  \big) -  \int_{\Sigma} \bs d\Tr\big(  \bs{d\gamma\gamma}\-\, D^Ae\big).    \label{Field-depGT-Theta-3D-grav2}
\end{align}
Here again, only the piece of the field equations linear in $\bs dA$ can contribute to the  transformation formulae, since the 1-form $ \bs{d\gamma\gamma}\-$ is Lie$\bs\H$-valued. 
These can also be verified algebraically using \eqref{GT-dA-Cartan} in \mbox{$\bs\theta_\Sigma^{\bs \gamma}= \int_{\Sigma}\Tr\big( \bs dA^{\bs \gamma} e^{\bs \gamma} \big)$} and $\bs \Theta_\Sigma^{\bs \gamma}=- \int_{\Sigma}\Tr\big( \bs d A^{\bs \gamma} \bs d e^{\bs \gamma}\big)$. 

The normal Cartan connection $\b A=\b A(e)$ in this case must simply be torsion-free, $D^Ae=0$. Which means that $A=A(e)$ is the Levi-Civita connection.  This is enforced by the field equations. But one may choose to start with a normal connection from the onset, so that the field equation reduces to $\bs E=\Tr\left( \bs d e  \big(R- \tfrac{\epsilon}{\ell^2} ee \big)\right)$. In which case even off-shell we have 
\begin{align*}
\qquad Q_\Sigma(\chi; \b A)= \int_{\d \Sigma} \Tr\big(\chi\, e \big)_{\ |N} \quad \text{and} \quad    \iota_{\chi^v} \bs \Theta_\Sigma =-\int_{\d\Sigma} \Tr\big(\chi \bs de \big)_{\ |N},
\end{align*}
and the $\bs\H$-gauge transformations of the presymplectic potential and 2-form are
\begin{align*}
\bs\theta_\Sigma^{\bs \gamma}= \bs \theta_\Sigma + \int_{\d\Sigma} \Tr\big(\bs{d\gamma\gamma}\- e \big)_{\ |N}  \qquad \text{and} \qquad 
\bs\Theta_\Sigma^{\bs\gamma} = \bs\Theta_\Sigma + \int_{\d\Sigma} \bs d\Tr\big( \bs{d\gamma\gamma}\- e  \big)_{\ |N}.
\end{align*}
Only proper boundary conditions are needed for these to be basic and descend, off-shell, as forms on $\A/\H$. On-shell they give symplectic potential and 2-form on the physical phase space $\S/\H$. 
Finally, by \eqref{Field-depGT-FieldEq}
\begin{align}
 \bs E^{\bs \gamma}= \bs E + dE \big(  \bs{d\gamma\gamma}\- ;  A \big) = \bs E + d\Tr\big(  \bs{d\gamma\gamma}\- D^Ae  \big)
													      = \bs E _{\ |N}, \notag
\end{align}
 which is verified algebraically by $\bs E^{\bs \gamma}=E\big( \bs d A^{\bs \gamma}; A^{\bs \gamma}\big)$, again with the help of  \eqref{GT-dA-Cartan}.


\subsubsection{4D Einstein-Cartan gravity}
\label{4D EC gravity}

Before giving applications to 4D gravity in this section and the next, we must say a word about the invariant multilinear map we will use to build the Lagrangians in an index-free way. Consider $P : \otimes^k M(2k, \mathbb{K}) \rarrow \mathbb{K}$ given by
\begin{align}
\label{Polyn-gravity}
P\big(A_1, \ldots, A_k \big)= A_1 \bullet\, \ldots\, \bullet A_k\defeq A^{i_1i_2}_1\, A^{i_3i_4}_2 \ldots \, A^{i_{2k-1}i_{2k}}_k \, \epsilon_{i_1 \ldots i_{2k}}, 
\end{align}
where the second equality defines the notation. Given $G \in GL(2k, \mathbb K)$, it satisfies the identity
\begin{align}
\label{Prop1-P}
P\big(G^TA_1G, \ldots, G^TA_k G\big)&= G^TA_1G \bullet\, \ldots\, \bullet G^TA_kG, \notag\\
							   &= {G^{i_1}}_{j_1}A^{j_1j_2}_1 {G_{j_2}}^{i_2} \  {G^{i_3}}_{j_3}A^{j_3j_4}_2   {G_{j_4}}^{i_4} \ldots \    {G^{i_{2k-1}}}_{j_{2k-1}} A^{j_{2k-1}j_{2k}}_k   {G_{j_{2k}}}^{i_{2k}}\ \epsilon_{i_1 \ldots i_{2k}}, \notag \\
							   &= \det(G)\ A^{j_1j_2}_1\, A^{j_3j_4}_2 \ldots \  A^{j_{2k-1}j_{2k}}_k\, \epsilon_{j_1 \ldots j_{2k}}, \notag\\
							   &=  \det(G)\  A_1 \bullet\, \ldots\, \bullet A_k =  \det(G)\  P\big(A_1, \ldots, A_k \big).
\end{align}
Then, $P$ is $S\!O(2k)$-invariant, since for $S\in S\!O(2k)$, $S^T=S\-$, we have $P\big(S\-A_1S, \ldots, S\-A_k S\big)=P\big(A_1, \ldots, A_k \big)$. 
Also, given some matrix $M \in M(2k, \mathbb{K})$ decomposed as the sum of its symmetric and antisymmetric parts as $M=\sfrac{1}{2}(M+M^T) + \sfrac{1}{2}(M-M^T)\rdefeq {\sf S}+  {\sf  A}$, we have
\begin{align}
\label{Prop2-P}
M\bullet A_2 \bullet \ldots \bullet A_k = \big(   \cancel{ {\sf S}^{i_1i_2}}_{} \!+ {\sf A}^{i_1i_2} \big) \, A^{i_3i_4}_2 \ldots \, A^{i_{2k-1}i_{2k}}_k \, \epsilon_{i_1 i_2 \ldots i_{2k}}= {\sf A} \bullet A_2 \ldots \bullet A_k. 
\end{align}
We have then actually a $\Ad\big(S\!O(2k)\big)$-invariant map $P: \otimes^k \so(2k) \rarrow \mathbb{K}$.\footnote{Remark that the diagonal combination $P(A, \ldots, A)=\text{Pf}(A)$ is the Pfaffian of the $2k \times 2k$ antisymmetric matrix $A$, which is the square root of its determinant $\text{Pf}(A)^2=\det(A)$. Conversely, $P$ is the polarisation of the Pfaffian polynomial. } We can use it to write the Lagrangians of even dimensional gravity theories. 

In particular the gauge formulation of 4D gravity with $\Lambda=0$, that we will call Einstein-Cartan (EC) gravity, is based on a Cartan geometry modelled on $(G, H)=\big(S\!O(1,3) \ltimes \RR^4, S\!O(1,3)\big)$ with underlying homogeneous space $G/H \sim \RR^4$. Equipped with the Minkowski metric $\eta\!=\!\eta_{ab}$, the latter is Minkowski space ${\sf M}=\big\{\RR^4, \eta \big\}$. 
This is a reductive geometry, the Cartan connection thus splits as $\b A =A+e$ 
and so does its curvature $\b F= R+T=(dA+A^2)  + D^Ae$. 
 The~gauge group is $\H=\SO(1,3)$, and acts on the connection as $A^\gamma = \gamma\- A\gamma + \gamma\-d\gamma$ and  $e^\gamma=\gamma\- e$, so that $R^\gamma=\gamma\- R\gamma$ and $T^\gamma=\gamma\- T$. 

Remembering that for $X \in \so(1,3)$ we have by definition $X\eta\- \in \so(4)$, we get that $R\eta\-$ is $\so(4)$-valued. And since $\gamma \eta\- = \eta\- \gamma^{-1T}$, we have $(R\eta\-)^\gamma=\gamma\- R \gamma\eta\- = \gamma\- R \eta\- \gamma^{-1T}$.
Also, since $e$ is a $\RR^4$-valued 1-form, $e \w e^T$ is a antisymmetric matrix as we have  $\big(e \w e^T\big)^T= -e\w e^T$. It transforms as $\big(e \w e^T\big)^\gamma = e^\gamma \w (e^\gamma)^T=\gamma\- e \w e^T \gamma^{-1T}$.
The Lagrangian of  4D EC gravity $\Lambda=0$ can then be written 
\begin{align}
\label{Lagrangian-EC}
L_\text{\tiny EC}(\b A)=L_\text{\tiny EC}(A, e)= P\big(R\eta\-, e\w e^T\big)= R \bullet e \w e^T= R^{ab}e^c e^d \epsilon_{abcd},
\end{align}
where $\eta\-$ behind $R$ is tacit in the third equality, as it will be from now on in front of any $\so(1,3)$-valued variable when using the `bullet' notation for $P$. It is manifest that $R^\star_\gamma L_\text{\tiny EC}=L_\text{\tiny EC}$, with $\gamma \in \SO(1,3)$.

If we notice that $\bs d(e \w e^T)= \bs d e \w e^T + e \w \bs d e^T$ is the antisymmetric part of $2\bs d e \w e^T$, and that in the same way $D^Ae\w e^T - e \w (D^Ae)^T$ is the antisymmetric part of $2 D^Ae \w e^T$, since again $\bs d \b A = \bs d A + \bs d e$, it is easily found from $\bs d L_\text{\tiny EC}$ that the field equations and presymplectic potential current are
\begin{align}
 \bs E_\text{\tiny EC}= E_\text{\tiny EC}\big(\bs d \b A; \b A \big)= 2\left( \bs d A \bullet D^Ae \w e^T + \bs d e \w e^T \bullet R  \right), \quad \text{and} \quad \bs \theta_\text{\tiny EC}= \theta_\text{\tiny EC}(\bs d \b A; \b A)= \bs dA \bullet e \w e^T. 
\end{align}
The ground state of the theory is the homogeneous space of the underlying Cartan geometry, i.e. Minskowski space $G/H \sim {\sf M}$. Given $\chi \in$ Lie$\SO(1,3)$, by \eqref{Noether-charge-invariant} the Noether charge is 
\begin{align}
\label{Noether-charge-4D-EC}
Q^\text{\tiny EC}_\Sigma(\chi; \b A)&= \int_{\d\Sigma} \theta_\text{\tiny EC}(\chi; \b A) - \int_\Sigma E_\text{\tiny EC}(\chi; \b A) 
			  = \int_{\d\Sigma}\chi \bullet e \w e^T -2 \int_\Sigma \chi \bullet  D^Ae \w e^T.           
\end{align}
Here again, only the piece of $\bs E_\text{\tiny EC}$ linear in  $\bs d A$ can  contribute to the result. 
By \eqref{presympl-2form-Noether-charge}, the presymplectic 2-form $\bs\Theta^\text{\tiny EC}_\Sigma=\int_\Sigma \bs{d\theta}_\text{\tiny EC}=-2 \int_\Sigma \bs dA \bullet  \bs d e \w e^T$ relates to the charge as 
\begin{align}
 \iota_{\chi^v} \bs \Theta^\text{\tiny EC}_\Sigma = - \bs d Q^\text{\tiny EC}_\Sigma(\chi; \b A) 
 										           = -2\int_{\d\Sigma} \chi \bullet \bs d e \w e^T    + 2\int_\Sigma  \chi \bullet \bs d  \big( D^Ae \w e^T \big), 
 \end{align} 
 and generates by \eqref{Poisson-bracket} the Poisson bracket of Lorentz charges $\big\{ Q^\text{\tiny EC}_\Sigma(\chi; \b A) ,  Q^\text{\tiny EC}_\Sigma(\eta; \b A)\big\}=Q^\text{\tiny EC}_\Sigma([\chi, \eta]; \b A)$.
 From \eqref{Field-depGT-presymp-pot} and  \eqref{Field-depGT-presymp-form}, the $\bs \SO(1,3)$-gauge transformations of the presymplectic potential and 2-form are
\begin{align}
(\bs\theta_\Sigma^\text{\tiny EC})^{\bs \gamma}&= \bs \theta_\Sigma^\text{\tiny EC} + \int_{\d\Sigma}  \theta_\text{\tiny EC} \big(  \bs{d\gamma\gamma}\- ;  \b A \big) - \int_{\Sigma} E_\text{\tiny EC} \big( \bs{d\gamma\gamma}\- ; \b A \big), \notag\\
				  &= \bs \theta_\Sigma^\text{\tiny EC} + \int_{\d\Sigma} \bs{d\gamma\gamma}\- \bullet  e \w e^T   -  2\int_{\Sigma} \bs{d\gamma\gamma}\-\bullet D^Ae \w e^T ,                           \label{SO-GT-presymp-pot-EC}  \\[1mm]
(\bs\Theta_\Sigma^\text{\tiny EC})^{\bs\gamma} &= \bs\Theta_\Sigma^\text{\tiny EC} +  \int_{\d\Sigma} \bs d  \theta_\text{\tiny EC}\big(\bs{d\gamma\gamma}\-; \b A\big) -  \int_{\d\Sigma} \bs d E_\text{\tiny EC}\big(\bs{d\gamma\gamma}\-; \b A\big), \notag\\
				    &= \bs\Theta_\Sigma^\text{\tiny EC} +  \int_{\d\Sigma}  \bs d  \big(\bs{d\gamma\gamma}\- \bullet  e \w e^T  \big) -  2  \int_{\Sigma} \bs d (\bs{d\gamma\gamma}\-\bullet D^Ae \w e^T).         \label{SO-GT-presymp-form-EC}
\end{align}
Still, only the piece of  $\bs E_\text{\tiny EC}$ linear in $\bs dA$ contributes since $\bs{d\gamma\gamma}\-$ is Lie$\bs \SO(1,3)$-valued. 
These can also be verified algebraically using \eqref{GT-dA-Cartan} in $(\bs\theta_\Sigma^\text{\tiny EC})^{\bs \gamma}= \int_{\Sigma} \bs dA^{\bs \gamma} \bullet e^{\bs \gamma} \w (e^{\bs \gamma})^T$ and 
$(\bs\Theta_\Sigma^\text{\tiny EC})^{\bs \gamma}=-\int_\Sigma \bs dA^{\bs \gamma} \bullet 2 \bs d e^{\bs \gamma} \w (e^{\bs \gamma})^T$. 
 The same comments apply for the result given by   \eqref{Field-depGT-FieldEq} 
\begin{align}
 \bs E_\text{\tiny EC}^{\bs \gamma}= \bs E_\text{\tiny EC} + dE_\text{\tiny EC} \big(  \bs{d\gamma\gamma}\- ;  A \big) = \bs E_\text{\tiny EC} + 2 d \big(  \bs{d\gamma\gamma}\-  \bullet D^Ae \w e^T  \big).
\end{align}

The normal Cartan connection $\b A=\b A(e)$ is simply torsion-free, $D^Ae=0$, so that  $A=A(e)$ is the Levi-Civita connection. To have this enforced by the field equation maybe seen as too strong a constraint, and instead we could admit normality from the beginning so that the field equation reduces to $\bs E_\text{\tiny EC}= \bs d e \w e^T \bullet R_{ \ |N} $, and $\bs E_\text{\tiny EC}^{\bs \gamma}= {\bs E_\text{\tiny EC} }_{\ |N}$. 
In~which case, off-shell we get
\begin{align*}
 \qquad Q^\text{\tiny EC}_\Sigma(\chi; \b A)= \int_{\d \Sigma}  \chi\bullet e\w e^T_{\ \ |N} \quad \text{and} \quad     \iota_{\chi^v} \bs \Theta^\text{\tiny EC}_\Sigma = -2\int_{\d\Sigma} \chi \bullet \bs d e \w e^T _{\ \ |N} 
\end{align*}
and the $\bs\H$-gauge transformation of the presymplectic potential and 2-form are
\begin{align*}
(\bs\theta^\text{\tiny EC}_\Sigma)^{\bs \gamma}= \bs \theta^\text{\tiny EC}_\Sigma + \int_{\d\Sigma} \bs{d\gamma\gamma}\- \bullet  e \w e^T_{\ \ |N}  \qquad \text{and} \qquad 
(\bs\Theta^\text{\tiny EC}_\Sigma)^{\bs\gamma} = \bs\Theta^\text{\tiny EC}_\Sigma + \int_{\d\Sigma} \bs d\big( \bs{d\gamma\gamma}\- \bullet  e \w e^T \big)_{\ |N}.
\end{align*}
We then only need adequate boundary conditions for these to be off-shell basic forms on  $\A$, and then of course give symplectic potential and 2-form on the physical phase space $\S/\H$.

\bigskip

We can add a  cosmological constant term to   $L_\text{\tiny EC}$ so as to obtain the new $\SO(1,3)$-invariant Lagrangian
\begin{align}
\label{Lagrangian-EC-2}
L_\text{\tiny EC-$\Lambda$}(\b A)=L_\text{\tiny EC-$\Lambda$}(A, e)&= P\big(R\eta\-, e\w e^T\big) - \tfrac{\epsilon}{2\ell^2} P(e\w e^T ,  e\w e^T), \notag \\
													&= R \bullet e \w e^T -  \tfrac{\epsilon}{2\ell^2} e\w e^T \bullet e\w e^T = \left( R^{ab}e^c e^d -  \tfrac{\Lambda}{6} e^a e^be^ce^d\right)\epsilon_{abcd},
\end{align}
with  $\tfrac{1}{\ell^2}= \tfrac{2|\Lambda|}{(n-1)(n-2)}=\tfrac{|\Lambda|}{3}$ for $n=4=$ dim$\M$, and $\epsilon=\pm$ is the sign of $\Lambda$. The field equations change into 
\begin{align}
 \bs E_\text{\tiny EC-$\Lambda$}= E_\text{\tiny EC-$\Lambda$}\big(\bs d \b A; \b A \big)= 2\left( \bs d A \bullet D^Ae \w e^T + \bs d e \w e^T \bullet \big( R -\tfrac{\epsilon}{\ell^2} e \w e^T  \big) \right), 
\end{align}
which changes the ground state of the theory, which is no more the homogeneous space of the underlying Cartan geometry, but  de Sitter ($\epsilon=+1$) or anti-de Sitter ($\epsilon=-1$) space.
The presymplectic potential current remains unchanged: $\bs \theta_\text{\tiny EC-$\Lambda$}=\bs \theta_\text{\tiny EC}= \bs dA \bullet e \w e^T $.\footnote{In that respect the cosmological constant acts like a mass term in massive YM theory which also changes the field equations of $m=0$ YM theory but doesn't affect the presymplectic potential. While such a mass term compromises the gauge invariance of the Lagrangian in massive YM theory, the cosmological constant term does not in gravity.}
Thus, 
  the Noether  charge is the same as for $L_\text{\tiny EC}$ \eqref{Noether-charge-4D-EC}, and 
  $\bs \Theta_\Sigma^\text{\tiny EC-$\Lambda$}\!=\bs \Theta_\Sigma^\text{\tiny EC}$ so the Poisson bracket of charge is identical as well. 
  As~furthermore $\bs E_\text{\tiny EC-$\Lambda$}$ and $\bs E_\text{\tiny EC}$ share the piece linear in the Lorentz parameter,
  the $\bs\SO(1,3)$-gauge transformation formulae for $\bs \theta_\Sigma^\text{\tiny EC-$\Lambda$}$ and $\bs \Theta_\Sigma^\text{\tiny EC-$\Lambda$}$ are identical to  \eqref{SO-GT-presymp-pot-EC}-\eqref{SO-GT-presymp-form-EC}. 
\bigskip

We observe that for neither $L_\text{\tiny EC}$ nor $L_\text{\tiny EC-$\Lambda$}$ does the Noether charge vanish on the ground state of the theory, which sets the mass-energy reference. 
Also,  for solutions of the field equations that decay asymptotically to the ground state (e.g. isolated star systems, black holes ...)  $\bs\theta_\Sigma^\text{\tiny EC}$ and $\bs\Theta_\Sigma^\text{\tiny EC}$ are not $\bs\SO(1,3)$-invariant, thus do not induce a symplectic structure on $\S/\H$, unless boundary conditions on the gauge parameter are specified.

\subsubsection{4D MacDowell-Mansouri gravity}
\label{4D MM gravity}

As a manner of preparation, we first consider the topological theory given by $L_\text{\tiny Euler}(A)=\tfrac{1}{2}R \bullet R$, with $A$ a connection on $\P\big(\M, S\!O(1,3)\big)$ and $R=dA+\sfrac{1}{2}[A, A]$ the Riemann curvature.  This Lagrangian is (proportional to) the Euler density of $\M$, $\text{e}(\M)\defeq \tfrac{1}{(2\pi)^4}\text{Pf}(R)$. It  clearly  has trivial $\H=\SO(1,3)$-equivariance, $R^\star_\gamma L_\text{\tiny Euler} = L_\text{\tiny Euler}$. From $\bs dL_\text{\tiny Euler}$ we~find that $\bs E_\text{\tiny Euler}=\bs d A \bullet D^AR\equiv 0$ by the Bianchi identity, and the presymplectic potential current is $\bs \theta_\text{\tiny Euler}=\bs dA \bullet R$. All the quantities to follow are therefore automatically on-shell. For $\chi \in$ Lie$\SO(1,3)$, the Noether  charge is
\begin{align}
 Q^\text{\tiny Euler}_\Sigma(\chi; A) = \int_{\d\Sigma} \chi \bullet R, 
\end{align} 
and it relates by \eqref{presympl-2form-Noether-charge} to the presymplectic 2-form  $\bs\Theta_\Sigma^\text{\tiny Euler}= -\int_\Sigma\bs dA \bullet \bs d R= \tfrac{1}{2} \int_{\d\Sigma} \big( \bs d A \bullet \bs d A\big)$, so that  by \eqref{Poisson-bracket} the latter generates the Poisson bracket of charges.
We have the $\bs\SO(1,3)$-gauge transformations given by \eqref{Field-depGT-presymp-pot} and  \eqref{Field-depGT-presymp-form}
\begin{align}
(\bs\theta_\Sigma^\text{\tiny Euler})^{\bs \gamma} = \bs\theta_\Sigma^\text{\tiny Euler}+ \int_{\d \Sigma} \big( \bs{d\gamma\gamma}\- \bullet R \big), \quad \text{and} \quad
(\bs\Theta_\Sigma^\text{\tiny Euler})^{\bs \gamma} = \bs\Theta_\Sigma^\text{\tiny Euler} + \int_{\d \Sigma}   \bs d \big( \bs{d\gamma\gamma}\- \bullet R \big). 
\end{align}

\medskip

We now define 4D MacDowell-Mansouri (MM) gravity as the gauge theory of gravity based on the Cartan geometry $(\P, \b A)$ modeled on either $(G, H)=\big(S\!O(1,4), S\!O(1,3)\big)$ for $\Lambda \geq 0$, or $(G, H)=\big(S\!O(2,3), S\!O(1,3)\big)$ for $\Lambda \leq 0$.\footnote{See \cite{Wise09, Wise10} for a discussion of the link between Cartan geometry and MM gravity.}
 In the first case the homogeneous space is $G/H \sim dS_4$ the de Sitter space, and in the second it is anti-de Sitter space $G/H\sim AdS_4$.
The geometry being reductive, the Cartan connection splits as $\b A=A + \tfrac{1}{\ell} e$, and correspondingly the curvature is $\b F=F+ \tfrac{1}{\ell} T = \big( R-\tfrac{\epsilon}{\ell^2}e e^t\big) + \tfrac{1}{\ell}D^A e $, where $e^t\defeq e^T \eta= e^a \eta_{ab}$. 
Or in matrix form, 
\begin{align*}
\b A = \begin{pmatrix} A & \tfrac{1}{\ell}e \\ \tfrac{-\epsilon}{\ell}e^t & 0 \end{pmatrix}, \qquad \b F = d\b A + \b A ^2= \begin{pmatrix} F & \tfrac{1}{\ell}T \\ \tfrac{-\epsilon}{\ell}T^t & 0 \end{pmatrix}=\begin{pmatrix} R-\tfrac{\epsilon}{\ell^2}e e^t & \tfrac{1}{\ell}D^A e \\ \tfrac{-\epsilon}{\ell}(D^Ae)^t & 0 \end{pmatrix}. 
\end{align*}
The homogeneous space $G/H$ is by definition Cartan flat, $\b F=0$, so it is torsion-free and satisfies $F=0 \rarrow R=\tfrac{\epsilon}{\ell}ee^t$. 
 The~normal Cartan connection $\b A=\b A(e)$ is  simply defined by $D^Ae=0$, making $A=A(e)$ the Levi-Civita connection. 
 
 The gauge group is  $\H=\SO(1,3)$ and acts on $A$, $e$, $R$ and $T$ as in EC gravity. 
So, defining the Lagrangian of 4D MM gravity as 
\begin{align}
L_\text{\tiny MM}(\b A)=\tfrac{1}{2} F \bullet F= L_\text{\tiny Euler}(A) -\tfrac{\epsilon}{\ell^2}\, L_\text{\tiny EC-$\Lambda$}(\b A) 
				  = \tfrac{1}{2}R \bullet R -\tfrac{\epsilon}{\ell^2} \left(  R \bullet e \w e^T -  \tfrac{\epsilon}{2\ell^2} e\w e^T \bullet e\w e^T  \right),
\end{align}
its trivial $\SO(1,3)$-equivariance  is manifest, $R^\star_\gamma L_\text{\tiny MM}=L_\text{\tiny MM}$.\footnote{For arbitrary choices of  coefficients in front of the three terms, this is otherwise known as the Lagrangian of  4D Zumino-Lovelock gravity. See \cite{Zumino1986}, \cite{Hassaine-Zanelli2016} section 4.1, or \cite{Kurihara2017} and references therein.} 
From $\bs d L_\text{\tiny MM}=\bs d L_\text{\tiny Euler} -\tfrac{\epsilon}{\ell^2} \,\bs d L_\text{\tiny EC-$\Lambda$}$ we find the field equations and presymplectic potential to be respectively, 
\begin{align}
\bs E_\text{\tiny MM}&= \cancel{\bs E_\text{\tiny Euler}} -\tfrac{\epsilon}{\ell^2}\, \bs E_\text{\tiny EC-$\Lambda$} =  -\tfrac{\epsilon}{\ell^2} \, 2 \left\{     \bs d A \bullet D^Ae \w e^T + \bs d e \w e^T \bullet \big( R -\tfrac{\epsilon}{\ell^2} e \w e^T  \big)   \right\}, \\[1mm]
\bs\theta_\text{\tiny MM}&= \bs\theta_\text{\tiny Euler} -\tfrac{\epsilon}{\ell^2}\, \bs \theta_\text{\tiny EC-$\Lambda$} = \bs\theta_\text{\tiny Euler} -\tfrac{\epsilon}{\ell^2}\, \bs \theta_\text{\tiny EC} = \bs d A \bullet \big(  R -\tfrac{\epsilon}{\ell^2} e \w e^T \big).
\end{align}
Given $\chi \in$ Lie$\SO(1,3)$, by \eqref{Noether-charge-invariant}  the Noether charge is found to be
%
\begin{align}
Q^\text{\tiny MM}_\Sigma(\chi; \b A)= \int_{\d\Sigma} \theta_\text{\tiny MM}(\chi; \b A) - \int_\Sigma E_\text{\tiny MM}(\chi; \b A) 
			   &= \int_{\d\Sigma}\chi \bullet \big( R -\tfrac{\epsilon}{\ell^2} e \w e^T  \big) + \tfrac{\epsilon}{\ell^2}\,2\! \int_\Sigma \chi \bullet  D^Ae \w e^T.            \label{Noether-charge-4D-MM} \\
			   &=\int_{\d\Sigma}\chi \bullet \big( R -\tfrac{\epsilon}{\ell^2} e \w e^T  \big) _{\, |N}, \notag
\end{align}
and of course $Q_\Sigma^\text{\tiny MM}(\chi;\b A)= Q_\Sigma^\text{\tiny Euler}(\chi; A) - \tfrac{\epsilon}{\ell^2}Q_\Sigma^\text{\tiny EC}(\chi;\b A)$. 
The presymplectic 2-form of 4D MM gravity 
\begin{align}
\bs\Theta_\Sigma^\text{\tiny MM}=\int_\Sigma \bs{d\theta}\text{\tiny MM} =\bs\Theta_\Sigma^\text{\tiny Euler} -\tfrac{\epsilon}{\ell^2}\, \bs\Theta_\Sigma^\text{\tiny EC} = -\int_\Sigma \bs d A \bullet \bs d  \big(  R -\tfrac{\epsilon}{\ell^2} e \w e^T \big)
\end{align}
is  such that by  \eqref{presympl-2form-Noether-charge} 
\begin{align}
\iota_{\chi^v} \bs \Theta^\text{\tiny MM}_\Sigma = - \bs d Q^\text{\tiny MM}_\Sigma(\chi; A) 
 										           &= -\int_{\d\Sigma} \chi \bullet \bs d \big(R -\tfrac{\epsilon}{\ell^2} e \w e^T\big)   -\tfrac{\epsilon}{\ell^2}\,2\! \int_\Sigma  \chi \bullet \bs d  \big( D^Ae \w e^T \big), \\
										           &=-\int_{\d\Sigma} \chi \bullet \bs d \big(R -\tfrac{\epsilon}{\ell^2} e \w e^T\big)  _{\ |N}.
\end{align} 
 One thus verifies that, in accord with \eqref{Poisson-bracket}, 
 it generates the Poisson bracket $\big\{ Q^\text{\tiny MM}_\Sigma(\chi; \b A) ,  Q^\text{\tiny MM}_\Sigma(\eta; \b A)\big\}=Q^\text{\tiny MM}_\Sigma([\chi, \eta]; \b A)$.
 From \eqref{Field-depGT-presymp-pot} and  \eqref{Field-depGT-presymp-form}, the $\bs \SO(1,3)$-gauge transformations of the presymplectic potential and 2-form are
\begin{align}
(\bs\theta_\Sigma^\text{\tiny MM})^{\bs \gamma}&= \bs \theta_\Sigma^\text{\tiny MM} + \int_{\d\Sigma}  \theta_\text{\tiny MM} \big(  \bs{d\gamma\gamma}\- ;  \b A \big) - \int_{\Sigma} E_\text{\tiny MM} \big( \bs{d\gamma\gamma}\- ; \b A \big), \notag\\
				  &= \bs \theta_\Sigma^\text{\tiny MM} + \int_{\d\Sigma} \bs{d\gamma\gamma}\- \bullet  \big(  R -\tfrac{\epsilon}{\ell^2} e \w e^T \big)   + \tfrac{\epsilon}{\ell^2}\,2\!  \int_{\Sigma} \bs{d\gamma\gamma}\-\bullet D^Ae \w e^T ,                           \label{SO-GT-presymp-pot-MM} \\
				  &=\bs \theta_\Sigma^\text{\tiny MM} + \int_{\d\Sigma} \bs{d\gamma\gamma}\- \bullet  \big(  R -\tfrac{\epsilon}{\ell^2} e \w e^T \big) _{\ |N},  \notag\\[2mm]
(\bs\Theta_\Sigma^\text{\tiny MM})^{\bs\gamma} &= \bs\Theta_\Sigma^\text{\tiny MM} +  \int_{\d\Sigma} \bs d  \theta_\text{\tiny MM}\big(\bs{d\gamma\gamma}\-; \b A\big) -  \int_{\d\Sigma} \bs d E_\text{\tiny MM}\big(\bs{d\gamma\gamma}\-; \b A\big), \notag\\
				    &= \bs\Theta_\Sigma^\text{\tiny MM} +  \int_{\d\Sigma}  \bs d  \left(\bs{d\gamma\gamma}\- \bullet  \big(  R -\tfrac{\epsilon}{\ell^2} e \w e^T \big) \right) +    \tfrac{\epsilon}{\ell^2}\,2\!    \int_{\Sigma} \bs d (\bs{d\gamma\gamma}\-\bullet D^Ae \w e^T),         \label{SO-GT-presymp-form-MM} \\
				    &=\bs\Theta_\Sigma^\text{\tiny MM} +  \int_{\d\Sigma}  \bs d  \left(\bs{d\gamma\gamma}\- \bullet  \big(  R -\tfrac{\epsilon}{\ell^2} e \w e^T \big) \right)_{\ |N}. \notag
\end{align}
Here again, only the piece of  $\bs E_\text{\tiny MM}$ linear in the Lorentz parameter contributes since $\bs{d\gamma\gamma}\-$ is Lie$\bs \SO(1,3)$-valued. 
These formulas are verified algebraically using \eqref{GT-dA-Cartan}.
 The same goes for the result given by   \eqref{Field-depGT-FieldEq} 
\begin{align}
 \bs E_\text{\tiny MM}^{\bs \gamma}= \bs E_\text{\tiny MM} + dE_\text{\tiny MM} \big(  \bs{d\gamma\gamma}\- ;  \b A \big) &= \bs E_\text{\tiny MM} - \tfrac{\epsilon}{\ell^2}\, 2\, d \big(  \bs{d\gamma\gamma}\-  \bullet D^Ae \w e^T  \big), \\
 																						   &= {\bs E_\text{\tiny MM}}_{\ |N}. \notag
\end{align}

It is noteworthy that in 4D MM gravity - and contrary to EC gravity - the Noether charges do indeed vanish on the  ground state of the theory, i.e. the homogeneous (anti-) de Sitter space $G/H$, which sets the zero mass-energy reference. 
Also, for solutions of the theory that asymptotically decay to the ground state, both $\bs\theta_\Sigma^\text{\tiny MM}$ and $\bs\Theta_\Sigma^\text{\tiny MM}$ are $\bs\SO(1,3)$-invariant, and thus induce respectively a symplectic potential and 2-form on the physical phase space $\S/\H$,  without restrictive boundary conditions. This shows there are benefits to taking Cartan geometry seriously.

\subsection{Presymplectic structure for non-invariant (c-equivariant) Lagrangians}
\label{Presymplectic structure for non-invariant Lagrangians}

We now consider Lagrangians that are $c$-equivariant 0-forms on $\A$, so s.t. $R^\star_\gamma L=L+ c(\ \, ; \gamma)$ for $\gamma \in \H$. 
\mbox{Infinitesimally} this means the classical anomaly is given by 
\begin{align}
\label{def_class_anomaly}
\bs L_{\chi^v} L_{|A}=\iota_{\chi^v} \bs d L_{|A}=\tfrac{d}{d\tau} c(A; e^{\tau \chi})\big|_{\tau=0}\rdefeq \alpha(\chi; A),
\end{align}
 where $\chi^v \in \Gamma(V\A)$ and $\chi \in \text{Lie}\H$.
We can no longer say that $L$ factors through the curvature map $F$, so we will only assume that it is  of the form $L(A)=\t L([F]; A)$, with $\t L$ a   $\Ad(H)$-invariant symmetric multilinear map (as is virtually always the case) and $[F]$ denoting $F$-dependent terms. Then we can write 
\begin{align*}
\bs dL_{|A}=&\,\t L(\bs d[F]; A)+ \t L([F]; A; \bs dA)= \t L \big(D^A(\bs d A); [F]; A \big)+ \t L(\bs dA; [F]; A),  \\
		=&\, d\t L\big( \bs d A; [F]; A \big) + \t L\big( \bs d A; D^A\{[F];A\}\big) + \t L(\bs dA; [F];A), \\
		 \rdefeq&\, d\theta(\bs d A; A)+ E(\bs d A;A) = d\bs \theta_{|A} + \bs E_{|A}. 
\end{align*}
where in the third equality the Leibniz identity and $Ad(H)$-invariance of $\t L$ was used. We thus see that this time $\bs E$ contains a piece that is not coming from the integration by parts of the term linear in $D^A(\bs d A)$.

As we've seen, the de Rham derivative doesn't preserve the space of $c$-equivariant forms, we have that $\bs d L=\bs E + d \bs \theta$ has equivariance 
\begin{align*}
R^\star_\gamma \bs dL &=  \bs d R^\star_\gamma L = \bs d L + \bs d c(\ \, ; \gamma).\\
 R^\star_\gamma \bs E + R^\star_\gamma d \bs \theta &=\bs E + d\bs \theta + \bs d c(\ \, ; \gamma).
\end{align*}
A priori, one can imagine that the respective $\H$-equivariances of $\bs E$ and $\bs \theta$ can be anything that add up to $\bs d c(\ \, ; \gamma)$. 
To~say something general we must make  assumptions. One can for example entertain  two extreme possibilities. 

The first would be to assume $R^\star_\gamma \bs E= \bs E + \bs d c(\ \, ; \gamma)$ and   $R^\star_\gamma  \bs \theta= \bs \theta$. In this case the presymplectic structure is obtained as in section~\ref{Presymplectic structure for invariant Lagrangians}, but the theory is quite badly behaved as a $\H$-gauge transformation brings us off-shell. Such a case is of limited interest, so we will not spent too much time on it and defer its treatment to appendix \ref{Presymplectic structure for non-invariant Lagrangians: the other extreme case}.

In the following, we will rather focus on the more subtle  case of theories satisfying 
\begin{align}
\label{Hyp0}
R^\star_\gamma \bs E= \bs E \quad \text{so that} \quad R^\star_\gamma d \bs \theta= d\bs \theta + \bs d c(\ \, ; \gamma). 
\end{align}
Call this our \emph{hypothesis 0}.
The theory is better behaved and its presymplectic structure is richer: we will  need further restricting hypothesis  to obtain results of some generality. 
\medskip

\noindent {\bf Noether current and charge: } 
Of course it is still true that $\bs d ^2L=0=\bs {dE}+d \bs \Theta$ with $\bs \Theta=\bs{d\theta}$, so that $\bs{d\Theta}=0$ and $\bs\Theta$ is still the candidate presymplectic form. From \eqref{Hyp0} follows that we must have $\bs d c(\ \, ,\gamma)=d \bs{b}(\gamma)$ where $\bs b(\gamma)$ is a variational 1-form (notice this doesn't mean that $c(\ \, ,\gamma)$ is $d$-exact). The latter is obviously linear in $\bs dA$, but it can furthermore be written (up to $d$-exact terms) as linear in \emph{underived} $\bs d A$. 
The $\H$-equivariance of the presymplectic potential current is thus 
\begin{align}
\label{H-eq-theta-non-inv}
R^\star_\gamma \bs \theta &= \bs \theta + \bs b(\gamma),
\qquad\text{infinitesimally this is} \quad
\bs L_{\chi^v} \bs \theta = \tfrac{d}{d\tau} \bs b(e^{\tau \chi})\big|_{\tau=0} \rdefeq \bs\alpha(\chi), \\
&= \bs \theta +  b(\bs d A; \gamma), \notag
\end{align}
By Cartan formula, the latter equation is also
\begin{align}
\label{def_bs_alpha}
\iota_{\chi^v} \bs \Theta = - \bs d (\iota_{\chi^v} \bs \theta) + \bs \alpha(\chi). 
\end{align}
Hypothesis 0 gives a constraint on $\bs \alpha(\chi)$. Indeed $R^\star_\gamma \bs E = \bs E$ is infinitesimally $\bs L_{\chi^v} \bs E=0$, which gives
\begin{align}
&\iota_{\chi^v}\bs d \bs E + \bs d \iota_{\chi^v} \bs E =
- \iota_{\chi^v} d \bs \Theta + \bs d \big( - \iota_{\chi^v} d\bs \theta + \alpha(\chi; A) \big)=0, \quad \text{by \eqref{def_class_anomaly}.} \notag \\[1.5mm]
&\hookrightarrow\quad   d\left( \iota_{\chi^v} \bs \Theta + \bs d (\iota_{\chi^v} \bs \theta) \right) =\bs d \alpha(\chi; A), \quad \text{which is } \quad
d \bs \alpha(\chi) = \bs d \alpha(\chi; A), \quad \text{by \eqref{def_bs_alpha}.} \label{alpha-bs-alpha}
\end{align}
This last relation will be used later on. 

Remark that since by \eqref{def_class_anomaly} we have on-shell $d(\iota_{\chi^v} \bs \theta)=\alpha(\chi; A)$, if the Noether current was defined the usual way it would have an anomalous conservation law. As things stand, we couldn't say more. But suppose the classical anomaly is $d$-exact
\vspace{-3mm}
\begin{align}
\label{Hyp1}
\alpha(\chi; A) = d \beta(\chi; A), 
\end{align}
call this our \emph{hypothesis 1} (it is independent of  \eqref{Hyp0}, hypothesis 0),
then we can define a conserved Noether current~by 
\begin{align}
\label{J-non-inv}
J(\chi; A) = \iota_{\chi^v} \bs \theta -  \beta(\chi; A) - dQ(\chi;A). 
\end{align}
But if this current is to generate $\H$-gauge transformations it must be related to $\bs \Theta$ in such a way that the exact term cannot be arbitrary. We must indeed have, $\iota_{\chi^v} \bs \Theta = -\bs d J(\chi;A)$, that is $-\bs d (\iota_{\chi^v} \bs \theta) + \bs \alpha(\chi) = -\bs d \left( \iota_{\chi^v} \bs \theta -  \beta(\chi; A) - dQ(\chi;A) \right)$. This gives a  constraint that we take to be the \emph{definition} of $dQ(\chi;A)$:
\begin{align}
\label{def-dQ}
\bs d d Q(\chi; A) = \bs\alpha(\chi) - \bs d \beta(\chi; A). 
\end{align}

As we are ultimately interested in on-shell quantities, we can refine our definition of the current so as to express it in terms of the field equations. Since by hypothesis we have   $R^\star_\gamma \bs E =\bs E$, we can write $\bs E= E\big(\bs d A; A \big)=~\t L\big(\bs d A; [\Omega] \big)$ where $[\Omega]$ denotes collectively tensorial terms. Then $\iota_{\chi^v}\bs E=E\big(D^A\chi; A\big)=\t L\big(D^A\chi; [\Omega] \big)=d\t L(\chi; [\Omega])- \t L\big(\chi; D^A[\Omega]\big)=d E(\chi; A)-\t L\big(\chi; D^A[\Omega]\big) $. But under  hypothesis 1, \eqref{def_class_anomaly} is: $-\iota_{\chi^v}\bs E = d\big(\iota_{\chi^v}\bs \theta - \beta(\chi;A) \big)$. Thus $\t L\big(\chi; D^A[\Omega]\big) \equiv 0$ since it is linear in \emph{underived}
$\chi$. We have then $-d E(\chi; A)=d\big(\iota_{\chi^v}\bs \theta - \beta(\chi;A) \big)$, which integrates into
\begin{align}
\label{def-dQ-tilde}
\iota_{\chi^v}\bs \theta - \beta(\chi;A)= - E(\chi; A) +  d \t Q(\chi; A).
\end{align}
We take this equation to \emph{define} $d \t Q(\chi; A)$. Then, we have the alternative form of the Noether current, 
\begin{align}
\label{J-non-inv-bis}
J(\chi; A) =   d \t Q(\chi; A) - dQ(\chi;A) - E(\chi; A),
\end{align}
from which appears immediately that it is $d$-exact on-shell. The corresponding Noether charge is then, 
\begin{align}
\label{Noether-Charge-non-inv}
Q_\Sigma(\chi; A) \defeq \int_\Sigma J(\chi;A)=  \int_{\d \Sigma} \big( \t Q(\chi; A) - Q(\chi;A) \big) - \int_\Sigma E(\chi; A). 
\end{align}
It is immediate that in the absence of anomaly, \eqref{J-non-inv-bis}-\eqref{Noether-Charge-non-inv} reduce to 
\eqref{Noether-current-invariant}-\eqref{Noether-charge-invariant}.
\medskip

\noindent {\bf Poisson bracket: } Defining as usual the presymplectic form as $\bs\Theta_\Sigma\defeq\int_\Sigma \bs \Theta$, we have by construction of our current and charge: $\iota_{\chi^v} \bs \Theta_\Sigma = -\bs d Q_\Sigma(\chi; A)$. We can thus define the Poisson bracket of charges the usual way: 
$\big\{Q_\Sigma(\chi; A),  Q_\Sigma(\eta; A)\big\}\defeq \bs\Theta_\Sigma \big(\chi^v, \eta^v \big)=\int_\Sigma \bs \Theta\big(\chi^v, \eta^v \big)$. 
To see if it is well-behaved, let us compute
\begin{align}
\bs \Theta\big( \chi^v, \eta^v\big)&= \iota_{\eta^v}\big( \iota_{\chi^v} \bs \Theta \big) = - \iota_{\eta^v} \left( \bs d (\iota_{\chi^v} \bs \theta) - \bs \alpha(\chi) \right) 
					= -\bs L_{\eta^v} \iota_{\chi^v} \bs \theta + \iota_{\eta^v} \bs \alpha(\chi), \notag \\
					&= -\iota_{\chi^v} \bs L_{\eta^v} \bs \theta - \iota_{[\eta^v, \,\chi^v]}\bs \theta + \iota_{\eta^v} \bs \alpha(\chi)
					=-\iota_{\chi^v} \bs \alpha(\eta) + \iota_{\eta^v} \bs \alpha(\chi) + \iota_{[\chi, \eta]^v}\bs \theta, \qquad \text{by \eqref{H-eq-theta-non-inv}}. \label{PB-intermediaire}
\end{align}
To make progress, we can take advantage of elements of twisted geometry.
Adapting the discussion of  section \ref{Quantum gauge anomalies} to classical anomalies, were we to endow $\A$ with a twisted connection $\bs{\t A}$, it would satisfy by definition $\bs{\t A}_{|A}(\chi^v_{|A})=\alpha(\chi, A)$.  Its curvature $\bs{\t F}$ would give us a consistency condition 
\begin{align*}
\bs{\t F}_{|A} \big(\chi^v_{A}, \eta^v_{A} \big)=\bs d \bs{\t A}_{|A}\big(\chi^v_{A}, \eta^v_{A} \big) 
				&=\chi^v \big( \bs{\t A}_{|A}(\eta^v_{A}) \big) -{\eta^v} \big( \bs{\t A}_{|A}(\chi^v_{A}) \big) - \iota_{[\chi^v,\, \eta^v]}\bs{\t A}, \\
				0&=\iota_{\chi^v} \bs d \big( \bs{\t A}_{|A}(\eta^v_{|A}) \big) -\iota_{\eta^v} \bs d \big( \bs{\t A}_{|A}(\chi^v_{|A}) \big) - \iota_{[\chi,\, \eta]^v}\bs{\t A}.\\[1.5mm]
				\hookrightarrow&\quad \iota_{\chi^v} \bs d \alpha(\chi; A) -\iota_{\eta^v} \bs d \alpha(\eta; A) =\alpha\big([\chi, \eta];A \big).
\end{align*}
By the identity \eqref{alpha-bs-alpha} and hypothesis 1, eq.\eqref{Hyp1}, this gives us $d\left( \iota_{\chi^v} \bs\alpha(\eta) - \iota_{\eta^v} \bs\alpha(\chi)  \right)=d\beta\big([\chi, \eta]; A \big)$ which integrates~as
\begin{align}
\label{def-dA}
 \iota_{\chi^v} \bs\alpha(\eta) - \iota_{\eta^v} \bs\alpha(\chi)= \beta\big([\chi, \eta]; A \big) - d \mathscr{A}\big( \lfloor\, \chi, \eta \rfloor ;A\big),
\end{align}
where $d\mathscr{A}\big( \lfloor\, \chi, \eta \rfloor ;A\big)$ is antisymmetric in $\chi$ and $\eta$, as the notation suggests, and we take this equation as defining it.
Injecting \eqref{def-dA} into \eqref{PB-intermediaire} and using our first definition \eqref{J-non-inv} of the Noether current, we get
\begin{align}
\label{vert-double-Theta}
\bs \Theta\big( \chi^v, \eta^v\big)=d \mathscr{A}\big( \lfloor\, \chi, \eta \rfloor ;A\big) - \beta\big([\chi, \eta]; A \big) + \iota_{[\chi, \eta]^v}\bs \theta 
						 = d \mathscr{A}\big( \lfloor\, \chi, \eta \rfloor ;A\big) + J\big([\chi, \eta] ;A\big) + dQ\big([\chi, \eta];A \big).
\end{align}
The Poisson bracket of Noether charges induced by the presymplectic 2-form $\bs \Theta_\Sigma$ is then
\begin{align}
\label{PB-anomalous}
\big\{ Q_\Sigma(\chi; A),  Q_\Sigma(\eta; A) \big\} &=  Q_\Sigma([\chi, \eta] ; A) + \mathscr{C}(\chi, \eta)  \\[1mm]
									&\text{with } \quad  \mathscr{C}(\chi, \eta) \defeq \int_{\d\Sigma}  \mathscr{A}\big( \lfloor\, \chi, \eta \rfloor ;A\big) + Q\big([\chi, \eta];A \big). \notag
\end{align}
It is thus anomalous, and the anomaly is a boundary term (in coherence with \eqref{Hyp1}). More is true: the Lie algebra of Noether charges $\big( Q_\Sigma(\ \, ; A), \big\{\, ,\, \big\}\big)$ is a central extension of Lie$\H$. We collect  basic definitions about Lie algebra extensions in appendix \ref{Lie algebras extensions}. As reminded there, to prove the above statement one just need to show that $\mathscr C$ is a 2-cocycle on Lie$\H$, i.e. that in addition to being antisymmetric in $\chi$ and $\eta$ - which it obviously is - it satisfies $ \mathscr{C}(\chi, [\eta, \zeta])+\mathscr{C}(\eta, [\zeta, \chi]) +\mathscr{C}(\zeta, [\chi, \eta]) \equiv 0$. This we show in appendix \ref{PB central extension}.

It is still the case that through the action of the Poisson bracket \eqref{PB-anomalous}, Noether charges are  generators of $\H$-gauge transformations: $\big\{Q_\Sigma(\chi; A),  \ \ \big\}$ acting on  functionals on $\A$ generates the infinitesimal action by $\chi^v$, see appendix \ref{Noether charges as generators of gauge transformations}. 

\medskip

\noindent To recapitulate, given the $c$-equivariant Lagrangian $L$ of a theory, the different stages of  the algorithm are:
\begin{enumerate}[label=\roman*, leftmargin=*, itemsep=1pt]
\item Check hypothesis 0, eq.  \eqref{Hyp0}.
\item Check hypothesis 1, eq.  \eqref{Hyp1}, and determine $\beta(\chi; A)$.
\item Determine $\bs\alpha(\chi)$ by \eqref{H-eq-theta-non-inv}. 
\item Determine $dQ(\chi; A)$ by \eqref{def-dQ} and $d\t Q(\chi; A)$ by \eqref{def-dQ-tilde} $\Rightarrow$ build $J(\chi; A)$ and $Q_\Sigma(\chi; A)$.
\item Determine $d \mathscr{A}\big( \lfloor\, \chi, \eta \rfloor ;A\big)$ by \eqref{def-dA} $\Rightarrow$ obtain the anomalous Poisson bracket of Noether charges. 
\end{enumerate}
Remark that actually we don't necessarily need the finite $c$-equivariances of the Lagrangian $L$  and $\bs\theta$, we only need the infinitesimal versions. 
\bigskip

\noindent {\bf Field-dependent gauge transformations: }  As always, to get the $\bs\H$-gauge transformations of $\bs E$, $\bs\theta$ and $\bs\Theta$ we only need their $\H$-equivariance and verticality properties. For $\bs E$ the situation is exactly as in section \ref{Presymplectic structure for invariant Lagrangians}, and the result identical to eq. \eqref{Field-depGT-FieldEq}: 
 \begin{align}
\label{Field-depGT-E-non-inv}
{\bs E^{\bs\gamma}}_{|A}(\bs X_A) \defeq&\ \big( \bs\Psi^\star \bs E \big)_{|A}(\bs X_A) = \ldots  
					=  \bs E_{|A} \big( \bs{X}_A \big) + d E \big( \{ \bs{d\gamma\gamma}_{|A}\-(\bs X_A) \}; A \big)  \notag \\[1mm]
	 \bs E^{\bs \gamma}=&\ \bs E + dE \big(  \bs{d\gamma\gamma}\- ;  A \big).
\end{align}
This is confirmation that the theory is well-behaved, as the action of $\bs\H$ does not takes us off-shell. 

Now, under hypothesis 0 and 1, $\bs\theta$ has $\H$-equivariance given by \eqref{H-eq-theta-non-inv} and its verticality can be taken to be determined by \eqref{def-dQ-tilde}. Thus, 
\begin{align}
\label{Field-depGT-theta-non-inv}
{\bs\theta^{\bs\gamma}}_{|A}(\bs X_A) &\defeq \big( \bs\Psi^\star \bs \theta \big)_{|A}(\bs X_A) = \bs \theta_{A^{\bs\gamma}}\big(\bs\Psi_\star \bs X_A\big)
					=   \bs \theta_{|A^{\bs\gamma}} \left(R_{\bs{\gamma}(A)\star} \left( \bs{X}_A + \left\{ \bs{d}\bs{\gamma} {\bs{\gamma}\- }_{|A}(\bs{X}_A)\right\}^v_A \right) \right) \quad \text{ by \eqref{Pushforward-X-inf}}, \notag \\
			&= R^\star_{\bs{\gamma}(A)} \bs \theta_{|A^{\bs\gamma}}  \left( \bs{X}_A + \left\{ \bs{d}\bs{\gamma} {\bs{\gamma}\- }_{|A}(\bs{X}_A)\right\}^v_A \right),  \notag \\
			&= \left(\, \bs \theta_{|A} + \bs b\big(\bs\gamma(A)\big)^{}_{}\, \right)_{|A}\left( \bs{X}_A + \left\{ \bs{d}\bs{\gamma} {\bs{\gamma}\- }_{|A}(\bs{X}_A)\right\}^v_A \right) \quad \text{ by \eqref{H-eq-theta-non-inv}}, \notag \\
			&=  \bs \theta_{|A} \big( \bs{X}_A \big) - E\big( \bs{d\gamma\gamma}\-_{|A}(\bs X_A); A \big)+ d \t Q \big( \{ \bs{d\gamma\gamma}_{|A}\-(\bs X_A) \}; A \big) + \beta \big( \{ \bs{d\gamma\gamma}_{|A}\-(\bs X_A) \}; A \big) 
			 \notag \\ 
			& \hspace{4cm} + \bs b\big(\bs\gamma(A)\big)_{|A}(\bs X_A) + b\left( D^A \big\{ \bs{d\gamma\gamma}\-_{|A}(\bs X_A )\big\}; \bs\gamma(A) \right)                     \quad \text{ by \eqref{def-dQ-tilde}}, \notag \\[1.5mm]
		\text{that is \quad} \bs\theta^{\bs \gamma}&= \bs \theta + d\t Q \big(  \bs{d\gamma\gamma}\- ;  A \big) - E \big( \bs{d\gamma\gamma}\- ; A \big) 
												+  \beta \big(  \bs{d\gamma\gamma}\- ; A \big) + \bs b\big(\bs\gamma(A)\big)_{|A} +  b\big( D^A \big\{ \bs{d\gamma\gamma}\-\big\}; \bs\gamma(A) \big). 
\end{align}
It is clear how, in the absence of anomaly, this result reduces to \eqref{Field-depGT-presymp-pot-current}. From this, the $\bs\H$-gauge transformation of the presymplectic potential follows, 
\begin{align}
\label{Field-depGT-presym pot-non-inv}
\bs\theta_\Sigma^{\bs \gamma}&= \bs \theta_\Sigma + \int_{\d\Sigma}\t Q \big(  \bs{d\gamma\gamma}\- ;  A \big) + \int_\Sigma - E \big( \bs{d\gamma\gamma}\- ; A \big) 
												+  \beta \big(  \bs{d\gamma\gamma}\- ; A \big) + \bs b\big(\bs\gamma(A)\big)_{|A} +  b\left( D^A \big\{ \bs{d\gamma\gamma}\-\big\}; \bs\gamma(A) \right). 
\end{align}
Unfortunately we see that even on-shell and with boundary conditions, it is not basic because of the anomalous terms. So it does not induce a well-defined symplectic potential on the physical phase space $\S/\H$. 
\medskip

\clearpage

Turning our attention to $\bs\Theta$, under hypothesis 0 and 1 its verticality is determined by definition (of the current) to be $\iota_{\chi^v} \bs \Theta = -\bs d J(\chi;A)$. 
It follows that $\bs L_{\chi^v} \bs \Theta = \iota_{\chi^v}  \cancel{\bs d \bs \Theta} + \bs d \iota_{\chi^v} \bs \Theta = - \bs d^2 J(\chi; A) =0$. From which we infer the $\H$-equivariance $R^\star_\gamma \bs \Theta = \bs \Theta$. Then,
\begin{align}
\label{1-bis}
{\bs\Theta^{\bs\gamma}}_{|A}\big(\bs X_A, \bs Y_A \big)\defeq&\, \big(  \bs\Psi^\star \bs\Theta \big)_{|A}(\bs X_A, \bs Y_A) = \bs\Theta_{|A^{\bs\gamma}} \left(\bs\Psi_\star \bs X_A, \bs\Psi_\star\bs Y_A  \right), \notag\\*[1mm]
			=&\, \bs\Theta_{|A^{\bs\gamma}} \left( R_{\bs{\gamma}(A)\star} \left( \bs{X}_A + \left\{ \bs{d}\bs{\gamma} {\bs{\gamma}\- }_{|A}(\bs{X}_A)\right\}^v_A   \right) ,  R_{\bs{\gamma}(A)\star} \left( \bs Y_A + \left\{ \bs{d}\bs{\gamma} {\bs{\gamma}\- }_{|A}(\bs Y_A)\right\}^v_A   \right)  \right), \notag \\[1mm]
			=&\, R^\star_{\bs\gamma(A)} \bs \Theta_{|A^{\bs\gamma}} \left( \bs{X}_A + \left\{ \bs{d}\bs{\gamma} {\bs{\gamma}\- }_{|A}(\bs{X}_A)\right\}^v_A,     \bs{Y}_A + \left\{ \bs{d}\bs{\gamma} {\bs{\gamma}\- }_{|A}(\bs{Y}_A)\right\}^v_A \right) ,\notag\\[1mm]
			=&\, \bs\Theta_{|A} \big(\bs X_A, \bs Y_A \big) + \bs\Theta_{|A}\left( \bs X_A,  \left\{ \bs{d}\bs{\gamma} {\bs{\gamma}\- }_{|A}(\bs{Y}_A)\right\}^v_A\right) +  \bs\Theta_{|A} \left(  \left\{ \bs{d}\bs{\gamma} {\bs{\gamma}\- }_{|A}(\bs{X}_A)\right\}^v_A, \bs Y_A  \right)   \notag\\
			&\hspace{5cm} +  \bs\Theta_{|A} \left(  \left\{ \bs{d}\bs{\gamma} {\bs{\gamma}\- }_{|A}(\bs{X}_A)\right\}^v_A,  \left\{ \bs{d}\bs{\gamma} {\bs{\gamma}\- }_{|A}(\bs{Y}_A)\right\}^v_A \right). 
\end{align}
To recognise some structure, let us consider intermediary results.
Using the second definition \eqref{J-non-inv-bis} of the current, we can rewritte \eqref{vert-double-Theta} as, 
\begin{align*}
\bs \Theta\big( \chi^v, \eta^v\big) &= d \mathscr{A}\big( \lfloor\, \chi, \eta \rfloor ;A\big) + J\big([\chi, \eta] ;A\big) + dQ\big([\chi, \eta];A \big), \\
						 &=  d \mathscr{A}\big( \lfloor\, \chi, \eta \rfloor ;A\big) + d\t Q\big([\chi, \eta];A \big) - E\big([\chi, \eta] ;A\big). 
\end{align*}
So the last term of \eqref{1-bis} is 
\begin{align}
\bs\Theta_{|A} \left(  \left\{ \bs{d}\bs{\gamma} {\bs{\gamma}\- }_{|A}(\bs{X}_A)\right\}^v_A,  \left\{ \bs{d}\bs{\gamma} {\bs{\gamma}\- }_{|A}(\bs{Y}_A)\right\}^v_A \right)&=
				 d \mathscr{A}\left( \lfloor\, {\bs{d\gamma\gamma}\-}_{|A}(\bs X_A), {\bs{d\gamma\gamma}\-}_{|A}(\bs Y_A) \rfloor ;A\right)  \notag\\
					& \quad      + d\t Q\left(\big[{\bs{d\gamma\gamma}\-}_{|A}(\bs X_A), {\bs{d\gamma\gamma}\-}_{|A}(\bs Y_A) \big];A \right) \notag\\
      					& \quad	  - E\left(\big[{\bs{d\gamma\gamma}\-}_{|A}(\bs X_A), {\bs{d\gamma\gamma}\-}_{|A}(\bs Y_A) \big] ;A\right). 
\end{align}
We remind that 
\begin{align}
\label{identity}
\big[ \bs{d\gamma\gamma}\-_{|A}(\bs X_A), \bs{d\gamma\gamma}\-_{|A}(\bs Y_A) \big] &= \bs d \big( \bs{d\gamma\gamma}\-\big)_{|A} (\bs X_A, \bs Y_A),  \notag\\
																  &= \bs X \cdot \big\{  \bs{d\gamma\gamma}\-_{|\munderline{red}{A}} (\bs Y_{\munderline{red}{A}})\big\} 
																  - \bs Y \cdot \big\{  \bs{d\gamma\gamma}\-_{|\munderline{red}{A}} (\bs X_{\munderline{red}{A}})\big\}   
																  -  \bs{d\gamma\gamma}\-_A\big( \big[\bs X, \bs Y\big]_A \big),
\end{align}
Also, by Kozsul formula, 
\begin{align*}
\bs d \left( E \big( \bs{d\gamma\gamma}\-_{|A}; A \big) \right)(\bs X_A, \bs Y_A)&= \bs X \cdot  E \big(\bs{d\gamma\gamma}\-_{|\munderline{blue}{A}}(\bs Y_{\munderline{blue}{A}}); \munderline{blue}{A} \big) 
															    - \bs Y \cdot  E \big(\bs{d\gamma\gamma}\-_{|\munderline{blue}{A}}(\bs X_{\munderline{blue}{A}}); \munderline{blue}{A} \big) 
															- E\big( \bs{d\gamma\gamma}\-_{|A}([\bs X_A, \bs Y_A]); A\big), \\
				&= \bs X \cdot  E \big(\bs{d\gamma\gamma}\-_{|A}(\bs Y_A); \munderline{ForestGreen}{A} \big)  +  E \big( \bs X \cdot  \bs{d\gamma\gamma}\-_{|\munderline{red}{A}}(\bs Y_{\munderline{red}{A}}); A \big) \\
				&\quad - \bs Y \cdot  E \big(\bs{d\gamma\gamma}\-_{|A}(\bs X_A); \munderline{ForestGreen}{A} \big)  +  E \big( \bs Y \cdot  \bs{d\gamma\gamma}\-_{|\munderline{red}{A}}(\bs X_{\munderline{red}{A}}); A \big) \\
				& \hspace{4cm}    - E\big( \bs{d\gamma\gamma}\-_{|A}([\bs X_A, \bs Y_A]); A\big), 
\end{align*}
where the second equality follows from the linearity of E in its first argument. In exact analogy we have
\begin{align*}
\bs d \left( d\t Q \big( \bs{d\gamma\gamma}\-_{|A}; A \big) \right)(\bs X_A, \bs Y_A)&= 
			\bs X \cdot  d\t Q \big(\bs{d\gamma\gamma}\-_{|A}(\bs Y_A); \munderline{ForestGreen}{A} \big)  +  d\t Q \big( \bs X \cdot  \bs{d\gamma\gamma}\-_{|\munderline{red}{A}}(\bs Y_{\munderline{red}{A}}); A \big) \\
			&\quad - \bs Y \cdot  d\t Q \big(\bs{d\gamma\gamma}\-_{|A}(\bs X_A); \munderline{ForestGreen}{A} \big)  +  d\t Q \big( \bs Y \cdot  \bs{d\gamma\gamma}\-_{|\munderline{red}{A}}(\bs X_{\munderline{red}{A}}); A \big) \\
				& \hspace{4cm}    - d\t Q\big( \bs{d\gamma\gamma}\-_{|A}([\bs X_A, \bs Y_A]); A\big). 
\end{align*}
And finally, using \eqref{J-non-inv-bis}, the second term in \eqref{1-bis} is
\begin{align*}
\bs\Theta_{|A}\left( \bs X_A,  \left\{ \bs{d}\bs{\gamma} {\bs{\gamma}\- }_{|A}(\bs{Y}_A)\right\}^v_A\right) &= \iota_{\bs X} \bs d \, J\big( \bs{d\gamma\gamma}\-_{|A}(\bs Y_A); \munderline{ForestGreen}{A} \big) 
																			     = \bs X \cdot J\big( \bs{d\gamma\gamma}\-_{|A}(\bs Y_A); \munderline{ForestGreen}{A} \big), \\
															&=\bs X \cdot \left(  d\t Q \big( \bs{d\gamma\gamma}\-_{|A}(\bs Y_A); \munderline{ForestGreen}{A} \big) 
															                               - E\big( \bs{d\gamma\gamma}\-_{|A}(\bs Y_A); \munderline{ForestGreen}{A} \big)
															                               - dQ\big( \bs{d\gamma\gamma}\-_{|A}(\bs Y_A); \munderline{ForestGreen}{A} \big) \right). 
\end{align*}
Idem for the third term in \eqref{1-bis}. 
Collecting all results, we get 

\begin{align}
{\bs\Theta^{\bs\gamma}}_{|A}\big(\bs X_A, \bs Y_A \big) &= \bs\Theta_{|A}\big(\bs X_A, \bs Y_A \big)  \notag \\*
						&\quad + \bs X \cdot \bigg(  \underline{{d\t Q \big( \bs{d\gamma\gamma}\-_{|A}(\bs Y_A); \munderline{ForestGreen}{A} \big)}  }_{\,_\text I}
															                               - \underline{{ E\big( \bs{d\gamma\gamma}\-_{A}(\bs Y_A); \munderline{ForestGreen}{A} \big) }}_{\,_\text{II}}
															                               - dQ\big( \bs{d\gamma\gamma}\-_{|A}(\bs Y_A); \munderline{ForestGreen}{A} \big) \bigg)  \notag \\*
						&\quad - \bs Y \cdot \bigg( \underline{{ d\t Q \big( \bs{d\gamma\gamma}\-_{|A}(\bs X_A); \munderline{ForestGreen}{A} \big)  }}_{\,_\text I}
															                               - \underline{{ E\big( \bs{d\gamma\gamma}\-_{|A}(\bs X_A); \munderline{ForestGreen}{A} \big) }}_{\,_\text{II}}
															                               - dQ\big( \bs{d\gamma\gamma}\-_{|A}(\bs X_A); \munderline{ForestGreen}{A} \big) \bigg)	\notag \\*
						&\quad + d \mathscr{A}\left( \lfloor {\bs{d\gamma\gamma}\-}_{|A}(\bs X_A), {\bs{d\gamma\gamma}\-}_{|A}(\bs Y_A) \rfloor ;A\right) 	\notag\\*
						&\quad+ \underline{ d\t Q\big( \bs X \cdot \big\{  \bs{d\gamma\gamma}\-_{|\munderline{red}{A}} (\bs Y_{\munderline{red}{A}})\big\} 
							 - \bs Y \cdot \big\{  \bs{d\gamma\gamma}\-_{|\munderline{red}{A}} (\bs X_{\munderline{red}{A}})\big\}    \big)
					   		           - d\t Q\left( \bs{d\gamma\gamma}\-_A\big( \big[\bs X, \bs Y\big]_A \big) \right)	 }_{\,_\text I}  \notag\\
						&\quad- \underline{ E\big( \bs X \cdot \big\{  \bs{d\gamma\gamma}\-_{|\munderline{red}{A}} (\bs Y_{\munderline{red}{A}})\big\} 
							 - \bs Y \cdot \big\{  \bs{d\gamma\gamma}\-_{|\munderline{red}{A}} (\bs X_{\munderline{red}{A}})\big\}    \big)
					   		           + E \left( \bs{d\gamma\gamma}\-_A\big( \big[\bs X, \bs Y\big]_A \big) \right) }_{\,_\text{II}},	 \notag \\[1.5mm]
				&= 	\bs\Theta_{|A}\big(\bs X_A, \bs Y_A \big) + \bs d \bigg(  \  \underline{{d\t Q \big( \bs{d\gamma\gamma}\-_{|A}; A \big)}}_{\,_\text I} 
															-  \underline{{E \big( \bs{d\gamma\gamma}\-_{|A}; A \big)}} _{\,_\text{II}}        \bigg)\, (\bs X_A, \bs Y_A)	\notag\\	
				&\hspace{3cm} 		+ d \mathscr{A}\left( \lfloor {\bs{d\gamma\gamma}\-}_{|A}(\bs X_A), {\bs{d\gamma\gamma}\-}_{|A}(\bs Y_A) \rfloor ;A\right) \notag	\\
				 &\hspace{3cm}              - \bs X \cdot dQ\big( \bs{d\gamma\gamma}\-_{|A}(\bs Y_A); \munderline{ForestGreen}{A} \big)		
				                        		      +  \bs Y \cdot dQ\big( \bs{d\gamma\gamma}\-_{|A}(\bs X_A); \munderline{ForestGreen}{A} \big). 			              \label{2-bis}       
\end{align}
It is possible to further simplify this result by noticing that, 
\begin{align*}
\bs d \left( dQ \big( \bs{d\gamma\gamma}\-_{|A}; A \big) \right)(\bs X_A, \bs Y_A)&= \bs X \cdot  dQ \big(\bs{d\gamma\gamma}\-_{|\munderline{blue}{A}}(\bs Y_{\munderline{blue}{A}}); \munderline{blue}{A} \big) 
															    - \bs Y \cdot  dQ \big(\bs{d\gamma\gamma}\-_{|\munderline{blue}{A}}(\bs X_{\munderline{blue}{A}}); \munderline{blue}{A} \big) 
															- dQ\big( \bs{d\gamma\gamma}\-_{|A}([\bs X_A, \bs Y_A]); A\big), \\
				&= \bs X \cdot  dQ \big(\bs{d\gamma\gamma}\-_{|A}(\bs Y_A); \munderline{ForestGreen}{A} \big)  +  dQ \big( \bs X \cdot  \bs{d\gamma\gamma}\-_{|\munderline{red}{A}}(\bs Y_{\munderline{red}{A}}); A \big) \\
				&\quad - \bs Y \cdot  dQ \big(\bs{d\gamma\gamma}\-_{|A}(\bs X_A); \munderline{ForestGreen}{A} \big)  +  dQ \big( \bs Y \cdot  \bs{d\gamma\gamma}\-_{|\munderline{red}{A}}(\bs X_{\munderline{red}{A}}); A \big) \\
				& \hspace{4.5cm}    - dQ\big( \bs{d\gamma\gamma}\-_{|A}([\bs X_A, \bs Y_A]); A\big), \\[1mm]
				&=\bs X \cdot  dQ \big(\bs{d\gamma\gamma}\-_{|A}(\bs Y_A); \munderline{ForestGreen}{A} \big) - \bs Y \cdot  dQ \big(\bs{d\gamma\gamma}\-_{|A}(\bs X_A); \munderline{ForestGreen}{A} \big) 
				    + dQ\left(\bs d \big(\bs{d\gamma\gamma}\-_{|A} (\bs X_A, \bs Y_A)\big) ; A\right), 
\end{align*}
by \eqref{identity} in the last step. This allows to re-express the last two terms in \eqref{2-bis}, so that
\begin{align*}
{\bs\Theta^{\bs\gamma}}_{|A}\big(\bs X_A, \bs Y_A \big) &= \bs\Theta_{|A}\big(\bs X_A, \bs Y_A \big) + \bs d \left(  \, d\t Q \big( \bs{d\gamma\gamma}\-_{|A}; A \big)
															-  E \big( \bs{d\gamma\gamma}\-_{|A}; A \big)   \right)\, (\bs X_A, \bs Y_A)	\notag\\	
				&\hspace{3cm} 		+ d \mathscr{A}\left( \lfloor {\bs{d\gamma\gamma}\-}_{|A}(\bs X_A), {\bs{d\gamma\gamma}\-}_{|A}(\bs Y_A) \rfloor ;A\right) \notag	\\
				&\hspace{3cm} 		+  dQ\left(\bs d \big(\bs{d\gamma\gamma}\-_{|A} \big) ; A\right) (\bs X_A, \bs Y_A) - \bs d \left( dQ \big( \bs{d\gamma\gamma}\-_{|A}; A \big) \right)(\bs X_A, \bs Y_A).
\end{align*}
That is finally, 
\begin{align}
\bs\Theta^{\bs\gamma} &= \bs\Theta + \bs d \left(  \, d\t Q \big( \bs{d\gamma\gamma}\-; A \big)  -  dQ \big( \bs{d\gamma\gamma}\-; A \big)    -  E \big( \bs{d\gamma\gamma}\-; A \big)   \right) \notag \\[1mm]
					& \hspace{3cm} + d \mathscr{A}\left( \lfloor {\bs{d\gamma\gamma}\-}, {\bs{d\gamma\gamma}\-} \rfloor ;A\right) + dQ\left(\bs d \big(\bs{d\gamma\gamma}\- \big) ; A\right). 
\end{align}
We recognise the expression of the Noether current (as given by \eqref{J-non-inv-bis}) and the anomalous terms appearing in the Poisson bracket \eqref{PB-anomalous}. It is manifest how this result generalises \eqref{Field-depGT-Theta}. it follows that the $\bs\H$-gauge transformation of the presymplectic 2-form is
\begin{align}
\label{Field-depGT-Theta-non-inv}
\bs\Theta_\Sigma^{\bs\gamma} &= \bs\Theta_\Sigma   +  \bs d\, Q_\Sigma \big( \bs{d\gamma\gamma}\-; A \big) + \mathscr{C}\big( \bs{d\gamma\gamma}\-, \bs{d\gamma\gamma}\-\big), \\[1mm]
						&= \bs\Theta_\Sigma  \, + \int_{\d\Sigma}  \bs d \left(\t Q \big( \bs{d\gamma\gamma}\-; A \big)  -  Q \big( \bs{d\gamma\gamma}\-; A \big)  \right) 
								+ \mathscr{A}\big( \lfloor \bs{d\gamma\gamma}\-, \bs{d\gamma\gamma}\- \rfloor ;A\big) + Q\big(\bs d (\bs{d\gamma\gamma}\- ) ; A\big) 
							\,    - \int_\Sigma \bs d E \big( \bs{d\gamma\gamma}\-; A \big), \notag
\end{align}
which clearly generalises \eqref{Field-depGT-presymp-form}.
We have the pleasant surprise that on-shell and given appropriate boundary conditions, $\bs\Theta_\Sigma$ is basic, $\bs\H$-invariant, and thus descends as a symplectic form on the physical phase space $\S/\H$. This,~despite the failure of the presymplectic potential $\bs\theta_\Sigma$ to do so. 
 \bigskip
 
 We now apply the above general results to obtain  the presymplectic structures of  Non-Abelian 3D Chern-simons theory and  3D-$\CC$-gravity when translation are gauged. 
 Here again  the repetitiveness of the exposition is designed to illustrate the systematicity of the proposed algorithm, and so that each example can be appreciated independently of the others. 


\subsubsection{3D non-Abelian Chern-Simons theory}
\label{3D-non-Abelian Chern-Simons theory}

The Lagrangian of the theory is $L_\text{{\tiny CS}}(A)=\Tr\big(AdA + \tfrac{2}{3}A^3\big)$, and as previewed in paragraph \ref{3D-CS-theory}, it is $c$-equivariant
 \begin{align}
 \big(R^\star_\gamma L_\text{{\tiny CS}}\big)(A) &= L_\text{{\tiny CS}}(A) + c(A,\gamma),  \notag\\
 						   &= L_\text{{\tiny CS}}(A) + \Tr \left(  d\big(  \gamma d\gamma\- A  \big) - \tfrac{1}{3}\big(  \gamma\-d\gamma \big)^3  \right), \quad \gamma \in \H=\SU(n).
 \end{align}
To find the presymplectic structure of the theory we deploy the algorithm detailed above. First, one shows that 
\begin{align}
\bs d L_\text{{\tiny CS}}= \bs E_\text{{\tiny CS}} + d\bs \theta_\text{{\tiny CS}}&=  E_\text{{\tiny CS}}(\bs d A; A) + d  \theta_\text{{\tiny CS}}(\bs d A; A), \notag\\*
														   &= 2\Tr(\bs dA\, F) + d\Tr(\bs d A\, A).  
\end{align}
It is manifest that $R^\star_\gamma \bs E_\text{{\tiny CS}} = \bs E_\text{{\tiny CS}}$, so hypothesis 0 
is checked. 
It then follows that we must have  $\bs d c(\ \, ,\gamma)=d \bs{b}(\gamma)$ with $\bs{b}(\gamma)$ linear in $\bs d A$ so that, 
\begin{align}
R^\star_\gamma d \bs \theta_\text{{\tiny CS}} &=  d\bs \theta_\text{{\tiny CS}} + \bs d c(A; \gamma) = d\bs \theta_\text{{\tiny CS}} + \Tr \left(d \big( \gamma d\gamma\- \bs d A\big) \right), \notag\\[1mm]
\hookrightarrow 
R^\star_\gamma  \bs \theta_\text{{\tiny CS}} &=  \bs \theta_\text{{\tiny CS}} + \bs b (\gamma) = \bs \theta_\text{{\tiny CS}} + b (\bs d A; \gamma) = \bs \theta_\text{{\tiny CS}} + \Tr  \big( \gamma d\gamma\- \bs d A\big).
\end{align}
As is easily checked directly. From this we find 
\begin{align}
\bs \alpha(\chi) \defeq \bs L_{\chi^v} \bs \theta_\text{{\tiny CS}} = \tfrac{d}{d\tau} \bs b(e^{\tau \chi})\big|_{\tau=0}= \Tr(\bs d A d \chi). 
\end{align}
 The classical anomaly is
 \begin{align}
\alpha(\chi, A) \defeq \bs{L}_{\chi^v} L_\text{{\tiny CS}}=\tfrac{d}{d\tau} c\big(A; e^{\tau \chi}\big)\big|_{\tau=0}= d\Tr(A d\chi) \rdefeq d \beta(\chi; A).
\end{align}
It is $d$-exact, so  hypothesis 1 is verified. We also checked that we have indeed $d\bs \alpha(\chi)= \bs d \alpha(\chi; A)$.  We next identify the quantity $dQ(\chi; A)$, which is by \eqref{def-dQ},
\begin{align}
\bs d dQ(\chi; A) \defeq \bs \alpha(\chi) - \bs d\beta(\chi; A)= \Tr(\bs d A d\chi) - \bs d \Tr(A d\chi) =0.
\end{align}
And the quantity $d\t Q(\chi; A)$ is by \eqref{def-dQ-tilde},
\begin{align}
d\t Q(\chi; A)\defeq&\, \iota_{\chi^v} \bs \theta - \beta(\chi; A) + E_\text{{\tiny CS}}(\chi; A), \notag\\
			=&\, \Tr(D^A\chi\, A) - \Tr(A d\chi) + 2\Tr(\chi F), \notag\\
			=& \Tr\big( 2d\chi\, A +\cancel{ [A, \chi ]A} + 2\chi dA - \cancel{\chi [A, A]}  \big) = 2d\Tr(\chi A). 
\end{align}
Finally, by \eqref{def-dA} we obtain, 
\begin{align}
d \mathscr{A}\big( \lfloor\, \chi, \eta \rfloor ;A\big)\defeq&\, - \iota_{\chi^v} \bs\alpha(\eta) + \iota_{\eta^v} \bs\alpha(\chi)+ \beta\big([\chi, \eta]; A \big), \notag \\
										=&\, -\Tr(D^A\chi d\eta) + \Tr(D^A\eta d\chi) + \Tr(Ad[\chi, \eta]), \notag\\
										=&\, -\Tr\big( d\chi d\eta + [A, \chi]d\eta - d\eta d\chi - [A,\eta]d\chi - A d [\chi, \eta] \big), \notag\\
										=&\, -\Tr\big( 2d\chi d\eta - \underline{A[d\eta, \chi] + A[d\chi, \eta]} - \cancel{A d [\chi, \eta]} \big) = d 2\Tr(d\chi\, \eta). 
\end{align}
We have now all the necessary quantities  to obtain the Noether current, which by  \eqref{J-non-inv-bis} is 
\begin{align}
\label{current-CS}
J_\text{{\tiny CS}}(\chi; A) &=  d \t Q(\chi; A) - dQ(\chi;A) - E_\text{{\tiny CS}}(\chi; A), \notag\\
				      &=  2d\Tr(\chi A) - 2\Tr(\chi F). 
\end{align}
It is clearly $d$-exact on-shell. By \eqref{Noether-Charge-non-inv}, the corresponding Noether charge is 
\begin{align}
\label{charge-CS}
Q^\text{{\tiny CS}}_\Sigma(\chi; A) = \int_\Sigma  J_\text{{\tiny CS}}(\chi; A) &= \int_{\d\Sigma} 2\Tr(\chi A) - \int_\Sigma 2\Tr(\chi F), \\
						     								  &= \int_{\d\Sigma} 2\Tr(\chi A) _{\ |\S}. \notag
\end{align}
This, we notice, is the ``extended generator'' proposed in \cite{Geiller2017} for Chern-Simons theory, see eq.(3.29)-(3.33) and eq.(D.16) there.
The Poisson bracket of charges is by \eqref{PB-anomalous},
\begin{align}
\label{PB-CS}
\big\{ Q^\text{{\tiny CS}}_\Sigma(\chi; A),  Q^\text{{\tiny CS}}_\Sigma(\eta; A) \big\} &=  Q^\text{{\tiny CS}}_\Sigma([\chi, \eta] ; A) +\int_{\d\Sigma}  \mathscr{A}\big( \lfloor\, \chi, \eta \rfloor ;A\big) + Q\big([\chi, \eta];A \big), \notag\\
									   &=Q^\text{{\tiny CS}}_\Sigma([\chi, \eta] ; A) + \int_{\d\Sigma}  2\Tr(d\chi\, \eta). 
\end{align}
This reproduces the results eq.(3.32) and  eq.(D.13) of \cite{Geiller2017}. 
\medskip

Considering now the question of field-dependent $\bs \SU(n)$ gauge transformations, by \eqref{Field-depGT-E-non-inv} we have that
\begin{align}
 \bs E_\text{{\tiny CS}}^{\, \bs  \gamma}&= \bs E_\text{{\tiny CS}} + dE_\text{{\tiny CS}} \big(  \bs{d\gamma\gamma}\- ;  A \big),\notag \\*
 							&= \bs E_\text{{\tiny CS}} + 2d\Tr\big( \bs{d\gamma\gamma}\- F\big),
\end{align}
as is easily checked.
By \eqref{Field-depGT-presym pot-non-inv}, we can write the $\bs\SU(n)$-gauge transformation of  the presymplectic potential 
\begin{align*}
(\bs\theta^\text{{\tiny\,\! CS}}_\Sigma)^{\bs \gamma}&= \bs \theta^\text{{\tiny\,\!  CS}}_\Sigma + \int_{\d\Sigma} \t Q \big(  \bs{d\gamma\gamma}\- ;  A \big) + \int_\Sigma - E_\text{{\tiny CS}} \big( \bs{d\gamma\gamma}\- ; A \big) 
												+  \beta \big(  \bs{d\gamma\gamma}\- ; A \big) + \bs b\big(\bs\gamma(A)\big)_{|A} +  b\left( D^A \big\{ \bs{d\gamma\gamma}\-\big\}; \bs\gamma(A) \right), \notag  \\
									&=\bs \theta^\text{{\tiny\,\!  CS}}_\Sigma + \int_{\d\Sigma} 2\Tr\big(\bs{d\gamma\gamma}\- A\big) +  \int_\Sigma - 2 \Tr\big(\bs{d\gamma\gamma}\-F\big) 
											+ \Tr\left(A d(\bs{d\gamma\gamma}\-)\right) + \Tr\left( \bs \gamma d \bs \gamma\- \bs d A\right) + \Tr\left( \bs \gamma d \bs \gamma\- D^A\big\{ \bs{d\gamma\gamma}\- \big\} \right).
\end{align*}
We can work a bit on the last three terms, 
\begin{align*}
&\Tr\left(A d(\bs{d\gamma\gamma}\-)+ \bs \gamma d \bs \gamma\- \bs d A +  \bs \gamma d \bs \gamma\- D^A\big\{ \bs{d\gamma\gamma}\- \big\} \right) \\
&\hspace{2cm}	=\Tr \bigg(	 A d(\bs d \bs \gamma)\bs \gamma\- \!+ A \bs{d\gamma}d\gamma\-	\ +\  \bs \gamma d \bs \gamma\- \bs d A 
							\ +\  	\bs \gamma d \bs \gamma\- d\big( \bs{d\gamma\gamma}\-\big)	
							  +\hspace{-1.1cm} \underbrace{\bs \gamma d \bs \gamma\-  [A, \bs{d\gamma\gamma}\-]}_{\hspace{2mm} -[	\bs \gamma d \bs \gamma\-, \, \bs{d\gamma\gamma}\-]A\, = -d\bs \gamma \bs{d\gamma}\-A + \bs{d\gamma}d\bs \gamma\- A}	\hspace{-1.1cm} \bigg), \\
&\hspace{2cm}= \Tr \bigg( \bs \gamma d \bs \gamma\- d\big( \bs{d\gamma\gamma}\-\big)	+ \bs \gamma d \bs \gamma\- \bs d A  
				- \underbrace{   \left( d(\bs d \bs \gamma)\bs \gamma\- +  \bs{d\gamma}d\bs \gamma\-  + d\bs \gamma \bs{d\gamma}\-\right)}_{=\bs \gamma \bs d d \bs\gamma\-, \text{ by } \bs d d(\bs{\gamma\gamma}\-)=0 }A +	\bs{d\gamma}d\bs \gamma\- A \bigg),\\
&\hspace{2cm}= \Tr \bigg( \bs \gamma d \bs \gamma\- d\big( \bs{d\gamma\gamma}\-\big)	+ \bs \gamma d \bs \gamma\- \bs d A  + \bs d(\bs \gamma d \bs \gamma\-)A  \bigg), \\
&\hspace{2cm}= \Tr \bigg( \bs \gamma d \bs \gamma\- d\big( \bs{d\gamma\gamma}\-\big)	+ \bs d \big( \bs \gamma d \bs \gamma\-  A \big)  \bigg).
\end{align*}
So the final expression for the field-dependent gauge transformation of the CS presymplectic potential is, 
\begin{align}
\label{Field-depGT-theta-CS}
(\bs\theta^\text{{\tiny \,\! CS}}_\Sigma)^{\bs \gamma}=\bs \theta^\text{{\tiny\,\!  CS}}_\Sigma + \int_{\d\Sigma} 2\Tr\big(\bs{d\gamma\gamma}\- A\big) +  \int_\Sigma - 2 \Tr\big(\bs{d\gamma\gamma}\-F\big) 
											+\Tr \left( \bs \gamma d \bs \gamma\- d\big( \bs{d\gamma\gamma}\-\big)	+ \bs d \big( \bs \gamma d \bs \gamma\-  A \big)  \right).
\end{align}
This reproduces the results eq.(3.42) and eq.(C.5)/eq.(D.25) in \cite{Geiller2017}. In this case, even with boundary conditions and on-shell, $\bs\theta_\Sigma$ is not $\bs \H$-invariant - it is never basic - so it cannot induce a well-define symplectic potential on $\S/\H$. 

The presymplectic 2-form is $\bs\Theta_\Sigma\defeq \bs d \bs \theta_\Sigma= -\int_\Sigma \Tr\big( \bs d A\, \bs d A \big)$, and by \eqref{Field-depGT-Theta-non-inv} its $\bs\H$-gauge transformation  is 
\begin{align}
\label{Field-depGT-Theta-CS}
(\bs\Theta_\Sigma^\text{{\tiny CS}})^{\bs\gamma} &= \bs\Theta^\text{{\tiny CS}}_\Sigma  \, + \int_{\d\Sigma}  \bs d \left(\t Q \big( \bs{d\gamma\gamma}\-; A \big)  -  Q \big( \bs{d\gamma\gamma}\-; A \big)  \right) 
								+ \mathscr{A}\big( \lfloor \bs{d\gamma\gamma}\-, \bs{d\gamma\gamma}\- \rfloor ;A\big) + Q\big(\bs d (\bs{d\gamma\gamma}\- ) ; A\big) 
							\notag\\[-0.5mm]
							& \hspace{12cm} - \int_\Sigma \bs d E_\text{{\tiny CS}} \big( \bs{d\gamma\gamma}\-; A \big), \notag \\[-0.5mm]
						&= \bs\Theta^\text{{\tiny CS}}_\Sigma  \, + \int_{\d\Sigma}  \bs d\left(  2 \Tr\big(  \bs{d\gamma\gamma}\- A \big) \right)  - 2\Tr\big( d(\bs{d\gamma\gamma}\-) \bs{d\gamma\gamma}\- \big)   - \int_\Sigma  2\bs d\Tr\big( \bs{d\gamma\gamma}\- F \big),
\end{align}
which  already is  the most compact form, as can be checked explicitly either from \eqref{Field-depGT-theta-CS} using $\bs\Theta^\text{{\tiny CS}}_\Sigma =\bs d \bs\theta^\text{{\tiny CS}}_\Sigma$, or from $(\bs\Theta^\text{{\tiny CS}}_\Sigma)^{\, \bs \gamma}=-\int_\Sigma \Tr\big( \bs d A^{\bs \gamma} \bs d A^{\bs \gamma} \big)$ and using \eqref{GT-dA}. As in the general case, we see that given adequate boundary conditions $\bs\Theta^\text{{\tiny CS}}_\Sigma$ is on-shell $\bs \H$-invariant and descends as a well-defined symplectic 2-form on the physical phase space $\S/\H$.

The next example is complex 3D gravity without cosmological constant. In this example the symmetry algebra is more subtle, thus so is the presymplectic structure. Also, in this case we can write finite gauge transformations associated to gauge translations. 


\subsubsection{3D-$\mathbb{C}$-gravity $\Lambda = 0$ (with translations)}
\label{3D gravity I (translations)}

The theory has been described in section \eqref{3D-C-grav-noLambda} in terms of the underlying Cartan geometry $(\P, \b A)$ based on the groups  $(G, H)=\big(S\!U(2)\ltimes \text{Herm}(2, \CC), S\!U(2)\big)$ in Euclidean signature, or $\big(S\!U(1, 1) \ltimes \text{Herm}(2, \CC), S\!U(1,1)\big)$ in Lorentzian signature. As stressed there, from a Cartan theoretic point of view there is no "translational gauge transformations", only the gauge group $\H$ of $\P$. 

Geometrically, the only  meaningful way to speak of "gauged translations" is to consider the associated bundle $\Q\defeq \P\times_H G$ with gauge group  $\G=\H \ltimes \T$ with  $\T\defeq \left\{ \, t: \Q \rarrow \text{Herm}(2, \CC)\, |\, \ldots\, \right\}$, and then lift $\b A$ to an Ehresmann connection on $\Q$.
The latter splits  as $\b A=A+e$, and its curvature is $\b F= R+T= dA+\tfrac{1}{2}[A,A]\ +\ de+[A, e]$.
In~the~case $\Lambda = 0$, it is easy to write the finite $\G$-gauge transformations:  given $\gamma \in \H$ and $t \in \T$ we have
\begin{align}
e^t&= e+ D^At, \qquad e^\gamma =\gamma\- e \gamma,\notag \\*
A^t&= A, \qquad A^\gamma=\gamma\- A \gamma + \gamma\- d\gamma.  
\end{align}
Correspondingly, for $\chi \in$ Lie$\H$ and $\tau \in $ Lie$\T$, we have, 
\begin{align}
\iota_{\tau^v}\bs d \b A &= \iota_{\tau^v}\bs d  A + \iota_{\tau^v}\bs d e = 0 + D^a\tau, \notag\\
 \iota_{\chi^v}\bs d \b A &= \iota_{\chi^v}\bs d  A + \iota_{\chi^v}\bs d e = D^A \chi +   [e, \chi].
\end{align}
The Lagrangian of the theory is $L(\b A)= L(A, e)= \Tr(eR)$, 
so that
\begin{align}
\bs d L= \bs E + d\bs \theta &= E(\bs d \b A ; \b A) + d \theta(\bs d \b A; \b A), \notag\\
					 &= \Tr\big( \bs d e\, R + \bs d A\, D^Ae \big) +d\Tr\big( \bs d A\, e \big).
\end{align}
We can analyse the presymplectic structure in three steps. First consider the Lorentz sector, then the translations sector, and finally find the Poisson bracket of Lorentz and translation charges. 
\medskip

\noindent {\bf Lorentz sector:}  We have remarked that the $\Lambda$ term in the theory does not affect its presymplectic structure (much like the mass term in massive YM theory), and does not compromise the $\H$-invariance (unlike the mass term in massive YM theory). So all results to be found here were already derived in section \ref{3D gravity}: in particular, for $\chi, \eta \in$ Lie$\H$, the Noether charges and Poisson bracket thereof are
\begin{align}
\label{Noether-charges-Lorentz-3DGrav-flat}
Q_\Sigma(\chi, \b A) =\int_{\d \Sigma} \Tr(\chi\, e) - \int_\Sigma \Tr(\chi \, D^Ae), \qquad \text{and} \qquad  \big\{ Q_\Sigma(\chi, \b A), Q_\Sigma(\eta, \b A)\big\}= Q_\Sigma([\chi, \eta], \b A).
\end{align}
The reader can nonetheless check that these results are also found  via the algorithm that we now deploy to analyse gauge translations.
\medskip

\noindent {\bf Translations sector:} The first thing to observe is that translational transformations are anomalous.  As seen in section \ref{3D-C-grav-noLambda}, eq.  \eqref{3D-grav-cocycle}, for $t \in \T$ we have
\begin{align}
\big(R^\star_t L\big)(\b A)= L(\b A) + c(\b A; t)
				      = L(A, e) + d\Tr\big( t\,R \big).
\end{align}
The cocycle is $d$-exact, the theory is quasi-invariant. 
We start by checking hypothesis 0:
\vspace{-6mm}
\begin{align}
R^\star_t \bs E= \Tr\big( (\bs de + \bs d D^At) R + \bs d A(e + D^At) \big) = \Tr\big( \bs d e\, R + \cancelto{\,_\text{I}}{[\bs d A, t] R} + \bs d A\, D^A e + \cancelto{\,_\text{I}}{\bs d A\, \overbrace{D^A(D^At)}^{ [R, t]} }\big) = \bs E.
\end{align}
It follows, and is clear in this case, that $\bs d c(\b A; t)= d\bs b(t)$ so that 
\begin{align}
R^\star_t d \bs \theta =  d\bs \theta + \bs d c(\b A; t) &= d\bs \theta  + d\Tr \big( t\, \bs d R\big), \notag \\
												&= 	d\bs \theta  + d\Tr \big( t\, D^A(\bs d A) \big) = 	d\bs \theta  + d\Tr \big( \cancel{d(t\, \bs dA)} - D^At \bs dA \big), \notag\\[1mm]
\hookrightarrow 
R^\star_t  \bs \theta =  \bs \theta + \bs b (t) &= \bs \theta + b (\bs d \b A; t) = \bs \theta + \Tr  \big(  \bs d A \, D^At  \big).
\end{align}
Notice we have been mindful of the fact that $\bs b(t)$ should be linear in $\bs d\b A$ ($\bs d A$). From this we obtain, for $\tau \in $ Lie$\T$,
\begin{align}
\label{bs-alpha-3D-grav-no-lambda}
\bs \alpha(\tau) \defeq \bs L_{\tau^v} \bs \theta = \tfrac{d}{ds} \bs b(e^{s \tau})\big|_{s=0}= \Tr\big(\bs d A D^A \tau\big). 
\end{align}
The classical anomaly is 
 \begin{align}
\alpha(\tau, \b A) \defeq \bs{L}_{\tau^v} L=\tfrac{d}{ds} c\big(A; e^{s \tau}\big)\big|_{s=0}= d\Tr(\tau\, R ) \rdefeq d \beta(\tau; \b A).
\end{align}
It is $d$-exact so hypothesis 1 is verified, and we check that $d\bs \alpha(\tau)= \bs d \alpha(\tau; \b A)$.  We next identify the quantity $dQ(\tau; \b A)$, which is by \eqref{def-dQ},
\begin{align}
\label{dQ-3D-grav-no-Lambda}
\bs d dQ(\tau; \b A) \defeq \bs \alpha(\tau) - \bs d\beta(\tau; \b A)= \Tr(\bs d A D^A\tau) -   \Tr(\tau \bs dR) =  -\Tr\big(d (\tau \bs d A) \big)= \bs d\big(  -d\Tr(\tau A) \big).
\end{align}
By \eqref{def-dQ-tilde}, the quantity $d\t Q(\tau; \b A)$ is
\begin{align}
d\t Q(\tau; \b A)\defeq&\, \iota_{\tau^v} \bs \theta - \beta(\tau; \b A) + E(\tau; \b A), \notag \\
			       =&\, 0  -\Tr\big( \tau R \big) + \Tr\big( \tau R \big) =0. 
\end{align}
Remark that only the Lie$\T$-part of $\bs E$ contributes.
Finally by \eqref{def-dA}, for $\tau, \tau' \in$ Lie$\T$, we obtain
\begin{align}
d \mathscr{A}\big( \lfloor\, \tau, \tau' \rfloor ; \b A\big)\defeq - \iota_{\tau^v} \bs\alpha(\tau') + \iota_{{\tau'}^v} \bs\alpha(\tau)+ \beta\big([\tau, \tau]; A \big) =0,					
\end{align}
because  Lie$\T$ is abelian and $\iota_{\tau^v} \bs dA =0$. 

We have now all the necessary quantities  to obtain the Noether current, which by  \eqref{J-non-inv-bis} is 
\begin{align}
\label{current-3D-grav-no-Lambda}
J(\tau; \b A) &=  d \t Q(\tau; \b A) - dQ(\tau; \b A) - E(\tau; \b A), \notag\\
				      &=  d\Tr(\tau A) - \Tr(\tau R) = -\Tr\big( Ad\tau +\tfrac{1}{2} \tau [A, A] \big).
\end{align}
Again, only the Lie$\T$-part of $\bs E$ contributes. This  is  $d$-exact on-shell. By \eqref{Noether-Charge-non-inv}, the corresponding Noether charge~is 
\begin{align}
\label{charge-3D-grav-no-Lambda}
Q_\Sigma(\tau; \b A) = \int_\Sigma  J(\tau; \b A) &= \int_{\d\Sigma} \Tr(\tau A) - \int_\Sigma\Tr(\tau R), \\
						     								  &= \int_{\d\Sigma} \Tr(\tau A) _{\ |\S}. \notag
\end{align}
This is the ``extended generator'' proposed in \cite{Geiller2017} for 3D-$\CC$-gravity with $\Lambda =0$, see eq.(4.28) there.
The Poisson bracket of charges induced by the presymplectic 2-form $\bs\Theta_\Sigma\defeq \bs d \bs \theta_\Sigma= -\int_\Sigma \Tr\big( \bs d A\, \bs d e \big)$ is by \eqref{PB-anomalous},
\begin{align}
\label{PB-3D-grav-no-Lambda-1}
\big\{ Q_\Sigma(\tau; \b A),  Q_\Sigma(\tau'; \b A) \big\} =  Q_\Sigma([\tau, \tau'] ; \b A) + \int_{\Sigma}  d\mathscr{A}\big( \lfloor\, \tau, \tau' \rfloor ; \b A\big) + dQ\big([\tau, \tau'];A \big)=0. 							   
\end{align}
This is as expected, as it reflects the fact that Lie$\T$ is abelian. 
\medskip

Finally we want to find the Poisson bracket of Lorentz and translation charges. By  \eqref{PB-anomalous} still, for $\chi \in$ Lie$\H$ and $\tau \in$ Lie$\T$, 
\begin{align*}
\big\{ Q_\Sigma(\chi; \b A),  Q_\Sigma(\tau; \b A) \big\} =  Q_\Sigma([\chi, \tau] ; \b A) + \int_{\Sigma}  d\mathscr{A}\big( \lfloor\, \chi, \tau \rfloor ; \b A\big) + dQ\big([\chi, \tau];A \big). 							   
\end{align*}
Since $\G=\H \ltimes \T$ is a semi-direct product we have  $[\chi, \tau] \in $ Lie$\T$, so the first term on the right is a translation charge
\begin{align*}
Q_\Sigma([\chi, \tau] ; \b A)=  \int_{\d\Sigma} \Tr([\chi, \tau] \, A) - \int_\Sigma\Tr([\chi, \tau]\, R), \qquad \text{and by \eqref{dQ-3D-grav-no-Lambda}} \qquad  dQ\big([\chi, \tau];A \big)=-d\Tr\big( [\chi, \tau]\, A\big).
\end{align*}
By \eqref{def-dA} again,  and given \eqref{bs-alpha-3D-grav-no-lambda}, we have 
\begin{align*}
d \mathscr{A}\big( \lfloor\, \chi, \tau \rfloor ; \b A\big)\defeq&\, - \iota_{\chi^v} \bs\alpha(\tau) - \cancel{\iota_{\tau^v} \bs\alpha(\chi)} + \beta\big([\chi, \tau]; \b A \big),  \quad \text{as there is no $\bs\alpha$-term for Lorentz symmetry,}\\
											=&\  -\Tr\big(D^A\chi D^A\tau \big) + \Tr\big( [\chi, \tau]\, R \big),\\
											=&\  -\Tr\left( d\chi d \tau + [A, \chi]d\tau + d\chi[A, \tau]  +  [A, \chi][A, \tau]	  -   [\chi, \tau]\, dA -  [\chi, \tau]\, \tfrac{1}{2} [A, A] \right).
\end{align*}
Now, using the graded Jacobi identity $[A, [A, \chi]] - [A, [\chi, A]]+ [\chi, [A, A]]=0$, the fourth term is 
\begin{align*}
 \Tr\big( [A, \chi][A, \tau]\big) = -\Tr\big( [A, [A, \chi]] \tau \big) = -\Tr\big( \tfrac{1}{2}[\chi, [A, A]] \tau \big) = \tfrac{1}{2} \Tr\big( [A, A][\chi, \tau] \big),
\end{align*}
and cancels the last one. So, 
\begin{align*}
d \mathscr{A}\big( \lfloor\, \chi, \tau \rfloor ; \b A\big)&=  -\Tr\big( d\chi d \tau + [A, \chi]d\tau + d\chi[A, \tau]     -   [\chi, \tau]\, dA \big), \\
									     &= -\Tr\bigg(   d\chi d \tau   \underbrace{ -A\, [d\tau, \chi] -  [d\chi, \tau]\,A}_{\Tr\big(-d[\chi, \tau]\, A\big)}   -   [\chi, \tau]\, dA  \bigg)
									     =  - \Tr\left( d\chi d \tau \right) + d\Tr \big(  [\chi, \tau]\, A\big) .
\end{align*}
Then $d \mathscr{A}\big( \lfloor\, \chi, \tau \rfloor ; \b A\big) + dQ\big([\chi, \tau];A \big)  = d\Tr\big(d\chi\, \tau \big)$, so the Poisson bracket of Lorentz and translation charges  is finally
\begin{align}
\label{PB-3D-grav-no-Lambda-2}
\big\{ Q_\Sigma(\chi; \b A),  Q_\Sigma(\tau; \b A) \big\} =  Q_\Sigma([\chi, \tau] ; \b A) + \int_{\d\Sigma}  \Tr\big(d\chi\, \tau \big). 							   
\end{align}
This reproduces the result eq.(4.29) in \cite{Geiller2017}. The Poisson algebra of charges is a central extension of Lie$\G$. 
\medskip

Considering now field-dependent translational gauge transformation, for $\bs t \in \bs\T$ and $\bs{dt} \in$ Lie$\bs\T$  we have~by~\eqref{Field-depGT-E-non-inv} 
\begin{align}
 \bs E^{\, \bs  t}= \bs E + dE\big(  \bs{dt} ; \b A \big) 
 				 = \bs E + 2d\Tr\big( \bs{dt} R\big). 
\end{align}
as is easily checked. Notice that only the Lie$\T$-valued piece of the field equations contributes. By \eqref{Field-depGT-presym pot-non-inv}, the $\bs\T$-gauge transformation of  the presymplectic potential  is
\begin{align}
\label{Field-depGT-theta-transl-3D-grav}
\bs\theta_\Sigma^{\, \bs t}&= \bs \theta_\Sigma + \int_{\d\Sigma} \cancel {\t Q \big(  \bs{dt} ;  \b A \big)} + \int_\Sigma - E \big( \bs{dt} ; \b A \big) 
												+  \beta \big(  \bs{dt} ; \b A \big) + \bs b\big(\bs t(\b A)\big)_{|\b A} +  b\left(  \cancel{\iota_{\bs t^v} \bs d A} ; \bs t(\b A) \right), \notag  \\
						&=\bs \theta_\Sigma +   \int_\Sigma -  \Tr\big(\bs{dt}\, R\big) + \Tr\big(\bs{dt}\, R \big) + \Tr\big( \bs d A \, D^A \bs{t}\big), \notag \\
						&=\bs \theta_\Sigma -   \int_{\d\Sigma} \Tr(\bs d A\,  \bs t)  +  \int_\Sigma \Tr\big( \bs d  R \,\bs t \big),
\end{align}
since $\Tr\big( \bs d A\, D^A \bs{t} \big) = -d\Tr(\bs d A\,  \bs t) + \Tr\big(D^A(\bs d A)\, \bs t\big)$, and $\bs d R=D^A(\bs d A)$. A result easily checked explicitly, which happens to reproduce eq.(4.43) of \cite{Geiller2017}. In~this case, contrary to the general situation, it appears that given adequate boundary conditions, $\bs\theta_\Sigma$ can be on-shell $\bs\T$-invariant and thus descend as a well-defined symplectic potential on the physical phase space $\S/\H$.

By \eqref{Field-depGT-Theta-non-inv} its $\bs\T$-gauge transformation of the presymplectic 2-form $\bs\Theta_\Sigma$ is 
\begin{align}
\label{Field-depGT-Theta-3D-grav}
\bs\Theta_\Sigma^{\, \bs t} &= \bs\Theta_\Sigma  \, + \int_{\d\Sigma}  \bs d \left( \cancel{ \t Q \big( \bs{dt}; \b A \big)}  -  Q \big( \bs{dt}; \b A \big)  \right) 
								+ \cancel{ \mathscr{A}\big( \lfloor \bs{dt}, \bs{dt} \rfloor ;A\big) } + Q\big(\cancel{\bs d (\bs{dt}} ) ; \b A\big)  
								- \int_\Sigma \bs d E \big( \bs{dt}; \b A \big), \notag \\
						&= \bs\Theta_\Sigma  \, + \int_{\d\Sigma}  \bs d \Tr \big(  \bs{dt}\,  A \big)  - \int_\Sigma  \bs d \Tr \big( \bs{dt} R \big),
\end{align}
as can be checked either from \eqref{Field-depGT-theta-transl-3D-grav} using $\bs\Theta_\Sigma =\bs d \bs\theta_\Sigma$, or from $\bs\Theta_\Sigma^{\, \bs t}=-\int_\Sigma \Tr\big( \bs d A^{\bs t} \bs d e^{\bs t} \big)$ and using   \eqref{GT-dA-Cartan}. 
Illustrating again the general case, we see  that given adequate boundary conditions $\bs\Theta_\Sigma$ is on-shell $\bs \H$-invariant and thus descends as a well-defined symplectic 2-form on the physical phase space $\S/\H$.
\medskip

As a final example, we give the presymplectic structure of complex 3D gravity with cosmological constant. In this case we do not write finite translational gauge transformations and we will be only concerned with deriving Noether currents and charges and the Poisson bracket structure. 


\subsubsection{3D-$\mathbb{C}$-gravity $\Lambda \neq 0$ (with translations)}
\label{3D gravity II (translations)}

The theory has been described in section \ref{3D gravity} in terms of the underlying Cartan geometry $(\P, \b A)$ based either on
 $(G, H)=\big(S\!U(2) \times S\!U(2), S\!U(2)\big)$ for $\Lambda > 0$ and $\big(S\!L(2, \CC), S\!U(2)\big)$ for  $\Lambda < 0$ in Euclidean signature, or on  $(G, H)=\big( S\!L(2, \CC), S\!U(1,1) \big)$ for $\Lambda >0$ and  $(G, H)=\big(S\!U(1,1)\times S\!U(1,1), S\!U(1,1)  \big)$  for  $\Lambda <0$ in Lorentzian signature.
 The Cartan connection splits as $\b A=A +\tfrac{1}{\ell}e$, with $\tfrac{1}{\ell^2}= \tfrac{2|\Lambda|}{(n-1)(n-2)}=|\Lambda|$ for $n=3=$ dim$\M$, and correspondingly the curvature splits as 
 $\b F=F+\tfrac{1}{\ell}T = R - \tfrac{\epsilon}{\ell^2}ee \ + \ \tfrac{1}{\ell} \big(de + [A,e]\big)$, with $\epsilon=\pm$ is the sign of $\Lambda$.

Again, from the Cartan theoretic point of view the only gauge transformations come from the gauge group $\H$ of $\P$. But if we consider the Cartan connection as coming from an Ehresmann connection on the associated bundle 
$\Q \defeq \P \times_H G$, gauge translation transformations $\T=\G/\H$ become geometrically meaningful.
The algebra Lie$\G$ being, for $\chi, \eta \in$ Lie$\H$ and $\tau, \tau' \in$ Lie$\T$, 
 \begin{align}
[\chi, \eta] \in \text{Lie}\H, \quad [\chi, \tau] \in \text{Lie}\T, \quad \text{and} \quad-\tfrac{\epsilon}{\ell^2} [\tau, \tau'] \in \text{Lie}\H,
 \end{align}
  the infinitesimal gauge transformations of the connection are encoded as
   \begin{align}
   \label{infGT-3D-grav-Lambda}
  \iota_{\chi^v}\bs{d} \b A = \iota_{\chi^v}\bs{d} A + \iota_{\chi^v} \bs{d}e = D^A\chi + [e, \chi],  \notag\\
  \iota_{\tau^v}\bs{d} \b A = \iota_{\tau^v}\bs{d} A + \iota_{\tau^v}\bs{d}e = -\tfrac{\epsilon}{\ell^2} [e, \tau] + D^A\tau.
  \end{align}
  Also, for future use, notice that 
 \begin{align}
 \iota_{\tau^v} \bs dR= \iota_{\tau^v} D^A(\bs d A)= D^A\left(- \tfrac{\epsilon}{\ell^2}[e, \tau] \right) = - \tfrac{\epsilon}{\ell^2}\left( [D^Ae, \tau] - [e, D^A\tau] \right).
 \end{align} 

  The Lagrangian is $L(\b A)=L(A, e)= \Tr\big( eF\big) = \Tr\left\{ e\big(R-\tfrac{\epsilon}{3\ell^2}ee\big) \right\}$, so that
  \begin{align}
  \label{dL-3D-grav-Lambda}
\bs d L=\bs E + d\bs \theta &= E(\bs d \b A; \b A) + d \theta(\bs d \b A; \b A),\notag \\
					 &= \Tr\big(\bs d e\, F+ \bs d A\, T \big)= \Tr\left( \bs d e  \big(R- \tfrac{\epsilon}{\ell^2} ee \big) + \bs d A\, D^Ae  \right) \ + \ d\Tr\big(  \bs d A \, e  \big). 
  \end{align}
  The  $\Lambda$-term does not affect $\bs\theta$, so the presymplectic 2-form current is still $\bs\Theta=-\Tr\big( \bs d A \, \bs de \big)$. 
 As we did above, we  can analyse the presymplectic structure in three steps. First consider the Lorentz sector, then the translations sector, and finally find the Poisson bracket of Lorentz and translation charges. 
\medskip

\noindent {\bf Lorentz sector:} The Lagrangian is $\H$-invariant, so the Lorentz charges and their Poisson bracket are as described in section \ref{3D gravity}, reiterated in section \ref{3D gravity I (translations)}:
 \begin{align}
 \label{Lorentz-charge-3Dgrav}
Q_\Sigma(\chi, \b A) =\int_{\d \Sigma} \Tr(\chi\, e) - \int_\Sigma \Tr(\chi \, D^Ae), \qquad \text{and} \qquad  \big\{ Q_\Sigma(\chi, \b A), Q_\Sigma(\eta, \b A)\big\}= Q_\Sigma([\chi, \eta], \b A).
\end{align}
Again, one can nonetheless check that these results are also found  via the algorithm that we now deploy to analyse infinitesimal gauge translations.
 \medskip
 
 \noindent {\bf Translations sector:}  The first thing to observe is that translational gauge transformations are anomalous. Infinitesimally, using    \eqref{infGT-3D-grav-Lambda} and   \eqref{dL-3D-grav-Lambda},
 \begin{align}
 \bs L_{\tau^v} L &=   \iota_{\tau^v} \bs d L = \Tr\left(  D^A\tau\, \big( R  -\tfrac{\epsilon}{\ell^2}ee \big)   -\tfrac{\epsilon}{\ell^2}[e,\tau]\, D^Ae \right) - d\Tr \left(  -\tfrac{\epsilon}{\ell^2}[e,\tau]\, e\right), \notag\\
 			 &= \Tr\bigg(  d\left(\tau \big( R -\tfrac{\epsilon}{\ell^2} ee\big) \right) - \tau \, D^A\big( \cancel{R} -\underline{\tfrac{\epsilon}{\ell^2}ee}_{\,_\text{I}} \big)  
			 		+ \underline{\tfrac{\epsilon}{\ell^2} \tau\, [e, D^Ae]}_{\,_\text{I}} + d\left(  \tfrac{\epsilon}{\ell^2}  \tau \,[ee]\right) \bigg), \notag\\
			&= d\Tr\left( \tau \, \big( R  + \tfrac{\epsilon}{\ell^2} ee \big)\right). \qquad \Rightarrow \qquad \alpha(\tau; \b A)= d\beta(\tau; \b A). 
 \end{align}
 This shows us that hypothesis 1 is satisfied, the classical anomaly is $d$-exact (the Lagrangian is quasi-invariant). 
 We~must  then check hypothesis 0, the translational invariance of the field equations,
  \begin{align}
\bs L_{\tau^v} \bs E& = \iota_{\tau^v} \bs d \bs E + \bs d \iota_{\tau^v} \bs E,\notag \\
			     &=\iota_{\tau^v}  \Tr\left( -\bs de\,\bs d \big( R -\tfrac{\epsilon}{\ell^2}ee\big) - \bs d A\, \bs d D^Ae \right) + \bs d\left( D^A\tau \big(R -\tfrac{\epsilon}{\ell^2}ee \big)  -\tfrac{\epsilon}{\ell^2} [e, \tau ] D^Ae \right), \notag \\[-2mm]
			     &= \Tr\bigg(  \underline{-D^A\tau \ \bs d \big( R -\tfrac{\epsilon}{\ell^2}ee \big)}_{\,_\text{I}}  + \bs d e\ \iota_{\tau^v} \big(  \bs d R -\tfrac{\epsilon}{\ell^2} [\bs d e, e]  \big)   
			    	   + \underline{ \tfrac{\epsilon}{\ell^2} [e, \tau ]\, \bs d D^Ae}_{\,_\text{II}} + \ \bs d A\ \iota_{\tau^v} \hspace{-8mm}\overbrace{\bs d D^Ae}^{\qquad D^A(\bs de) + [\bs dA, e]} \notag\\
			      & \hspace{3cm}        + \underbrace{\bs d D^A\tau}_{[\bs dA, \tau]} \ \,  \big( R  -\tfrac{\epsilon}{\ell^2}ee \big) + \underline{D^A\tau\  \bs d \big( R-\tfrac{\epsilon}{\ell^2}ee \big)}_{\,_\text{I}} -\tfrac{\epsilon}{\ell^2}[\bs de, \tau]\, D^Ae - \underline{\tfrac{\epsilon}{\ell^2} [e, \tau]\, \bs d D^Ae}_{\,_\text{II}} \ \bigg), \notag\\[1mm]
			      &= \Tr\bigg( \bs de \left(  -\tfrac{\epsilon}{\ell^2}\big(\underline{ [D^Ae, \tau]}_{\,_\text{A}} - \cancel{[e,  D^A\tau]}\big) -\tfrac{\epsilon}{\ell^2}\cancel{[D^A\tau, e]}  \right)\ + \ \bs d A\bigg( \overbrace{ D^A\big( D^A \tau\big)}^{ \underline{[R, \tau]}_{\,_\text{B}} } - \underline{\tfrac{\epsilon}{\ell^2}[[e, \tau], e] }_{\,_\text{C}} \bigg) \notag\\
			      & \hspace{3cm}        + \underline{[\bs dA, \tau]\, R}_{\,_\text{B}}  -\underline{\tfrac{\epsilon}{\ell^2}[\bs dA, \tau]\, \tfrac{1}{2}[e,e]}_{\,_\text{C}}   -\underline{\tfrac{\epsilon}{\ell^2}[\bs de , \tau]\, D^Ae}_{\,_\text{A}}    \bigg)
			      =0.
 \end{align}
 The terms A, B and C vanishe by  \eqref{usefull-identity} and the graded Jacobi identity.\begin{samepage} This means $\bs\theta$ carries the non-invariance, and we must have
  \vspace{-2mm}
 \begin{align}
\bs\alpha(\tau) \defeq&\ \bs L_{\tau^v} \bs \theta = \iota_{\tau^v} \bs d \bs \theta + \bs d \iota_{\tau^v} \bs \theta= \iota_{\tau^v} \Tr\left( -\bs d A \bs d e \right) + \bs d \Tr\left( -\tfrac{\epsilon}{\ell^2} [e, \tau]\, e \right),\notag \\
			      =&\ \Tr\left(   \cancel{\tfrac{\epsilon}{\ell^2}[e\, \tau]\, \bs de} + \bs d A\, D^A\tau - \tfrac{\epsilon}{\ell^2}[\bs de, \tau]\, e - \cancel{\tfrac{\epsilon}{\ell^2}[e, \tau]\, \bs de} \right), \notag\\
			      =&\ \Tr\left(   \bs d A\, D^A\tau + \tfrac{\epsilon}{\ell^2} \tau\, [\bs de, e] \right).
 \end{align}~One checks that $d\bs \alpha(\tau)= \bs d \alpha(\tau; \b A)$.  
\end{samepage}
 We next identify the quantity $dQ(\tau; \b A)$, which is by \eqref{def-dQ},
\begin{align}
\label{dQ-3D-grav-Lambda}
\bs d dQ(\tau; \b A) \defeq&\ \bs \alpha(\tau) - \bs d\beta(\tau; \b A)= \Tr\left( \bs d A D^A\tau + \tfrac{\epsilon}{\ell^2}\tau\, [\bs de, e]  \right) -   \Tr\left( \tau\, \left(\bs dR + \tfrac{\epsilon}{\ell^2} [\bs de, e] \right) \right), \notag \\
				      =&\  \Tr\left( -d\big(\bs d A\,\tau \big) + D^A(\bs dA)\, \tau  - \tau\, D^A(\bs dA)\right)
				      = \bs d\big(  -d\Tr(\tau A) \big).
\end{align}
This is the same result as in the $\Lambda=0$ case. By \eqref{def-dQ-tilde}, the quantity $d\t Q(\tau; \b A)$ is
\begin{align}
d\t Q(\tau; \b A)\defeq&\, \iota_{\tau^v} \bs \theta - \beta(\tau; \b A) + E(\tau; \b A), \notag\\
			       =&\, \Tr\left(  - \tfrac{\epsilon}{\ell^2} [e, \tau]\,e \right)  -\Tr\left(  \tau\, \big( R+ \tfrac{\epsilon}{\ell^2}ee \big) \right) + \Tr \left( \tau\, \big( R-\tfrac{\epsilon}{\ell^2}ee \big)\right), \notag\\
			       =&\ \Tr\left( \tfrac{\epsilon}{\ell^2}\tau\, 2ee - \tfrac{\epsilon}{\ell^2} \tau\, 2ee \right)=0. 
\end{align}
Remark that only the Lie$\T$-part of $\bs E$ contributes. Now, since for  $\tau, \tau' \in$ Lie$\T$, $-\tfrac{\epsilon}{\ell^2}[\tau, \tau'] \in$ Lie$\H$, by \eqref{def-dA} we get 
\begin{align}
d \mathscr{A}\big( \lfloor\, \tau, \tau' \rfloor ; \b A\big)\defeq&\ - \iota_{\tau^v} \bs\alpha(\tau') + \iota_{{\tau'}^v} \bs\alpha(\tau)+ \cancel{\beta\big(-\tfrac{\epsilon}{\ell^2}[\tau, \tau]; A \big)}, \notag\\
										      =&\  -\Tr\left( -\tfrac{\epsilon}{\ell^2}[e, \tau]\, D^A\tau' + \tfrac{\epsilon}{\ell^2}\tau'[D^A\tau, e]\right)	
										      	     + 	\Tr\left( -\tfrac{\epsilon}{\ell^2}[e, \tau']\, D^A\tau + \tfrac{\epsilon}{\ell^2}\tau\, [D^A\tau', e]\right)	=0. 
\end{align}
The $\beta$-term cancels because there is no Lorentz anomaly, the Lagrangian being $\H$-invariant. 

We have now all the necessary quantities  to obtain the Noether current, which by  \eqref{J-non-inv-bis} is 
\begin{align}
\label{current-3D-grav-Lambda}
J(\tau; \b A) &=  d \t Q(\tau; \b A) - dQ(\tau; \b A) - E(\tau; \b A), \notag\\
		   &=  d\Tr\big(\tau A\big) - \Tr\left( \tau\, \big( R -\tfrac{\epsilon}{\ell^2} ee \big) \right). 
\end{align}
Again, only the Lie$\T$-part of $\bs E$ contributes. Compare with  \eqref{current-3D-grav-no-Lambda}. On-shell it is  $d$-exact  and identical to the current of the  $\Lambda=0$ case.  By \eqref{Noether-Charge-non-inv}, the corresponding Noether charge is 
\begin{align}
\label{charge-3D-grav-Lambda}
Q_\Sigma(\tau; \b A) = \int_\Sigma  J(\tau; \b A) &= \int_{\d\Sigma} \Tr(\tau A) - \int_\Sigma \Tr\left( \tau\, \big( R -\tfrac{\epsilon}{\ell^2} ee \big) \right), \\
						     								  &= \int_{\d\Sigma} \Tr(\tau A) _{\ |\S}. \notag
\end{align}
This is the ``extended generator'' proposed in \cite{Geiller2017} for 3D-$\CC$-gravity with $\Lambda \neq0 $, see eq.(4.21) and eq.(4.24) there.

The~Poisson bracket of translation charges induced by the presymplectic 2-form $\bs\Theta_\Sigma\defeq \bs d \bs \theta_\Sigma= -\int_\Sigma \Tr\big( \bs d A\, \bs d e \big)$ is by \eqref{PB-anomalous},
\begin{align}
\label{PB-3D-grav-no-Lambda-1}
\big\{ Q_\Sigma(\tau; \b A),  Q_\Sigma(\tau'; \b A) \big\} &=  Q_\Sigma(-\tfrac{\epsilon}{\ell^2}[\tau, \tau'] ; \b A) + \int_{\Sigma}  d\mathscr{A}\big( \lfloor\, \tau, \tau' \rfloor ; \b A\big) + \cancel{dQ\big(-\tfrac{\epsilon}{\ell^2}[\tau, \tau'];A \big)}, \notag\\
										    &= -\tfrac{\epsilon}{\ell^2} Q_\Sigma( [\tau, \tau'] ; \b A), 
\end{align}
where the $dQ$-term cancels because there is no Lorentz anomaly, and the final term is a Lorentz charge 
\begin{align*}
Q_\Sigma( [\tau, \tau'] ; \b A) &= \int_{\d \Sigma} \Tr\big( [\tau, \tau']\, e\big) -\int_\Sigma \Tr\big( [\tau, \tau'] \, D^Ae\big), \\
					    &= \int_{\d \Sigma} \Tr\big( [\tau, \tau']\, e\big) _{\ |N},
\end{align*}
as expected. This result can be cheked explicitly by proving $\bs\Theta_\Sigma(\tau^v, {\tau'}^v)= -\iota_{{\tau'}^v}\bs d Q_\Sigma\big( \tau;\b A\big)= -\tfrac{\epsilon}{\ell^2} Q_\Sigma\big( [\tau, \tau'] ;\b A\big)$, with the latter charge given by \eqref{Noether-charge-3D-grav}. 

Finally we want to find the Poisson bracket of Lorentz and translation charges. By  \eqref{PB-anomalous} still, for $\chi \in$ Lie$\H$ and $\tau \in$ Lie$\T$, 
\begin{align*}
\big\{ Q_\Sigma(\chi; \b A),  Q_\Sigma(\tau; \b A) \big\} =  Q_\Sigma([\chi, \tau] ; \b A) + \int_{\Sigma}  d\mathscr{A}\big( \lfloor\, \chi, \tau \rfloor ; \b A\big) + dQ\big([\chi, \tau];A \big). 							   
\end{align*}
Since we have  $[\chi, \tau] \in $ Lie$\T$, the first term on the right is a translation charge
\begin{align*}
Q_\Sigma([\chi, \tau] ; \b A)=  \int_{\d\Sigma} \Tr([\chi, \tau] \, A)  - \int_\Sigma \Tr\left( [\chi, \tau]\, \big( R -\tfrac{\epsilon}{\ell^2} ee \big) \right)
, \qquad \text{and by \eqref{dQ-3D-grav-Lambda}} \qquad  dQ\big([\chi, \tau];A \big)=-d\Tr\big( [\chi, \tau]\, A\big).
\end{align*}
By \eqref{def-dA} again, we get 
\begin{align*}
d \mathscr{A}\big( \lfloor\, \chi, \tau \rfloor ; \b A\big)\defeq&\ - \iota_{\chi^v} \bs\alpha(\tau) - \cancel{\iota_{\tau^v} \bs\alpha(\chi)} + \beta\big([\chi, \tau]; \b A \big),  \quad \text{as there is no $\bs\alpha$-term for Lorentz symmetry,}\\*
											=&\  -\Tr\left(D^A\chi D^A\tau   + \tfrac{\epsilon}{\ell^2} \tau\, [ [e, \chi], e]   \right) \ 
													+\  \Tr\left( [\chi, \tau] \big( R  + \tfrac{\epsilon}{\ell^2} ee\big)\right), \\*
										 	=&\ -\Tr\left(D^A\chi D^A\tau +  [\chi, \tau]\,  R \right)= \ldots =  - \Tr\left( d\chi d \tau \right) + d\Tr \big(  [\chi, \tau]\, A\big).
\end{align*}
Which is the same result as in the case $\Lambda=0$. So here also we have $d \mathscr{A}\big( \lfloor\, \chi, \tau \rfloor ; \b A\big) + dQ\big([\chi, \tau];A \big) = d\Tr\big(d\chi\, \tau \big)$.
Then the Poisson bracket of Lorentz and translation charges is, 
\begin{align}
\label{PB-3D-grav-no-Lambda-3}
\big\{ Q_\Sigma(\chi; \b A),  Q_\Sigma(\tau; \b A) \big\} =  Q_\Sigma([\chi, \tau] ; \b A) + \int_{\d\Sigma}  \Tr\big(d\chi\, \tau \big). 							   
\end{align}
Equations \eqref{Lorentz-charge-3Dgrav}, \eqref{PB-3D-grav-no-Lambda-1} and \eqref{PB-3D-grav-no-Lambda-3} reproduce the results eq.(4.25)-(4.26) in \cite{Geiller2017}. The Poisson algebra of Noether charges is a central extension of Lie$\G$. 
\medskip

This conclude our illustrations of the general results of section \ref{Presymplectic structure for non-invariant Lagrangians}, and more generally our analysis of the presymplectic structure of both invariant and $c$-equivariant gauge theories. 
In the next section we turn to the problem of defining a symplectic structure for gauge theories over bounded regions.


\subsection{Boundaries and dressed presymplectic structure}
\label{Boundaries and dressed presymplectic structure}

In the case of invariant theories, in view of \eqref{Field-depGT-presymp-pot} and \eqref{Field-depGT-presymp-form}, for $\bs\theta_\Sigma$ and $\bs\Theta_\Sigma$ to be basic, $\bs\H$-invariant, and descend as a well-defined symplectic structure on the reduced phase space $\S/\H$, one has to assume either $\d\Sigma=\emptyset$  or adequate boundary or fall-off conditions.
In the case of $c$-equivariant theories, in view of  \eqref{Field-depGT-presym pot-non-inv}  generically the presymplectic potential $\bs\theta_\Sigma$ is not basic, no matter what - even though by chance it might be in specific situations, as is the case in 3D gravity, section \ref{3D gravity I (translations)}. Unexpectedly, in view of \eqref{Field-depGT-Theta-non-inv} the corresponding presympletic 2-form $\bs\Theta_\Sigma$ can be basic when either  
$\d\Sigma=\emptyset$ or adequate boundary or fall-off conditions are specified. In which case the latter still descends as a well-defined symplectic 2-form on $\S/\H$. 

A difficulty arises when the region under consideration has a boundary,  $\d\Sigma\neq\emptyset$, and there is no good reason to assume the fields vanish there. This may happen with a physical boundary, but the problem arises also in a more conceptual way: Suppose an \emph{arbitrary} partition of the region D under study into two subregions D' and D'' sharing a fictitious boundary,  one would  expect to be able to meaningfully resolve the symplectic structure associated to D into well-defined symplectic structures associated to D' and D'' (modulo compatibility conditions along the boundary). Replace `symplectic structure' by `Hilbert space' and you get the quantum counterpart of this puzzle, which relates e.g. to considerations about entanglement entropy. A good solution to the boundary problem should therefore solve not only the case of a region with a physical boundary, but also the  case of an arbitrary partition with fictitious boundaries that in principle can be placed anywhere.

One obvious strategy is to build a presymplectic structure on $\A$ that is basic on-shell, without any restrictions on the fields or the region. This is basically the option entertained in \cite{DonnellyFreidel2016} which first introduced the so-called ``edge modes" in YM theory and metric gravity. The DFM offers a systematic framework to assess this strategy (as pointed out first in \cite{Teh-et-al2020, Teh2020}). We thus apply it, using the general results of sections \ref{The dressing field method} and \ref{A-dependent dressing fields and basic variational forms on A}, to build a \emph{dressed} presymplectic structure on $\A$, first for invariant theories in section \ref{Dressed presymplectic structure for invariant Lagrangians}, then for $c$-equivariant theories in section \ref{Dressed presymplectic structure for non-invariant Lagrangians}.

\subsubsection{Dressed presymplectic structure for invariant Lagrangians}
\label{Dressed presymplectic structure for invariant Lagrangians}

In section \ref{The dressing field method, complement} we have argued that given a variational form $\bs\alpha = \alpha\big(\Lambda^\bullet \bs d A; A\big) \in \Omega^\bullet(\A)$ whose $\bs\H$-gauge transformation is $\bs\alpha^{\bs \gamma}= \alpha(\Lambda^\bullet \bs d A^{\bs \gamma}; A^{\bs \gamma})$ (by  \eqref{GT-var-form}), and a field-dependent dressing field $\bs u \in \D r[G, K]$ - with $K \triangleleft H$ so that $H/K=J$ is a group\footnote{This is the favorable case where is it meaningful to speak of residual gauge transformations of the first kind discussed in section \ref{Residual gauge transformations (first kind)}. } - such that $R^\star_\gamma \bs u = \gamma\- \bs u $ for $\gamma \in \K \triangleleft  \H$,  one can build a corresponding $\K$-basic form given by equation  \eqref{Dressed-variational-form}: $\bs\alpha^{\bs u}= \alpha\big(\Lambda^\bullet \bs d A^{\bs u}; A^{\bs u}\big)$. As very elementary examples we discussed there the dressing of Lagrangians (and actions). We here reiterate this discussion, but further discuss the corresponding dressing of the presymplectic structure derived from such dressed Lagrangian.

For an invariant Lagrangian, $R^\star_\gamma L=L$ with $\gamma \in \H$, which implies $L^{\bs \gamma}=L$, i.e. $L^{\bs \gamma}(A)= L(A^{\bs \gamma})=L(A)$, for $\bs \gamma \in \bs\H$, the corresponding dressed Lagrangian is $L^{\bs u}(A)\defeq L(A^{\bs u})$.\footnote{Remind that in case $G\supset H$, we may need to factor in the fact that the polynomial $P$ on which $L$  is based must be extended to (or is the restriction of)  a \mbox{$G$-invariant} polynomial $\b P$ on Lie$G$-valued variables. A fact that we could keep track of notationally as e.g. $L^{\bs u}(A)= \b L (A^{\bs u})$, idem for other dressed functionals/forms. We avoid this in the following, but keep it in mind as it will be relevant to the case of 4D gravity. \label{note1}} Notice that since the invariance of $L$ holds as a formal functional property, we have $L^{\bs u}=L$, which means that the Lagrangian can be rewritten in terms of the $\K$-invariant variable $A^{\bs u}$ and that if the theory given by $L(A)$ appeared to be a $\H$-gauge theory, it is actually a $\J$-gauge theory written in a ``Ockamized'' way as $L(A^{\bs u})$.\footnote{We must qualify this statement by stressing that it is true if $\bs u$ is  \emph{local}  in the sense of field theory. In such a case (part of) the gauge symmetry can be killed without losing the locality of field variables. Such a gauge symmetry is said \emph{artificial}. Gauge symmetries that can only be killed at the price of the locality of the theory are called \emph{substantial}. For a detailed discussion of this distinction, see \cite{Francois2018} and references therein. }

What about the presymplectic structure derived from $L^{\bs u}$? It is quite obvious that we must have as usual, 
\begin{align}
\bs d L^{\bs u} = \bs E^{\bs u} + d\bs\theta^{\bs u} = E\big(\bs d A^{\bs u}; A^{\bs u}\big) + d\theta \big(\bs dA^{\bs u}; A^{\bs u}\big),
\end{align}
where $\bs E^{\bs u}$ and $\bs \theta^{\bs u}$ are $\K$-basic. One still defines $\bs\Theta^{\bs u}=\bs d \bs \theta^{\bs u}$. The dressed presymplectic potential is $\bs \theta^{\bs u}_\Sigma=\int_\Sigma \bs \theta^{\bs u} $ and the corresponding dressed presymplectic 2-form is  $\bs \Theta^{\bs u}_\Sigma=\int_\Sigma \bs \Theta^{\bs u}$. 
Furthermore, since we have derived the respective $\bs\H$-gauge transformations of $\bs E$ \eqref{Field-depGT-FieldEq}, $\bs \theta_\Sigma$ \eqref{Field-depGT-presymp-pot}, and $\bs \Theta_\Sigma$ \eqref{Field-depGT-presymp-form},  we have by \eqref{Dressed-variational-form}: 
\begin{align}
\bs E^{\bs u}&= E(\bs d A^{\bs u}; A^{\bs u})= \bs E + dE \big(  \bs{duu}\- ;  A \big),    \label{Field-Eq-dressed} \\[1.5mm]
\bs\theta_\Sigma^{\bs u} &=\int_\Sigma \theta(\bs d A^{\bs u}; A^{\bs u})=  \bs \theta_\Sigma + \int_{\d\Sigma} \theta( \bs{duu}\- ; A) - \int_\Sigma E(\bs{duu}\-; A),   \label{theta-dressed}  \\[1.5mm]
\bs\Theta_\Sigma^{\bs u} &= \int_\Sigma\Theta(\bs \Lambda^2 d A^{\bs u}; A^{\bs u})=\bs\Theta_\Sigma + \int_{\d\Sigma}   \bs d \theta\big(\bs{duu}\-; A\big) - \int_\Sigma \bs d E\big(\bs{duu}\-; A\big).         \label{Theta-dressed}
\end{align}
 We notice  that $A$ and $A^{\bs u}$ satisfy the same field equations, if $\bs E=0$ then $\bs E^{\bs u}=0$. By definition, the above are off-shell $\K$-basic, thus $\bs\K$-invariant, and remain so on-shell. 
  So that $\bs\theta_\Sigma^{\bs u}$ and $\bs\Theta_\Sigma^{\bs u}$ descend as well-defined~forms~on~$\S/\K$.
  The~$\K$-basicity of $\bs\theta_\Sigma^{\bs u}$ also means that  dressed Noether charges associated to Lie$\K$ vanish identically: for $\chi \in$~Lie$\K$, $Q_\Sigma\big(\chi; A^{\bs u}\big)\defeq \iota_{\chi^v} \bs\theta_\Sigma^{\bs u} \equiv 0$. Relatedly, the $\K$-basicity of $\bs\Theta_\Sigma^{\bs u}$ means that the Poisson bracket of such dressed charges is identically $0$: 
  $\big\{ Q_\Sigma\big(\chi; A^{\bs u}\big) ; Q_\Sigma\big(\chi'; A^{\bs u}\big) \big\} \defeq \bs\Theta_\Sigma^{\bs u} \big( \chi^v, {\chi'}^v \big) \equiv 0$.

  In particular if  we have a $H$-dressing field $\bs u  \in \D r[G, H]$,  then $\bs\theta_\Sigma^{\bs u}$ and $\bs\Theta_\Sigma^{\bs u}$ are off-shell and on-shell $\H$-basic and $\bs\H$-invariant, and may descend as a well-defined symplectic structure on $\S/\H$. And this without any conditions on the fields and/or the boundary. Clearly, the dressed $\H$-Noether charges and their Poisson bracket vanish identically.

Since on-shell we have
\begin{align}
\label{theta-Theta-dressed-on-shell}
\bs\theta_\Sigma^{\bs u} =  \bs \theta_\Sigma + \int_{\d\Sigma} \theta( \bs{duu}\- ; A) \qquad \text{and} \qquad 
\bs\Theta_\Sigma^{\bs u} &= \bs\Theta_\Sigma + \int_{\d\Sigma}   \bs d \theta\big(\bs{duu}\-; A\big), 
\end{align}
it may seem as if the dressing field needs only to exist at the boundary $\d\Sigma$, hence the name ``edge mode"  the dressing field was given in this context \cite{DonnellyFreidel2016, Geiller2017, Geiller2018, Speranza2018, Geiller2019}. This is understandable when considering a genuine physical boundary, but if we are concerned with arbitrary partition of a region into subregions, then the fictitious boundary can be anywhere in the bulk of the region and it is more natural to admit that the dressing field is defined everywhere and not only at the boundary. Furthermore, since it is field-dependent, $\bs u =\bs u(A)$, and the gauge field $A$ is a priori supposed to be defined across the bulk as well as at the boundary, it is again natural to assume that so is the dressing field built from it. 

In the literature, equations  \eqref{theta-dressed}-\eqref{Theta-dressed}  - or rather their restriction on-shell  - are referred to as  the \emph{extended} presymplectic structure brought about by edge modes, and to be associated to a gauge field theory over a bounded region. For obvious reasons, we may rather call these a \emph{dressed} presymplectic structure. 

\medskip

\noindent{{\bf Residual gauge transformations of the first kind:}} In  case $\bs u  \in \D r[G, K]$ and $K \triangleleft H$ as above, there are residual $\J$-transformations. Suppose $\bs u$ satisfies Proposition \ref{Residual1} of section \ref{Residual gauge transformations (first kind)}, 
so that
\begin{align}
R^\star_\eta \bs{u} = \eta\- \bs{u}\eta \quad \text{with} \quad \eta \in \J, \qquad \text{and} \qquad \bs{u^\eta}=\bs{\eta}\- \bs{u \eta}\quad \text{with} \quad\bs{\eta} \in \bs{\J}.
\end{align}
Then $R_\eta \, A^{\bs u}\defeq (A^{\bs u})^\eta = \eta\- A^{\bs u} \eta+ \eta\- d\eta$ and we have $R^\star_\eta L^{\bs u}=L^{\bs u}$. From there, the whole analysis of section \ref{Presymplectic structure for invariant Lagrangians}  regarding Noether currents and charges and their Poisson bracket holds for  $\J$-transformations. Given $\lambda \in$ Lie$\J$, the dressed Noether charge is 
\begin{align}
\label{dressed-Noether-charge-invariant-1st-kind}
Q_\Sigma(\lambda;A^{\bs u})= \int_{\d\Sigma} \theta(\lambda; A^{\bs u}) - \int_\Sigma E(\lambda; A^{\bs u}), \quad \text{and s.t.} \quad \iota_{\lambda^v}\bs\Theta^{\bs u}_\Sigma = -\bs dQ_\Sigma(\lambda; A^{\bs u}).
\end{align}
The dressed presymplectic 2-form induces a Poisson bracket for those charges by
\begin{align}
\label{dressed-Poisson-bracket-1st-kind}
\big\{ Q_\Sigma(\lambda; A^{\bs u}) ,  Q_\Sigma(\lambda'; A^{\bs u})\big\}\defeq\, \bs\Theta^{\bs u}_\Sigma\big(\lambda^v, {\lambda'}^v\big) =	 Q_\Sigma\big([\lambda, \lambda']; A^{\bs u}\big),
\end{align}
so that the Poisson algebra of dressed charges is isomorphic to Lie$\J$. The dressed charges generate infinitesimal $\J$-transformations via $\big\{ Q_\Sigma(\lambda; A^{\bs u}) , \  \ \ \ \big\}$, as in appendix \ref{Noether charges as generators of gauge transformations}. 

 Now, as seen in section \ref{Residual transformations and the bundle structure of the space of dressed connections}, the residual $\bs\J$-gauge transformation of the $\K$-basic dressed form $\bs\alpha^{\bs u}$ is given by 
\eqref{ResidualGT-dressed-alpha}: $\big( \bs{\alpha}^{\bs u} \big)^{\bs \eta} = \alpha \left( \Lambda^\bullet \big(\bs{d}A^{\bs u}\big)^{\bs \eta}; (A^{\bs u})^{\bs \eta} \right)$. Thus the field-dependent $\bs\J$-gauge transformations of 
$\bs E^{\bs u} $, $\bs\theta_\Sigma^{\bs u}$ and $\bs\Theta_\Sigma^{\bs u}$ are 
\begin{align}
  \big( \bs E^{\bs u}\big)^{\bs \eta}&= \bs E^{\bs u} + dE\big(  \bs{d\eta\eta}\- ;  A^{\bs u} \big),  \label{ResidualGT-FieldEq-1st-kind-inv}\\[1.5mm]
 \big(  \bs\theta^{\bs u}_\Sigma \big)^{\bs \eta}& =  \bs \theta^{\bs u}_\Sigma + \int_{\d\Sigma} \theta( \bs{d\eta\eta}\- ; A^{\bs u}) - \int_\Sigma E(\bs{d\eta\eta}\-; A^{\bs u}),  \label{ResidualGT-theta-1st-kind-inv}\\[1.5mm]
 \big(\bs\Theta^{\bs u}_\Sigma\big)^{\bs \eta} &= \bs\Theta^{\bs u}_\Sigma + \int_{\d\Sigma}   \bs d \theta\big(\bs{d\eta\eta}\-; A^{\bs u}\big) - \int_\Sigma \bs d E\big(\bs{d\eta\eta}\-; A^{\bs u}\big).    \label{ResidualGT-Theta-1st-kind-inv}
\end{align}
Here the reduced phase space is $\S^{\bs u}/\J$, and we see that for $\bs\theta_\Sigma^{\bs u}$ and $\bs\Theta_\Sigma^{\bs u}$ to be on-shell $\J$-basic and to thus induce a well-defined symplectic structure on  $\S^{\bs u}/\J$, one needs again to stipulate adequate boundary conditions. 
\smallskip

One may wonder what happens if $\bs u$ satisfies Proposition \ref{Residual2} of section \ref{Residual gauge transformations (first kind)}, so that $A^{\bs u}$  is a twisted connection. As we have briefly commented in section \ref{Residual transformations and the bundle structure of the space of dressed connections}, we would need to generalise what has been done in section \ref{Bundle geometry of A} and now in section \ref{Presymplectic structure for invariant Lagrangians} (and \ref{Presymplectic structure for non-invariant Lagrangians}) to analyse the bundle geometry and presymplectic structures of the space of twisted connections $\t \A$.
As hinted in \cite{Francois2019_II}, this is needed to properly understand the presymplectic structure of conformal gravity - which turns out to be a twisted gauge theory - a topic we may address in another paper. 
\smallskip

As we have already noticed, in case $\bs u$ is a $H$-dressing field, $\bs u  \in \D r[G, H]$, there are no residual gauge transformations of the first kind:   $\bs\theta_\Sigma^{\bs u}$ and $\bs\Theta_\Sigma^{\bs u}$ are $\H$-basic (so in particular $Q_\Sigma\big(\chi; A^{\bs u}\big) \defeq \iota_{\chi^v}\bs\theta_\Sigma^{\bs u} \equiv 0$ for $\chi \in $ Lie$\H$) and may give a symplectic structure on $\S/\H$. In this case it is particularly relevant to consider residual transformations of the second kind. 
\medskip

\noindent{{\bf Residual  transformations of the second kind:}} In sections \ref{Residual transformations (second kind) : ambiguity in choosing a dressing field} and \ref{Residual transformations and the bundle structure of the space of dressed connections} we have shown how  an ambiguity in choosing a dressing field implies a priori  that the space of dressed connections $\A^{\bs u}$ is a $\t \G$-principal bundle, with $\t\G\defeq \left\{ \xi :\P \rarrow G\, |\, R^*_h\xi = \xi \right\}$\footnote{Even though it is not necessary, and contrary to the general case treated in section \ref{Residual transformations and the bundle structure of the space of dressed connections}, here we will assume that $\bs u  \in \D r[G, H]$ so that $\H$-transformations are fully killed and there are no residual gauge transformations of the first kind. It makes the discussion slightly simpler.} acting as 
\begin{align}
\label{residual-2nd-kind-action-2}
\bs u^\xi \defeq \bs u\xi, \qquad \text{and} \qquad A^\xi=A,
\end{align}
so that $R_\xi A^{\bs u} \defeq (A^{\bs u})^\xi=\xi \- A^{\bs u} \xi + \xi\-d\xi$. 
Then we have $R^\star_\xi L^{\bs u} = L^{\bs u}$, i.e. $L\big( (A^{\bs u})^\xi\big)=L(A^{\bs u})$,\footnote{In case $G \supset H$,  as remarked in footnote \footref{note1},  we would write $\b L\big( (A^{\bs u})^\xi\big)=\b L(A^{\bs u})$. \label{note2}} and again the whole analysis of section \ref{Presymplectic structure for invariant Lagrangians}  regarding Noether currents and charges and their Poisson bracket is available on the $\t\G$-bundle $\A^{\bs u}$. So, for $\alpha\in$ Lie$\t\G$ we get the dressed Noether charges
\begin{align}
\label{dressed-Noether-charge-invariant-2nd-kind}
Q_\Sigma(\alpha; A^{\bs u})= \int_{\d\Sigma} \theta(\alpha; A^{\bs u}) - \int_\Sigma E(\alpha; A^{\bs u}), \quad \text{which are s.t.} \quad \iota_{\alpha^v}\bs\Theta^{\bs u}_\Sigma = -\bs dQ_\Sigma(\alpha; A^{\bs u}).
\end{align}
The dressed presymplectic 2-form induces a Poisson bracket for these dressed charges by
\begin{align}
\label{dressed-Poisson-bracket-2nd-kind}
\big\{ Q_\Sigma(\alpha; A^{\bs u}) ,  Q_\Sigma(\alpha'; A^{\bs u})\big\}\defeq\, \bs\Theta^{\bs u}_\Sigma\big(\alpha^v, {\alpha'}^v\big) =	 Q_\Sigma\big([\alpha, \alpha']; A^{\bs u}\big),
\end{align}
so that the Poisson algebra of dressed charges is isomorphic to Lie$\t\G$, and as in appendix \ref{Noether charges as generators of gauge transformations}  they generate infinitesimal \mbox{$\t\G$-transformations} via 
$\big\{ Q_\Sigma(\alpha; A^{\bs u}) , \  \ \ \ \big\}$. 

In section \ref{Residual transformations and the bundle structure of the space of dressed connections} we showed that the field-dependent $\bs{\t\G}$-transformation of the dressed form $\bs\alpha^{\bs u}$ is given by 
\eqref{Dressed-alpha-residual-2nd-kind}: $ \big( \bs{\alpha}^{\bs u} \big)^{\bs \xi} = \alpha \left( \Lambda^\bullet \big(\bs{d}A^{\bs u}\big)^{\bs \xi}; (A^{\bs u})^{\bs \xi} \right)$. The $\bs{\t\G}$-gauge transformations of 
$\bs E^{\bs u} $,$\bs\theta_\Sigma^{\bs u}$ and $\bs\Theta_\Sigma^{\bs u}$ are  then
\begin{align}
  \big( \bs E^{\bs u}\big)^{\bs \xi}&= \bs E^{\bs u} + dE\big(  \bs{d\xi\xi}\- ;  A^{\bs u} \big), \\[1.5mm]
 \big(  \bs\theta^{\bs u}_\Sigma \big)^{\bs \xi}& =  \bs \theta^{\bs u}_\Sigma + \int_{\d\Sigma} \theta( \bs{d\xi\xi}\- ; A^{\bs u}) - \int_\Sigma E(\bs{d\xi\xi}\-; A^{\bs u}),  \label{GT-2nd-kind-theta}\\[1.5mm]
 \big(\bs\Theta^{\bs u}_\Sigma\big)^{\bs\xi} &= \bs\Theta^{\bs u}_\Sigma + \int_{\d\Sigma}   \bs d \theta\big(\bs{d\xi\xi}\-; A^{\bs u}\big) - \int_\Sigma \bs d E\big(\bs{d\xi\xi}\-; A^{\bs u}\big).   \label{GT-2nd-kind-Theta}
\end{align}
In view of the SES  \eqref{SESgroups-dressed} and \eqref{SESLieAlg-dressed}, we have determined that $\A/\H \simeq \A^{\bs u}/\t\G$.  It then follows that the reduced phase space is $\S/\H \simeq \S^{\bs u}/\t\G$,  and we see that for $\bs\theta_\Sigma^{\bs u}$ and $\bs\Theta_\Sigma^{\bs u}$ to be on-shell $\t\G$-basic and to thus induce a well-defined symplectic structure on  $\S^{\bs u}/\t\G$, one needs  to stipulate adequate boundary conditions. 
The situation exactly parallels what happened for $\bs\theta_\Sigma$ and $\bs\Theta_\Sigma$ w.r.t. $\S/\H$. 
\bigskip

In the literature, $\t\G$ is often called ``surface symmetry" or ``boundary symmetry" - e.g. in \cite{DonnellyFreidel2016, Geiller2017, Speranza2018} - as $\bs u$ is believed to live on  $\d\Sigma$ only (for reasons evoked below \eqref{theta-Theta-dressed-on-shell}), and it is claimed to be a new \emph{physical} symmetry arising from the introduction of dressing fields. From our perspective, we see this assertion as misguided. 

As the analysis of transformations of the second kind in section \ref{Residual transformations and the bundle structure of the space of dressed connections} made clear, in view of the  SES  \eqref{SESgroups-dressed}-\eqref{SESLieAlg-dressed} associated to the $\t\G$-bundle $\A^{\bs u}$, the group $\t\G$ obviously does not permute points of $\S^{\bs u}/\t\G$ which is isomorphic to the physical phase space,  $\S^{\bs u}/\t\G\simeq \S/\H$. Therefore,  $\t\G$ can never be a physical transformation group. \mbox{A genuine} physical transformation, a Hamiltonian flow, belongs to $\Diff\big(\S/\H \big) \simeq \Diff\big(\S^{\bs u}/\t\G \big)$, or infinitesimally to $\Gamma \big(T\S/\H \big) \simeq \Gamma\big( T\S^{\bs u}/\t\G \big) $. 

In particular, as already remarked in section \ref{Residual transformations (second kind) : ambiguity in choosing a dressing field}, when a \emph{local} dressing field in $\D r[H, H]$ is introduced  \emph{by fiat}  in a theory it amounts to the tacit assumption that the underlying bundle is trivial \cite{FLM2015_I}, which implies  $\t\G=\t\H \simeq \H$, i.e. the ``new symmetry" is just  the initial gauge symmetry in another guise. In this case, the gauge symmetry $\H$ is arguably \emph{artificial} to begin with, and thus dispensable: The dressing operation,  by rewriting the theory in terms of invariant local variables $A^{\bs u}$ representing the physical d.o.f., simply rids it of a `fake' gauge symmetry \cite{Jackiw-Pi2015} that plays no relevant physical role (the theory is ``Ockhamised"). See  \cite{Francois2018} for a deeper discussion of this point.

In case a local dressing field is not  introduced by \emph{fiat} but built from the field content of the theory, here $A$, then the constructive procedure may be s.t.  $\t\G$, while still not a physical transformation group, is either  ``small'' compared to $\H$ or an interesting new gauge symmetry. The dressing operation thus results in the switching from one gauge symmetry to another. This latter point is illustrated by the case of 4D gravity below. Unfortunately, the problem of defining a symplectic structure on the reduced phase space faces us again w.r.t. $\t\G$, as we see from  \eqref{GT-2nd-kind-theta}-\eqref{GT-2nd-kind-Theta}.

We conclude that the only hope for the DFM 
 to help conclusively with the boundary problem in a (pure) gauge theory rests on the possibility  1)  to construct a $H$-dressing field from the gauge potential $A$, and 2) to provide good reasons that this constructive procedure is free of ambiguities so that $\t\G$ is essentially reduced to the trivial group. 
\smallskip

Remark that for theories where  $\H$ is a \emph{substantial} gauge symmetry, there are good reasons to believe that no \emph{local} $H$-dressing field can be built - see again \cite{Francois2018}. Which means that even if the above two conditions were met, the basic (dressed) variable $A^{\bs u}$ as well as all derived objects, such as $\bs\theta^{\bs u}_\Sigma$ and $\bs\Theta^{\bs u}_\Sigma$, would be gauge-invariant but non-local. The boundary problem would be solved at the price of the locality of the theory. It is not clear how satisfying such a resolution is. 
But as the trade-off locality/gauge-invariance is a well-known hallmark of substantial gauge symmetries, it may be that this result is unavoidable. The boundary problem could then be seen as one more instance of technical and conceptual difficulty posed by the intrinsic non-local way in which physical d.o.f. are encoded in (substantial) gauge theories \cite{Lyre2004,Guay2008, Healey2009, Dougherty2017, Nguyen-et-al2017}. 
Relatedly, for some it signals the necessity of gauge symmetries to dynamically couple local systems, i.e. the \emph{relational} nature of the physics of gauge fields \cite{Rovelli, Rovelli2020, Gomes2019, Gomes2019-bis}.
\smallskip

Let us  illustrate this discussion by the application of the above scheme to the cases of  YM theory and 4D gravity.

\subsubsubsection{Yang-Mills theory}
\label{Yang-Mills theory}

In YM theory the Lagrangian is $L_\text{\tiny YM}(A)=\tfrac{1}{2}\Tr (F \, *\!F)$, the gauge group is $\H=\SU(n)$,  and $R^\star_\gamma L_\text{\tiny YM}=L_\text{\tiny YM}$ for $\gamma \in \H$. The associated field equations are $\bs E = E(\bs d A; A) = \Tr\big( \bs d A \, D^A\!*\!F\big)$ and the presymplectic potential current is $\bs \theta =   \theta(\bs d A; A) = \Tr\big( \bs d A *\!F\big)$, while the presymplectic 2-form current is $\bs\Theta= -\Tr\big(\bs d A *\! \bs d F \big)$.

Admitting there is a $S\!U(n)$-dressing field $\bs u$, the dressed Lagrangian is $L^{\bs u}_\text{\tiny YM}(A)=L_\text{\tiny YM}(A^{\bs u})=\tfrac{1}{2}\Tr (F^{\bs u} *\!F^{\bs u})$, 
 and by  \eqref{Field-Eq-dressed} the  field equations are $\bs E^{\bs u}= \Tr\big( \bs d A^{\bs u} \, D^{A^{\bs u}}\!*\!F^{\bs u}\big)= \Tr\big( \bs d A \, D^A\!*\!F\big) + d\Tr\big( \bs{duu}\- \, D^A\!*\!F\big)$.
By \eqref{theta-dressed} and \eqref{Theta-dressed}, the  dressed presymplectic potential and 2-form are, 
\begin{align}
\bs\theta_\Sigma^{\bs u} = \int_\Sigma\Tr\big( \bs d A^{\bs u} \, *\!F^{\bs u}\big) &= \bs\theta_\Sigma +  \int_{\d _\Sigma }\Tr\big(\bs{duu}\- *\!F \big) - \int_\Sigma\Tr\big( \bs{duu}\-\, D^A\!*\!F\big),  \notag\\*
					&= \bs\theta_\Sigma +  \int_{\d _\Sigma }\Tr\big(\bs{duu}\- *\!F \big) _{\ |\S},  \label{dressed-theta-YM}\\[1mm]
\bs\Theta_\Sigma^{\bs u} = - \int_\Sigma \Tr\big(\bs d A^{\bs u} *\! \bs d F^{\bs u} \big)&= \bs\Theta_\Sigma +  \int_{\d \Sigma} \bs d \Tr\big( \bs{duu}\- *\!F \big)  - \int_\Sigma \bs d  \Tr\big(  \bs{duu}\-\, D^A\!*\!F \big), \notag\\
					 &= \bs\Theta_\Sigma +  \int_{\d \Sigma} \bs d \Tr\big( \bs{duu}\- *\!F \big) _{\ |\S}, \\
					 &= \bs\Theta_\Sigma -  \int_{\d \Sigma}  \Tr\big( \bs{duu}\- *\!\bs d F  -\tfrac{1}{2} [\bs{duu}\-, \bs{duu}\-] *\!F\big) _{\ |\S}. \notag
\end{align}
Using $\bs d \big(\bs{duu}\-\big) -\sfrac{1}{2}\big[\bs{duu}\-, \bs{duu}\- \big]=0$  in the last step. This reproduces eq (2.19) and eq.(2.22)-(2.23) in \cite{DonnellyFreidel2016}. 

Since $\bs u$ kills all of $\H$, there are no residual gauge transformations of the first kind. In particular,  for $\chi \in$ Lie$\H$, $Q_\Sigma(\chi; A^{\bs u}) \defeq \iota_{\chi^v}\bs\theta^{\bs u}_\Sigma  \equiv 0$, which - using $\iota_{\chi^v}\bs{duu}\-\!=-\chi$ in  \eqref{dressed-theta-YM} - reproduces eq.(2.29) of   \cite{DonnellyFreidel2016}. But there is a priori an ambiguity in the choice of the dressing field, so we have  residual transformations of the second kind embodied by the group $\t\H\defeq \left\{ \xi :\P \rarrow S\!U(n)\, |\, R^*_h\xi = \xi \right\}$ acting as $\bs u^\xi = \bs u \xi$ and $A^\xi=A$. Then $R^\star_\xi L^{\bs u}_\text{\tiny YM}(A) = L^{\bs u}_\text{\tiny YM}(A)$, and for $\alpha \in$ Lie$\t\H$ the dressed Noether charge is, by \eqref{dressed-Noether-charge-invariant-2nd-kind},
\begin{align}
\label{dressed -Noether-charge-YM}
Q_\Sigma(\alpha; A^{\bs u}) &= \int_{\d\Sigma} \Tr\big(\alpha *\!F^{\bs u} \big) - \int_\Sigma \Tr\big( \alpha\, D^{A^{\bs u}}\!*\!F^{\bs u} \big), \\
					   &=	\int_{\d\Sigma} \Tr\big(\alpha *\!F^{\bs u} \big)_{\ |\S}. \notag
\end{align}
This reproduces eq.(2.35) of \cite{DonnellyFreidel2016}, 
and can also  be checked directly via $Q_\Sigma(\alpha; A^{\bs u})= \iota_{\alpha^v} \bs\theta_\Sigma^{\bs u} $, using $\iota_{\alpha^v} \bs d A=0$ and $\iota_{\alpha^v}\bs{duu}\-\!= \bs u\alpha \bs u\-$ in  \eqref{dressed-theta-YM}. 
Also, $\iota_{\alpha^v}\bs\Theta^{\bs u}_\Sigma = -\bs dQ_\Sigma(\alpha; A^{\bs u})$ reproduces eq.(2.36) of \cite{DonnellyFreidel2016}, while the Poisson bracket of dressed charges induced by  $\bs\Theta_\Sigma^{\bs u}$ via \eqref{dressed-Poisson-bracket-2nd-kind} reproduces eq.(2.38) in the same reference. 

Finally, by  \eqref{GT-2nd-kind-theta} and  \eqref{GT-2nd-kind-Theta} the $\bs{\t\H}$-transformations of the dressed presymplectic potential and 2-form are, 
\begin{align}
(\bs\theta^{\bs u}_\Sigma)^{\bs \xi} &= \bs \theta^{\bs u}_\Sigma + \int_{\d _\Sigma }\Tr\big(\bs{d\xi\xi}\- *\!F^{\bs u} \big) - \int_\Sigma\Tr\big( \bs{d\xi\xi}\-\, D^{A^{\bs u}}\!*\!F^{\bs u}\big),     \label{GT-2nd-kind-thetaYM} \\[1mm]
(\bs\Theta^{\bs u}_\Sigma)^{\bs\xi} &= \bs\Theta^{\bs u}_\Sigma + \int_{\d \Sigma} \bs d \Tr\big( \bs{d\xi\xi}\- *\!F^{\bs u} \big)  - \int_\Sigma \bs d  \Tr\big(  \bs{d\xi\xi}\-\, D^{A^{\bs u}}\!*\!F^{\bs u} \big),   \label{GT-2nd-kind-ThetaYM}
\end{align}
in complete analogy with  \eqref{GT-thetaYM}-\eqref{GT-ThetaYM}. We see how the  boundary problem reemerges. 
\smallskip

As here the dressing field is introduced essentially by hand, we have that $\t\H \simeq\H$, and the Poisson algebra of dressed charges is isomorphic to Lie$\H$. 
As stressed in the general case, $\t\H$ doesn't permute points in the physical phase space $\S/\H \simeq \S^{\bs u}/\t\H$, it is thus never a physical transformation group - contrary to the assertion in~\cite{DonnellyFreidel2016}. 

For the edge mode strategy to solve the boundary problem in (pure) YM theory, one needs 1) to show that it is possible to construct a $S\!U(n)$-dressing field from the gauge potential $A$ and 2) to provide good reasons that this constructive procedure is s.t. $\t\H$  reduces to the trivial group. Remark that if $\SU(n)$ is a substantial gauge symmetry, then no \emph{local} $S\!U(n)$-dressing field can be built in YM theory. Which means that even if the above conditions were met, the dressed variable $A^{\bs u}$ would be gauge-invariant but non-local, and so would be the dressed symplectic structure induced by $\bs\theta^{\bs u}_\Sigma$ and $\bs\Theta^{\bs u}_\Sigma$ .


\subsubsubsection{4D gravity}
\label{4D gravity}

In this section we will take the viewpoint that $\b \A$ is the space of local representatives on $U\subset \M$ of Cartan connections on $\P$, which changes nothing of substance to the general results derived up to now.  
We will consider in turn 4D Einstein-Cartan gravity with $\Lambda \neq 0$ and 4D McDowell-Mansouri gravity.
\medskip

\noindent {\bf 4D Einstein-Cartan gravity $\bs{\Lambda\neq0}$:} 
 As seen in section \ref{4D EC gravity}, the underlying Cartan geometry is reductive and the Cartan connection splits as $\b A= A + e$, where $A$ is $\so(1,3)$-valued and the soldering form $e$ is $\RR^4$-valued. The gauge group is $\H=\SO(1,3)$ and acts as $R_\gamma A \defeq A^\gamma = \gamma\- A \gamma + \gamma \- d\gamma$ and $R_\gamma e\defeq e^\gamma= \gamma\- e$. 
 
 From this we see that a $S\!O$-dressing is readily extracted from the soldering form, i.e. the Cartan connection: Given a coordinate system $\{ x^{\, \mu}\}$ on $U$ the soldering is $e={ e^a}_\mu\, dx^{\,\mu}$, so the map $\bs e \defeq {e^a}_\mu: U \rarrow GL(4)$  is s.t. $\bs e^\gamma  = \gamma\- \bs e$. We thus have indeed a field-dependent local dressing field
$ \bs u : \b \A \rarrow D r\big[GL, S\!O\big]$, 
 $           \b A  \mapsto	\bs u(\b A)=\bs e$,
s.t. $R^\star _\gamma \bs u(\b A)= \bs u(R_\gamma \b A)=\bs u(\b A^\gamma)= \bs e^\gamma=\gamma\- \bs e=\gamma\- \bs u(\b A)$. 
Said otherwise, the tetrad field is a Lorentz dressing field. 
The~$\SO$-invariant dressed Cartan connection is then 
\begin{align}
\label{dressed Cartan connection}
\b A^{\bs u}= \bs u\- \b A \bs u + \bs u\- d\bs u \rdefeq \b\Gamma \quad \Rightarrow \quad  \left\{ \begin{array}{l} A^{\bs u} =  \bs e\- A \bs e + \bs e \- d\bs e  \rdefeq \Gamma,  \\ e^{\bs u}=\bs e \- e = dx, \end{array} \right.
\end{align}
where  $dx= {\delta^{\,\mu}}_\rho\, dx^{\, \rho}$ and $\Gamma={\Gamma^\mu}_\nu ={\Gamma^\mu}_{\nu, \,\rho}\,dx^{\,\rho}$ has values in $M(4, \RR)=$ Lie$GL(4)$. The latter is the familiar linear connection. Correspondingly the dressed Cartan curvature is 
\begin{align}
\label{dressed Cartan curvature}
\b F^{\bs u}= \bs u\- \b F \bs u  \Rightarrow  \left\{ \begin{array}{l} R^{\bs u} =  \bs e\- R \bs e   \rdefeq {\sf R}= d\,\Gamma +\tfrac{1}{2}[\Gamma, \Gamma],  \\ T^{\bs u}=\bs e \- T \rdefeq {\sf T} = \Gamma \w dx, \end{array} \right.
\end{align}
where ${\sf T} = {\sf T}^{\,\mu}=\tfrac{1}{2}{{\sf T}^{\,\mu}}_{\rho\sigma} \, dx^{\,\rho} \w dx^{\,\sigma} = {\Gamma^\mu}_{\rho\sigma}\, dx^{\,\rho}\w dx^{\,\sigma}$ and ${\sf R}=\tfrac{1}{2} {{\sf R}^{\,\mu}}_{\nu, \, \rho\sigma} \, dx^{\,\rho} \w dx^{\,\sigma}$ is $M(4, \RR)$-valued. 

The geometry being reductive, $\bs d \b A^{\bs u} = \bs d \b \Gamma$ splits and by   \eqref{Dressed-dA-Cartan} we have 
\begin{align}
\label{dressed-fields-EC}
\bs dA^{\bs u}=\bs d\ \! \Gamma = \bs e\-  \left( \bs{d}A  + D^A\left\{ \bs{dee}\- \right\} \right)  \bs e, \qquad \text{and} \qquad \bs de^{\bs u} = \bs{e}\-   \left( \bs{d}e   -  \bs{d}\bs{e} {\bs{e}\- e} \right) \equiv 0,
\end{align}
where we remark that  $\bs{dee}\- =\bs d { e^a}_\mu { (e\-)^{\,\mu}}_b  \in $ Lie$\SO$, and the second equality follows from the fact that $\bs{dee}\- e =\bs d { e^a}_\mu { (e\-)^{\,\mu}}_b\ e = \bs d { e^a}_\mu \, dx^{\, \mu} = \bs d e$ (or indeed simply from $\bs d\, dx =0$).

As reminded in section \ref{Ehresmann and Cartan connections}, given a non-degenerate bilinear form $\eta$ on Lie$G/$Lie$H$, the soldering of a Cartan connection induces a metric on $M$ (or a class thereof, depending on the action of $H$-action on $\eta$) by $g\defeq \eta(e, e)$. Here, the bilinear form is the Minkowski metric $\eta$ on $\RR^4$, the metric is  in components $\bs g\defeq \bs e^T \eta\, \bs e$, i.e $g_{\mu\nu}={e_\mu}^a\eta_{ab}\,{e^{b}}_\nu$, and it is another $\SO$-invariant field arising naturally. Remark that the metricity condition is automatic, as we have  $\nabla \bs g\defeq d\bs g - \Gamma^T\bs g - \bs g \Gamma = -\bs e^T\big( A^T\eta + \eta\, A \big)\bs e =0$.
\medskip

The polynomial \eqref{Polyn-gravity} with which we wrote the Lagrangian for 4D gravity in sections \ref{4D EC gravity} and \ref{4D MM gravity} is $S\!O$-invariant by \eqref{Prop1-P}. It is the restriction of the polynomial $\b P : \otimes^k M(2k, \mathbb{K}) \rarrow \mathbb{K}$ given by
\begin{align}
\label{Polyn-gravity2}
\b P\big(M_1, \ldots, M_k \big)= \sqrt{|\det(\bs g)|}\ M_1 \bullet\, \ldots\, \bullet M_k\defeq \sqrt{|\det(\bs g)|}\ M^{\mu_1 \mu_2}_1\, M^{\mu_3\mu_4}_2 \ldots \, M^{\mu_{2k-1}\mu_{2k}}_k \, \epsilon_{\mu_1 \ldots \mu_{2k}}.
\end{align}
Clearly only the antisymmetric part  {\sf  A} of a variable $M= {\sf S}+  {\sf  A}$ contributes.  
Under the substitution 
$\bs g \rarrow G^{T} \bs g\, G$ and $M\rarrow G^{-1T} M\,G\-$, with $G={G^{\alpha}}_{\beta} \in GL(4)$, by a computation analogue to  \eqref{Prop1-P} we have
\begin{align}
\label{Polyn-gravity3}
\b P\big(G^{-1T}M_1G^{-1}, \ldots, G^{-1T}M_kG^{-1} \big) 
											&= \sqrt{|\det(\bs g)|} \det(G) \det(G\-)\ M^{\mu_1 \mu_2}_1\, M^{\mu_3\mu_4}_2 \ldots \, M^{\mu_{2k-1}\mu_{2k}}_k \, \epsilon_{\mu_1 \ldots \mu_{2k}}, \notag \\
											&= \b P\big(M_1, \ldots, M_k \big).
\end{align}
In this sense,  $\b P$ is  thus $GL$-invariant. One obtains the $S\!O$-invariant polynomial $P$ by the substitution  $\bs g \rarrow \eta$. 
Conversely, if in $P$ one plugs variables $\bs e\, M\,\bs e^T $ then by \eqref{Prop1-P} again we get
\begin{align}
\label{Polyn-gravity4}
P\big(\bs e\, M_1 \,\bs e^T, \ldots, \bs e \,M_k\, \bs e^T  \big) = \bs e\, M_1 \,\bs e^T \bullet  \ldots \bullet \bs e \,M_k\, \bs e^T  = \det(\bs e)\  M_1 \,\bs \bullet  \ldots \bullet M_k =\b P\big(M_1, \ldots, M_k \big).
\end{align}
As we are about to see, this situation illustrates the caveat expressed in footnote \footref{note1} and \footref{note2} about the necessary extension of the polynomial on which are based the Lagrangian and all derived variational forms, when the structure group is a subgroup of the target group of the dressing field, $G \supset H$. 
\medskip

The  Lagrangian of the theory is 
$L(A, e)= R \eta\- \bullet e \w e^T -  \tfrac{\epsilon}{2\ell^2} e\w e^T \bullet e\w e^T$, see~\eqref{Lagrangian-EC-2},
and is $\SO$-invariant. The associated field equations  and presymplectic potential current are 
\begin{align}
 \bs E = E\big(\bs d \b A; \b A \big)= 2\left( \bs d A \bullet D^Ae \w e^T + \bs d e \w e^T \bullet \big( R -\tfrac{\epsilon}{\ell^2} e \w e^T  \big) \right) \quad \text{and} \quad   \bs \theta= \theta(\bs d \b A; \b A)= \bs dA \bullet e \w e^T,
\end{align}
while the presymplectic 2-form current is $\bs\Theta=-2\, \bs dA \bullet  \bs d e \w e^T$.

Noticing that $e\w e^T = \bs e\, dx \w dx^T\bs e^T $ and $R\,\eta\- = \bs e\, {\sf R}\,\bs e\- \eta\- = \bs e\,  {\sf R} \bs g\- \bs e^T$,\footnote{Remark also that $ {\sf R}\bs g^{-1} + \bs g^{-1}  {\sf R}^T=\bs e^{-1} (R\eta^{-1} + \eta^{-1} R^T) \,\bs e^{-1T}=0$, so ${\sf R}\bs g^{-1}={\sf R}^{\,\mu\nu}$ is a  2-form with values in antisymmetric matrices. }
 the  dressed Lagrangian is found by
 \begin{align}
 \label{dressed-Lagrangian-EC}
 L^{\bs u}(A, e) &= P\big(  \bs e\,  {\sf R} \bs g\- \bs e^T,\  \bs e\, dx \w dx^T\bs e^T \big) - \tfrac{\epsilon}{2\ell^2}  P(\bs e\, dx \w dx^T\bs e^T, \ \bs e\, dx \w dx^T\bs e^T), \notag\\
 								&=\bs e\,  {\sf R} \bs g\- \bs e^T  \bullet  \bs e\, dx \w dx^T\bs e^T - \tfrac{\epsilon}{2\ell^2}  \bs e\, dx \w dx^T\bs e^T\bullet  \bs e\, dx \w dx^T\bs e^T,  \notag\\[1mm]
								&= \sqrt{|\det(\bs g)|} \ \left( \ {\sf R} \bs g\- \bullet dx \w dx^T  - \tfrac{\epsilon}{2\ell^2} dx \w dx^T \bullet dx \w dx^T \ \right), \notag\\ 
  &= \b P\big(  {\sf R} \bs g\-, \ dx \w dx^T\big) - \tfrac{\epsilon}{2\ell^2}  \b P(dx \w dx^T, \  dx \w dx^T) \rdefeq \b L(\Gamma, \bs g).
  \end{align}
 Developing the expression, with $\tfrac{\epsilon}{\ell^2}=\tfrac{\Lambda}{3}$, this is of course the Lagrangian of GR in the metric formulation,
  \begin{align*}
  \b L(\Gamma, \bs g) = \sqrt{|\det(\bs g)|}  \ \big( {\sf R}^{\, \mu\nu} dx^{\,\alpha} dx^{\,\beta} - \tfrac{\epsilon}{2\ell^2}dx^{\,\mu} dx^{\,\nu} dx^{\,\alpha} dx^{\,\beta} \big) \epsilon_{\mu\nu\alpha\beta}= 2 \sqrt{|\det(\bs g)|} \,d^4\!x \ \big( Ric - 2\Lambda \big). 
 \end{align*}
 
 The dressed presymplectic potential associated to \eqref{dressed-Lagrangian-EC} is thus simply the presymplectic potential of metric GR.
 By \eqref{theta-dressed} it is, 
\begin{align}
\bs\theta_\Sigma^{\bs u} =\int_\Sigma \b\theta\big(\bs d \b A^{\bs u}; \b A^{\bs u}\big) =\int_\Sigma \theta\big(\bs e \bs d\ \!\b\Gamma\, \bs e\-; \b A \big)
					&=  \bs \theta_\Sigma + \int_{\d\Sigma} \theta( \bs{duu}\- ; \b A) - \int_\Sigma E(\bs{duu}\-; \b A),   \notag\\
					&=  \bs \theta_\Sigma + \int_{\d\Sigma} \bs{dee}\- \bullet e\w e^T - \int_\Sigma 2\, \bs{dee}\-  \bullet D^A e \w e^T,   \label{dressed-theta-EC} \\
					&= \bs\theta_\Sigma + \int_{\d\Sigma} \bs{dee}\- \bullet e\w e^T \,  _{\ | N}.  \notag
\end{align}
This can be checked explicitly by plugging $\bs d A = \bs e \bs d\ \!\Gamma\, \bs e\- - D^A\big\{ \bs{dee}\- \big\}$ (obtained from \eqref{dressed-fields-EC}) in $\bs\theta_\Sigma = \theta_\Sigma(\bs d\b A; \b A)$. 
Notice that only the Lie$\SO$-part of the field equations $\bs E$ contributes to this expression. In components we have  
\begin{align}
\label{dressed-theta-components}
\bs\theta_\Sigma^{\bs u} &= \int_\Sigma \theta \big(\bs e \bs d\ \!\b\Gamma\, \bs e\-; \b A \big) = \int_\Sigma \ \bs e \bs d\ \!\Gamma \,\bs e\- \bullet   e\w e^T =   \int_\Sigma \ \sqrt{|\det(\bs g)|}\ \bs d\ \!\Gamma\, \bs g\- \bullet dx \w dx^T,  \notag\\
				       &= \int_\Sigma \ \sqrt{|\det(\bs g)|} \, d^3\!x\,  \delta^\bullet_\mu\ \  2\,  \bs d\ \!{\Gamma_{\alpha \beta}}^{[\mu} g^{\alpha]\beta}. 
\end{align}
which is indeed the presymplectic potential of metric GR, as given e.g. in  \cite{De-Paoli-Speziale2018} eq.(2.14) or   \cite{Oliveri-Speziale2020} eq.(2.8).

The boundary term in  \eqref{dressed-theta-EC} is the - aptly named - ``dressing 2-form" proposed in \cite{De-Paoli-Speziale2018} (see eq(2.32)-(2.34)) and \cite{Oliveri-Speziale2020} (see eq.(2.11)-(2.12)) so as to make $\bs\theta_\Sigma$ ``\emph{fully gauge invariant}"$\,$\footnote{\emph{Horizontal} is what is actually meant there. Even though indeed the conjunction of trivial $\SO$-equivariance (gauge invariance in the standard sense) and horizontality implies basicity, and therefore field-dependent $\bs \SO$-invariance.  } and equivalent to the presymplectic potential of the metric formulation in case of vanishing torsion (i.e. when $\b A$ is the normal Cartan connection).\footnote{Actually \cite{De-Paoli-Speziale2018, Oliveri-Speziale2020} addressed this problem for the classically equivalent theory including the Holst term, which is the starting point of the Loop Quantum Gravity program. It is very easy to take this contribution into account, and we do so in appendix \ref{G}.}
 In~the framework of the DFM, this result and its generalisation are naturally delivered as a matter of course.  

 By  \eqref{Theta-dressed} the  dressed presymplectic  2-forms is
 \begin{align}
\bs\Theta_\Sigma^{\bs u} &=\bs\Theta_\Sigma + \int_{\d\Sigma}   \bs d \theta\big(\bs{duu}\-; \b A\big) - \int_\Sigma \bs d E\big(\bs{duu}\-; \b A\big).    \notag  \\   
					&=  \bs \Theta_\Sigma + \int_{\d\Sigma} \bs d\left(  \bs{dee}\- \bullet e\w e^T  \right) - \int_\Sigma 2\, \bs d\left( \bs{dee}\-  \bullet D^A e \w e^T \right) ,   \label{dressed-Theta-EC} \\
					&= \bs \Theta_\Sigma + \int_{\d\Sigma}  \bs d\left( \bs{dee}\- \bullet e\w e^T\right)\,  _{\ | N}.  \notag
\end{align}
This last result reproduces eq.(6.19) of \cite{Oliveri-Speziale2020}. Using the alternative expression $\bs\Theta_\Sigma^{\bs u} =\bs\Theta_\Sigma + \int_{\Sigma} \bs d \theta\big(D^A \{\bs{duu}\-\}; \b A\big) $ (see \eqref{Field-depGT-Theta} with the substitution $\bs\gamma \rarrow \bs u$) it is easy to check that $\bs\Theta_\Sigma^{\bs u} = \bs d \bs\theta_\Sigma^{\bs u}= \int_\Sigma \bs d \, \left(  \bs e \bs d\ \!\Gamma\, \bs e\- \bullet   e\w e^T \right)$. 
Using \eqref{dressed-theta-components},  in components this is 
\begin{align}
\bs\Theta_\Sigma^{\bs u} = \bs d \bs\theta_\Sigma^{\bs u}= \int_\Sigma \ \sqrt{|\det(\bs g)|} \, d^3\!x\,  \delta^\bullet_\mu\ \  2\, \left\{ \tfrac{1}{2}g^{\alpha\beta} \bs d g_{\alpha\beta} \, \bs d \ \! {\Gamma^{[\mu}}_{\gamma\nu} g^{\,\gamma]\nu} - \bs d \ \! {\Gamma^{[\mu}}_{\gamma\nu}\, \bs d g^{\,\gamma]\nu} \right\}.
\end{align}
It takes some work to show that in the torsion-free case  this reproduces the standard result, 
\begin{align}
\bs\Theta_\Sigma^{\bs u} = \int_\Sigma \ \sqrt{|\det(\bs g)|} \, d^3\!x\,  \delta^\bullet_\mu\ \,
 						& \left\{ g^{\alpha\beta} \bs d g_{\alpha\beta}  \left(  g^{\lambda [\gamma}\, \nabla_\lambda (\bs d g_{\gamma\nu})\, g^{\,\mu]\nu} \right) \right. \notag \\
					        &\qquad \left. - 2\, g^{\lambda\mu}\, \nabla_\gamma( \bs d g_{\nu\lambda})\, \bs d g^{\gamma\nu}
						            	         	     + g^{\lambda\mu}\, \nabla_\lambda(\bs d g_{\gamma\nu})\, \bs d g^{\gamma\nu}
							         	             + g^{\lambda\gamma} \,\nabla_\nu(\bs d g_{\gamma\lambda})\, \bs d g^{\,\mu\nu} \right\},
\end{align}
as given e.g. in  \cite{Hollands-Wald2012} eq.(22)-(23) or \cite{DonnellyFreidel2016}  eq.(33). 

Remark that despite the last equalities in \eqref{dressed-theta-EC} and \eqref{dressed-Theta-EC} -  in case $\b A$ is normal - one is not tempted to think of the tetrad $\bs e$ as a edge mode living only at the boundary $\d\Sigma$. This highlights again that the notion of edge modes is a special case of the more  general and systematic DFM framework.  
\medskip

Since the dressing $\bs u = \bs e$ kills all of $\SO$, there are no residual gauge transformations of the first kind. In particular,  for $\chi \in$ Lie$\SO$, $Q_\Sigma(\chi; \b A^{\bs u}) \defeq \iota_{\chi^v}\bs\theta^{\bs u}_\Sigma  \equiv 0$, which   reproduces eq.(6.26)-(6.28) of   \cite{Oliveri-Speziale2020}, i.e. there are no Lorentz charges. But there is an ambiguity in the choice of the dressing field: the choice of coordinate system! 

Indeed the soldering form $e$ is a coordinate invariant object, but we identified its components $\bs e ={e^a}_{\,\mu}$ in a given coordinate system $\{ x^{\,\mu}\}$ as a good candidate $S\!O$-dressing field. In another coordinate system the components of $e$ are  $\bs e ' = \bs e\, \xi$, i.e. ${{e'}^a}_{\,\nu} = {e^a}_{\,\mu}\, {\xi^{\,\mu}}_\nu$, where  $\xi={\xi^{\,\mu}}_\nu \in GL(4)$ is the Jacobian of the coordinate change. The residual transformations of the second kind are thus here the group of local coordinate transformations $\t\G\defeq \left\{ \xi :U \rarrow GL(4)\, |\, \xi^\gamma=\xi \right\}$ acting on $S\!O$-dressing fields (the tetrad fields) as $\bs u^\xi = \bs u \xi$ and of course trivially on the Cartan connection $\b A^\xi=\b A$. 
 The space of dressed Cartan connections $\b \A^{\bs u}$ is thus a $\t\G$-principal bundle, with a right action of  $\t\G$ given by 
 \begin{align}
 R_\xi {\b A}^{\bs u} = (\b A^{\bs u})^\xi \defeq \xi \- \b A^{\bs u} \xi + \xi\- d\xi \quad \Rightarrow \quad \left\{ \begin{array}{l} (A^{\bs u})^\xi = \Gamma^\xi =  \xi \-   \Gamma \xi + \xi\- d\xi ,  \\[1mm] (e^{\bs u})^\xi= (dx)^\xi = \xi\- dx. \end{array} \right.
 \end{align}
 By \eqref{Polyn-gravity3} the dressed Lagrangian  \eqref{dressed-Lagrangian-EC} has trivial $\t\G$-equivariance $R^\star_\xi \b L=\b L$, so for $\alpha={\alpha^{\, \mu}}_{\!\nu} \in $ Lie$\t\G$ and by  \eqref{dressed-Noether-charge-invariant-2nd-kind}, the dressed Noether charge is
\begin{align}
\b Q_\Sigma(\alpha; \b A^{\bs u}) &=  \int_{\d\Sigma} \b \theta \big(\alpha; \b A^{\bs u}\big) - \int_\Sigma \b E\big(\alpha; \b A^{\bs u}\big) 
					       =  \int_{\d\Sigma}  \theta \big(\bs e\, \alpha \, \bs e\-; \b A\big) - \int_\Sigma  E\big(\bs e\, \alpha \, \bs e\-; \b A\big),  \notag\\
					       &=  \int_{\d\Sigma} \bs e\, \alpha \, \bs e\- \bullet e \w e^T - \int_\Sigma  2\ \bs e\, \alpha \, \bs e\- \bullet D^Ae \w e^T ,  \notag\\
					       &=  \int_{\d\Sigma} \sqrt{|\det(\bs g)|} \ \ \alpha \bs g\-  \bullet dx \w dx^T - \int_\Sigma  2\,\sqrt{|\det(\bs g)|} \ \ \alpha  \bs g\- \bullet {\sf T} \w dx^T,   \label{dressed -Noether-charge-EC} \\
					       &=  \int_{\d\Sigma} \sqrt{|\det(\bs g)|} \  \ \alpha \bs g\-  \bullet dx \w {{dx^T} }_{\ | N}, \notag \\
					       &=  \int_{\d\Sigma} \sqrt{|\det(\bs g)|} \ \ {\alpha^{\,\mu}}_{\!\lambda}\, g^{\lambda \nu} dx^{\,\alpha} dx^{\,\beta} \, {\epsilon_{\mu\nu\alpha\beta}} _{\ | N}.  \label{Pre-Komar-mass}
\end{align}
We notice that in case $\xi$ is interpreted as manifestation  of an active diffeomorphism viewed in a given coordinate system, so that $\alpha = \d \zeta = \d_{\,\nu}\zeta^{\mu}$ with $\zeta$ the components of the vector field generating the diffeomorphism, then the above dressed Noether charge in the absence of torsion is 
\begin{align}
\label{Komar-mass}
\b Q_\Sigma(\d \zeta; \b A^{\bs u}) =  \int_{\d\Sigma} \sqrt{|\det(\bs g)|} \ \d \zeta\, \bs g\-  \bullet dx \w {dx^T}   
						    &=\int_{\d\Sigma} \sqrt{|\det(\bs g)|} \  \  \d^{[\,\nu } \zeta^{\,\mu]} dx^{\,\alpha} dx^{\,\beta} \, {\epsilon_{\mu\nu\alpha\beta}}  _{\ | N}\,  ,  \notag  \\
						     &=\int_{\d\Sigma} \sqrt{|\det(\bs g)|} \  \  \nabla^{[\,\nu } \zeta^{\,\mu]} dx^{\,\alpha} dx^{\,\beta} \, {\epsilon_{\mu\nu\alpha\beta}}  _{\ | N} \, .
\end{align}
If $\zeta$ is a Killing vector field, this is exactly the Komar mass as given  in \cite{Choquet-Bruhat2009} (definition 4.6, eq.(4.8) p. 460). It~is~known to coincide with the Newtonian mass and ADM mass for (stationnary) asymptotically flat spacetimes $\M$, and  to vanish if and only if $\M$ is  flat (Lemma 4.10, Theorem 4.13 and Theorem 4.11 in \cite{Choquet-Bruhat2009}). This is satisfying when $\Lambda=0$, in which case the groundstate of the theory is Minkowski space, but in case $\Lambda\neq0$ this means that we have a non-vanishing Komar mass associated to the (A)dS groundstate. As we will see shortly, this problem is solved in 4D MacDowell-Mansouri gravity. 

The dressed Noether charges \eqref{dressed -Noether-charge-EC} satisfy  $\iota_{\alpha^v}\bs\Theta^{\bs u}_\Sigma = -\bs d\b Q_\Sigma(\alpha; \b A^{\bs u})$ so that the dressed presymplectic 2-form  $\bs\Theta_\Sigma^{\bs u}$  induces via \eqref{dressed-Poisson-bracket-2nd-kind} the Poisson bracket 
\begin{align}
\label{dressed-Poisson-bracket-EC}
\big\{ \b Q_\Sigma(\alpha; \b A^{\bs u}) ,  \b Q_\Sigma(\alpha'; \b A^{\bs u})\big\}\defeq\, \bs\Theta^{\bs u}_\Sigma\big(\alpha^v, {\alpha'}^v\big) =	 \b Q_\Sigma\big([\alpha, \alpha']; \b A^{\bs u}\big). 
\end{align}
The Poisson algebra of dressed Noether charges is thus isomorphic to the Lie algebra of coordinate changes Lie$\t\G$.

Finally, by  \eqref{GT-2nd-kind-theta}  the field-dependent coordinate transformation of the dressed presymplectic potential  is, 
\begin{align}
 \big(  \bs\theta^{\bs u}_\Sigma \big)^{\bs \xi} &=  \bs \theta^{\bs u}_\Sigma + \int_{\d\Sigma} \b \theta( \bs{d\xi\xi}\- ; \b A^{\bs u}) - \int_\Sigma \b E(\bs{d\xi\xi}\-; \b A^{\bs u}) 
 								       =  \bs \theta^{\bs u}_\Sigma + \int_{\d\Sigma}  \theta( \bs e\, \bs{d\xi\xi}\- \bs e \- ; \b A) - \int_\Sigma  E( \bs e\, \bs{d\xi\xi}\- \bs e \- ; \b A),  \notag\\
								     &=  \bs \theta^{\bs u}_\Sigma + \int_{\d\Sigma} \bs e\, \bs{d\xi\xi}\-  \bs e\- \bullet e \w e^T - \int_\Sigma  2\ \bs e\, \bs{d\xi\xi}\-  \bs e\- \bullet D^Ae \w e^T ,  \notag\\
		 &= \bs \theta^{\bs u}_\Sigma +  \int_{\d\Sigma} \sqrt{|\det(\bs g)|} \ \ \bs{d\xi\xi}\- \bs g\-  \bullet dx \w dx^T - \int_\Sigma  2\,\sqrt{|\det(\bs g)|} \ \ \bs{d\xi\xi}\-  \bs g\- \bullet {\sf T} \w dx^T, \label{GT-2nd-kind-thetaEC} \\
					         &=  \bs \theta^{\bs u}_\Sigma + \int_{\d\Sigma} \sqrt{|\det(\bs g)|} \  \ \bs{d\xi\xi}\- \bs g\-  \bullet dx \w {{dx^T}} _{\ | N}, \notag \\
					       &=  \bs \theta^{\bs u}_\Sigma + \int_{\d\Sigma} \sqrt{|\det(\bs g)|} \ \ \bs d{\bs{\xi}^{\,\mu}}_{\sigma}{(\bs{\xi}\-)^\sigma}_{\!\!\lambda}\, g^{\lambda \nu} dx^{\,\alpha} dx^{\,\beta} \, {\epsilon_{\mu\nu\alpha\beta}} _{\ | N}.  \notag
\end{align}
in complete analogy with  \eqref{SO-GT-presymp-pot-EC}. 
By  \eqref{GT-2nd-kind-Theta} the $\bs{\t\G}$-transformation of the presymplectic 2-form is,
\begin{align}
 \big(  \bs\Theta^{\bs u}_\Sigma \big)^{\bs \xi} &=  \bs \Theta^{\bs u}_\Sigma + \int_{\d\Sigma} \bs d  \b \theta( \bs{d\xi\xi}\- ; \b A^{\bs u}) - \int_\Sigma  \bs d \b E(\bs{d\xi\xi}\-; \b A^{\bs u}) 
 								       =  \bs \Theta^{\bs u}_\Sigma + \int_{\d\Sigma}   \bs d \theta( \bs e\, \bs{d\xi\xi}\- \bs e \- ; \b A) - \int_\Sigma \bs d   E( \bs e\, \bs{d\xi\xi}\- \bs e \- ; \b A),  \notag\\
								     &=  \bs \Theta^{\bs u}_\Sigma + \int_{\d\Sigma} \bs d \left(  \bs e\, \bs{d\xi\xi}\-  \bs e\- \bullet e \w e^T \right)  - 2\int_\Sigma  \bs d \left( \bs e\, \bs{d\xi\xi}\-  \bs e\- \bullet D^Ae \w e^T \right) ,  \notag\\
		 &= \bs \Theta^{\bs u}_\Sigma +  \int_{\d\Sigma} \bs d \left(  \sqrt{|\det(\bs g)|} \ \ \bs{d\xi\xi}\- \bs g\-  \bullet dx \w dx^T \right) - 2\int_\Sigma   \bs d \left( \sqrt{|\det(\bs g)|} \ \ \bs{d\xi\xi}\-  \bs g\- \bullet {\sf T} \w dx^T \right), \label{GT-2nd-kind-ThetaEC} \\
					         &=  \bs \Theta^{\bs u}_\Sigma + \int_{\d\Sigma} \bs d \left( \sqrt{|\det(\bs g)|} \  \ \bs{d\xi\xi}\- \bs g\-  \bullet dx \w {dx^T}\right) _{\ | N},   \notag
\end{align}
in analogy with \eqref{SO-GT-presymp-form-EC}. We see how the  boundary problem reemerges, intact, for coordinate transformations. 

In the case at hand - in good illustration of the general discussion -  we have no trouble appreciating that $\t\G$ is not a physical transformation group, and doesn't permute points in the physical phase space $\S/\SO \simeq \S^{\bs u}/\t\G$. The dressing operation did not solve the boundary problem, as it simply traded the Lorentz gauge symmetry for coordinate transformations, which is indeed a relevant (very likely, substantial) symmetry of the theory. 
\medskip

We've commented that if a gauge symmetry is substantial it cannot be killed unless one sacrifices the locality of the theory, yet here the tetrad plays the role of a dressing field killing Lorentz, and it is local. So are we to conclude that Lorentz is an artificial gauge symmetry, without physical relevance? It depends. In the pure theory, yes. In gravity coupled to scalar fields (fluid, or dust), the answer would be yes too. But were we  to couple gravity with spinorial matter, then the answer would be no. 

Indeed, in this case the tetrad is mapped to a Hermitian matrix-valued 1-form, $e^a \rarrow e^{AA'}=e^a \sigma_a^{\,AA'}$ with $\{\sigma_a\}$ the Pauli matrices,  on which $S\!L(2, \CC)$ acts by conjugation. In this representation, the tetrad is unfit to serve as a $S\!L(2, \CC)$-dressing for spinors that would make them Lorentz invariant (this phenomenon we will encounter in \emph{free} 3D-$\CC$-gravity!). Lorentz symmetry is unavoidable when gravity is coupled to spinorial matter fields, and in this more realistic theoretical framework, it is a substantial symmetry that likely cannot be killed without losing the locality of the theory (as a gravitational AB effect would suggest \cite{Hohensee-et-al2012}). See again \cite{Francois2018} for further elaboration on this point, as well as some caveats. The verdict on the substantiality of a gauge symmetry or lack thereof, and thus on its physical relevance, is crucially dependent on the field content of the theory (of course). This goes to show that one should thread carefully when interpreting simplified or idealised models, even seemingly very compelling or well-motivated ones.  
\bigskip


\noindent {\bf 4D MacDowell-Mansouri gravity:}  We rely on the previous case to streamline the presentation of the results in the case of 4D MM gravity. The underlying geometry being reductive, the Cartan connection splits as $\b A=A + \tfrac{1}{\ell} e$, and correspondingly the curvature is $\b F=F+ \tfrac{1}{\ell} T = \big( R-\tfrac{\epsilon}{\ell^2}e e^t\big) + \tfrac{1}{\ell}D^A e $, where $e^t\defeq e^T \eta= e^a \eta_{ab}$.
the $\SO$-invariant Lagrangian of the theory is 
$L(\b A) = \tfrac{1}{2}R \bullet R -\tfrac{\epsilon}{\ell^2} \left(  R \bullet e \w e^T -  \tfrac{\epsilon}{2\ell^2} e\w e^T \bullet e\w e^T  \right)$.
 The associated field equations  and presymplectic potential current are 
\begin{align}
\bs E =  -\tfrac{\epsilon}{\ell^2} \, 2 \left\{     \bs d A \bullet D^Ae \w e^T + \bs d e \w e^T \bullet \big( R -\tfrac{\epsilon}{\ell^2} e \w e^T  \big)   \right\},\quad \text{and} \quad
\bs\theta =  \bs d A \bullet \big(  R -\tfrac{\epsilon}{\ell^2} e \w e^T \big).
\end{align}
while the presymplectic 2-form current is $\bs\Theta=- \bs dA \bullet  \bs d\big(  R -\tfrac{\epsilon}{\ell^2} e \w e^T \big)$. 
\medskip

The dressing fiel is again the tetrad field $\bs u = \bs e$, and the dressed Lagrangian is, 
\begin{align}
\b L(\Gamma, \bs g)  = \sqrt{|\det(\bs g)|} \  \ \tfrac{1}{2}\,  {\sf R} \bs g\- \bullet  {\sf R} \bs g\-  -  \tfrac{\epsilon}{\ell^2}  \left( {\sf R} \bs g\- \bullet dx \w dx^T  - \tfrac{\epsilon}{2\ell^2} dx \w dx^T \bullet dx \w dx^T \right). 
\end{align}
 The dressed presymplectic potential is by \eqref{theta-dressed}, 
\begin{align}
\bs\theta_\Sigma^{\bs u}  =\int_\Sigma \theta\big(\bs e \bs d\ \!\b\Gamma\, \bs e\-; \b A \big)
					&=  \bs \theta_\Sigma + \int_{\d\Sigma} \theta( \bs{duu}\- ; \b A) - \int_\Sigma E(\bs{duu}\-; \b A),   \notag\\
					&=  \bs \theta_\Sigma + \int_{\d\Sigma} \bs{dee}\- \bullet \big( R -\tfrac{\epsilon}{\ell^2} e \w e^T  \big) + \tfrac{\epsilon}{\ell^2} \int_\Sigma 2\, \bs{dee}\-  \bullet D^A e \w e^T,   \label{dressed-theta-MM} \\
					&= \bs\theta_\Sigma + \int_{\d\Sigma} \bs{dee}\- \bullet \big( R -\tfrac{\epsilon}{\ell^2} e \w e^T  \big) \,  _{\ | N}.  \notag
\end{align}
The boundary term generalises the ``dressing 2-form'' of  \cite{De-Paoli-Speziale2018} and \cite{Oliveri-Speziale2020}. 
 By  \eqref{Theta-dressed} the associated dressed presymplectic  2-forms is
 \begin{align}
\bs\Theta_\Sigma^{\bs u} 
					&=  \bs \Theta_\Sigma + \int_{\d\Sigma} \bs d\left(  \bs{dee}\- \bullet \big( R -\tfrac{\epsilon}{\ell^2} e \w e^T  \big)  \right) + \tfrac{\epsilon}{\ell^2} \int_\Sigma 2\, \bs d\left( \bs{dee}\-  \bullet D^A e \w e^T \right) ,   \label{dressed-Theta-MM} \\
					&= \bs \Theta_\Sigma + \int_{\d\Sigma}  \bs d\left( \bs{dee}\- \bullet \big( R -\tfrac{\epsilon}{\ell^2} e \w e^T  \big) \right)\,  _{\ | N}.  \notag
\end{align}
Again, the dressing $\bs u = \bs e$ kills the Lorentz symmetry, such that there are no residual gauge transformations of the first kind. But as before, there are  residual transformations of the second kind embodied by the group $\t\G$ of local coordinate transformations, which leaves the dressed Lagrangian invariant, $R^\star_\xi \b L=\b L$. The associated dressed Noether charges, for  $\alpha={\alpha^{\, \mu}}_{\!\nu} \in$ Lie$\t \G$,  are by \eqref{dressed-Noether-charge-invariant-2nd-kind}
\begin{align}
\b Q_\Sigma(\alpha; \b A^{\bs u})  &=  \int_{\d\Sigma} \bs e\, \alpha \, \bs e\- \bullet\big( R -\tfrac{\epsilon}{\ell^2} e \w e^T  \big) + \tfrac{\epsilon}{\ell^2} \int_\Sigma  2\ \bs e\, \alpha \, \bs e\- \bullet D^Ae \w e^T ,  \notag\\
					            &=  \int_{\d\Sigma} \sqrt{|\det(\bs g)|} \ \ \alpha \bs g\-  \bullet \left(  {\sf R} \bs g\- -  \tfrac{\epsilon}{\ell^2} dx \w dx^T \right) + \tfrac{\epsilon}{\ell^2} \int_\Sigma  2\,\sqrt{|\det(\bs g)|} \ \ \alpha  \bs g\- \bullet {\sf T} \w dx^T,   \label{dressed -Noether-charge-MM} \\
					            &=  \int_{\d\Sigma} \sqrt{|\det(\bs g)|} \  \ \alpha \bs g\-  \bullet \left(  {\sf R} \bs g\- -  \tfrac{\epsilon}{\ell^2} dx \w dx^T \right)_{\ | N}, \notag \\
					            &=  \int_{\d\Sigma} \sqrt{|\det(\bs g)|} \ \ \left( \epsilon_{\, \mu\nu\sigma\rho} \, \alpha^{\,\mu\nu} \, \tfrac{1}{2} {{\sf R}^{\sigma\rho}}_{\alpha\beta}-  \tfrac{\epsilon}{\ell^2}\, \alpha^{\,\mu\nu} \, {\epsilon_{\mu\nu\alpha\beta}} \right) dx^{\,\alpha}\!\! \w\!{dx^{\,\beta} }_{\ | N}.  \label{Gen-Komar-mass}
\end{align}
In case $\alpha = \d \zeta = \d_{\,\nu}\zeta^{\mu}$ with $\zeta$ the components of a Killing vector field generating an isometry of spacetime, the above expression is a generalised Komar integral. 
 It gives a good notion of Komar mass in gravity with $\Lambda \neq0$ as it vanishes on the (A)dS groundstate of the theory, which is also the homogeneous space of the underlying Cartan geometry (showing that there are benefits to writing theories/Lagrangians that `respect' this geometry). In~particular, \eqref{Gen-Komar-mass} reproduces the result for 4D Zumino-Lovelock theory gravity  (Or Gauss-Bonnet gravity) obtained in  \cite{Kastor2008} eq.(17)-(20). 

The general dressed Noether charges \eqref{dressed -Noether-charge-MM} satisfy  $\iota_{\alpha^v}\bs\Theta^{\bs u}_\Sigma = -\bs d\b Q_\Sigma(\alpha; \b A^{\bs u})$ so that the dressed presymplectic 2-form  $\bs\Theta_\Sigma^{\bs u}$  induces via \eqref{dressed-Poisson-bracket-2nd-kind} the Poisson bracket 
\begin{align}
\label{dressed-Poisson-bracket-EC}
\big\{ \b Q_\Sigma(\alpha; \b A^{\bs u}) ,  \b Q_\Sigma(\alpha'; \b A^{\bs u})\big\}\defeq\, \bs\Theta^{\bs u}_\Sigma\big(\alpha^v, {\alpha'}^v\big) =	 \b Q_\Sigma\big([\alpha, \alpha']; \b A^{\bs u}\big),
\end{align}
so that, again, the Poisson algebra of dressed  charges is  isomorphic to the Lie algebra of coordinate changes Lie$\t\G$. 

We refrain from giving the formula of $\bs{\t\G}$-gauge transformation for $\bs\theta_\Sigma^{\bs u}$ and $\bs\Theta_\Sigma^{\bs u}$ which are easily guessed generalisations of \eqref{GT-2nd-kind-thetaEC}-\eqref{GT-2nd-kind-ThetaEC}. From these would be clear that the boundary problem resurfaces w.r.t. $\t\G$, which is clearly seen not to be a physical transformation group, as it doesn't  act on the physical phase space $\S/\SO \simeq \S^{\bs u}/\t\G$.


\subsubsection{Dressed presymplectic structure for $c$-equivariant Lagrangians}
\label{Dressed presymplectic structure for non-invariant Lagrangians}

We rely on the template of section \ref{Dressed presymplectic structure for invariant Lagrangians} and generalise it to obtain the dressed presymplectic structure for \mbox{$c$-equivariant} theories. 
For a $c$-equivariant Lagrangian, $R^\star_\gamma L=L + c(\ \, ; \gamma)$ with $\gamma \in \H$, which implies  $L^{\bs \gamma}(A)= L(A^{\bs \gamma})=L(A)+ c(A; \bs \gamma)$, for $\bs \gamma \in \bs\H$, the corresponding dressed Lagrangian is $L^{\bs u}(A)\defeq L(A^{\bs u})= L(A)+ c(A; \bs u)$.
As usual,
\begin{align}
\bs d L^{\bs u} = \bs E^{\bs u} + d\bs\theta^{\bs u} = E\big(\bs d A^{\bs u}; A^{\bs u}\big) + d\theta \big(\bs dA^{\bs u}; A^{\bs u}\big), 
\end{align}
and we still define $\bs\Theta^{\bs u}\defeq \bs d \bs \theta^{\bs u}$. 
From now on we operate under hypothesis 0 \eqref{Hyp0} and hypothesis 1 \eqref{Hyp1}, so that we are in the perimeter of validity of the analysis of section \ref{Presymplectic structure for non-invariant Lagrangians}.  This allows us to obtain by  \eqref{Dressed-variational-form} the explicit expressions of the above basic forms in terms of $\bs u$. Applying the rule of thumb $\bs \gamma \rarrow \bs u$ and using \eqref{Field-depGT-E-non-inv}, \eqref{Field-depGT-presym pot-non-inv} and \eqref{Field-depGT-Theta-non-inv}, we get
\begin{align}
\bs E^{\bs u}&= E(\bs d A^{\bs u}; A^{\bs u})= \bs E + dE \big(  \bs{duu}\- ;  A \big),    \label{Field-Eq-dressed-non-inv} \\[1.5mm]
\bs\theta_\Sigma^{\bs  u}&= \bs \theta_\Sigma + \int_{\d\Sigma}\t Q \big(  \bs{duu}\- ;  A \big) + \int_\Sigma - E \big( \bs{duu}\- ; A \big) 
												+  \beta \big(  \bs{duu}\- ; A \big) + \bs b\big(\bs u(A)\big)_{|A} +  b\left( D^A \big\{ \bs{duu}\-\big\}; \bs u(A) \right),   \label{theta-dressed-non-inv} \\[1.5mm]
\bs\Theta_\Sigma^{\bs u} 
						&= \bs\Theta_\Sigma  \, + \int_{\d\Sigma}  \bs d \left(\t Q \big( \bs{duu}\-; A \big)  -  Q \big( \bs{duu}\-; A \big) \right)   \label{Theta-dressed-non-inv}
								+ \mathscr{A}\big( \lfloor \bs{duu}\-, \bs{duu}\- \rfloor ;A\big) + Q\big(\bs d (\bs{duu}\- ) ; A\big)  
							\\& \hspace{12cm}    - \int_\Sigma \bs d E \big( \bs{duu}\-; A \big). \notag     
\end{align}
Notice  that $A$ and $A^{\bs u}$ satisfy the same field equations, if $\bs E=0$ then $\bs E^{\bs u}=0$. By definition, the above are off-shell $\K$-basic, thus $\bs\K$-invariant, and remain so on-shell. 
  So that $\bs\theta_\Sigma^{\bs u}$ and $\bs\Theta_\Sigma^{\bs u}$ descend as well-defined forms~on~$\S/\K$. 
    The~$\K$-basicity of $\bs\theta_\Sigma^{\bs u}$ means that  dressed Noether charges associated to Lie$\K$ vanish identically: for $\chi \in$~Lie$\K$, $Q_\Sigma\big(\chi; A^{\bs u}\big)\defeq \iota_{\chi^v} \bs\theta_\Sigma^{\bs u} \equiv 0$. And of course, consistently, the $\K$-basicity of $\bs\Theta_\Sigma^{\bs u}$ means that the Poisson bracket of such dressed charges is identically $0$: 
  $\big\{ Q_\Sigma\big(\chi; A^{\bs u}\big) ; Q_\Sigma\big(\chi'; A^{\bs u}\big) \big\} \defeq \bs\Theta_\Sigma^{\bs u} \big( \chi^v, {\chi'}^v \big) \equiv 0$.  
  
  In particular, if $\bs u  \in \D r[G, H]$ then $\bs\theta_\Sigma^{\bs u}$ and $\bs\Theta_\Sigma^{\bs u}$ are off-shell and on-shell $\H$-basic and $\bs\H$-invariant, and may thus induce a symplectic structure on $\S/\H$ without any conditions on the fields and/or the boundary. Obviously, the $\H$-dressed Noether charges and their Poisson bracket vanish identically. 
  
  In view of the on-shell restriction of \eqref{Theta-dressed-non-inv}, one may be tempted to think that $\bs u$ needs only to exist at the boundary $\d\Sigma$, but this inclination is dispelled when considering the on-shell restriction of \eqref{theta-dressed-non-inv} where clearly a bulk term involving the dressing field remains and cannot be neglected. It is then most clear in this  case  that the dressing field cannot be mistaken for an edge mode. 
  
  We now turn our attention to the residual transformations that may operate on these dressed objects. 
\medskip

\noindent{{\bf Residual gauge transformations of the first kind:}} In  case $\bs u  \in \D r[G, K]$ and $K \triangleleft H$ so that $H/K=J$, there are residual $\J$-transformations. We suppose $\bs u$ satisfies Proposition \ref{Residual1} of section \ref{Residual gauge transformations (first kind)}, 
so that
\begin{align}
R^\star_\eta \bs{u} = \eta\- \bs{u}\eta \quad \text{with} \quad \eta \in \J, \qquad \text{and} \qquad \bs{u^\eta}=\bs{\eta}\- \bs{u \eta}\quad \text{with} \quad\bs{\eta} \in \bs{\J}.
\end{align}
Then $R_\eta \, A^{\bs u}\defeq (A^{\bs u})^\eta = \eta\- A^{\bs u} \eta+ \eta\- d\eta$ and we thus have $R^\star_\eta L(A^{\bs u}) =L(A^{\bs u}) + c(A^{\bs u}; \eta )$. Then, as we are under Hypothesis 0 and 1, the analysis of section \ref{Presymplectic structure for non-invariant Lagrangians}  regarding Noether currents and charges and their Poisson bracket holds for  $\J$-transformations. Given $\lambda \in$ Lie$\J$, the dressed Noether charge is 
\begin{align}
\label{dressed-Noether-charge-non-invariant-1st-kind-non-inv}
Q_\Sigma\big(\lambda; A^{\bs u}\big)= \int_{\d \Sigma} \big( \t Q\big(\lambda; A^{\bs u}\big) - Q\big(\lambda; A^{\bs u}\big) \big) - \int_\Sigma E\big(\lambda; A^{\bs u}\big), \quad \text{and s.t.} \quad \iota_{\lambda^v}\bs\Theta^{\bs u}_\Sigma = -\bs d Q_\Sigma(\lambda; A^{\bs u}),
\end{align}
The dressed presymplectic 2-form induces a Poisson bracket for those charges by
\begin{align}
\label{dressed-Poisson-bracket-1st-kind-non-inv}
\big\{ Q_\Sigma\big(\lambda; A^{\bs u}\big) ,  Q_\Sigma\big(\lambda'; A^{\bs u}\big)\big\}\defeq&\, \bs\Theta^{\bs u}_\Sigma\big(\lambda^v, {\lambda'}^v\big) =	Q_\Sigma\big([\lambda, \lambda'] ; A^{\bs u}\big) + \mathscr{C}(\lambda, \lambda')  \\[1mm]
									&\text{with } \quad  \mathscr{C}(\lambda, \lambda') \defeq \int_{\d\Sigma}  \mathscr{A}\big( \lfloor\, \lambda, \lambda' \rfloor ;A^{\bs u}\big) + Q\big([\lambda, \lambda'];A^{\bs u} \big), \notag
\end{align}
so that the Poisson algebra of dressed charges is a central extansion of Lie$\J$. The dressed charges generate infinitesimal $\J$-transformations via $\big\{ Q_\Sigma\big(\lambda; A^{\bs u}\big) , \  \ \ \ \big\}$. 

 Following section \ref{Residual transformations and the bundle structure of the space of dressed connections}, the residual $\bs\J$-gauge transformation of the $\K$-basic dressed form $\bs\alpha^{\bs u}$ is given by 
\eqref{ResidualGT-dressed-alpha}: $\big( \bs{\alpha}^{\bs u} \big)^{\bs \eta} = \alpha \left( \Lambda^\bullet \big(\bs{d}A^{\bs u}\big)^{\bs \eta}; (A^{\bs u})^{\bs \eta} \right)$. Thus the field-dependent $\bs\J$-gauge transformations of 
$\bs E^{\bs u} $, $\bs\theta_\Sigma^{\bs u}$ and $\bs\Theta_\Sigma^{\bs u}$ are here
\begin{align}
  \big( \bs E^{\bs u}\big)^{\bs \eta}&= \bs E^{\bs u} + dE\big(  \bs{d\eta\eta}\- ;  A^{\bs u} \big), \label{GT-1stkind-FieldEq-non-inv} \\[1.5mm]
\big(\bs\theta_\Sigma^{\bs  u}\big)^{\bs \eta}&= \bs \theta^{\bs  u}_\Sigma + \int_{\d\Sigma}\t Q \big(  \bs{d\eta\eta}\- ;  A^{\bs  u} \big) + \int_\Sigma - E \big( \bs{d\eta\eta}\- ; A^{\bs  u} \big) 
												+  \beta \big(  \bs{d\eta\eta}\- ; A^{\bs  u} \big) + \bs b\big(\bs \eta\big)_{|A^{\bs u}} +  b\left( D^{A^{\bs  u}} \big\{ \bs{d\eta\eta}\-\big\}; \bs \eta \right),   \label{GT-1stkind-theta-non-inv} \\[1.5mm]
\big(\bs\Theta_\Sigma^{\bs u}\big)^{\bs \eta} 
						&= \bs\Theta^{\bs  u}_\Sigma  \, + \int_{\d\Sigma}  \bs d \left(\t Q \big( \bs{d\eta\eta}\-; A^{\bs  u} \big)  -  Q \big( \bs{d\eta\eta}\-; A^{\bs  u} \big)  \right)   \label{GT-1stkind-Theta-non-inv}
								+ \mathscr{A}\big( \lfloor \bs{d\eta\eta}\-, \bs{d\eta\eta}\- \rfloor ;A^{\bs  u}\big) + Q\big(\bs d (\bs{d\eta\eta}\- ) ; A^{\bs  u}\big)  
							\\& \hspace{12cm}    - \int_\Sigma \bs d E \big( \bs{d\eta\eta}\-; A^{\bs  u} \big). \notag       
\end{align}
The reduced phase space being $\S^{\bs u}/\J$,  for $\bs\theta_\Sigma^{\bs u}$ and $\bs\Theta_\Sigma^{\bs u}$ to be on-shell $\J$-basic and to induce a well-defined symplectic structure on  $\S^{\bs u}/\J$, one needs again to stipulate adequate boundary conditions. 
\smallskip

When $\bs u$ is a $H$-dressing field, $\bs u  \in \D r[G, H]$, there are no residual gauge transformations of the first kind:   $\bs\theta_\Sigma^{\bs u}$ and $\bs\Theta_\Sigma^{\bs u}$ are $\H$-basic (so in particular $Q_\Sigma\big(\chi; A^{\bs u}\big) \defeq \iota_{\chi^v}\bs\theta^{\bs u}_\Sigma \equiv 0$ for $\chi \in $ Lie$\H$) and may give a symplectic structure on $\S/\H$. In this case we must consider residual transformations of the second kind. 
\medskip

\noindent{{\bf Residual  transformations of the second kind:}} According to sections \ref{Residual transformations (second kind) : ambiguity in choosing a dressing field} and \ref{Residual transformations and the bundle structure of the space of dressed connections} any a priori ambiguity in choosing a dressing field turns  $\A^{\bs u}$ into a $\t \G$-principal bundle, with $\t\G\defeq \left\{ \xi :\P \rarrow G\, |\, R^*_h\xi = \xi \right\}$ 
acting as $\bs u^\xi \defeq \bs u\xi$ and $A^\xi=A$,
so that $R_\xi A^{\bs u} \defeq (A^{\bs u})^\xi=\xi \- A^{\bs u} \xi + \xi\-d\xi$. 
Then we have  $L\big( (A^{\bs u})^\xi\big)=L(A^{\bs u}) + c(A^{\bs u}; \xi)$, 
so that again the analysis of section \ref{Presymplectic structure for invariant Lagrangians}  regarding Noether currents and charges and their Poisson bracket is available on the $\t\G$-bundle $\A^{\bs u}$. For $\alpha\in$ Lie$\t\G$ we then get the dressed Noether charges
\begin{align}
\label{dressed-Noether-charge-non-invariant-2nd-kind}
Q_\Sigma\big(\alpha; A^{\bs u}\big)= \int_{\d \Sigma} \big( \t Q\big(\alpha; A^{\bs u}\big) - Q\big(\alpha; A^{\bs u}\big) \big) - \int_\Sigma E\big(\alpha; A^{\bs u}\big), \quad \text{and s.t.} \quad \iota_{\alpha^v}\bs\Theta^{\bs u}_\Sigma = -\bs d Q_\Sigma(\alpha; A^{\bs u}),
\end{align}
The dressed presymplectic 2-form induces a Poisson bracket for these dressed charges by
\begin{align}
\label{dressed-Poisson-bracket-non-inv-2nd-kind}
\big\{ Q_\Sigma\big(\alpha; A^{\bs u}\big) ,  Q_\Sigma\big(\alpha'; A^{\bs u}\big)\big\} &= Q_\Sigma\big([\alpha, \alpha'] ; A^{\bs u}\big) + \mathscr{C}(\alpha, \alpha')  \\[1mm]
									&\text{with } \quad  \mathscr{C}(\alpha, \alpha') \defeq \int_{\d\Sigma}  \mathscr{A}\big( \lfloor\, \alpha, \alpha' \rfloor ;A^{\bs u}\big) + Q\big([\alpha, \alpha'];A^{\bs u} \big), \notag
\end{align}
so that the Poisson algebra of  $\t\G$-Noether charges is a central extension of  Lie$\t\G$, and as in appendix \ref{Noether charges as generators of gauge transformations}  they generate infinitesimal $\t\G$-transformations via 
$\big\{ Q_\Sigma(\alpha; A^{\bs u}) , \  \ \ \ \big\}$.

From section \ref{Residual transformations and the bundle structure of the space of dressed connections} we have that the field-dependent $\bs{\t\G}$-transformation of the dressed form $\bs\alpha^{\bs u}$ is given by 
\eqref{Dressed-alpha-residual-2nd-kind}: $ \big( \bs{\alpha}^{\bs u} \big)^{\bs \xi} = \alpha \left( \Lambda^\bullet \big(\bs{d}A^{\bs u}\big)^{\bs \xi}; (A^{\bs u})^{\bs \xi} \right)$. The $\bs{\t\G}$-gauge transformations of 
$\bs E^{\bs u}$, $\bs\theta_\Sigma^{\bs u}$ and $\bs\Theta_\Sigma^{\bs u}$ are  then
\begin{align}
  \big( \bs E^{\bs u}\big)^{\bs \xi}&= \bs E^{\bs u} + dE\big(  \bs{d\xi\xi}\- ;  A^{\bs u} \big), \\[1.5mm]
\big(\bs\theta_\Sigma^{\bs  u}\big)^{\bs \xi}&= \bs \theta^{\bs  u}_\Sigma + \int_{\d\Sigma}\t Q \big(  \bs{d\xi\xi}\- ;  A^{\bs  u} \big) + \int_\Sigma - E \big( \bs{d\xi\xi}\- ; A^{\bs  u} \big) 
												+  \beta \big(  \bs{d\xi\xi}\- ; A^{\bs  u} \big) + \bs b\big(\bs \xi\big)_{|A^{\bs u}} +  b\left( D^{A^{\bs  u}} \big\{ \bs{d\xi\xi}\-\big\}; \bs \xi \right),   \label{GT-2ndkind-theta-non-inv} \\[1.5mm]
\big(\bs\Theta_\Sigma^{\bs u}\big)^{\bs \xi} 
						&= \bs\Theta^{\bs  u}_\Sigma  \, + \int_{\d\Sigma}  \bs d \left(\t Q \big( \bs{d\xi\xi}\-; A^{\bs  u} \big)  -  Q \big( \bs{d\xi\xi}\-; A^{\bs  u} \big)  \right)  \label{GT-2ndkind-Theta-non-inv}
								+ \mathscr{A}\big( \lfloor \bs{d\xi\xi}\-, \bs{d\xi\xi}\- \rfloor ;A^{\bs  u}\big) + Q\big(\bs d (\bs{d\xi\xi}\- ) ; A^{\bs  u}\big)  
							\\& \hspace{12cm}    - \int_\Sigma \bs d E \big( \bs{d\xi\xi}\-; A^{\bs  u} \big). \notag       
\end{align}
From the SES  \eqref{SESgroups-dressed}-\eqref{SESLieAlg-dressed}, and since  $\A/\H \simeq \A^{\bs u}/\t\G$ , we have that the reduced phase space is $\S/\H \simeq \S^{\bs u}/\t\G$.  Then, for $\bs\theta_\Sigma^{\bs u}$ and $\bs\Theta_\Sigma^{\bs u}$ to be on-shell $\t\G$-basic and to thus induce a well-defined symplectic structure on  $\S^{\bs u}/\t\G$, one needs  to stipulate adequate boundary conditions.  
The situation parallels what happens for $\bs\theta_\Sigma$ and $\bs\Theta_\Sigma$ w.r.t. $\S/\H$. 
\medskip

The comments made just before \ref{Yang-Mills theory} hold equally here. The group $\t\G$ is not a physical transformation group as it acts trivially on the physical phase space $\S/\H\simeq \S^{\bs u}/\t\G$. So the viability of the DFM as a solution to the boundary problem rests on the possibility to show 1) that a $H$-dressing can indeed be extracted from the connection $A$, and 2) that this constructive procedure is s.t. $\t\G$ is reduced to a discrete group, or the trivial group. 

In case the $\H$-gauge symmetry of the theory is substantial, there is very probably no local $H$-dressing field. So that even with the above two conditions satisfied, one solves the boundary problem for non-invariant theories via the DFM only at the cost of locality. 

We can nonetheless, at least formally, write down the dressed presymplectic structure for 3D non-Abelian Chern-Simons theory and 3D-$\CC$-gravity, as illustrations of the above general framework.

\subsubsubsection{3D non-Abelian Chern-Simons theory}
\label{Non-Abelian Chern-Simons theory}

The Lagrangian of CS theory is $L(A)=\Tr\big(AdA + \tfrac{2}{3}A^3\big)$, the gauge group is $\H=\SU(n)$  and the $\H$-equivariance of the Lagrangian is 
 $R^\star_\gamma L(A)= L(A) + \Tr \left(  d\big(  \gamma d\gamma\- A  \big) - \tfrac{1}{3}\big(  \gamma\-d\gamma \big)^3  \right)$ for $\gamma \in \H$. The associated field equations are $\bs E = E(\bs d A; A) = 2\Tr(\bs dA\, F)$ and the presymplectic potential current is $\bs \theta =   \theta(\bs d A; A) = \Tr(\bs d A\, A)$, while the presymplectic 2-form current is $\bs\Theta= -\Tr\big(\bs d A \, \bs d A \big)$.

Admitting there is a $S\!U(n)$-dressing field $\bs u$, the dressed Lagrangian is 
\begin{align}
L^{\bs u}(A)=L(A^{\bs u})=L(A)+\Tr \left(  d\big(  \bs u d{\bs u}\- A  \big) - \tfrac{1}{3}\big(  {\bs u}\-d\bs u \big)^3  \right), 
\end{align}
which generalises eq.(3.45) in \cite{Geiller2017}. By  \eqref{Field-Eq-dressed-non-inv} the associated field equations are $\bs E^{\bs u}= \Tr\big( \bs d A^{\bs u} \, F^{\bs u}\big)= \Tr\big( \bs d A \, F\big) + d\Tr\big( \bs{duu}\- \, F\big)$.
By \eqref{theta-dressed-non-inv}, or using  \eqref{Field-depGT-theta-CS} for our rule of thumb, the  dressed presymplectic potential is
\begin{align}
\bs\theta_\Sigma^{\bs u} = \int_\Sigma\Tr\big( \bs d A^{\bs u} \, A^{\bs u}\big) &= \bs\theta_\Sigma + \int_{\d\Sigma} 2\Tr\big(\bs{duu}\- A\big) +  \int_\Sigma - 2 \Tr\big(\bs{duu}\-F\big) 
															+\Tr \left( \bs u d \bs u\- d\big( \bs{duu}\-\big)	+ \bs d \big( \bs u d \bs u\-  A \big)  \right),  \label{dressed-theta-CS}\\
													             &= \bs\theta_\Sigma + \int_{\d\Sigma} 2\Tr\big(\bs{duu}\- A\big) +  \int_\Sigma 
																\Tr \left( \bs u d \bs u\- d\big( \bs{duu}\-\big)	+ \bs d \big( \bs u d \bs u\-  A \big)  \right) _{\ |\S}.  \notag
\end{align}
This generalises eq (3.43) in \cite{Geiller2017} (dealing with Abelian CS theory). 
By \eqref{Theta-dressed-non-inv}, or using \eqref{Field-depGT-Theta-CS}, the associated dressed presymplectic 2-forms is, 
\begin{align}
\bs\Theta_\Sigma^{\bs u} = - \int_\Sigma \Tr\big(\bs d A^{\bs u} \, \bs d A^{\bs u} \big)&=   \bs\Theta_\Sigma  + \int_{\d\Sigma}  \bs d\left(  2 \Tr\big(  \bs{duu}\- A \big) \right) 
																		- 2\Tr\big( d(\bs{duu}\-) \bs{duu}\- \big)   - \int_\Sigma  2\bs d\Tr\big( \bs{duu}\- F \big), \\
															        &= \bs\Theta_\Sigma  + \int_{\d\Sigma}  \bs d\left(  2 \Tr\big(  \bs{duu}\- A \big) \right) 
	 																	- 2\Tr\big( d(\bs{duu}\-) \bs{duu}\- \big)  _{\ |\S}.  \notag
\end{align}
This generalises eq (3.64) in \cite{Geiller2017}. 

The dressing $\bs u$ kills all of $\H$, so there are no residual gauge transformations of the first kind. In particular,  for $\chi \in$ Lie$\H$, $Q_\Sigma(\chi; A^{\bs u}) \defeq \iota_{\chi^v}\bs\theta^{\bs u}_\Sigma  \equiv 0$ and there is no Poisson bracket to speak about, as obviously the basicity of $\bs\Theta_\Sigma^{\bs u}$ implies $\big\{ Q_\Sigma(\chi; A^{\bs u}), Q_\Sigma(\chi'; A^{\bs u}) \big\}\equiv0$ - which generalises and trivialises eq.(3.66)-(3.67) in  \cite{Geiller2017}. 

But there is a priori an ambiguity in the choice of the dressing field, so we have  residual transformations of the second kind embodied by the group $\t\H\defeq \left\{ \xi :\P \rarrow S\!U(n)\, |\, R^*_h\xi = \xi \right\}$ acting as $\bs u^\xi = \bs u \xi$ and $A^\xi=A$. Then the $\t\H$-equivariance of the dressed Lagrangian is $R^\star_\xi L(A^{\bs u}) = L(A^{\bs u}) + \Tr \left(  d\big(  \xi d\xi\- A^{\bs u}  \big) - \tfrac{1}{3}\big( \xi\-d\xi \big)^3  \right)$, and for $\alpha \in$ Lie$\t\H$ the dressed Noether charge is by \eqref{dressed-Noether-charge-non-invariant-2nd-kind}, or from \eqref{charge-CS},
\begin{align}
\label{dressed -Noether-charge-YM}
Q_\Sigma(\alpha; A^{\bs u}) &= \int_{\d\Sigma} 2\Tr(\alpha A^{\bs u}) - \int_\Sigma 2\Tr(\alpha F^{\bs u}),  \quad \text{s.t. } \quad \iota_{\alpha^v}\bs\Theta^{\bs u}_\Sigma = -\bs dQ_\Sigma(\alpha; A^{\bs u}) \\
					   &= \int_{\d\Sigma} 2\Tr(\alpha A^{\bs u}) _{\ |\S}. \notag
\end{align}
This generalises eq.(3.69) of \cite{Geiller2017}.
The Poisson bracket of dressed charges induced by  $\bs\Theta_\Sigma^{\bs u}$ via \eqref{dressed-Poisson-bracket-non-inv-2nd-kind} is, via \eqref{PB-CS},  
\begin{align}
\big\{ Q_\Sigma(\alpha; A^{\bs u}),  Q_\Sigma(\alpha'; A^{\bs u}) \big\} &=Q_\Sigma([\alpha, \alpha'] ; A^{\bs u}) + \int_{\d\Sigma}  2\Tr(d\alpha\, \alpha'), 
\end{align}
which generalises eq.(3.70) in \cite{Geiller2017}. 
The Poisson algebra of  $\t\H$-Noether charges is then a central extension of Lie$\t\H$.

Finally, by  \eqref{GT-2ndkind-theta-non-inv} and  \eqref{GT-2ndkind-Theta-non-inv}, and using \eqref{Field-depGT-theta-CS}-\eqref{Field-depGT-Theta-CS}, we get the $\bs{\t\H}$-transformations of the dressed presymplectic potential and 2-form, 
\begin{align}
(\bs\theta^{\bs u}_\Sigma)^{\bs \xi}&=\bs \theta^{\bs u}_\Sigma + \int_{\d\Sigma} 2\Tr\big(\bs{d \xi \xi}\- A^{\bs u}\big) +  \int_\Sigma - 2 \Tr\big(\bs{d \xi \xi}\-F^{\bs u}\big) 
											+\Tr \left( \bs  \xi d \bs  \xi\- d\big( \bs{d \xi \xi}\-\big)	+ \bs d \big( \bs  \xi d \bs  \xi\-  A^{\bs u} \big)  \right),    \label{GT-2nd-kind-thetaCS} \\
(\bs\Theta^{\bs u}_\Sigma)^{\bs\xi} &= \bs\Theta^{\bs u}_\Sigma + \int_{\d\Sigma}  \bs d\left(  2 \Tr\big(  \bs{d\xi\xi}\- A^{\bs u} \big)\right)  - 2\Tr\big( d(\bs{d\xi\xi}\-) \bs{d\xi\xi}\- \big)   - \int_\Sigma  2\bs d\Tr\big( \bs{d\xi\xi}\- F^{\bs u} \big),
   \label{GT-2nd-kind-ThetaCS}
\end{align}
We see how the  boundary problem reemerges w.r.t. the $\t\H$-residual symmetry.

The dressing field is here introduced by hand, so $\t\H$ is $\H$ in another guise. It is certainly not a physical symmetry,  as is clear from the fact that it acts trivially on the physical phase space $\S/\H \simeq \S^{\bs u}/\t\H$.
 As in the case of YM theory, for the DFM to be of any help for the boundary problem in CS theory, one would have to prove that a $S\!U$-dressing field $\bs u$ can be extracted from $A$ in such a way that $\t\H$ can in effect be neglected. Even if this can be done, the $\SU(n)$ symmetry of CS theory being (probably) substantial, a gauge-invariant symplectic structure would be non-local.

\subsubsubsection{3D-$\CC$-gravity $\Lambda=0$ (Lorentz+translations)}
\label{3D-gravity  (Lorentz+translations)}

This example is a good one to conclude the paper, as it allows to illustrate how to perform multiple dressings.
The~Lagrangian of the theory is $L(\b A)=L(A, e)= \Tr \big( e\, R\big)$, with $R=dA+\tfrac{1}{2}[A,A]$. The associated field equations are $\bs E = E\big(\bs d \b A; \b A\big)= \Tr\big(\bs d e\, R + \bs d A\, D^A e \big)$, the presymplectic potential current is $\bs \theta = \theta\big(\bs d \b A; \b A\big)= \Tr\big(\bs dA \, e \big)$, so that $\bs \Theta= -\Tr\big( \bs dA\, \bs d e \big)$. See details in sections \eqref{3D gravity} and \eqref{3D gravity I (translations)}. 

We will again depart from the Cartan theoretic point of view, and start by considering the Lie$G$-valued connection $\b A$ as an Ehresmann connection on a $G$-bundle $\Q$. As detailed in section \eqref{3D gravity I (translations)}, the gauge group is, like $G$, a semi-direct product $\G=\H \ltimes  \T$ in either Euclidean or Lorentzian signature. We remind that, given $\gamma \in \H$ and $t \in \T$,  its action on the connection, 
\begin{align}
e^t&= e+ D^At, \qquad e^\gamma =\gamma\- e \gamma,\notag \\
A^t&= A, \qquad A^\gamma=\gamma\- A \gamma + \gamma\- d\gamma.  
\end{align}
We will  refer to $\H$ as the Lorentz gauge group, and to $\T$ as the translation gauge group. The Lagrangian has trivial $\H$-equivariance $R^\star_\gamma L(\b A)=L(\b A)$, 
but  non-trivial $\T$-equivariance $R^\star_t L(\b A)=L(\b A) + c(\b A; t)=L(\b A) + d\Tr( t\, R)$. Yet, for the latter sector Hypothesis 0 ($R^\star_t \bs E =\bs E$) and 1 are satisfied.  

We have two gauge sectors. Were we to perform dressings, where should we start? Well, if  a dressing operation must leave us with a well defined residual gauge symmetry, according to section \ref{Residual gauge transformations (first kind)} we must start by killing the gauge subgroup that is normal in the full gauge group. Since here we have a semi-direct product, this is easy: we have  $\T \!\triangleleft \G$, so we must start by dressing for gauge translations. 
\medskip

\noindent {\bf First dressing for translational invariance:}\ \ Suppose  that we have a $\T$-dressing $\bs v$ which, since $\T$ is additive abelian, is s.t. $R^\star_t \bs v= \bs v -t$.   We have the dressed Cartan connection  and curvature 
\begin{align}
\b A^{\bs v} &=  A^{\bs v} + e^{\bs v}  = A + ( e+ D^A\bs v ), \\
\b F^{\bs v} &=  R^{\bs v} + T^{\bs v}  = R + D^A( e+ D^A\bs v )= R+ (T+ [R, \bs v]), 
\end{align}
and the dressed Lagrangian is thus
\begin{align}
L^{\bs v}(\b A)=L(\b A^{\bs v})=\Tr\big( e^{\bs v} R \big)=L(\b A)+ d\Tr\big( \bs v \, R\big).
\end{align}
By  \eqref{Field-Eq-dressed-non-inv} the associated field equations  are $\bs E^{\bs v}=E\big( \bs d \b A^{\bs v}; \b A ^{\bs v}\big)=\Tr\big( \bs d e^{\bs v} R + \bs d A\, D^A(e^{\bs v}) \big) =\bs E + \Tr\big( \bs{dv}\, R\big)$. By \eqref{theta-dressed-non-inv} and \eqref{Theta-dressed-non-inv}, or using  \eqref{Field-depGT-theta-transl-3D-grav} and  \eqref{Field-depGT-Theta-3D-grav} with our rule of thumb, the  dressed presymplectic potential and 2-form are
\begin{align}
\label{Dressed-theta-transl-3D-grav}
\bs\theta_\Sigma^{\, \bs v}&= \int_\Sigma \theta \big( \bs d \b A^{\bs v}; \b A ^{\bs v}\big) = \int_\Sigma \Tr\big( \bs dA^{\bs v}\, e^{\bs v} \big)=\bs \theta_\Sigma -   \int_{\d\Sigma} \Tr(\bs d A\,  \bs v)  +  \int_\Sigma \Tr\big( \bs d  R \,\bs v \big), \\[1mm]
\label{Dressed-Theta-transl-3D-grav}
\bs\Theta_\Sigma^{\, \bs v} &= \int_\Sigma\Theta\big( \Lambda^2\bs d \b A^{\bs v}; \b A ^{\bs v}\big)= -\int_\Sigma \Tr\big( \bs dA^{\bs v}\, \bs d e^{\bs v} \big)= \bs\Theta_\Sigma  \, + \int_{\d\Sigma}  \bs d \Tr \big(  \bs{dv}\,  A \big)  - \int_\Sigma  \bs d \Tr \big( \bs{dv} R \big).
\end{align}
We notice that \eqref{Dressed-theta-transl-3D-grav} reproduces eq.(4.44) in \cite{Geiller2017}. 
These are by construction $\T$-basic both off-shell and on-shell, which means in particular that for $\tau \in$ Lie$\T$ we have vanishing dressed Noether charges, $Q_\Sigma\big( \tau; \b A^{\bs v}\big)\defeq \iota_{\tau^v} \bs \theta^{\bs v}_\Sigma \equiv 0 $, and that their Poisson bracket is obviously trivial $\big\{ Q_\Sigma\big( \tau; \b A^{\bs v}\big), Q_\Sigma\big( \tau'; \b A^{\bs v}\big) \big\}\defeq  \bs\Theta_\Sigma^{\, \bs v} \big( \tau^v, {\tau'}^v \big) \equiv 0$.  

Consider how, looking at the on-shell restriction ($R=0$) of \eqref{Dressed-theta-transl-3D-grav}-\eqref{Dressed-Theta-transl-3D-grav}, one may be tempted to think that the dressing $\bs v$ need only exist at the boundary $\d\Sigma$.
\medskip
 
\noindent {\bf Residual (Lorentz) transformations of the fist kind: }\ \ As $\T \!\triangleleft \G$ we have $\H$ as our residual symmetry of the first kind, which by the way acts by conjugation on $\T$. Meaning that our translation-valued dressing field is s.t $R^\star_\gamma \bs v = \gamma\- \bs v \gamma$. This is precisely the condition of validity of Proposition \ref{Residual1} in section \ref{Residual gauge transformations (first kind)}, by which we can conclude that 
\begin{align}
R_\gamma \, \b A^{\bs v}\defeq (\b A^{\bs v})^\gamma = \gamma\- \b A^{\bs v} \gamma+ \gamma\- d\gamma \quad \Rightarrow \quad 
					\left\{ \begin{array}{l} (A^{\bs v})^\gamma =A^\gamma= \gamma\-  A \gamma+ \gamma\- d\gamma, \\[1mm]  (e^{\bs v})^\gamma= \gamma\- e^{\bs v} \gamma, \end{array} \right.
 \end{align}
 i.e. the dressed fields are genuine $\H$-gauge fields.  Then, since the initial Lagrangian has trivial $\H$-equivariance, so does the dressed Lagrangian: $R^\star_\gamma L(\b A^{\bs v})=L(\b A^{\bs v})$. 
 There is no Lorentz anomaly, but we may still use the general results of section \ref{Dressed presymplectic structure for invariant Lagrangians} 
(even though those   of section \ref{Dressed presymplectic structure for non-invariant Lagrangians} would work as well).

By \eqref{dressed-Noether-charge-invariant-1st-kind} (or  \eqref{dressed-Noether-charge-non-invariant-1st-kind-non-inv}), for $\chi \in$ Lie$\H$ the dressed Lorentz charges are
\begin{align}
Q_\Sigma\big(\chi, \b A^{\bs v}\big) =\int_{\d \Sigma} \Tr\big(\chi\, e^{\bs v}\big) - \int_\Sigma \Tr\big(\chi \, D^Ae^{\bs v}\big), \qquad \text{s.t.}  \qquad  \iota_{\chi^v} \bs\Theta_\Sigma^{\, \bs v} = - \bs d Q_\Sigma\big(\chi, \b A^{\bs v}\big), 
\end{align}
as is easily checked from $\iota_{\chi v} \bs d A = D^A\chi$, $\iota_{\chi v} \bs d R = [R, \chi]$, and \eqref{Dressed-theta-transl-3D-grav}. Their Poisson bracket is by \eqref{dressed-Poisson-bracket-1st-kind-non-inv} (or just \eqref{dressed-Poisson-bracket-1st-kind}) 
$\big\{ Q_\Sigma\big(\chi, \b A^{\bs v}\big), Q_\Sigma\big(\chi', \b A^{\bs v}\big)\big\} \defeq\bs\Theta_\Sigma^{\, \bs v}\big(\chi^v, {\chi'}^v \big) = Q_\Sigma\big([\,\chi, \chi'], \b A^{\bs v}\big)$, so that the Poisson algebra of dressed Lorentz charges is isomorphic to Lie$\H$. These dressed charges generate infinitesimal Lorentz gauge transformations by $\big\{ Q_\Sigma\big(\chi, \b A^{\bs v}\big), \ \ \ \ \big\} $.
 
 
 By  \eqref{ResidualGT-FieldEq-1st-kind-inv}-\eqref{ResidualGT-Theta-1st-kind-inv} (or  \eqref{GT-1stkind-FieldEq-non-inv}-\eqref{GT-1stkind-Theta-non-inv}) the residual $\bs \H$-gauge transformations of the dressed field equations, the dressed presymplectic potential and 2-form are
 \begin{align}
(\bs E^{\bs v})^{\bs \gamma}&= \bs E^{\bs v} + E\big( \bs{d\gamma\gamma}\-, \b A^{\bs v}\big) = \bs E^{\bs v} + \Tr\big( \bs{d\gamma\gamma}\- \, D^A(e^{\bs v}) \big) \\[1.5mm]
(\bs\theta_\Sigma^{\bs v})^{\bs \gamma}&= \bs \theta_\Sigma^{\bs v} + \int_{\d_\Sigma}\theta \big(  \bs{d\gamma\gamma}\- ; \b A^{\bs v} \big) - \int_\Sigma E \big( \bs{d\gamma\gamma}\- ; \b A^{\bs v} \big) , \notag\\[1mm]
				 & = \bs \theta^{\bs v}_\Sigma +  \int_{\d \Sigma}\Tr\big(\bs{d\gamma\gamma}\- e^{\bs v} \big) -  \int_{\Sigma}\Tr\big( \bs{d\gamma\gamma}\-\, D^Ae^{\bs v}\big). \\[1.5mm]
(\bs\Theta_\Sigma^{\bs v})^{\bs\gamma} &= \bs\Theta_\Sigma^{\bs v} +  \int_{\d \Sigma}\bs d \theta\big(\bs{d\gamma\gamma}\-; \b A^{\bs v}\big) - \int_\Sigma \bs d E\big(\bs{d\gamma\gamma}\-; \b A^{\bs v}\big), \notag\\[1mm]
				   & = \bs\Theta^{\bs v}_\Sigma +  \int_{\d \Sigma} \bs d \Tr\big( \bs{d\gamma\gamma}\- e^{\bs v}  \big) -  \int_{\Sigma} \bs d\Tr\big(  \bs{d\gamma\gamma}\-\, D^Ae^{\bs v}\big).
\end{align}
Notice how only the Lorentz part of $\bs E^{\bs v}$ contributes.
 \bigskip
 
 These are the exact same results as those found in section \ref{3D gravity} (given that $\Lambda$, like a mass term, does not affect the presympletic structure) adopting a strictly Cartan geometric viewpoint.
 One may indeed notice that what is effectively achieved geometrically, after this first dressing eliminating translational gauge transformations,  is that the underlying $G$-bundle $Q$ with gauge group $\G$ on which $\b A$ is an Ehresmann connection has been factorised to a   $H$-subbundle $\P$ with gauge group $\H$ on which $\b A^{\bs v}$ is a Cartan connection. That is, we end-up where it would have been natural to start, from a Cartan theoretic point of view on gravitational gauge theories.
 
 But then we see that the boundary problem is still with us w.r.t. Lorentz gauge transformations: Even restricting the above results on-shell, we see that to get a symplectic structure on the physical phase space $\S^{\bs v}/\H \simeq \S/\G$, we must impose boundary conditions. 
  \medskip

\noindent {\bf Second dressing for Lorentz invariance:}\ \ Given that we have a well-behaved $\H$-gauge theory given by $L^{\bs v}$, we may try to perform a second dressing operation, this time to erase Lorentz symmetry. 
In section \ref{4D gravity} we had found that in 4D (real) gravity, the tetrad field extracted from the soldering form  could play that role. But notice that in the $\CC$-representation at hand we have $e^\gamma= \gamma\- e\gamma$, for $\gamma \in \H$, so that no Lorentz dressing field can be extracted from the soldering form.\footnote{See the comment at the end of our treatment of EC gravity in section \ref{4D gravity} regarding the coupling to spinors and the ensuing change of status of Lorentz symmetry.}
Nonetheless we can formally write down what the fully invariant theory and associated presymplectic structure would look like. 

Suppose then that we have a $\H$-dressing field $\bs u$, s.t. $R^\star_\gamma \bs u = \gamma\- \bs u$. As we have stressed in sections  \ref{Residual gauge transformations (first kind)} and \ref{Residual transformations and the bundle structure of the space of dressed connections}, in order not to spoil the $\T$-invariance gained in the previous operation, this new dressing field must satisfy $R^\star_t \bs u= \bs u $, for $t \in \T$
(this is the last of the compatibility conditions  stipulated by equation  \eqref{CompCond}). 
 We then have the dressed Cartan connection and curvature, 
\begin{align}
(\b A^{\bs v})^{\bs u}= ( A^{\bs v})^{\bs u} + (e^{\bs v})^{\bs u} &= \big( {\bs u}\-  A^{\bs v} \bs u + {\bs u}\- d\bs u \big) + {\bs u}\- e^{\bs v}\, \bs u,  \notag\\
											    &=\big( {\bs u}\-  A \, \bs u + {\bs u}\- d\bs u \big) + {\bs u}\- \big(e+ D^A{\bs v}\big)\, \bs u,  \label{fully-dressed-Cartan-connection-3D-grav}\\[1mm]
(\b F^{\bs v})^{\bs u}= ( R^{\bs v})^{\bs u} + (T^{\bs v})^{\bs u} &=  {\bs u }\-  R^{\bs v}  \bs u + \bs u \- T^{\bs v}\, \bs u,  \notag\\
											       &= {\bs u}\-  R \, \bs u + D^{A^{\bs u}} \big( {\bs u }\- e^{\bs v} \bs u\big).
\end{align}
Equation \eqref{fully-dressed-Cartan-connection-3D-grav} reproduces the (aptly named) ``dressed fields" defined below eq.(4.54) in \cite{Geiller2017}. The dressed Lagrangian  is, 
\begin{align}
L^{\bs u}\big(\b A^{\bs v}\big)= L\big( (\b A^{\bs v})^{\bs u}\big) = \Tr \big( (e^{\bs v})^{\bs u}\, R^{\bs u} \big),
\end{align}
and is manifestly $\H$-invariant (thus $\G$-invariant) because its variables are. The corresponding field equations are, by \eqref{Field-Eq-dressed}, 
\begin{align}
(\bs E^{\bs v})^{\bs u}= \Tr\left( \bs d (e^{\bs v })^{\bs u} R^{\bs u} + \bs dA^{\bs u} D^{A^{\bs u}}(e^{\bs v })^{\bs u} \right) = \bs E^{\bs v} + \Tr\big( \bs{duu}\- D^A(e^{\bs v}) \big). 
\end{align}
By \eqref{theta-dressed}, or using  \eqref{Field-depGT-theta-3D-grav} with the rule of thumb, the associated fully  dressed presymplectic potential is 
\begin{align}
(\bs\theta_\Sigma^{\bs v})^{\bs u} 
						&=\int_\Sigma  \Tr\left(  \bs d A^{\bs u} \, (e^{\bs v})^{\bs u} \right) ,  \notag \\
			          		 &= \bs \theta_\Sigma^{\bs v} + \int_{\d\Sigma} \theta( \bs{duu}\- ; \b A^{\bs v}) - \int_\Sigma E(\bs{duu}\-; \b A^{\bs v}), \notag \\
						&=   \bs \theta_\Sigma^{\bs v} + \int_{\d\Sigma} \Tr\big( \bs{duu}\-  e^{\bs v} \big) - \int_\Sigma \Tr \big(\bs{duu}\- D^A( e^{\bs v}) \big), \notag\\
						&= \bs \theta  -   \int_{\d\Sigma} \Tr(\bs d A\,  \bs v)  +  \int_\Sigma \Tr\big( \bs d  R \,\bs v \big) +  \int_{\d\Sigma} \Tr\big( \bs{duu}\-  (e+D^A{\bs v}) \big) - \int_\Sigma \Tr \big(\bs{duu}\- D^A( e+D^A{\bs v}) \big), \notag \\
						&=\bs \theta  +   \int_{\d\Sigma} \Tr\left(\bs{duu}\-  (e+D^A{\bs v}) - \bs d A\,  \bs v \right)  +   \int_\Sigma \Tr\left( \bs d  R \,\bs v - \bs{duu}\- ( D^A e+ [R, \bs v] )   \right). 
\end{align}
Where  we used \eqref{Dressed-theta-transl-3D-grav}. This reproduces eq.(4.53)-(4.54) in \cite{Geiller2017}.  By \eqref{Theta-dressed}, or using  \eqref{Field-depGT-Theta-3D-grav2} with the rule of thumb, the fully dressed presymplectic 2-form is, 
\begin{align}
(\bs\Theta_\Sigma^{\bs v})^{\bs u} &= - \int_\Sigma\Tr\left(  \bs d A^{\bs u} \, \bs d(e^{\bs v})^{\bs u} \right), \notag \\
 					              & =\bs\Theta_\Sigma^{\bs v} + \int_{\d\Sigma}   \bs d \theta\big(\bs{duu}\-; \b A^{\bs v}\big) - \int_\Sigma \bs d E\big(\bs{duu}\-; \b A^{\bs v}\big), \notag\\
					              &= \ldots  \notag\\
					              &= \bs \Theta +  \int_{\d\Sigma} \bs d \Tr\left(  \bs{duu}\-  (e+D^A{\bs v}) + A \bs {dv} \right)  -  \int_\Sigma  \bs d \Tr\left(   \bs{dv}\,  R +   \bs{duu}\- ( D^A e+ [R, \bs v] )  \right), 
\end{align}
using  \eqref{Dressed-Theta-transl-3D-grav}. This last result generalises eq.(4.56)-(4.58) of \cite{Geiller2017}. Now, as these are $\H$-basic, there remains no Lorentz charges, $Q_\Sigma\big(\chi; (\b A^{\bs v})^{\bs u} \big) \equiv0$, and that there is no Poisson bracket to speak of. 
\medskip

If it may seem that the above dressed presymplectic structure finally solves the boundary problem and gives a well behave symplectic structure on the reduced phase space $\S/\G$ of the theory, it is not so.
Indeed, here again, as both the dressings $\bs v$ and $\bs u$ have been introduced by hand, a priori there are corresponding ambiguities so that we have residual symmetries of the second kind: $\t\H$ replicating $\H$ and $\t\T$ replicating $\T$. We have corresponding dressed $\t\H$ charges and $\t\T$ charges exactly analogous to $Q_\Sigma\big(\chi; \b A\big)$ and $Q_\Sigma\big(\tau; \b A\big)$ - $\chi \in$ Lie$\H$ and $\tau \in $ Lie$\T$ - satisfying exactly the same Poisson bracket (\eqref{Noether-charges-Lorentz-3DGrav-flat}, \eqref{PB-3D-grav-no-Lambda-1} and \eqref{PB-3D-grav-no-Lambda-2}). These dressed charges are not observables and the group $\t\G \defeq \t\H \ltimes \t\T$ is not a physical transformation group as it acts trivially on the physical phase space $(S^{\bs v})^{\bs u}/\t\G \simeq \S/\G$. The boundary problem reappears intact w.r.t. $\t\G$. 

As we have now commented several times, the DFM strategy will fail unless one put forth a principled way to build the dressing from the original field content of the theory so that $\t\G$ reduces to something negligible. If the $\G$-symmetry of 3D-$\CC$-gravity is believed to be substantial, no \emph{local} such dressing can be found.

\section{Conclusion}
\label{Conclusion}


The first concern of the covariant Hamiltonian formalism is to associate a symplectic (physical) phase space to a gauge field theory, with gauge group $\H$, on a region of spacetime. In pure gauge theories, the starting point is the  space of connections (gauge potentials) $\A$ seen as a configuration space. 
A choice of a Lagrangian functional $L$ on $\A$, i.e. of a theory, provides the means to determine both the phase space $\S$ (via the field equations $\bs E$) and a symplectic structure via the presymplectic potential  $\bs \theta_\Sigma$ and 2-form $\bs \Theta_\Sigma= \bs{d\theta}_\Sigma$. The main preoccupation is to show that the latter, especially  $\bs \Theta_\Sigma$, decend as well-behaved objects on the physical/reduced phase space $\S/\H$.

 But one may notice that $\A$ has a $\H$-principal bundle geometry preexisting to the choice of $L$, the latter  being seen as (equivalent to) a section of some bundle associated to $\A$. We have  argued here that paying close attention to the bundle geometry of $\A$ allows to clarify several notions and to systematise many results appearing in the literature on the covariant Hamiltonian formalism. Let us  sumarize, in a non-chronological order, what has been done in this paper:
\medskip

$\bullet$\ \  Exposing the bundle geometry of $\A$ (section \ref{Bundle geometry of A}) we could appreciate how the  gauge group $\bs \H$ of $\A$ gives geometric substance to the often encountered notion of \emph{field dependent} gauge   transformations - either explicitly e.g. in \cite{Lee-Wald1990, Geiller2017, Geiller2018, De-Paoli-Speziale2018, Oliveri-Speziale2020}, or tacitly e.g. in \cite{DonnellyFreidel2016}. As the gauge transformation of a form on a bundle is dictated by its  equivariance and verticality properties, we could produce geometrically the field-dependent gauge transformation $\bs d A^{\bs \gamma}$, as well as $\bs d F^{\bs \gamma}$ (usually derived in a more heuristic fashion).
\smallskip

$\bullet$\ \ We also draw attention (section \ref{Twisted bundle geometry}) to a new kind of bundles one can associate to a principal bundle $\P$: \emph{twisted} bundles $E^C$, built not from representations of the structure group $H$, as standard associated bundles are, but from 1-cocycles $C$ for the action of $H$ on $\P$. A generalisation of Ehresmann connections, \emph{twisted} connections, are needed on $\P$ to produce a covariant derivative on the space of sections  $\Gamma\big(E^C \big)$ of such bundles. 

We have shown  (section \ref{Anomalies in gauge theories and twisted geometry on A}) that this twisted geometry, first developed in  \cite{Francois2019_II}, has natural applications in gauge theory. Anomalous quantum action functionals on the $\H$-bundle $\A$ are indeed twisted sections, and so are classically non-invariant Lagrangians/actions (e.g. non-Abelian 3D Chern-Simons theory). In both cases, by definition a variational twisted connection on $\A$ reproduces the quantum/classical anomaly via its verticality property, and the horizontality of the  twisted curvature encodes the associated Wess-Zumino consistency condition. Invariant Lagrangians are contained as a special subclass with trivial anomaly. We have also shown how the twisted covariant derivative of a non-invariant classical action reproduces the Wess-Zumino improved (i.e. invariant) action. 
\smallskip

$\bullet$\ \  Taking advantage of these insights about the bundle structure of $\A$ and associated twisted structures, we have obtained from first principles the general field-dependent gauge transformations of the presymplectic potentials $\bs \theta_\Sigma$ and presymplectic 2-forms $\bs \Theta_\Sigma$ associated to invariant Lagrangians (section \ref{Presymplectic structure for invariant Lagrangians}, eq.\eqref{Field-depGT-presymp-pot}-\eqref{Field-depGT-presymp-form}). To do the same in the case of non-invariant Lagrangian (section \ref{Presymplectic structure for non-invariant Lagrangians}, eq.\eqref{Field-depGT-theta-non-inv}-\eqref{Field-depGT-Theta-non-inv}) we had to work under the assumption that the field equations are invariant (Hypothesis 0 \eqref{Hyp0}) and the classical anomaly is exact (Hypothesis 1 \eqref{Hyp1}). 
 These results allow to appreciate the generic fact that $\bs \theta_\Sigma$ and $\bs \Theta_\Sigma$ are basic and induce the sought-after symplectic structure on the physical phase space only when boundary conditions are imposed (either fall-off for $A$ or $\d\Sigma=\emptyset$).\footnote{Actually, in the non-invariant case even with such conditions $\bs \theta_\Sigma$ may never be basic. See eq.\eqref{Field-depGT-presym pot-non-inv}. This is notably the case for the presymplectic potential of non-Abelian 3D Chern-Simons theory.} 

We also proved that the Poisson algebra of Noether charges - with the Poisson bracket induced by $\bs \Theta_\Sigma$ -  is isomorphic to the Lie algebra of (field-independent) gauge transformations  in the invariant case (see eq.\eqref{Poisson-bracket}), while it is anomalous and a central extension of it in the non-invariant case (see eq.\eqref{PB-anomalous}). 
Applications of the general formalism to Yang-Mills theory, 3D-$\CC$ and 4D-$\RR$ gravity, and 3D Chern-Simons theory allowed to straightforwardly recover various results of the literature, e.g. \cite{CrnkovicWitten1986, Crnkovic1987, DonnellyFreidel2016, Geiller2017}.
\smallskip

$\bullet$\ \  The above general  results on the $\bs\H$-transformations of $\bs \theta_\Sigma$ and $\bs \Theta_\Sigma$ make very clear the challenge to  defining  well-behaved symplectic structures over bounded regions. The most obvious strategy to solve this boundary problem is to attempt to define a presymplectic structure on $\A$ strictly $\bs \H$-invariant, without any restrictions on $A$ or the region. We have thus pointed out (section \ref{The dressing field method}) the existence of a geometric framework, known as the dressing field method (DFM), that allows to build in a systematic way basic forms on a principal bundle, provided one identifies a so-called \emph{dressing field}. We showed how, applied on $\A$ (section \ref{A-dependent dressing fields and basic variational forms on A}), it  allows to define the crucial notion of \emph{field-dependent dressing fields} $\bs u$. 

Next relying on the result about $\bs\H$-gauge transformations of $\bs \theta_\Sigma$ and $\bs \Theta_\Sigma$, we defined \emph{dressed} presymplectic structures  $\bs \theta_\Sigma^{\bs u}$ and $\bs \Theta_\Sigma^{\bs u}$ for both invariant (section \ref{Dressed presymplectic structure for invariant Lagrangians}, eq.\eqref{theta-dressed}-\eqref{Theta-dressed}) and non-invariant theories (section \ref{Dressed presymplectic structure for non-invariant Lagrangians}, eq.\eqref{theta-dressed-non-inv}-\eqref{Theta-dressed-non-inv}). We also showed that these are obtained in the usual way from dressed Lagrangians $L^{\bs u}$. We showed why, when considering the restriction on-shell of the formulae for  $\bs \theta_\Sigma^{\bs u}$ and $\bs \Theta_\Sigma^{\bs u}$, one may think that $\bs u$ needs only to exist at the boundary, thereby indicating that - as first pointed out by \cite{Teh2020} - the DFM is the geometric framework underlying the so-called ``edge modes" as introduced by \cite{DonnellyFreidel2016}, and used in various contexts by several others since. Thus, applying our general results to the various examples already encountered above straightforwardly reproduced, or generalised, results of the recent literature on edge modes, e.g.  \cite{DonnellyFreidel2016, Geiller2017, Geiller2018}.\footnote{We could have given still other examples, such as Maxwell-Chern-Simons theory or BF theory (which would have reproduced and generalised some results of \cite{Geiller2019}), but the multiplication of examples was pointless given the main lesson drawn from the DFM. See next.}

The geometric insights given by the dressing field method helped correct a common misconception encountered in the edge modes  literature. Indeed it is often claimed that the introduction of edges modes entails a new \emph{physical} symmetry - or boundary symmetry - whose associated (dressed) Noether charges are observables, exactly similar to the original gauge Noether charges and with identical Poisson brackets. We have stressed (section \ref{Residual transformations and the bundle structure of the space of dressed connections}) that in the DFM framework, this `symmetry' $\t\G$ is known   to naturally stem from the a priori ambiguity in the choice of the dressing field, and we  further argued that this is never a physical symmetry as it acts trivially on the physical phase space $\S^{\bs u}/\t\G \simeq \S/\H$. Actually, when a dressing field  is introduced by \emph{fiat} in a theory, this `new' symmetry is simply a replica of the original gauge symmetry. 

We concluded (section \ref{Dressed presymplectic structure for invariant Lagrangians}) that unless one has a principled way to build a dressing field  from the original field content of the theory, such that the constructive procedure is essentially free of ambiguities, the DFM (a.k.a. edge mode strategy) cannot solve the boundary problem.\footnote{We point e.g. to \cite{Ibort-Spivak2017} and  \cite{Margalef-Bentabol-Villasenor2020}  for   mathematically  more sophisticated approaches to the boundary problem. See also \cite{Gomes-et-al2018, Gomes-Riello2018, Gomes2019, Riello2020} for an approach using an Ehresmann  connection on field~space.} We also stressed that even if these strictures are met,  for theories with substantial gauge symmetry only non-local dressing fields are likely to exist, so that the dressed symplectic structure obtained is non-local. We raised the possibility that the boundary problem is yet another instance of the trade-off gauge-invariance/locality by now strongly suspected to be characteristic of substantial gauge theories, which encodes physical d.o.f. in a non-local way (\cite{Francois2018} and references therein). This might be relevant to the new quantum gravity program based on edge modes recently initiated in \cite{Freidel-et-al2020-1, Freidel-et-al2020-2, Freidel-et-al2020-3}.  
\medskip

$\bullet $ \ \ The only truly physically relevant application of the DFM  was in 4D gravity (section \ref{4D gravity}), with or without $\Lambda$, approached via Cartan geometry. A local Lorentz dressing field could be extracted from the Cartan connection: the tetrad field (which no one is tempted to think of as a mere edge mode). Our result for $\bs \theta_\Sigma^{\bs u}$ in the Eintein-Cartan ($\Lambda \neq 0$) formulation, eq.\eqref{dressed-theta-EC}, immediately reproduces the ``dressing 2-form" of \cite{De-Paoli-Speziale2018, Oliveri-Speziale2020} introduced to restore the equivalence between the presymplectic potentials of GR in the tetrad and metric formulations and to properly derive the black hole first law. Our result for $\bs \theta_\Sigma^{\bs u}$ in the MacDowell-Mansouri ($\Lambda \neq 0$) formulation, eq.\eqref{dressed-theta-MM},  generalises this dressing 2-form. In this context $\t\G$ is the group of local coordinate transformations, a relevant gauge symmetry that we have yet no problem recognising as non physical and acting trivially on the physical phase space. So that dressing to kill Lorentz gauge symmetry did not solve the boundary problem, which reemerges intact w.r.t. local coordinate transformations. Furthermore, we  recover the Komar mass and obtain its generalisation to $\Lambda \neq 0$ as dressed $\t\G$-Noether charges, eq.\eqref{Komar-mass}-\eqref{Gen-Komar-mass}. 
See appendix \ref{G} for a similar analysis of the LQG Lagrangian. 
 \medskip

This work can undoubtedly be improved mathematically. A better geometrical approach would certainly have been to work on  $J^1(\A)$ or $J^r(\A)$, the first and $r^{\text th}$-jet spaces of $\A$. 
 In particular, it is worth comparing our results with  the proposition of \cite{Margalef-Bentabol-Villasenor2020}.
Taking for now this framework as a starting point, its domain of validity can be extended on several fronts, each will be the object of a forthcoming paper.
\smallskip

$\bullet$\ \ To obtain general results about the presymplectic structure of gauge theories including matter fields, the formalism described here must be extended from $\A$ to $\A \times \Gamma(E)$. This should pose no special challenge as the space of sections $\Gamma(E)$ of a (standard) bundle $E$ associated to $\P$ is  also an infinite dimensional $\H$-principal bundle, so the material of section \ref{The space of connections as a principal bundle} naturally exports to this (simpler) context.  It is again the case that non-invariant Lagrangians $L(A, \vphi)$, $\vphi \in \Gamma(E)$, or quantum actions $Z[A, \vphi]$ are $c$-equivariant functionals, i.e. sections of \emph{twisted} associated bundles to the $\H$-principal product bundle $\A\times \Gamma(E)$. A twisted variational connection on the latter still reproduces the quantum and classical anomalies, whose Wess-Zumino consistency conditions are encoded in the tensoriality of the twisted curvature. 
  This extension of the framework will also make possible to define field-dependent dressing fields extracted from the matter sector of a gauge theory, $\bs u: \Gamma(E) \rarrow \D r[G, K]$, $\vphi \mapsto \bs u(\vphi)$. 
  
  This extended framework will allow us to derive easily e.g. the presymplectic structure of 4D gravity coupled to spinors. Also, two  illustrations of how the DFM gives interesting results will be considered. First, we will provide the presymplectic structure of Maxwell theory coupled to a charged scalar field, and the corresponding \emph{local} $\U(1)$-invariant dressed symplectic structure. We will also discuss why in the case of Maxwell coupled to charged spinors, the dressed symplectic structure is non-local - and the dressed variables are those introduced by Dirac \cite{Dirac55, Dirac58}. Second, we will consider the case of the Electroweak model 
  where a  \emph{local} $S\!U(2)$-dressing can be built from the $\CC^2$-scalar (Higgs) field:\footnote{This, as  argued elsewhere \cite{Francois2018, Attard_et_al2017bis},  provides an alternative to the SSB narrative. See also \cite{Maas2019}.} we thus get a \emph{local} $\SU(2)$-invariant dressed presymplectic structure, leaving a residual substantial $\U(1)$-gauge symmetry of the first kind.
\smallskip

$\bullet$\ \  As alluded to in sections \ref{Residual transformations and the bundle structure of the space of dressed connections} and  \ref{Dressed presymplectic structure for invariant Lagrangians}, if a dressing field satisfies Proposition \ref{Residual2} then $A^{\bs u}$ is a \emph{twisted} connection. We have then a natural motivation to study the $\H$-bundle geometry of the space $\t\A$ of twisted connections, as we did in section \ref{The space of connections as a principal bundle} for $\A$. This would allow us to obtain general results on the presymplectic structure of twisted gauge theories, as we did here for standard gauge theories in sections \ref{Presymplectic structure for invariant Lagrangians} and \ref{Presymplectic structure for non-invariant Lagrangians}. 
This is no idle formal endeavor. Indeed, it turns out that it is needed to truly understand the presymplectic structure of conformal gravity. 

Starting from the  conformal Cartan geometry $(\P, \b A)$ \cite{Sharpe}, one writes the associated Yang-Mills-type Lagrangian $L(\b A)=\Tr\big( \b F\, *\! \b F\big)$.  As shown in \cite{Attard-Francois2016_I, Attard-Francois2016_II}, two dressing fields can be extracted from the conformal Cartan connection $\b A$, so that after the first dressing $\b A^{\bs u}$ remains a standard Cartan connection w.r.t. Lorentz transformations, but has become a twisted connection w.r.t. Weyl rescalings. Furthermore, the dressed Lagrangian reduces to the Lagrangian of conformal (or Weyl) gravity when evaluated on the dressed \emph{normal} conformal Cartan connection: $L(\b A^{\bs u}) =\Tr\big( \b F^{\bs u}\, *\! \b F^{\bs u}\big)= \Tr\big( W\, *\! W\big)$, with $W$ the Weyl 2-form (tensor). Conformal gravity is thus an example of (pure) twisted gauge theory.
One can of course anticipate that the results for the presymplectic potential and 2-form for conformal gravity closely parallel the expressions for YM theory, given in section \ref{The case of Yang-Mills theory}.

The dressing of sections of the $\RR^6$-bundle  and $\CC^4$-bundle associated to the conformal Cartan bundle $\P$, give respectively the so-called conformal tractors \cite{curry_gover_2018, Bailey-et-al94} and local twistors \cite{Penrose-Rindler-vol2}. The tractor and twistor bundles are then \emph{twisted} bundles associated  to the bundle $\P$ of conformal Cartan geometry (section \ref{Twisted bundle geometry}). Thus, connecting to the previous point, extending our framework to $\t\A\times  \Gamma\big( E^C\big)$ would automatically give us the (dressed) presymplectic structure of conformal gravity coupled to tractors and/or twistors as a special case. 
\smallskip

$\bullet$\ \ Finally, one may consider how the presymplectic structure of a theory behaves under the action of  \emph{field-dependent} diffeomorphisms of $\M$.  It is quite clear how the framework developed for $c$-equivariant theories 
is well adapted to give general results on this: Given a Lagrangian ${\sf L}(\phi^i)$ on a $n$-dimensional $\M$, $\phi^i$ a collection of fields,  its Lie derivative along a vector field $X \in \Gamma(TM)\simeq$ Lie$\Diff(\M)$ is $L_X {\sf L}(\phi^i) = \iota_X \cancel{d {\sf L}(\phi^i)} + d \iota_X{\sf L}(\phi^i)\rdefeq \alpha (X; \phi^i)$,  the first term cancelling on $\M$ because $\sf L$ is a $n$-form. In other words the classical $\Diff(\M)$-anomaly is necessary $d$-exact, satisfying hypothesis 1  \eqref{Hyp1}. With minor adaptations, the material of section  \ref{Presymplectic structure for non-invariant Lagrangians} can thus be made to encompass the case of (field-dependent) $\Diff(\M)$-transformations. 

At the moment the DFM has been developed to cover only internal gauge groups. We will show how it can be adapted to the case of $\Diff(\M)$, in which case it gives a systematic way to build $\Diff(\M)$-invariant quantities, and in particular the $\Diff(\M)$-invariant dressed presymplectic structure of a theory $\sf L$.  This will show that  $\Diff(\M)$-dressing fields are the `edge modes' introduced in  \cite{DonnellyFreidel2016} for metric GR and extended in \cite{ Speranza2018}  to an arbitray $\sf L$. But it can be anticipated that the general lesson drawn here will hold in this context as well: A local dressing field should be constructed from the initial field content of the theory in such a way that no ambiguity remains, otherwise one will obtain a non-physical residual symmetry of the second kind that will simply duplicate $\Diff(\M)$. Furthermore, in a  theory where $\Diff(\M)$ is a substantial symmetry, only non-local such dressing fields are likely to exist, so that $\Diff(\M)$-invariance (of the presymplectic structure)  costs locality. 

Reversing the viewpoint, we will also comment that a local $\Diff(\M)$-dressing field is quite literally a preferred choice of coordinates on $\M$ (a global chart) that artificially implements a $\Diff(\M)$-symmetry in a theory that lacks it, exactly like a Stueckelberg field\footnote{Which is also an instance of dressing field. See the end of section \ref{Residual transformations (second kind) : ambiguity in choosing a dressing field} and section 2 of \cite{GaugeInvCompFields} for a short demonstration.}  artificially enforces a gauge symmetry. Hidden preferred coordinates in theories with artificial $\Diff(\M)$-symmetries are known in some quarter of philosophy of physics as ``clock fields" \cite{Pitts2008, Pitts2009}. One may then conjecture that $\Diff(\M)$-dressing fields and clock fields are one and the same thing.

\medskip

\section*{Acknowledgment}  

This work was supported by the Fonds de la
Recherche Scientifique - FNRS under grants PDR No.\ F.4503.20
(``HighSpinSymm'') and grant MIS No. T.0022.19 (``Fundamental issues in extended
gravitational theories'').


\appendix
\section{Lie algebra (anti)-isomorphisms}
\label{Lie algebra (anti)-isomorphisms}

We reproduce here the proofs of the assertions of section \ref{Principal bundles and their smooth structure} that the map Lie$H \rarrow \Gamma(V\P)$, $X\mapsto X^v$, is a (injective) Lie algebra morphism, while the map Lie$\H \rarrow \Gamma_H(V\P)$, $\chi \mapsto \chi^v$, is a Lie algebra anti-isomorphism. 
\medskip

We remind that a  vertical vector at $p\in \P$ generated by $X \in$ Lie$H$ is defined as $X^v \defeq \tfrac{d}{d\tau} p\exp(\tau\, X)\big|_{\tau=0}$, with flow $\phi_\tau : \P \rarrow \P$, $p \mapsto \phi_\tau(p)=p\exp(\tau\, X)$. It is then easy to prove that the pushforward by the right action of the structure group $H$ of $\P$ is, for $h \in H$, 
\begin{align}
R_{h*}X^v_p &= \tfrac{d}{d\tau} R_h\, p\exp(\tau\, X)\big|_{\tau=0} = \tfrac{d}{d\tau} p\exp(\tau\, X) h \big|_{\tau=0}= \tfrac{d}{d\tau} ph\, h\-\!\exp(\tau\, X)h\big|_{\tau=0} , \notag \\[1mm]
		     &= \tfrac{d}{d\tau} ph\, \exp(\tau\, h\-Xh)\big|_{\tau=0}\rdefeq (h\-Xh)_p^v.
\end{align}
Such a vertical vector field is not $H$-right-invariant, $X^v \notin \Gamma_H(TP)$. 
But $\Gamma(V\P)$ is a Lie subalgebra of $\Gamma(T\P)$, and   Lie$H \rarrow\Gamma(V\P)$ is clearly injective. Using the definition of the Lie derivative  of a vector field $L_X Y = \tfrac{d}{d\tau} \phi\-_{\tau*} Y_{\phi_\tau(p)} \big|_{\tau=0}$, and with the help of the Baker-Campbell-Haussdorff formula which gives, for $A$ and $B$ matrices, 
\begin{align}
\label{BCH-formula}
\exp(-A)\exp(B)\exp(A)= \exp\big( B -[A, B] + \cdots  \big),
\end{align}
we have, for $X^v$ with flow $\phi_\tau$ and $Y^v$ with flow $\vphi_s$:
\begin{align}
\big[ X^v, Y^v\big]_p &= L_{X^V} Y^v_{\ |p}=  \tfrac{d}{d\tau} \phi\-_{\tau*} Y^v_{\phi_\tau(p)}\big|_{\tau=0},   \notag\\[1.5mm]
			    &=  \tfrac{d}{d\tau} \left(   \phi\-_{\tau*} \,  \tfrac{d}{ds} \vphi_s \big( \phi_\tau(p) \big) \big|_{s=0}   \right)\big|_{\tau=0} 
			    =  \tfrac{d}{d\tau}  \tfrac{d}{ds} \left( 	\phi\-_\tau \circ \vphi_s \circ \phi_\tau	\right)(p)\, \big|_{s=0}\,\big|_{\tau=0},  \notag\\[1.5mm]
			    &=   \tfrac{d}{d\tau}  \tfrac{d}{ds} 	    R_{\exp(-\tau X)} \circ 	R_{\exp(s Y)} \circ R_{\exp(\tau X)}\, p\ \big|_{s=0}\,\big|_{\tau=0} 
			    =   \tfrac{d}{d\tau}  \tfrac{d}{ds} 	    p \exp(\tau X) \exp(s Y) \exp(-\tau X)\ \big|_{s=0}\,\big|_{\tau=0},  \notag\\[1.5mm]
			    &=   \tfrac{d}{d\tau}  \tfrac{d}{ds} 	    p \exp\big( sY+ \tau\,s\, [X, Y] + \cdots   \big)\ \big|_{s=0}\,\big|_{\tau=0} 
			    =   \tfrac{d}{d\tau}      p \big( Y+ \tau\, [X, Y]   \big)\ \big|_{\tau=0} = p[X, Y]  \notag\\[1.5mm]
			    &=  \tfrac{d}{dt} p \exp(t\, [X, Y]) \big|_{t=0} \rdefeq [X,Y]^v_p.
\end{align}
So,  Lie$H \rarrow\Gamma(V\P)$ is indeed an injective morphism of Lie algebras.
Similarly, an element $\chi \in $ Lie$\H$, \mbox{satisfying} $R^*_h \, \chi = h\- \chi h$, generates a vertical vector field by $\chi^v \defeq \tfrac{d}{d\tau} p\exp\big(\tau\, \chi(p)\big)\big|_{\tau=0}$. 
 The pushforward by $R_h$ is, 
\begin{align}
R_{h*}\chi^v_p &= \tfrac{d}{d\tau} R_h\, p\exp\big(\tau\, \chi(p)\big)\, \big|_{\tau=0} = \tfrac{d}{d\tau} p\exp\big(\tau\, \chi(p)\big) h\, \big|_{\tau=0}= \tfrac{d}{d\tau} ph\, h\-\!\exp\big(\tau\, \chi(p)\big)h\, \big|_{\tau=0} , \notag \\[1mm]
		     &= \tfrac{d}{d\tau} ph\, \exp\big(\tau\, h\-\chi(p)h\big)\, \big|_{\tau=0} = \tfrac{d}{d\tau} ph\, \exp\big(\tau\, \chi(ph)\big)\,\big|_{\tau=0} \rdefeq \chi_{ph}^v.
\end{align}
Which proves that $\chi^v \in \Gamma_H(TP)$. Furthermore, for $\chi^v$ with flow $\phi_\tau$ and $\eta^v$ with flow $\vphi_s$,
\begin{align}
\big[ \chi^v, \eta^v\big]_p &= L_{\chi^V} \eta^v_{\ |p}=  \tfrac{d}{d\tau} \phi\-_{\tau*} \eta^v_{\phi_\tau(p)}\big|_{\tau=0},   \notag\\[1.5mm]
			    &=  \tfrac{d}{d\tau} \left(   \phi\-_{\tau*} \,  \tfrac{d}{ds} \vphi_s \big( \phi_\tau(p) \big) \big|_{s=0}   \right)\big|_{\tau=0} 
			    =  \tfrac{d}{d\tau}  \tfrac{d}{ds} \left( 	\phi\-_\tau \circ \vphi_s \circ \phi_\tau	\right)(p)\, \big|_{s=0}\,\big|_{\tau=0}
\end{align}
Now, we have that
\begin{align}
\big( \vphi_s \circ \phi_\tau \big)(p) &= \vphi_s \left( p \exp\big(\tau\, \chi(p)\big) \right) 
						      = p \exp\big(\tau\, \chi(p)\big)\  \exp\left(  s\ \eta\bigg(  p \exp\big(\tau\, \chi(p)\big)  \bigg) \right) , \notag\\[1.5mm]
						     &= p \exp\big(\tau\, \chi(p)\big)\ \exp\bigg( s\ \exp\big(-\tau\, \chi(p) \big) \,  \eta(p) \exp\big( \tau\, \chi(p) \big) \bigg), \notag\\[1.5mm]
						      &= p \exp\big(\tau\, \chi(p)\big)\  \exp\big(-\tau\, \chi(p) \big) \exp\bigg( s\  \eta(p) \bigg)  \exp\big( \tau\, \chi(p)\big) , \notag\\*[1.5mm]
						      &= p \exp\big( s\  \eta(p) \big)  \exp\big( \tau\, \chi(p)\big). 
\end{align}
So, 
\begin{align}
\phi_{-\tau} \, \big( \vphi_s \circ \phi_\tau \big)(p) &= \left[ p \exp\big( s\  \eta(p) \big)  \exp\big( \tau\, \chi(p)\big)\right]\ \, \exp \left(  	- \tau\, \chi \bigg(     p \exp\big( s\  \eta(p) \big)\exp\big( \tau\, \chi(p)\big)   \bigg) 	\right), \notag\\[1.5mm]
									&\hspace{-6mm}=  \left[ p \exp\big( s\  \eta(p) \big)  \exp\big( \tau\, \chi(p)\big)\right]\ \, \exp\big( -\tau\, \chi(p)\big) \exp\big( - s\  \eta(p) \big) \ \exp\bigg( -\tau\, \chi(p) \bigg)\  \exp\big( s\  \eta(p) \big)\exp\big( \tau\, \chi(p)\big), \notag\\[1.5mm]
									&= p \ \exp\big( -\tau\, \chi(p) \big)\  \exp\big( s\  \eta(p) \big)\exp\big( \tau\, \chi(p)\big), \notag\\[1.5mm]
									&= p \ \exp \bigg( s\, \eta(p) - s\, \tau\ \big[ \chi(p), \eta(p)\big] + \cdots  \bigg).
\end{align}
And finally we get, 
\begin{align}
\big[ \chi^v, \eta^v\big]_p &=  \tfrac{d}{d\tau}  \tfrac{d}{ds} \left( 	\phi\-_\tau \circ \vphi_s \circ \phi_\tau	\right)(p)\, \big|_{s=0}\,\big|_{\tau=0}
				      = p\ \left( - \big[ \chi(p), \eta(p)\big] \right), \notag\\[1.5mm]
				      &= \tfrac{d}{dt} p\ \exp\left(  -t\, \big[ \chi, \eta\big](p) \right) \rdefeq \big( -[\chi, \eta ]\big)^v_p.
\end{align}
Which proves that the map Lie$\H \rarrow \Gamma_H(V\P)$,  $\chi \mapsto \chi^v$, is a Lie algebra \emph{anti}-isomorphism. 
An obvious corollary is that by altering  the definition  to $\chi^v_p \defeq \tfrac{d}{d\tau} \ p\, \exp\big( -\tau\, \chi(p)\big)\, \big|_{\tau=0}$, this map becomes a Lie algebra isomorphism.


\section{Proofs of pushforward formulae for variational vector fields}
\label{Proofs of pushforward formulae for variational vector fields}

The pushforward of a \emph{vertical} variational vector field $\chi^v \in \Gamma(V\A)$, with $\chi \in$ Lie$\H$, by the right action of the structure group $\H$ of $\A$  is, for $\gamma \in \H$,
\begin{align}
\label{A1}
R_{\gamma\star} \chi^v_A\defeq &\  R_{\gamma\star}\tfrac{d}{d\tau}\, R_{\exp(\tau\, \chi)} \ A \big|_{\tau=0}  = \tfrac{d}{d\tau}\, R_\gamma\ R_{\exp(\tau\, \chi)} \ A \big|_{\tau=0} 
						     =  \tfrac{d}{d\tau} \, R_{\exp(\tau\, \chi) \gamma}\ A \big|_{\tau=0}  =  \tfrac{d}{d\tau} \, R_{\gamma\gamma\- \exp(\tau\, \chi) \gamma} \ A \big|_{\tau=0},  \notag\\*[1mm]
					           =&\ \tfrac{d}{d\tau} \, R_{\gamma\- \exp(\tau\, \chi) \gamma}\ R_\gamma \ A \big|_{\tau=0} = \tfrac{d}{d\tau} \, R_{ \exp(\tau\, \gamma\-\chi \gamma) } \ A^\gamma \big|_{\tau=0} \rdefeq \left( \gamma\-\chi \gamma\right)^v_{A^\gamma}. 
\end{align}
This is in exact analogy with the finite dimensional case. One proves that the map  Lie$\H \rarrow \Gamma(V\A)$ is an injective morphism of Lie algebras as in section \ref{Lie algebra (anti)-isomorphisms}.

By a computation analogue to \eqref{A1}, the action of $\H$ on vertical vector fields $\bs\chi^v$ induced by $\bs \chi \in $ Lie$\bs\H$ is
\begin{align}
R_{\gamma\star} \bs\chi^v_A\defeq &\   \tfrac{d}{d\tau}\, R_\gamma\ R_{\exp\big(\tau\, \bs\chi(A)\big)} \ A \big|_{\tau=0} 
						     =  \tfrac{d}{d\tau} \, R_{\exp\big(\tau\, \bs\chi(A)\big) \, \gamma}\ A \big|_{\tau=0}  =  \tfrac{d}{d\tau} \, R_{ \exp\big(\tau\, \gamma\-\bs\chi(A) \gamma\big) } \ A^\gamma \big|_{\tau=0},  \notag\\[1mm]
					           =&\  \tfrac{d}{d\tau} \, R_{ \exp(\tau\, \bs\chi(A^\gamma) ) } \ A^\gamma \big|_{\tau=0} \rdefeq \bs \chi ^v_{A^\gamma}. 
\end{align}
In exact analogy with the finite dimensional case. Which proves that $\bs\chi^v$ is a $\H$-rigth-invariant vector field. One proves that the map  Lie$\bs\H \rarrow \Gamma_{\H}(V\A)$ is a Lie algebra anti-isomorphism as in section \ref{Lie algebra (anti)-isomorphisms}.
\bigskip

The action of $\bs\Aut_v(\A) \simeq \bs\H$ on a variational vector field $\bs X \in \Gamma(T\A)$ is obtained as in the finite dimensional case. For $\bs \Psi \in \bs\Aut_v(\A)$, s.t. $\bs \Psi(A)=R_{\bs \gamma(A)} A$, and $\phi_\tau: \A \rarrow \A$ the flow of $\bs X$, we have
\begin{align*}
\bs \Psi_\star \bs X_{A}\defeq&\ \tfrac{d}{d\tau} \bs \Psi\big(  \phi_\tau(A)\big) \big|_{\tau=0} =   \tfrac{d}{d\tau} R_{\bs \gamma\big(  \phi_\tau(A) \big)} \phi_\tau(A)  \big|_{\tau=0}, \\[1mm]
					  =&\ 	 \tfrac{d}{d\tau} \bs\gamma\big(  \phi_\tau(A) \big)\- \ \phi_\tau(A) \ \bs\gamma\big(  \phi_\tau(A) \big) + \bs\gamma\big(  \phi_\tau(A) \big)\- d \bs\gamma\big(  \phi_\tau(A) \big) \big|_{\tau=0}, \\[1mm]
					  =&\ \underline{ \tfrac{d}{d\tau}\bs\gamma\big(  \phi_\tau(A) \big)\- \big|_{\tau=0}\ A\ \bs \gamma(A) }_{\,_\text I} 
					  	+  \bs\gamma(A)\-  \tfrac{d}{d\tau}  \phi_\tau(A) \big|_{\tau=0} \ \bs \gamma(A)
					  				+  \underline{ \bs\gamma(A)\- A \ \tfrac{d}{d\tau}\bs\gamma\big(  \phi_\tau(A) \big) \big|_{\tau=0}   }_{\,_\text{II}} \\[1mm]
					&\hspace{5cm}			+ \underline{ \tfrac{d}{d\tau}\bs\gamma\big(  \phi_\tau(A) \big)\- \big|_{\tau=0} \ d \bs\gamma(A) }_{\,_\text{III}}
										+ \underline{  \bs \gamma(A)\- d\  \tfrac{d}{d\tau}\bs\gamma\big(  \phi_\tau(A) \big)\big|_{\tau=0} }_{\,_\text{IV}}.
\end{align*}
Now, on the one hand
\begin{align*}
R_{\bs\gamma(A) \star} \bs X_A \defeq&\ \tfrac{d}{d\tau} R_{\bs \gamma(A)} \phi_\tau(A)  \big|_{\tau=0} =  \tfrac{d}{d\tau}  \bs \gamma(A)\- \phi_\tau(A) \ \bs \gamma(A) + \bs \gamma(A)\- d \bs \gamma(A)\big|_{\tau=0}, \\
						=&\  \bs \gamma(A)\-   \tfrac{d}{d\tau}\phi_\tau(A) \big|_{\tau=0} \bs \gamma(A), 
\end{align*}
and one the other hand, 
\begin{align*}
D^{A^{\bs\gamma(A)}}\left( \bs\gamma(A)\- \bs d \bs\gamma_{|A} (\bs X_A)\right) &= d\left( \bs\gamma(A)\- \bs d \bs\gamma_{|A} (\bs X_A) \right) + \left[   \bs \gamma(A)\- \phi_\tau(A) \ \bs \gamma(A) + \bs \gamma(A)\- d \bs \gamma(A) , \bs\gamma(A)\- \bs d \bs\gamma_{|A} (\bs X_A) \right], \\[1mm]
														      &= d\bs\gamma(A)\- \cdot \bs d \bs\gamma_{|A} (\bs X_A) + \bs\gamma(A)\- d\, \bs d \bs\gamma_{|A} (\bs X_A)   \\[-4mm]
											&\hspace{3.5cm}  + \bs\gamma(A)\- A \ \bs d \bs\gamma_{|A} (\bs X_A) - \overbrace{\bs\gamma(A)\- \bs d \bs\gamma_{|A} (\bs X_A)\ \bs\gamma(A)\-}^{-\bs d \bs\gamma(A)\- _{|A}(\bs X_A)} A\ \bs\gamma(A) \\[1mm]
&\hspace{3.5cm}  +  \underbrace{\bs\gamma(A)\- d \bs\gamma(A) \  \bs\gamma(A)}_{-d\bs\gamma(A)\-} \ \bs d \bs\gamma_{|A} (\bs X_A) - \underbrace{\bs\gamma(A)\- \bs d \bs\gamma_{|A} (\bs X_A)\, \bs\gamma(A)\-}_{-\bs d \bs\gamma(A)\- _{|A}(\bs X_A)}d \bs\gamma(A),  \\[1mm]
														    & \hspace{-1cm}= \underline{ \bs\gamma(A)\- d\, \bs d \bs\gamma_{|A} (\bs X_A) }_{\,_\text{IV}}
														           + \underline{ \bs\gamma(A)\- A \ \bs d \bs\gamma_{|A} (\bs X_A) }_{\,_\text{II}}
														           + \underline{ \bs d \bs\gamma(A)\- _{|A}(\bs X_A)\ A\ \bs\gamma(A) }_{\,_\text{I}}
														           + \underline{ \bs d \bs\gamma(A)\- _{|A}(\bs X_A)\ d \bs\gamma(A)}_{\,_\text{III}}.
\end{align*}
Since the above quantity is clearly the component of a vertical vector field at $A^{\bs\gamma(A)}$, 
\begin{align*}
D^{A^{\bs\gamma(A)}}\left( \bs\gamma(A)\- \bs d \bs\gamma_{|A} (\bs X_A) \right) = \tfrac{d}{ds}\ R_{\exp \big(s\   \bs\gamma(A)\- \bs d \bs\gamma_{|A} (\bs X_A) \big) } A^{\bs\gamma(A)} \big|_{s=0} \rdefeq \left\{ \bs\gamma(A)\- \bs d \bs\gamma_{|A} (\bs X_A) \right\}^v_{A^{\bs\gamma(A)}},
\end{align*}
which is also written as, 
\begin{align*}
\bs\gamma(A)\-  D^A\left( \bs d \bs\gamma_{|A} \bs\gamma(A)\-  (\bs X_A) \right)\bs\gamma(A) = \bs\gamma(A)\-  \tfrac{d}{ds}\ R_{\exp \big(s\   \bs d \bs\gamma\bs\gamma\-_{|A}  (\bs X_A) \big) }\ A \big|_{s=0} \ \bs\gamma(A) \rdefeq R_{\bs\gamma(A)\star} \left\{ \bs d \bs\gamma \gamma\-_{|A} (\bs X_A) \right\}^v_A,
\end{align*}
we get the final expression, 
\begin{align}
\bs \Psi_\star \bs X_{A} = R_{\bs\gamma(A) \star} \bs X_A  + \left\{ \bs\gamma(A)\- \bs d \bs\gamma_{|A} (\bs X_A) \right\}^v_{A^{\bs\gamma(A)}} 
				    &= R_{\bs\gamma(A) \star}  \left( \bs X_A + \left\{ \bs d \bs\gamma \bs\gamma\-_{|A} (\bs X_A) \right\}^v_A  \right), \\[1mm]
	\qquad \text{or symbolically, in `component',}	 \quad		    &= \bs\gamma(A)\- \left(  \bs X(A) + D^A \big\{ \bs{d\gamma\gamma}\-_{|A}  (\bs X_A) \big\} \right) \bs\gamma(A) \ \tfrac{\delta}{\delta A}. \notag
\end{align}
It's  the touchstone of the geometric derivation of (field-dependent) gauge transformations of variational forms on~$\A$, defined by $\bs \alpha^{\bs\gamma}(\bs X)\defeq \big(\bs\Psi^\star \bs\alpha \big)(\bs X)= \bs \alpha\big(\bs\Psi_\star \bs X \big)$, $\bs \gamma \in \bs \H$. 

\medskip

\section{Cocycle relations for c-equivariant theories}
\label{Cocycle relations for c-equivariant theories}

We prove here the claim of section \ref{Classical gauge anomalies} that massive Yang-Mills (mYM) theory and non-Abelian 3D Chern-Simon (CS) theory are sections of \emph{twisted} bundles associated to the $\H$-principal bundle $\A$, i.e. that their equivariance is s.t. $R^\star_\gamma L=L + c(\ \, ; \gamma)$ with $c(A; \gamma\alpha)= c(A: \gamma) + c\big(A^\gamma; \alpha \big)$, for $\gamma, \alpha \in \H$. 

In the case of mYM theory, the $\H$-equivariance of the Lagrangian is 
\begin{align}
R^\star_\gamma L_\text{{\tiny mYM}}(A)= L_\text{{\tiny mYM}}(A)+ c(A; \gamma), \quad \text{with}  \quad c(A; \gamma)\defeq  - m^2 \Tr\left( 2\, A\, *\!d\gamma\gamma\- - d\gamma\-\,*\!d\gamma \right). 
\end{align}
So, the (linearized) cocycle $c$ verifies
\begin{align}
c(A; \gamma\alpha) =&\  -m^2 \Tr\left( 2\, A\, *\!d(\gamma\alpha)(\gamma\alpha)\- - d(\gamma\alpha)\-\,*\!d(\gamma\alpha) \right), \notag \\
			       =&\  -m^2 \Tr\left(  2\, A\, *\!d\gamma\gamma\- + 2\, A\, *\!\gamma (d\alpha\alpha\-)\gamma\- - \big( d\alpha\-\gamma\- + \alpha\- d\gamma\- \big)\, *\!\big( d\gamma\alpha + \gamma d\alpha \big) \right), \notag \\
			       =&\  -m^2 \Tr\bigg( 2\, A\, *\!d\gamma\gamma\-  + 2(\gamma\- A\, \gamma)  \, *\!d\alpha\alpha\- - \ \underbrace{\alpha d\alpha\-}_{-d\alpha \alpha\-} \, *\gamma\-d\gamma - d\alpha\- \, *\!d\alpha - d\gamma\- \, *\!d\gamma - \underbrace{d\gamma\-\gamma}_{-\gamma\-d\gamma} \, *\ \!d\alpha \alpha\- \bigg), \notag\\[-1mm]
			       =&\  -m^2 \Tr\bigg( 2\, A\, *\!d\gamma\gamma\-  - d\gamma\-  \, *\!d\gamma \ +\  2\, \big( \gamma\- A\, \gamma + \gamma\-d\gamma \big)  \, *\! d\alpha\alpha\- - d\alpha\- \, *\!d\alpha \bigg), \notag\\
	  	       \rdefeq&\ c(A: \gamma) + c\big(A^\gamma; \alpha \big). 
\end{align}
Which proves the claim. Remark that in the Abelian limit, i.e. massive Maxwell theory, we have
\begin{align*}
R^\star_\gamma L_\text{{\tiny mM}}(A)= L_\text{{\tiny mM}}(A)+ c(A; \gamma), \quad \text{with}  \quad c(A; \gamma)\defeq  -m^2 \left( 2\, A\, *\!d\chi + d\chi\,*\!d\chi \right),
\end{align*}
where $\chi \in$ Lie$\U(1)$. So that 
\begin{align*}
c(A; \gamma\alpha) &= -m^2\ \left( 2\, A \, *\!d\chi + d\chi \, *\!d\chi \ +\ 2\big( A+ d\chi \big)\, *\!d\lambda + d\lambda \, *\!d\lambda   \right) = c(A; \gamma)+ c(A^\gamma; \alpha), \\
			       &= -m^2\ \left( 2\, A \, *\!d\lambda + d\lambda \, *\!d\lambda \ +\ 2\big( A+ d\lambda \big)\, *\!d\chi + d\chi \, *\!d\chi    \right) = c(A; \alpha)+ c(A^\alpha; \gamma) = c(A; \alpha\gamma),
\end{align*}
as indeed the gauge group is Abelian. Yet the cocycle is still non-trivial: $c(A; \gamma\alpha) \neq  c(A; \gamma) + c(A; \alpha)$.
\bigskip

In the case of CS theory, the  $\H$-equivariance of the Lagrangian is well-kown to be
\begin{align}
R^\star_\gamma L_\text{{\tiny CS}}(A)= L_\text{{\tiny CS}}(A)+ c(A; \gamma), \quad \text{with}  \quad c(A; \gamma)\defeq   \Tr\left( d\big( \gamma d\gamma\- A \big) -\tfrac{1}{3} \big( \gamma\-d\gamma\big)^3\right). 
\end{align}
Since we have on the one hand, 
\begin{align}
c\big(A^\gamma; \alpha\big) &= \Tr\left(  d\big( \alpha d\alpha\- [\gamma\- A \, \gamma+\gamma\-d\gamma] \big)  - \tfrac{1}{3} \big( \alpha\-d\alpha\big)^3 \right), \notag \\
					   &= \Tr\left(  d\big( \alpha d\alpha\-\, \gamma\- A \, \gamma \big) + d\big( \alpha d\alpha\-\, \gamma\-d\gamma \big)  - \tfrac{1}{3} \big( \alpha\-d\alpha\big)^3 \right), \notag\\
					   &=  \Tr\left(  d\alpha d\alpha\-\, \gamma\- A \, \gamma -  \alpha d\alpha\- d\gamma\- A\, \gamma -  \alpha d\alpha\-\gamma\-dA\, \gamma +  \alpha d\alpha\- \gamma\-A\, d\gamma \right. \notag \\
					   &\hspace{4cm} \left. + d \alpha d\alpha\-\,  \gamma\-d\gamma - \alpha d\alpha\-\, d\gamma\-d\gamma    - \tfrac{1}{3} \big( \alpha\-d\alpha\big)^3 \right). 
\end{align}
We find that, 
\begin{align}
c(A; \gamma\alpha)  &= \Tr\left(  d\left( \gamma\alpha d(\gamma\alpha)\-A \right)  - \tfrac{1}{3}\left(  (\gamma\alpha)\-d(\gamma\alpha) \right)^3\right), \notag\\
			        &= \Tr\left( d\left(  \big( \gamma\alpha d\alpha\- \gamma\- + \gamma d\gamma\- \big)A \right)  - \tfrac{1}{3}\left(  \alpha\- \gamma\-d\gamma\alpha + \alpha\-d\alpha \right) \right), \notag\\
			        &= \Tr\left( d\big( \gamma\alpha d\alpha\-\gamma\- A \big) + \underline{d\big( \gamma d\gamma\- A \big) -   \tfrac{1}{3}\big( \gamma\-d\gamma \big)^3}_{}
			        		- \alpha\-\gamma\- d\gamma d\alpha\, \alpha\-d\alpha  - \alpha\-\gamma\-d\gamma\,\gamma\-d\gamma d \alpha - \tfrac{1}{3} \big( \alpha\-d\alpha \big)^3 \right), \notag\\
				&= \underline{c(A; \gamma)}_{}  + \Tr\bigg( \left( d\gamma \alpha d\alpha\- \gamma\- A + \gamma d\alpha d\alpha\- \gamma\- A - \gamma \alpha d\alpha\-d\gamma\-A  - \gamma \alpha d\alpha\-\gamma\-dA \right)  \notag\\
				& \hspace{8cm }	+  \gamma\-d\gamma d\alpha d\alpha\-  + \underbrace{\alpha\-d\gamma\- d\gamma d\alpha}_{-\alpha d\alpha\- d\gamma\- d\gamma}  - \tfrac{1}{3} \big( \alpha\-d\alpha\big)^3 \bigg), \notag\\[-3mm]
				&=  c(A; \gamma)+  c(A^\gamma; \alpha).
\end{align}
Which proves the claim. Remark that in the Abelian limit we have
\begin{align*}
R^\star_\gamma L_\text{{\tiny Ab-CS}}(A)= L_\text{{\tiny Ab-CS}}(A)+ c(A; \gamma), \quad \text{with}  \quad c(A; \gamma)\defeq  d\big(\chi dA \big),
\end{align*}
where $\chi \in$ Lie$\U(1)$. So that 
\begin{align*}
c(A; \gamma\alpha) =   d\big((\chi+\lambda) dA \big) &=   d\big(\chi dA \big) +  d\big(\lambda dA \big) \\
			        &= c(A; \gamma)+ c(A; \alpha)= c(A; \alpha\gamma),
\end{align*}
which reflects that $\U(1)$ is abelian, as it should. But the case of Abelian 3D CS theory is quite degenerate as the cocycle is trivial, i.e. it is a simple group morphism. 

%


\section{Noether charges as generators of gauge transformations}
\label{Noether charges as generators of gauge transformations}

We show how, in our notations, the Noether charges $Q_\Sigma(\chi; A)$ associated to Lie$\H$  generate infinitesimal gauge transformations via the Poisson bracket defined by the presymplectic 2-form $\bs \Theta_\Sigma$.  
As we managed to defined charges and their Poisson bracket for invariant and non-invariant theories (in the latter case under the restriction of Hypothesis 0 \eqref{Hyp0} and 1 \eqref{Hyp1}), the following holds in both cases. 

Consider a functional $f : \A \rarrow \Omega^{n-1}(\P)$, $A \mapsto f(A)$. Define $V^f$, its associated variational Hamiltonian vector field, via $\iota_{V^f} \bs \Theta_\Sigma = -\int_\Sigma \bs d f$. The action of Lie$\H$ on $f$ is usually given by the Lie derivative. Using Cartan's formula, 
\begin{align}
\int_\Sigma \bs L_{\chi^v} f = \int_\Sigma \iota_{\chi^v}\bs d f + \bs d  \cancel{\iota_{\chi^v} f} =  \iota_{\chi^v} \left(- \iota_{V^f} \bs \Theta_\Sigma  \right) = \iota_{V^f} \iota_{\chi^v} \bs \Theta_\Sigma  \rdefeq&\ \big\{ Q_\Sigma(\chi; A), \ f \big\},   \\*
																					\hookrightarrow\  =&\ -  \iota_{V^f} \bs d Q_\Sigma(\chi; A). \notag
\end{align}
The first line shows why Noether charges generate  Lie$\H$-transformations via the Poisson bracket, the second line gives the explicit mean of computation: One must first determine the Hamiltonian vector field of $f$ via the symplectic 2-form $ \bs \Theta_\Sigma$, then feed it to the variational 1-form $\bs d Q_\Sigma(A; \chi)$.

In the case of the `identity functional` $A \mapsto \id(A)$, we define the associated Hamiltonian vector field $V^A$ via $\iota_{V^A} \bs \Theta_\Sigma = -\int_\Sigma \bs d \id(A)= -\int_\Sigma \id(\bs d A)$. So the action of a Noether charge is 
\begin{align}
\big\{ Q_\Sigma(\chi; A), \id(A) \big\} \defeq \bs \Theta_\Sigma\big(\chi^v, V^A \big)= -\iota_{ \chi^v} \iota_{ V^A}\bs \Theta_\Sigma = \iota_{ \chi^v} \int_\Sigma \id(\bs d A) = \int_\Sigma \id\big(D^A\chi\big),
\end{align}
which indeed reproduces the infinitesimal gauge transformation of $A$. 
The definition of  $\id(A)$ might need to be adapted according to the example treated. It must be built from the functional on which the Lagrangian is based (like all other quantities derived from the latter), so in most cases an invariant polynomial $P$ on Lie$\H$. As $A= A^a \tau_a \in \Omega^1(\P)\, \otimes$ Lie$H$, one may use $\tau_a \vol^a_{n-2} \in \Omega^{n-2}(\P) \, \otimes$ Lie$H$ to defined $\id(A)\defeq P\big(A, \tau_a \vol^a_{n-2}\big) \in \Omega^{n-1}(\P)$. 
Let us illustrate how this works in concrete situations. We take two examples of invariant theories, two examples of non-invariant theories. 
\bigskip

\noindent {\bf Yang-Mills theory:} We remind that the presymplectic 2-form is $\bs\Theta_\Sigma = -\int_\Sigma \Tr \big( \bs d A \, *\! \bs d F\big)$, and the Noether charge is $Q_\Sigma(\chi; A)= \int_\Sigma \Tr\big(D^A \chi \, *\!  F  \big)$, with $\chi \in $ Lie$\SU(n)$. In this case we define $\id(A)=\Tr\big( A\  \tau_a \vol^a_{n-2} \big)$
So, we get first, 
\begin{align}
\label{dQ-SigmaYM}
\bs d Q_\Sigma(\chi; A) = \int_\Sigma \Tr\big([\bs dA,  \chi] \, *\!  F + D^A\chi  \, *\!  \bs d F \big) = \int_\Sigma \Tr\big( - \bs d A \, *\![F, \chi] +  D^A\chi  \, *\!  \bs d F \big).
\end{align}
Then we derive the constraints on  $V^A$, 
\begin{align}
\iota_{V^A} \bs \Theta_\Sigma &= -\int_\Sigma \id(\bs d A),  \quad \Rightarrow   \quad    -\int_\Sigma \Tr \big( \iota_{V^A}\bs d A \, *\! \bs d F - \bs d A\, *\!    \iota_{V^A} \bs d F \big)  =- \int_\Sigma \Tr\big( \bs dA\  \tau_a \vol^a_{n-2} \big), \notag\\[2mm]
\hookrightarrow &\  \text{ so that} \quad   \iota_{V^A}\bs d A =0 \quad \text{and} \quad  *\iota_{V^A}\bs d F= - \tau_a \vol^a_{n-2}. 
\end{align}
Then the Poisson bracket is, using \eqref{dQ-SigmaYM}, 
\begin{align}
\big\{ Q_\Sigma(\chi; A), \id(A) \big\} =  \bs \Theta_\Sigma\big(\chi^v, V^A \big) = - \iota_{V^A} \bs d Q_\Sigma(\chi; A) = \int_\Sigma \Tr \big(D^A\chi \,  \tau_a \vol^a_{n-2} \big)\defeq \id\big( D^A\chi  \big).
\end{align}
This indeed reproduces the infinitesimal (field-independent) gauge transformation of $A$. 

To do the same for the curvature $F$, let us define $\id(F)=\Tr\big(  \tau_a \vol^a_1 \, * F \big)$ -  inspired by the expression of  $\bs \Theta_\Sigma$. We~have then
\begin{align}
\iota_{V^F} \bs \Theta_\Sigma &= -\int_\Sigma \id(\bs dF) \quad \Rightarrow   \quad    -\int_\Sigma \Tr \big( \iota_{V^F}\bs d A \, *\! \bs d F - \bs d A\, *\!    \iota_{V^F} \bs d F \big)  =- \int_\Sigma \Tr\big( \tau_a \vol^a_1\, * \bs d F \big), \notag\\[2mm]
\hookrightarrow &\  \text{ so that} \quad   \iota_{V^F}\bs d A = \tau_a \vol^a_1 \quad \text{and} \quad  \iota_{V^F}\bs d F= 0. 
\end{align}
The Poisson bracket is, 
\begin{align}
\big\{ Q_\Sigma(\chi; A), \id(F) \big\} =  \bs \Theta_\Sigma\big(\chi^v, V^F \big) = - \iota_{V^F} \bs d Q_\Sigma(\chi; A) = \int_\Sigma \Tr \big(\tau_a \vol^a_1 \,  *[F, \chi]  \big) \defeq \id\big( [F, \chi]  \big).
\end{align}
This  reproduces the infinitesimal (field-independent) gauge transformation of $F$. To finish we can find the Poisson bracket of $\id(A)$ and  $\id(F)$, 
\begin{align}
\big\{ id(A), \id(F) \big\} &= \bs \Theta_\Sigma\big(V^A, V^F \big) = - \iota_{V^F}  \int_\Sigma \Tr\big( \bs dA\  \tau_b \vol^b_{n-2} \big) = -\int_\Sigma \Tr\big(\tau_a \vol^a_1 \tau_b \vol^b_{n-2} \big), \notag\\
		   		    &= -\int_\Sigma \vol_{n-1}\, \Tr\big(\tau_a  \tau_b \big) = -\int_\Sigma \vol_{n-1}\, \tfrac{1}{2} \delta_{ab}, 
\end{align}
using a standard normalisation in the last step. Putting a distribution to get a finite result, this reproduces the familiar Poisson bracket in components: $\big\{ A^a_i(x), F^b_j(y) \big\} = - \tfrac{1}{2}\, \delta_{ij}\,\delta^{ab}\, \delta^{n-1}(x-y) $.
\bigskip

\noindent {\bf 4D Einstein-Cartan gravity $\bs{\Lambda=0}$:} We remind that the presymplectic 2-form is $\bs\Theta_\Sigma = - 2 \int_\Sigma  \bs d A \bullet \bs d e \w e^T $, and the Noether charge is $Q_\Sigma(\chi; \b A)= \int_\Sigma D^A \chi \bullet  e \w e^T  $, with $\chi \in $ Lie$\SO(1,3)$. In this case we define $\id(A)= A \bullet  \tau_a \vol^a_2 $ and $\id(e)=   \tau_a \vol^a_1 \bullet\, e \w e^T  $. 
Get first, 
\begin{align}
\label{dQ-SigmaEC}
\bs d Q_\Sigma(\chi; \b A) = \int_\Sigma [\bs dA,  \chi] \bullet  e \w e^T +  2\ D^A\chi \bullet  \bs d e \w e^T=  \int_\Sigma  - \bs dA \bullet  [e \w e^T, \chi] +  2\  \bs d e \w e^T  \bullet  D^A\chi.
\end{align}
Derive the constraints on  $V^A$ and $V^e$, 
\begin{align}
\iota_{V^A} \bs \Theta_\Sigma &=\int_\Sigma \id(\bs d A),  \quad \Rightarrow   \quad    -2 \int_\Sigma  \iota_{V^A}\bs d A\bullet \bs d e\w e^T - \bs d A\bullet  \iota_{V^A} \bs d e\w e^T = - \int_\Sigma \bs d A \bullet \tau_a \vol^a_2 , \notag\\[2mm]
\hookrightarrow &\  \text{ so that} \quad   \iota_{V^A}\bs d A =0 \quad \text{and} \quad  \iota_{V^A}\bs d e\w e^T = -\tfrac{1}{2} \tau_a \vol^a_2. \\[2mm]
\iota_{V^e} \bs \Theta_\Sigma &=\int_\Sigma \id(\bs d e),  \quad \Rightarrow   \quad    -2 \int_\Sigma  \iota_{V^e}\bs d A\bullet \bs d e\w e^T - \bs d A\bullet  \iota_{V^e} \bs d e\w e^T = - 2 \int_\Sigma     \tau_a \vol^a_1 \bullet\, \bs d e \w e^T , \notag\\[2mm]
\hookrightarrow &\  \text{ so that} \quad   \iota_{V^e}\bs d A =\tau_a \vol^a_1 \quad \text{and} \quad  \iota_{V^e}\bs d e = 0. 
\end{align}
Then the Poisson brackets are, 
\begin{align}
\big\{ Q_\Sigma(\chi; \b A), \id(A) \big\} &=  \bs \Theta_\Sigma\big(\chi^v, V^A \big) = - \iota_{V^A} \bs d Q_\Sigma(\chi; \b A) = \int_\Sigma   \tau_a \vol^a_2 \bullet  D^A\chi = \int_\Sigma  D^A\chi \bullet  \tau_a \vol^a_2 \defeq \id\big( D^A\chi  \big), \\[1mm]
\big\{ Q_\Sigma(\chi; \b A), \id(e) \big\} &=  \bs \Theta_\Sigma\big(\chi^v, V^e \big) = - \iota_{V^e} \bs d Q_\Sigma(\chi; \b A) = \int_\Sigma   \tau_a \vol^a_1 \bullet\, [e\w e^T ,\chi] \defeq \id\big(  [e\w e^T ,\chi] \big).
\end{align}
This reproduces the infinitesimal gauge transformations of $A$ and $e$. 
\bigskip

\noindent {\bf 3D non-Abelian Chern-Simons theory:} We remind that the presymplectic 2-form is $\bs\Theta_\Sigma = -  \int_\Sigma \Tr\big(  \bs d A\, \bs d A \big) $, and the Noether charge can be written  $Q_\Sigma(\chi; A)= \int_\Sigma \Tr\big( D^A \chi \ A - A\, d\chi\big)$, with $\chi \in $ Lie$\SU(n)$. Define $\id(A)= \Tr \big( A \,  \tau_a \vol^a_1\big)$.  
Get~first, 
\begin{align}
\label{dQ-SigmaCS}
\bs d Q_\Sigma(\chi; A) &= \int_\Sigma  \Tr\big( [\bs dA, \chi] \, A + D^A\chi  \, \bs d A - \bs d A\, d\chi \big) = \int_\Sigma \Tr\big( - \bs d A \, [A, \chi] - \bs d A\, d\chi  - \bs d A\,  D^A\chi   \big). \notag\\
				    &=-2\int_\Sigma \Tr\big(  \bs d A\,  D^A\chi  \big).
\end{align}
Constraints on  $V^A$ are, 
\begin{align}
\iota_{V^A} \bs \Theta_\Sigma &= -\int_\Sigma \id(\bs d A),  \quad \Rightarrow   \quad    2 \int_\Sigma \Tr \big(\bs d A \,  \iota_{V^A}\bs d A  \big)  =- \int_\Sigma \Tr\big( \bs dA\  \tau_a \vol^a_1 \big), \notag\\[2mm]
\hookrightarrow &\  \text{ so that} \quad   \iota_{V^A}\bs d A = -\tfrac{1}{2} \tau_a \vol^a_1. 
\end{align}
Then the Poisson bracket is
\begin{align}
\big\{ Q_\Sigma(\chi; A), \id(A) \big\} =  \bs \Theta_\Sigma\big(\chi^v, V^A \big) = - \iota_{V^A} \bs d Q_\Sigma(\chi; A) = - \int_\Sigma \Tr \big( \tau_a \vol^a_1 \ D^A\chi  \big)   \defeq \id\big( D^A\chi  \big),
\end{align}
Which is the expected result. 
\bigskip

\noindent {\bf 3D-$\bs \CC$-gravity $\bs{\Lambda \neq 0}$:} We remind that the presymplectic 2-form is $\bs\Theta_\Sigma = -  \int_\Sigma \Tr\big(  \bs d A\, \bs d e \big) $, while the Lorentz and translation Noether charges are respectively, 
\vspace{-2mm}
\begin{align}
   Q_\Sigma(\chi; \b A)= \int_\Sigma \Tr\big( D^A \chi \ e \big), \quad \text{with }  \chi \in \text{Lie}\H, \\[1mm]
   Q_\Sigma(\tau; \b A)= \int_\Sigma \Tr\left( -\tau \, \big( R- \tfrac{\epsilon}{\ell^2} ee \big)  + d(\tau\, A)\right), \quad \text{with }  \tau \in \text{Lie}\T. 
   \end{align}
 Define $\id(A)= \Tr \big( A \,  \tau_a \vol^a_1\big)$ and $\id(e)= \Tr \big(e\,  \tau_a \vol^a_1\big)$. 
Get~first, 
\begin{align}
\bs d Q_\Sigma(\chi; \b A) &= \int_\Sigma  \Tr\big( [\bs dA, \chi] \, e + D^A\chi  \, \bs d e  \big) = \int_\Sigma \Tr\big( - \bs d A \, [e, \chi]  - \bs d e\,  D^A\chi   \big). \label{dQ-Sigma3DGrav-Lorentz} \\[1.5mm]
\bs d Q_\Sigma(\chi; \b A) &= \int_\Sigma  \Tr\left(-\tau \, \big( \bs d R- \tfrac{\epsilon}{\ell^2} [\bs de, e] + d\big( \tau \bs d A \big) \right)
					=\int_\Sigma  \Tr\left(-\tau \,D^A(\bs d A) + 	 \tfrac{\epsilon}{\ell^2}\, \tau\,[\bs de, e]	+ d\big( \tau \bs d A \big)	\right),  \notag \\
					&=\int_\Sigma  \Tr\bigg( \underbrace{ -\tau \,[A, \bs d A]}_{[A, \tau]\, \bs d A} + 	 \tfrac{\epsilon}{\ell^2}\, \tau\,[\bs de, e]	-  \bs d A \, d \tau 	\bigg)
					=\int_\Sigma  \Tr\left( - \bs dA \, D^A\tau +  \tfrac{\epsilon}{\ell^2}\, \bs d e [e, \tau] \right).
\end{align}
The constraints on $V^A$ and   $V^e$ are, 
\begin{align}
\iota_{V^A} \bs \Theta_\Sigma &= -\int_\Sigma \id(\bs d A),  \quad \Rightarrow   \quad    - \int_\Sigma \Tr \big(\iota_{V^A}\bs d A\, \bs d e - \bs d A \,  \iota_{V^A}\bs d e \big)  =- \int_\Sigma \Tr\big( \bs dA\  \tau_a \vol^a_1 \big), \notag\\[2mm]
\hookrightarrow &\  \text{ so that} \quad   \iota_{V^A}\bs d A = 0 \quad  \text{and} \quad  \iota_{V^A}\bs d e = - \tau_a \vol^a_1. \\[2mm]
\iota_{V^e} \bs \Theta_\Sigma &= -\int_\Sigma \id(\bs d e),  \quad \Rightarrow   \quad    - \int_\Sigma \Tr \big(\iota_{V^A}\bs d A\, \bs d e - \bs d A \,  \iota_{V^A}\bs d e \big)  =- \int_\Sigma \Tr\big(\bs de \,  \tau_a \vol^a_1 \big), \notag\\[2mm]
\hookrightarrow &\  \text{ so that} \quad   \iota_{V^e}\bs d A = - \tau_a \vol^a_1 \quad  \text{and} \quad  \iota_{V^e}\bs d e = 0.
 \end{align}
The Poisson brackets are then, 
\begin{align}
\big\{ Q_\Sigma(\chi; \b A), \id(A) \big\} &=  \bs \Theta_\Sigma\big(\chi^v, V^A \big) = - \iota_{V^A} \bs d Q_\Sigma(\chi; \b A) =   \int_\Sigma \Tr\left( - \tau_a \vol^a_1\, D^A\chi \right)	 \defeq \id\big(D^A\chi\big)	,\\[2mm]
\big\{ Q_\Sigma(\chi; \b A), \id(e) \big\} &=  \bs \Theta_\Sigma\big(\chi^v, V^e \big) = - \iota_{V^e} \bs d Q_\Sigma(\chi; \b A) =  \int_\Sigma \Tr\left(  - \tau_a \vol^a_1\, [e, \chi]	\right)\defeq \id\big([e, \chi] \big), \\[3mm]
\big\{ Q_\Sigma(\tau; \b A), \id(A) \big\} &=  \bs \Theta_\Sigma\big(\tau^v, V^A \big) = - \iota_{V^A} \bs d Q_\Sigma(\tau; \b A) =    \int_\Sigma \Tr\left( -\tfrac{\epsilon}{\ell^2}\, [e, \tau]\,  \tau_a \vol^a_1 \right)= \id\big( -\tfrac{\epsilon}{\ell^2}\, [e, \tau] \big) , \\[2mm]
\big\{ Q_\Sigma(\tau; \b A), \id(e) \big\} &=  \bs \Theta_\Sigma\big(\tau^v, V^e \big) = - \iota_{V^e} \bs d Q_\Sigma(\tau; \b A) =    \int_\Sigma \Tr\left( 	D^A\tau\, \tau_a \vol^a_1 \right)\defeq  \id\big(D^A\tau\big).
\end{align}
This  reproduces the infinitesimal Lorentz and translational gauge transformations of $A$ and $e$ (compare with \eqref{infGT-3D-grav-Lambda}).


\section{Presymplectic structure for non-invariant Lagrangians: the other extreme case}
\label{Presymplectic structure for non-invariant Lagrangians: the other extreme case}

We here briefly treat the opposite alternative to Hypothesis 0 \eqref{Hyp0}, i.e. we consider theories satisfying 
\begin{align}
R^\star_\gamma  \bs \theta= \bs \theta\quad \text{and} \quad R^\star_\gamma  \bs E= \bs E + \bs d c(\ \, ; \gamma). 
\end{align}
Since $R^\star_\gamma  \bs \theta= \bs \theta$, we may write    $\bs\theta_{|A}=\theta(\bs d A; A)=\t L(\bs d A; [F])$. The Noether current and charge are then
\begin{align}
J(\chi;A)=\bs \theta_{|A} (\chi_A^v)&=\theta\big(D^A\chi; A\big)=\t L\big(D^A\chi; [F]\big)=d\t L\big(\chi; [F]\big)- \t L\big(\chi; D^A[F]\big), \notag\\
						    &= d\theta\big(\chi; A\big)- \t L\big(\chi; D^A[F]\big). \\[1mm]
Q_\Sigma(\chi;A)=\int_\Sigma \theta\big(D&^A\chi; A\big) =\int_{\d\Sigma}\theta\big(\chi; A\big) -  \int_\Sigma \t L\big(\chi; D^A[F]\big). \label{Charge-def-2}
\end{align}
So, unfortunately in this case we cannot write these in terms of the field equations. Furthermore, since we have  $\bs dL (\chi^v)=\alpha(\chi; A) =\bs E(\chi^v)+ d\bs\theta(\chi^v)$, we observe that even on-shell the current and charge satisfy an anomalous conservation law 
\begin{align}
\label{Q-on-shell}
dJ(\chi;A)=\alpha(\chi; A)_{\, |\S}, \qquad \text{so} \qquad     dQ_\Sigma(\chi;A)=\int_\Sigma \alpha(\chi; A)_{\, |\S}. 
\end{align}

Yet, it is still true by assumption that  $\bs L_{\chi^v} \bs \theta =0$, rewritten as  $ \iota_{\chi^v} \bs \Theta = - \bs d J(\chi; A)$ with $\bs \Theta \defeq \bs{d\theta}$. So that, as previously, the charge is related to the presymplectic 2-form $\bs\Theta_\Sigma\defeq\int_\Sigma \bs\Theta$ as $\iota_{\chi^v} \bs \Theta_\Sigma = - \bs d Q_\Sigma(\chi; A)$, and the latter generates via \eqref{verticality-Theta} a Poisson bracket: $\big\{ Q_\Sigma(\chi; A) ,  Q_\Sigma(\eta; A)\big\}\defeq\, \bs\Theta_\Sigma(\chi^v, \eta^v)  = Q_\Sigma([\chi, \eta]; A)$. So, here also the map Lie$\H \rarrow \big(Q_\Sigma(\, \, ; A), \big\{\, , \big\} \big)$ is a Lie algebra morphism. 
\medskip

The  $\bs \H$-gauge transformations of $\bs \theta$ and $\bs\Theta$ are obviously identical to the results already obtained in section \ref{Presymplectic structure for invariant Lagrangians}. For $\bs\theta$ we get,
\begin{align*}
\bs\theta^{\bs\gamma}_{|A}(\bs X_A) &\defeq \big( \bs\Psi^\star \bs \theta \big)_{|A}(\bs X_A) = \bs \theta_{A^{\bs\gamma}}\big(\bs\Psi_\star \bs X_A\big)
					=   \bs \theta_{|A^{\bs\gamma}} \left(R_{\bs{\gamma}(A)\star} \left( \bs{X}_A + \left\{ \bs{d}\bs{\gamma} {\bs{\gamma}\- }_{|A}(\bs{X}_A)\right\}^v_A \right) \right), \quad \text{ using \eqref{Pushforward-X-inf}} \notag \\
					&= R^\star_{\bs{\gamma}(A)} \bs \theta_{|A^{\bs\gamma}}  \left( \bs{X}_A + \left\{ \bs{d}\bs{\gamma} {\bs{\gamma}\- }_{|A}(\bs{X}_A)\right\}^v_A \right)
					= \bs \theta_{|A} \left( \bs{X}_A + \left\{ \bs{d}\bs{\gamma} {\bs{\gamma}\- }_{|A}(\bs{X}_A)\right\}^v_A \right)
					=  \bs \theta_{|A} \big( \bs{X}_A \big) +  \theta\left( D^A\{ \bs{d\gamma\gamma}_{|A}\-(\bs X_A) \}; A \right). 
\end{align*}
That is $ \bs\theta^{\bs \gamma}= \bs \theta + \theta \left(D^A \big\{\bs{d\gamma\gamma}\-\big\} ;  A \right)$,  
which is the same as \eqref{Field-depGT-presymp-pot-current} without using  \eqref{Noether-current-invariant}. For $\bs\Theta$ the result is  given by \eqref{Field-depGT-Theta}. 
The corresponding field-dependent gauge transformations of the presymplectic potential and 2-form are thus  
\begin{align}
\label{theta-Theta-GT}
\bs\theta_\Sigma^{\bs\gamma} = \bs\theta_\Sigma + \int_\Sigma   \theta\left( D^A\big\{ \bs{d\gamma\gamma}\- \big\}; A \right), \quad \text{and} \quad
\bs\Theta_\Sigma^{\bs\gamma} = \bs\Theta_\Sigma + \int_\Sigma \bs d\, \theta\left( D^A\big\{ \bs{d\gamma\gamma}\- \big\}; A \right).
\end{align}
It  appears that $\bs\theta_\Sigma$ and $\bs\Theta_\Sigma$ would be basic, and would induce a symplectic structure on $\S/\H$, only if severe restriction are imposed on the field-dependent gauge element $\bs\gamma \in \bs\H$, such as requiring $D^A\big\{ \bs{d\gamma\gamma}\- \}=0$. 

To be complete, we now derive the $\bs\H$-gauge transformation of $\bs E$. By assumption we have $R^\star_\gamma \bs E = \bs E + \bs d  c(\ \, ,\gamma)$, and $\bs E(\chi^v)= \alpha(\chi; A) - d \theta(D^A\chi; A)$. Then
\begin{align}
\label{FieldGT-E-2}
\bs E^{\bs \gamma}_{|A} (\bs X_A) \defeq&\ \big( \bs\Psi^\star \bs E\big)_{|A}  (\bs X_A) =  \bs E_{| \bs\Psi(A)} \big(\bs\Psi_\star \bs X_A \big) 
							       =  \bs E_{| A^{\bs \gamma}} \left(  R_{\bs\gamma\star} \bs X_A + \big\{ \bs \gamma\- \bs{d\gamma}_{|A} (\bs X_A)\big\}^v_{A^{\bs \gamma}}  \right) ,  \notag\\
							       &=  R^\star_{\bs\gamma}\bs E_{| A^{\bs \gamma}} ( \bs X_A) + \bs E_{| A^{\bs \gamma}}\left( \big\{\bs \gamma\- \bs{d\gamma}_{|A} (\bs X_A)\big\}^v_{A^{\bs \gamma}} \right), \notag\\
							       &= \left( \bs E_{|A} + \bs d c\big(\ \, ; \bs\gamma(A)\big)_{|A} \right) ( \bs X_A ) + \alpha\left( \big\{\bs\gamma\- \bs{d\gamma}_{|A}(\bs X_A) ; A^{\bs \gamma} \right) 
							       																	- d\theta\left( D^A\big\{ \bs \gamma\- \bs{d\gamma}_{|A} (\bs X_A)\big\} ; A^{\bs \gamma}\right).
 \end{align}
 Now we remark that 
 \begin{align}
 \bs d c\big(\ \, ; \bs\gamma)_{|A}(\bs X_A)&= \left( \bs d c\big(\ \, ; \bs\gamma(A) \big)_{|A}   + \bs d c\big(A; \bs \gamma  \big)_{|A} \right) (\bs X_A), \\
 								&= \bs d c\big(\ \, ; \bs\gamma(A) \big)_{|A} (\bs X_A)  + \bs d c\big(A; \ \,  \big)_{|\bs \gamma(A)} \big(  \bs{d\gamma}_{|A}(\bs X_A)\big), 
 \end{align}
(see  \cite{Francois2019_II} section 4.2 for further details - in the finite dimensional case)  and,
 \begin{align}
 \alpha\left( \big\{\bs\gamma\- \bs{d\gamma}\-_{|A}(\bs X_A)\big\} ; A^{\bs \gamma} \right) \defeq&\ \tfrac{d}{d\tau} c\left(A^{\bs \gamma} ;  \exp\big(  \tau \, \big\{\bs\gamma\- \bs{d\gamma}_{|A}(\bs X_A) \big\} \big)\right)\big|_{\tau=0}, \notag \\[2mm]
 											=&\ \tfrac{d}{d\tau} c\left(A^{\bs \gamma} ;  \bs\gamma\-\exp\big(  \tau \, \big\{ \bs{d\gamma\gamma}\-_{|A}(\bs X_A) \big\} \big) \, \bs\gamma\ \right)\big|_{\tau=0} 
											= \tfrac{d}{d\tau} c\left(A  ;  \exp\big(  \tau \, \big\{ \bs{d\gamma\gamma}\-_{|A}(\bs X_A) \big\} \big) \, \bs\gamma\ \right)\big|_{\tau=0}, \notag \\[2mm]
									                 =&\   \bs d c\big( A; \bs{d\gamma}_{|A}(\bs X_A) \big) = \bs d  c(A; \bs\gamma) _{|A} (\bs X_A). 
 \end{align}
 So finally, 
 \begin{align}
 \bs E^{\bs \gamma}_{|A} (\bs X_A) = \left( \bs E_{|A}  + \bs d c\big( \ \, ; \bs \gamma \big)_{|A}  - d\theta \big( D^A \big\{ \bs{d\gamma\gamma}\- \big\}; A\big) \right)( \bs X_A ).
 \end{align}
 An instance of the above framwork is massive Yang-Mills theory. 
 \bigskip

\noindent{\bf Massive Yang-Mills theory:} \ \ The Lagrangian is $L(A)= \tfrac{1}{2} \Tr\big( F \, *\! F\big)- \tfrac{1}{2} \, m^2\,\Tr\big( A \, *\! A\big)$, whose equivariance is
\begin{align}
R^\star_\gamma L(A) = L(A) + c(A,\gamma),  \quad \text{with}\quad 
 						   c(A,\gamma)=   -m^2\Tr \left(  A *\!d\gamma\gamma\- - \tfrac{1}{2} d\gamma\-*\!d\gamma  \right), \quad \gamma \in \H=\SU(n).
\end{align}
The presymplectic potential and 2-form are, as in the massless YM case, 
\begin{align}
\bs\theta_\Sigma = \int_\Sigma \theta(\bs d A; A)= \int_\Sigma \Tr \big( \bs d A\, *\! F\big),\quad \text{and} \quad \bs\Theta_\Sigma =\bs d \bs \theta_\Sigma=- \int_\Sigma \Tr \big( \bs d A\, *\! \bs d F\big). 
\end{align}
While the field equations are 
\begin{align}
\label{E-MYM}
\bs E=E(\bs d A; A)= \Tr\left(\bs d A \big( D^A*\!F - m^2 *\!A\big) \right).
\end{align}
From the cocycle we obtain, 
\begin{align}
\alpha(\chi; A) \defeq&\ \tfrac{d}{d\tau} c(A,e^{\tau\, \chi}) \big|_{\tau=0} = - m^2\, \Tr\big( A\, *\! d\chi \big), \label{Anomaly-MYM} \\[2mm] 
 \bs d  c(A,\gamma)=&\   -m^2\Tr \left( \bs d  A *\!d\gamma\gamma\-  \right), \\[2mm]
  \bs d  c(A,\bs\gamma)=&\  -m^2\Tr \left(\bs d  A *\!d\bs{\gamma\gamma}\-  + A *\! \bs d\big( d\bs{\gamma\gamma}\-\big) + \bs d \big( \bs\gamma\- d\bs\gamma\big) *\! \bs\gamma\-d\bs\gamma \right)
\end{align}
In accordance with  \eqref{Charge-def-2} and \eqref{Q-on-shell}  we get 
\begin{align}
Q_\Sigma\big(\chi; A \big)&=\int_\Sigma \Tr\big( D^A\chi *\! F\big)= \int_{\d\Sigma} \Tr \big( \chi *\!F \big)- \int_\Sigma \Tr\big(\chi \ D^A *\! F\big),  \\*[2mm]
\text{and } \ & \ dQ_\Sigma\big(\chi; A \big)= -m^2 \int_\Sigma \Tr\big( A*\!d\chi \big)_{\ |\S}= \alpha\big(\chi; A \big).
\end{align}
By application of the general formulae \eqref{theta-Theta-GT}-\eqref{FieldGT-E-2} we get the $\bs\H$-gauge transformations, 
\begin{align}
\bs\theta_\Sigma^{\bs \gamma} &= \bs\theta_\Sigma + \int_\Sigma \Tr\big(  D^A\big\{\bs {d\gamma\gamma}\- \big\}\, *\! F \big), \\[2mm]
\bs\Theta_\Sigma^{\bs \gamma} &= \bs\Theta_\Sigma + \int_\Sigma  \bs d \Tr\big(  D^A\big\{\bs {d\gamma\gamma}\- \big\}\, *\! F \big), \\[2mm]
\bs E^{\bs \gamma} &= \bs E  -m^2\Tr \left(\bs d  A *\!d\bs{\gamma\gamma}\-  + A *\! \bs d\big( d\bs{\gamma\gamma}\-\big) + \bs d \big( \bs\gamma\- d\bs\gamma\big) *\! \bs\gamma\-d\bs\gamma \right)
						 - d\Tr\big(  D^A\big\{\bs {d\gamma\gamma}\- \big\}\, *\! F \big).
\end{align}
The case of $\bs E^{\bs \gamma}$ takes some work to check explicitly, while it is easier for the first two. 

\section{Lie algebras extensions}
\label{Lie algebras extensions}

We here remind basic definitions about Lie algebra extensions. We refer e.g.  \cite{Schottenloher2008} Chap 4  for complements. 

Lie$G$ is an extension of Lie$H$ by Lie$A$ is  they can be organised as a SES of vector spaces and Lie algebras,
\begin{align}
\label{SES-Lie-alg-ext}
\makebox[\displaywidth]{
\hspace{-18mm}\begin{tikzcd}[column sep=large, ampersand replacement=\&]
\&0     \arrow [r]         \& \text{Lie}A    \arrow[r, "\iota"  ]          \&  \text{Lie}G    \arrow[r, "\pi"]      \&   \text{Lie}H       \arrow[r]      \& 0,
\end{tikzcd}}  \raisetag{3.4ex}
\end{align}
We have $ \text{Lie}G / \iota(\text{Lie}A) \simeq  \text{Lie}H$ as vectors spaces, but not necessarily as Lie algebras. For $a\in$ Lie$A$ and $g\in$ Lie$G$, 
\begin{align}
\pi \big[\iota(a), g\big]_{\text{{\tiny Lie$G$}}} = \big[ \underbrace{\pi\circ \iota(a)}_{=0}, \pi(g) \big]_{\text{{\tiny Lie$H$}}}=0,
\end{align}
as $\pi$ is a Lie algebra morphism. Thus Lie$A$ is an ideal in Lie$G$. 
\medskip

{\bf Split extensions:}  \ The SES \eqref{SES-Lie-alg-ext} is a split extension if there is a section of $\pi$, i.e. a linear map \mbox{$\ell: $ Lie$H \rarrow $ Lie$G$} s.t. $\pi \circ\ell=\id_{\text{{\tiny Lie$H$}}}$, so  Lie$G \simeq \iota(\text{Lie}A) + \ell(\text{Lie}H)$ as vector spaces, but $\ell(\text{Lie}H)$ is a priori not a subalgebra in Lie$G$. 
\smallskip

$\Rightarrow$ \ If $\ell$ is a Lie algebra morphism, then $\ell(\text{Lie}H)$ is a subalgebra of Lie$G$. 
\smallskip

$\Rightarrow$ \ If furthermore $\ell(\text{Lie}H)$ is an ideal in Lie$G$, then the extension is \emph{trivial}, i.e. Lie$G \simeq \iota(\text{Lie}A) \oplus \ell(\text{Lie}H)$. 
Indeed the bracket is 
\begin{align}
\big[\iota(a)+ \ell(h), \iota(a')+ \ell(h')\big]_{\text{{\tiny Lie$G$}}}  &= \big[\iota(a), \iota(a')\big]_{\text{{\tiny Lie$G$}}}  + \big[\ell(h), \ell(h')\big]_{\text{{\tiny Lie$G$}}} 
												+ \underbrace{ \big[\iota(a) ,\ell(h')\big]_{\text{{\tiny Lie$G$}}} + \big[ \ell(h), \iota(a')\big]_{\text{{\tiny Lie$G$}}} }_{\in\ \iota(\text{Lie}A)\ \cap\ \ell(\text{Lie}H)\, =\, \emptyset, \ \text{ both ideals.}} \notag \\[-3mm]
												&= \iota \left( \big[ a, a'\big]_{\text{{\tiny Lie$A$}}} \right) + \ell \left( \big[ h, h'\big]_{\text{{\tiny Lie$H$}}} \right).
\end{align}

{\bf Central extensions:} \  For central extension one requires that the SES \eqref{SES-Lie-alg-ext} be s.t. $\ker \pi \subset Z(\text{Lie}G)$ - the center of Lie$G$ -  so that Lie$A$ is an \emph{Abelian}~ideal.
\smallskip

$\Rightarrow$ \ If there is a section of $\pi$, $\ell: $ Lie$H \rarrow $ Lie$G$, which is a Lie algebra morphism, the central extension is automatically trivial. Indeed, since $\iota(\text{Lie}A) =\ker \pi \subset Z(\text{Lie}G)$, the bracket is
\begin{align}
\big[\iota(a)+ \ell(h), \iota(a')+ \ell(h')\big]_{\text{{\tiny Lie$G$}}}  &= \big[\iota(a), \iota(a')\big]_{\text{{\tiny Lie$G$}}}  + \big[\ell(h), \ell(h')\big]_{\text{{\tiny Lie$G$}}} 
												+ \underbrace{ \big[\iota(a) ,\ell(h')\big]_{\text{{\tiny Lie$G$}}} }_{=0} + \underbrace{\big[ \ell(h), \iota(a')\big]_{\text{{\tiny Lie$G$}}} }_{=0} \notag \\*[-3mm]
												&= \iota \left( \cancel{\big[ a, a'\big]_{\text{{\tiny Lie$A$}}}}\right) + \ell \left( \big[ h, h'\big]_{\text{{\tiny Lie$H$}}} \right).
\end{align}

$\Rightarrow$ \  If $\ell: $ Lie$H \rarrow $ Lie$G$ is only a \emph{linear} map (vector space morphism) s.t. $\pi \circ\ell=\id_{\text{{\tiny Lie$H$}}}$, then Lie$G \simeq \iota(\text{Lie}A) + \ell(\text{Lie}H)$ as vector spaces still. 
Now, one  defines the map 
\begin{align}
\label{2-cocycle-def}
 \mathscr{C}: \text{Lie}H \times \text{Lie}H &\rarrow \iota\big( \text{Lie}A\big), \notag\\
 					(X, Y)		&\mapsto   \mathscr{C}(X, Y)\defeq \big[ \ell(X), \ell(Y)\big]_{\text{{\tiny Lie$G$}}}  - \ell \big(   \big[ X, Y \big]_{\text{{\tiny Lie$H$}}} 	\big), 
\end{align} 
as indeed $\pi \, \mathscr{C}(X, Y) =0$, i.e. $ \mathscr{C}(X, Y) \in \ker \pi \simeq \iota\big( \text{Lie}A\big)$. This map is \\
1) clearly bilinear and antisymmetric, and \\
2) satisfies $ \mathscr{C}(X, [Y, Z])+  \mathscr{C}(Y, [Z, X])+  \mathscr{C}(Z, [X,Y])=0$.
 \begin{proof} By definition, 
\begin{align}
 \mathscr{C}(X, [Y, Z])+  \mathscr{C}(Y, [Z, X])+  \mathscr{C}(Z, [X,Y])&= \big[ \ell(X), \ell([Y,Z])\big]  - \ell \big(  \big[ X, [Y,Z] \big]  \big) \notag \\
 													 & \quad	 + \  \big[ \ell(Y), \ell([Z,X])\big]  - \ell \big(  \big[Y, [Z,X] \big]  \big) \notag\\
													 & \qquad 		\ +\  \big[ \ell(Z), \ell([X,Y])\big]  - \ell \big(  \big[ Z, [X,Y] \big]  \big), \notag\\
													 &\hspace{-4cm}= - \ell\,  \bigg( \underbrace{\big[ X,  [Y, Z]\big]  +  \big[ Y,  [Z, X] + \big[ Z,  [X, Y]}_{=0\ \text{ by Jacobi in Lie}H.} \big]\big]\bigg) 
													      + \big[ \ell(X),\  [\ell(Y), \ell(Z)]  - \cancel{\mathscr{C}(Y, Z) }\big] \notag\\[-5mm]
													 & \hspace{2.7cm}  + \big[ \ell(Y),\  [\ell(Z), \ell(X)]  - \cancel{\mathscr{C}(Z, Y) }\big]  \notag\\
													 &  \hspace{2.7cm} +  \big[ \ell(Z),\  [\ell(X), \ell(Y)]  - \cancel{\mathscr{C}(X, Y) } \big].   \notag
\end{align}
Which vanishes by the Jacobi identity in Lie$G$. The definition \eqref{2-cocycle-def} and the fact that $\mathscr{C} \in Z(\text{Lie}G)$ is used in the second equality. 
\end{proof}
The map $\mathscr{C}$ is called a $\iota\big( \text{Lie}A\big)$-valued 2-cocycle on Lie$H$. By the vector space isomorphism we have,
\begin{align*}
\text{Lie}A \times \text{Lie}H &\rarrow \text{Lie}G \simeq \iota\big( \text{Lie}A\big) + \ell\big( \text{Lie}H\big),  \\
		  (a, h)			  &\mapsto \iota(a) + \ell(h),
\end{align*}
so the bracket in Lie$G$ is, 
\begin{align}
\big[  \ell(X)+ \iota(a),  \ell(Y)+ \iota(b)\big]_{\text{{\tiny Lie$G$}}} &= \big[  \ell(X),  \ell(Y)\big]_{\text{{\tiny Lie$G$}}}  + 0, \notag\\
												  &= \ell\big( \big[X, Y\big]_{\text{{\tiny Lie$H$}}} \big) + \mathscr{C}(X, Y). 
\end{align}
Remark that if $\ell$ is a Lie algebra morphism, $\mathscr{C}\equiv0$ and we recover  trivial central extensions. 
\medskip

Conversely, given Lie$A$ Abelian and Lie$H$, one can \emph{define} the central extension Lie$G$ of Lie$H$ by Lie$A$ as being the vector space Lie$G=$ Lie$H+$ Lie$A$ equipped with the bracket,
\begin{align}
\label{Bracket-central-ext}
\big[ X', Y'\big]_{\text{{\tiny Lie$G$}}}= \big[ X+a, Y+b \big]_{\text{{\tiny Lie$G$}}} \defeq \big[ X, Y\big]_{\text{{\tiny Lie$H$}}} +  \mathscr{C}(X, Y), 
\end{align}
with $ \mathscr{C}$ a Lie$A$-valued 2-cocycle on Lie$H$. This  bracket  satisfies the Jacobi identity, and is thus a Lie bracket,  iff $ \mathscr{C}(X, [Y, Z])+  \mathscr{C}(Y, [Z, X])+  \mathscr{C}(Z, [X,Y])=0$. 
\medskip

Such a central extansion is trivial iff $\mathscr{C}(X, Y) = \mu \big( \big[ X, Y\big]_{\text{{\tiny Lie$H$}}} \big)$ with $\mu \in \text{Hom}\big(\text{Lie}H; \text{Lie}A\big)$. 
\smallskip 

$\bullet$\ $\ \mathscr{C}(X, Y) = \mu \big( \big[ X, Y\big]_{\text{{\tiny Lie$H$}}} \big)$ $\quad \Rightarrow \quad $ $\exists \ \ell: \text{Lie}H \rarrow \text{Lie}G$ section of $\pi$ and injective Lie algebra morphism. 

\noindent Defining $\ell(X)\defeq X+ \mu(X)$, one gets indeed $\pi \circ \ell =\id_{\text{{\tiny Lie$H$}}}$. Then, on the one hand $\ell \big( [X, Y] \big) = [X,Y] + \mu \left([X, Y] \right)$,  
and on the other hand $[\ell(X), \ell(Y)]= [X+\mu(X), Y+\mu(Y)] = [X, Y] +\mathscr{C}(X, Y)$, by \eqref{Bracket-central-ext}. So  indeed $\ell \big( [X, Y]\big)=[\ell(X), \ell(Y)]$, i.e. $\ell$ is a Lie algebra morphism. 

\smallskip 

$\bullet$\ $\exists \ \ell: \text{Lie}H \rarrow \text{Lie}G$ section of $\pi$ and injective Lie algebra morphism $\quad \Rightarrow \quad$ $\mathscr{C}(X, Y) = \mu \big( \big[ X, Y\big]_{\text{{\tiny Lie$H$}}} \big)$.

\noindent Since $\pi \circ \ell(X)=X$ we must have $\ell(X)=X + \mu(X)$ for some $\mu \in \text{Hom}\big(\text{Lie}H; \text{Lie}A\big)$. Then on the one hand $[\ell(X), \ell(Y)]= [X+\mu(X), Y+\mu(Y)] = [X, Y] +\mathscr{C}(X, Y)$ by \eqref{Bracket-central-ext}, and on the other hand $[\ell(X), \ell(Y)]=\ell\big([X,Y] \big) = [X, Y] + \mu \big([X,Y] \big)$. Since $\ell$ is injective we get $\mathscr{C}(X, Y)= \mu \big([X,Y] \big)$. 
\smallskip

\noindent Thus, denoting the spaces of 2-cocycles and trivial 2-cocycles respectively as
\begin{align}
Z^2\big( \text{Lie}H; \text{Lie}A \big) \defeq&\ \big\{ \mathscr{C} : \text{Lie}H \times \text{Lie}H \rarrow \text{Lie}A\ |   \notag\\*[-.5mm]
									&\hspace{3cm}				 \text{bilin. antisym. and s.t. }\  \mathscr{C}(X, [Y, Z])+  \mathscr{C}(Y, [Z, X])+  \mathscr{C}(Z, [X,Y])=0 \big\}, \notag\\*[1mm]
B^2\big( \text{Lie}H; \text{Lie}A \big) \defeq&\ \big\{ \mathscr{C}\ | \ 	\mathscr{C}(X, Y) = \mu \big( \big[ X, Y\big]_{\text{{\tiny Lie$H$}}} \big), \ \text{ for } \ \mu \in \text{Hom}\big(\text{Lie}H; \text{Lie}A\big)	\big\}	,	\notag					
\end{align}
the cohomology group $H^2\big( \text{Lie}H; \text{Lie}A \big) \defeq Z^2\big( \text{Lie}H; \text{Lie}A \big) / B^2\big( \text{Lie}H; \text{Lie}A \big)$ of non-trivial 2-cocycles classifies the central extensions of Lie$H$. 

\subsection{The Poisson algebra of Noether charges as a central extension of Lie$\H$.} 
\label{PB central extension}

We here prove the claim of section \ref{Presymplectic structure for non-invariant Lagrangians}, that under hypothesis 0 and 1, the Poisson Lie algebra of Noether charges in $c$-equivariant theories,  noted here $\text{PAlg}[Q_\Sigma(\, ;A)]$,  is a central extension of Lie$\H$. The Poisson bracket is given by 
\begin{align*}
\big\{ Q_\Sigma(\chi; A),  Q_\Sigma(\eta; A) \big\} &=  Q_\Sigma([\chi, \eta] ; A) + \mathscr{C}(\chi, \eta) \\[1mm]
									&\text{ with } \quad  \mathscr{C}(\chi, \eta) \defeq \int_{\Sigma}  d\mathscr{A}\big( \lfloor\, \chi, \eta \rfloor ;A\big) + dQ\big([\chi, \eta];A \big).
\end{align*}
We remind that $d\mathscr{A}\big( \lfloor\, \chi, \eta \rfloor ;A\big) \defeq   \beta\big( [\chi, \eta]; A\big) - \iota_{\chi^v} \bs\alpha(\eta) + \iota_{\eta^v} \bs\alpha(\chi) $, and both  $\beta(\chi;A)$ and $dQ(\chi; A)$ are linear in~$\chi$.
To~prove the claim it is enough to show that the map $\mathscr{C}: \text{Lie}\H \times  \text{Lie}\H \rarrow \Omega^0(\A)$, which is obviously antisymmetric, satisfies $ \mathscr{C}(\chi, [\eta, \zeta])+\mathscr{C}(\eta, [\zeta, \chi]) +\mathscr{C}(\zeta, [\chi, \eta]) \equiv 0$. 
On the one hand, by linearity of $dQ$ in its first argument
\begin{align*}
 dQ([\chi, [\eta, \zeta]]; A) + dQ([\eta, [\zeta, \chi]]; A) + dQ([\zeta, [\chi, \eta]]; A) = dQ\big([\chi, [\eta, \zeta]] + [\eta, [\zeta, \chi]] + [\zeta, [\chi, \eta]]; A \big)=0,
\end{align*}
by Jacobi in Lie$\H$. On the other hand 
\begin{align}
\label{2nd-term}
d\mathscr{A}( \lfloor\, \chi, [\eta, \zeta] \rfloor; A) + d\mathscr{A}( \lfloor \eta, [\zeta, \chi]  \rfloor; A) + d\mathscr{A}( \lfloor \zeta, [\chi, \eta] \rfloor; A) &= 
                         \beta\big( [\chi, [\eta, \zeta]]; A\big) - \iota_{\chi^v} \bs\alpha([\eta, \zeta]) + \iota_{[\eta, \,\zeta]^v} \bs\alpha(\chi)  \notag\\
                       &+\beta\big( [\eta, [\zeta, \chi]]; A\big) - \iota_{\eta^v} \bs\alpha([\zeta, \chi]) + \iota_{[\zeta, \,\chi]^v} \bs\alpha(\eta) \notag\\
                       &+\beta\big( [\zeta, [\chi, \eta]]; A\big) - \iota_{\zeta^v} \bs\alpha([\chi, \eta]) + \iota_{[\chi,\, \eta]^v} \bs\alpha(\zeta). 
\end{align}
As for $dQ$, the $\beta$ terms vanish by linearity in the first argument and Jacobi in Lie$\H$. Now, we must use a consistency condition on $\bs \alpha$: By definition $\bs L_{\chi^v} \bs \theta \rdefeq \bs \alpha(\chi)$, and since $[\bs L_{\chi^v}, \bs L_{\eta^v}]= \bs L_{[\chi^v, \,\eta^v]}= \bs L_{[\chi, \,\eta]^v}$ we get 
\begin{align*}
\bs L_{\chi^v} \bs L_{\eta^v} \bs \theta - \bs L_{\eta^v} \bs L_{\chi^v} \bs \theta = \bs L_{[\chi, \eta]^v} \bs \theta \qquad \Rightarrow \qquad      \bs L_{\chi^v} \bs\alpha(\eta)  - \bs L_{\eta^v} \bs \alpha(\chi) = \bs \alpha([\chi, \eta]). 
\end{align*}
Using the latter consistency condition as well as $[\bs L_{\chi^v}, \iota_{\eta^v}]= \iota_{[\chi^v , \eta^v]}= \iota_{[\chi, \eta]^v}$, we can write the first term of each row of the  remaining six terms in \eqref{2nd-term} as, 
\begin{align}
\label{Aux}
- \iota_{\zeta^v} \bs \alpha([\chi, \eta]) &= -\iota_{\zeta^v} \bs L_{\chi^v} \bs\alpha(\eta) + \iota_{\zeta^v} \bs L_{\eta^v} \bs\alpha(\chi) \notag\\
							&= \underline{- \bs L_{\chi^v} \iota_{\zeta^v}  \bs \alpha(\eta) + \iota_{[\chi,\, \zeta]^v}  \bs\alpha(\eta) }_{\,_\text I}+  \underline{ \bs L_{\eta^v} \iota_{\zeta^v} \bs\alpha(\chi) }_{\,_\text{II}}- \iota_{[\eta, \,\zeta]^v}  \bs\alpha(\chi), \notag\\[1.5mm]
- \iota_{\eta^v} \bs \alpha([\zeta, \chi]) &= -\iota_{\eta^v} \bs L_{\zeta^v} \bs\alpha(\chi) + \iota_{\eta^v} \bs L_{\chi^v} \bs\alpha(\zeta) \notag\\
							&= \underline{- \bs L_{\zeta^v} \iota_{\eta^v}  \bs \alpha(\chi) + \iota_{[\zeta,\, \eta]^v}  \bs\alpha(\chi) }_{\,_\text{II}}+  \underline{ \bs L_{\chi^v} \iota_{\eta^v} \bs\alpha(\zeta) }_{\,_\text{III}}- \iota_{[\chi, \,\eta]^v}  \bs\alpha(\zeta),\notag \\[1.5mm]
- \iota_{\chi^v} \bs \alpha([\eta, \zeta]) &= -\iota_{\chi^v} \bs L_{\eta^v} \bs\alpha(\zeta) + \iota_{\chi^v} \bs L_{\zeta^v} \bs\alpha(\eta) \notag\\
							&= \underline{- \bs L_{\eta^v} \iota_{\chi^v}  \bs \alpha(\zeta) + \iota_{[\eta,\, \chi]^v}  \bs\alpha(\zeta) }_{\,_\text{III}}+  \underline{ \bs L_{\zeta^v} \iota_{\chi^v} \bs\alpha(\eta) }_{\,_\text{I}}- \iota_{[\zeta, \,\chi]^v}  \bs\alpha(\eta). 
\end{align}
The sums of terms I, II and III vanish separately, and by inspection of the  remaining terms in \eqref{2nd-term} and \eqref{Aux} it is clear that $d\mathscr{A}( \lfloor\, \chi, [\eta, \zeta] \rfloor; A) + d\mathscr{A}( \lfloor \eta, [\zeta, \chi]  \rfloor; A) + d\mathscr{A}( \lfloor \zeta, [\chi, \eta] \rfloor; A) =0$. 

It follows that $\mathscr{C}(\chi, [\eta, \zeta])+\mathscr{C}(\eta, [\zeta, \chi]) +\mathscr{C}(\zeta, [\chi, \eta]) \equiv 0$, and that $\mathscr{C}$ is a (non-trivial) $\Omega^0(\A)$-valued 2-cocycle on Lie$\H$, i.e.
$\text{PAlg}[Q_\Sigma(\, ;A)] $ is indeed a central extension of Lie$\H$, and we have the SES
\begin{align}
\label{PB-central-ext}
\makebox[\displaywidth]{
\hspace{-18mm}\begin{tikzcd}[column sep=large, ampersand replacement=\&]
\&0     \arrow [r]         \&  \Omega^0(\A)   \arrow[r, "\iota"  ]          \& \text{PAlg}[Q_\Sigma(\, ;A)]      \arrow[r, "\pi"]      \&      \text{Lie}\H    \arrow[l, bend left=25, end anchor={[xshift=1ex]}, "Q_\Sigma(\, \ ;\,A)"]      \arrow[r]      \& 0,
\end{tikzcd}}  \raisetag{3.4ex}
\end{align}
with $Q_\Sigma(\, \ ;A): \text{Lie}\H \rarrow \text{PAlg}[Q_\Sigma(\, ;A)]$, $\chi \mapsto Q_\Sigma(\chi ;A)$, a linear map s.t. $\pi \circ Q_\Sigma(\, \ ;A) =\id_{\text{Lie}\H}$.

\clearpage

\section{Holst Lagrangian}
\label{G}

A possible starting point of Loop Quantum Gravity (LQG) is the Lagrangian 
\begin{align}
\label{LQG Lagrangian}
L_{\text{{\tiny LQG}}}(\b A)=L_{\text{{\tiny EC-$\Lambda$}}}(\b A) +  \tfrac{1}{\upgamma}\, L_{\text{{\tiny Holst}}}(\b A) 
	 &= R \eta\- \bullet e \w e^T -  \tfrac{\epsilon}{2\ell^2} e\w e^T \bullet e\w e^T \ + \ \tfrac{1}{\upgamma}\,\Tr\left( R\, e\w e^T \eta \right) \\
	 &= \left( R^{ab}e^c e^d -  \tfrac{\Lambda}{6} e^a e^be^ce^d\right)\epsilon_{abcd}\ +\ \tfrac{1}{\upgamma}\,{R^a}_b\, e^b \w e_a.  \notag
\end{align}
where $\upgamma$ is  the Barbero-Immirzi parameter (\cite{Rovelli-Vidotto2014} section 3.3). It is classically equivalent to $L_{\text{{\tiny EC-$\Lambda$}}}(\b A) $ because the second term - known as the Holst term \cite{Holst1996} - does not affect the field equations, but only  the presymplectic potential of the theory. In that respect it is akin to the Euler density $L_{\text{{\tiny Euler}}}( A)$ - that we added to $L_{\text{{\tiny EC-$\Lambda$}}}(\b A)$ to get MM gravity $L_{\text{{\tiny MM}}}(\b A)$ - or to the $\uptheta$-term in QCD. Yet, contrary to the former situation, there is (yet) no Cartan geometric justification for the addition of the Holst term. 
Nonetheless, nothing prevent us from applying the results of \ref{Presymplectic structure for invariant Lagrangians} and \ref{Dressed presymplectic structure for invariant Lagrangians} to get the presymplectic structure of $L_{\text{{\tiny LQG}}}(\b A)$ and its dressing. 

As we already analysed the presymplectic structure of $L_{\text{{\tiny EC-$\Lambda$}}}(\b A)$ and its dressed version in sections \ref{4D EC gravity} and \ref{4D gravity}, we here focus on the Holst term. We have, 
\begin{align}
\bs d L_{\text{{\tiny Holst}}}(\b A) = \bs E + d \bs\theta &= E\left(\bs d \b A; \b A \right) + d\theta\left(\bs d \b A; \b A \right), \notag\\
						 				  &= 2\Tr\left(\bs dA\, D^Ae \w e^T\eta + R\,e \w \bs d e^T \eta \right)\ + \ d\Tr\left( \bs d A\, e\w e^T\eta \right).
\end{align}
Now, the Bianchi identity satisfied by the curvature of the Cartan connection $D^{\b A}\b F=0$ gives $D^A T=R\,e$.\footnote{This is true if we assume either Cartan geometries whose homogeneous spaces are (A)dS or Minkowski spaces, and where the Cartan curvatures are respectively $\b F= F+  \tfrac{1}{\ell} T = (R-\tfrac{\epsilon}{2\ell^2}e\w e^T\eta) + \tfrac{1}{\ell} T$ or $\b F= R + T$. In the former case the Bianchi identity gives $D^AT + F\, e =0$, but $F\,e = R\, e$ since $e \w e^T\eta e = e\w g_{\mu\nu}\, dx^{\,\mu} \w dx^\nu\equiv 0$.  } So if we restrict the discussion to the case where $\b A$ is the normal Cartan connection, $T=D^Ae=0$ and  $\bs E\equiv 0$. In any case, the field equations of $L_{\text{{\tiny EC-$\Lambda$}}}(\b A)$ are not modified by $L_{\text{{\tiny Holst}}}(\b A)$. Clearly the latter has trivial equivariance, $R^\star_\gamma L_{\text{{\tiny Holst}}}= L_{\text{{\tiny Holst}}}$ for $\gamma \in \SO(1,3)$. Then, by \eqref{Noether-charge-invariant} we find the associated Noether charges to be, for $\chi \in$ Lie$\SO$, 
\begin{align}
Q_\Sigma(\chi; \b A)&= \int_{\d\Sigma} \theta(\chi; \b A) - \int_\Sigma E(\chi; \b A). \notag\\
			  &= \int_{\d\Sigma} \Tr\big(  \chi\, e\w e^T\eta\big) - 2 \int_\Sigma \Tr\big(  \chi\, D^Ae \w e^T\eta\big). 
\end{align}
These are often called ``dual" charges (see e.g. \cite{Oliveri-Speziale2020-II}). The presymplectic 2-form is $\bs \Theta_\Sigma = - \int_\Sigma \Tr \left( \bs d A\, \bs d \big( e\w e^T \big) \eta \right)$ and satisfies $\iota_{\chi^v} \bs\Theta_\Sigma = -\bs d Q_\Sigma(\chi; \b A)$, so that it induces a Poisson bracket of charges via \eqref{Poisson-bracket}.

The field-dependent $\bs\SO$-gauge transformations of the presymplectic potential and 2-form are, by \eqref{Field-depGT-presymp-pot} and \eqref{Field-depGT-presymp-form}, 
\begin{align}
\bs\theta_\Sigma^{\bs\gamma} &= \bs\theta_\Sigma + \int_{\d\Sigma}    \theta\big(\bs{d\gamma\gamma}\-; \b A\big) - \int_\Sigma  E\big(\bs{d\gamma\gamma}\-; \b A\big), \notag\\[1mm]
						&=\bs\theta_\Sigma +  \int_{\d\Sigma}    \Tr\big(  \bs{d\gamma\gamma}\- e\w e^T\eta\big) - 2 \int_\Sigma \Tr\big(  \bs{d\gamma\gamma}\- D^Ae \w e^T\eta\big). \\[2mm]
\bs\Theta_\Sigma^{\bs\gamma} &= \bs\Theta_\Sigma + \int_{\d\Sigma}   \bs d \theta\big(\bs{d\gamma\gamma}\-; \b A\big) - \int_\Sigma \bs d E\big(\bs{d\gamma\gamma}\-; \b A\big). \notag\\[1mm] 
						&=\bs\theta_\Sigma +  \int_{\d\Sigma}    \bs d \Tr\big(  \bs{d\gamma\gamma}\- e\w e^T\eta\big) - 2 \int_\Sigma   \bs d \Tr\big(  \bs{d\gamma\gamma}\- D^Ae \w e^T\eta\big). 
\end{align}
While the field dependent gauge transformation of the field equations are, by \eqref{Field-depGT-FieldEq},
\begin{align}
\bs E^{\bs \gamma}&= \bs E+  dE \big(  \bs{d\gamma\gamma}\- ;  A \big), \notag \\
			      &= \bs E+  2\,d \Tr\big(  \bs{d\gamma\gamma}\-  D^Ae \w e^T\eta\big).
\end{align}
These relations can be checked explicitly using \eqref{GT-dA-Cartan}. 

Using the results of section \ref{4D EC gravity},  we obtain that the presymplectic potential and 2-form of $L_{\text{{\tiny LQG}}}(\b A)$ are,
\begin{align}
\bs \theta_\Sigma^{\text{{\tiny LQG}}} &= \bs \theta_\Sigma^{\text{{\tiny EC-$\Lambda$}}} +\tfrac{1}{\upgamma}\, \bs \theta_\Sigma^{\text{{\tiny Holst}}}
							 =  \int_\Sigma \bs d A \bullet e \w e^T + \tfrac{1}{\upgamma}\, \Tr\big(  \bs d A\, e\w e^T\eta\big), \\[1mm]
\bs \Theta_\Sigma^{\text{{\tiny LQG}}} &= \bs \Theta_\Sigma^{\text{{\tiny EC-$\Lambda$}}} +\tfrac{1}{\upgamma}\, \bs \Theta_\Sigma^{\text{{\tiny Holst}}}
							 =  - \int_\Sigma \bs d A \bullet  \bs d \big(e \w e^T\big) + \tfrac{1}{\upgamma}\, \Tr\big(  \bs d A\, \bs d\big( e\w e^T\big)\eta\big).							 
\end{align}
The latter induces a Poisson bracket of Noether charges
\begin{align}
Q_\Sigma^{\text{{\tiny LQG}}}(\chi; \b A)&= \int_{\d\Sigma} \chi \bullet e \w e^T + \tfrac{1}{\upgamma} \Tr\big(  \chi\, e\w e^T\eta\big) - 2 \int_\Sigma \chi \bullet D^Ae\w e^T  + \tfrac{1}{\upgamma} \Tr\big(  \chi\, D^Ae \w e^T\eta\big), \\[1mm]
								    &= \int_{\d\Sigma} \chi \bullet e \w e^T + \tfrac{1}{\upgamma} \Tr\big(  \chi\, e\w e^T\eta\big)_{\ |N},
\end{align}
since $\iota_{\chi^v} \bs\Theta_\Sigma^{\text{{\tiny LQG}}} \!= -\bs d Q_\Sigma^{\text{{\tiny LQG}}}(\chi; \b A)$. These charges  generate infinitesimal $\SO(1,3)$ transformations by $\big\{Q_\Sigma^{\text{{\tiny LQG}}}(\chi; \b A); \ \ \, \big\}$. 
\bigskip

\noindent {\bf Dressing:} \ The dressing field in this context is the tetrad field, $\bs u =\bs u(\b A)\defeq \bs e={e^a}_\mu:\U \subset \M \rarrow GL(4)$, which satisfies $R^\star_\gamma \bs e =\gamma\- \bs e$ for $\gamma \in \SO(1,3)$. The dressed Cartan connection $\b A^{\bs u}$ is given by \eqref{dressed Cartan connection} and its dressed curvature is  \eqref{dressed Cartan curvature}. 
Since the target group of the dressing field, $G=GL(4)$, is also the invariance group of $\Tr$ from which the Holst term is built, it is easy to obtain its dressed version and the corresponding dressed presymplectic structure.  The dressed Holst term is 
\begin{align}
L_{\text{{\tiny Holst}}}^{\bs u}(\b A)=L_{\text{{\tiny Holst}}}(\b A^{\bs u}) = \Tr\left(  \bs e {\sf R} \bs e\- \, e\w e^T \eta  \right)
													  = \Tr\left( {\sf R} \cdot dx \w dx^T\! \cdot g \right),
\end{align}  
with the ``$\cdot$" notation meaning contraction on greek indices. This is  the ``metric" version, which reads explicitly, 
\begin{align}
L_{\text{{\tiny Holst}}}(\b A^{\bs u}) = {{\sf R}^\alpha}_\beta\, dx^{\, \beta} \w dx^{\, \mu} g_{\mu\alpha} = -\tfrac{1}{2}\, {\sf R}_{\,\beta \mu,\, \rho\sigma}\, \epsilon^{\,\beta \mu \rho\sigma}\, d^4\!x. 
\end{align}  
The corresponding dressed presymplectic potential is, by \eqref{theta-dressed},
\begin{align}
\bs\theta_\Sigma^{\bs u} &=  \bs \theta_\Sigma + \int_{\d\Sigma} \theta( \bs{duu}\- ; \b A) - \int_\Sigma E(\bs{duu}\-; \b A),    \notag \\[1mm]
				      &= \bs \theta_\Sigma + \int_{\d\Sigma} \Tr\big(  \bs{dee}\-e\w e^T\eta \big) -  2 \int_\Sigma \Tr\big(  \bs{dee}\-D^Ae\w e^T\eta \big),
\end{align}  
where the boundary term simplifies as $\int_{\d\Sigma} \Tr\big(  \bs d e\w e^T\eta \big)= \int_{\d\Sigma}  \bs d e^a\w e_a$. It is also, in ``metric" form,
\begin{align}
\label{dressed-pot-Holst}
\bs\theta_\Sigma^{\bs u}=\int_\Sigma \theta\left( \bs e \bs d \Gamma \bs e\-; \b A\right)= \int_\Sigma \Tr\big( \bs e \bs d \Gamma \bs e\-e\w e^T \eta \big) 
				     &=  \int_\Sigma \Tr\big( \bs d \Gamma \cdot dx \w dx^T\! \cdot g \big), \\
				     & = \int_\Sigma g_{\mu \alpha}\, \bs d{ \Gamma^\alpha}_{\beta\nu} \,dx^{\nu}\w dx^{\,\beta}\w dx^{\, \mu}.  \notag
\end{align}  
By \eqref{Theta-dressed}, and by definition, the dressed presymplectic 2-form is 
\begin{align}
\label{dressed-2form-Holst}
\bs \Theta_\Sigma^{\bs u} 
				      = \bs \Theta_\Sigma + \int_{\d\Sigma}  \bs d \Tr\big(  \bs{d}e\w e^T\eta \big) -  2 \int_\Sigma  \bs d \Tr\big(  \bs{dee}\-D^Ae\w e^T\eta \big),
\end{align}  
or in ``metric" form $\bs \Theta_\Sigma^{\bs u} = \bs d\bs \theta^{\bs u}_\Sigma = \int_\Sigma  \bs d g_{\mu \alpha}\, \bs d{ \Gamma^\alpha}_{\beta\nu} \,dx^{\nu}\w dx^{\,\beta}\w dx^{\, \mu}$.
\medskip

With this and the results of section \ref{4D gravity}, we have that the dressed/``metric" version of the LQG Lagrangian is 
\begin{align}
L_{\text{{\tiny LQG}}}^{\bs u}(\b A)=L_{\text{{\tiny LQG}}}(\b A^{\bs u}) 
	 & =  \sqrt{|\det(\bs g)|} \ \left( \ {\sf R} \bs g\- \bullet dx \w dx^T  - \tfrac{\epsilon}{2\ell^2} dx \w dx^T \bullet dx \w dx^T \ \right)\ + \ \tfrac{1}{\upgamma} \Tr\left( {\sf R} \cdot dx \w dx^T \! \cdot g \right), \notag \\
	 & = 2 \sqrt{|\det(\bs g)|} \,d^4\!x \ \big( Ric - 2\Lambda \big)  \ - \ \tfrac{1}{2\upgamma} \, {\sf R}_{\,\beta \mu,\, \rho\sigma}\, \epsilon^{\,\beta \mu \rho\sigma}\, d^4\!x.
\end{align}  
The associated dressed presymplectic potential is thus, 
\begin{align}
\big(\bs \theta_\Sigma^{\text{{\tiny LQG}}}\big)^{\bs u} & = \bs \theta_\Sigma^{\text{{\tiny LQG}}} + \int_{\d\Sigma} \bs{dee}\-\bullet e\w e^T  + \tfrac{1}{\upgamma} \Tr\big( \bs de \w e^T\eta \big) 
																	- 2 \int_\Sigma \bs{dee}\-\bullet D^A e\w e^T  + \tfrac{1}{\upgamma} \Tr\big( \bs{dee}\- \w D^A \w e^T\eta \big), \notag\\
								       & =  \bs \theta_\Sigma^{\text{{\tiny LQG}}} + \int_{\d\Sigma} \bs{dee}\-\bullet e\w e^T  + \tfrac{1}{\upgamma} \Tr\big( \bs de \w e^T\eta \big) _{\ |N}. 
\end{align}  
This reproduces eq.(8.14)-(8.16) in \cite{Oliveri-Speziale2020}. The expression of the dressed presymplectic 2-form $\big(\bs \Theta_\Sigma^{\text{{\tiny LQG}}}\big)^{\bs u}$ is clear. 
\clearpage


\noindent {\bf Residual transformations:} \ 
Since $\bs\theta_\Sigma^{\bs u}$ \eqref{dressed-pot-Holst} and $\bs\Theta_\Sigma^{\bs u}$ \eqref{dressed-2form-Holst} are basic by construction, the  Lorentz charges vanish identically $Q_\Sigma\big(\chi; \b A^{\bs u} \big)\defeq \iota_{\chi^v}\bs\theta_\Sigma^{\bs u}\equiv0$, and there is no Poisson bracket to speak of since $\iota_{\chi^v}\bs\Theta_\Sigma^{\bs u}\equiv0$. 

Yet, as we know, there are residual transformations of the second kind due to the ambiguity in the choice of the dressing field: coordinate transformations $\t\G=\GL(4)$. By \eqref{dressed-Noether-charge-invariant-2nd-kind} the corresponding dressed Noether charges are, for $\alpha= {\alpha^\mu}_\nu \in$ Lie$\t\G$,
\begin{align}
Q_\Sigma\big(\alpha; \b A^{\bs u} \big)\defeq&\, \iota_{\alpha^v} \bs\theta_\Sigma^{\bs u} = \int_{\d\Sigma} \theta\big( \alpha ; \b A^{\bs u} \big)    -  \int_{\Sigma} E\big( \alpha ; \b A^{\bs u} \big) 
															           =  \int_{\d\Sigma} \theta\big(\bs e \alpha \bs e\-; \b A \big)    -  \int_{\Sigma} E\big(\bs e \alpha \bs e \-  ; \b A \big), \notag\\
						 =&\,  \int_{\d\Sigma} \Tr\big(\bs e \alpha \bs e\- e\w e^T \eta \big)    - 2 \int_{\Sigma} \Tr\big(\bs e \alpha \bs e \-   D^Ae \w e^T\eta \big) \notag \\
					         =&\, \int_{\d\Sigma} \Tr\big( \alpha \cdot dx \w dx^T\!  \cdot g \big)    - 2 \int_{\Sigma} \Tr\big( \alpha \cdot {\sf T} \w dx^T\!  \cdot g \big), \\
					         =&\, \int_{\d\Sigma} \tfrac{1}{2}\,\alpha_{[\sigma\nu]}\, dx^{\nu}\w dx^\sigma -  2 \int_{\Sigma} \alpha_{\sigma\nu}\, {\sf T}^\nu\w dx^\sigma, \\
					          =&\, \int_{\d\Sigma} \tfrac{1}{2}\,\alpha_{[\sigma\nu]}\, dx^{\nu}\w dx^\sigma_{\ \ \ |N}. \notag
\end{align}

Notice that when the coordinate change is interpreted actively, as resulting from the action of a diffeomorphism of $\M$, so that $\alpha=\d \zeta$ for $\zeta^\mu$ the components of the generating vector field, the above result gives the (dual) Noether charge for diffeomorphisms. In particular, the integrand of the boundary term becomes $\tfrac{1}{2}\, \d_{[\nu\,}\zeta_{\sigma]}\, dx^{\nu}\w dx^\sigma = d \zeta^*$, where $\zeta^*\defeq g(\zeta, \ )= \zeta^*_\sigma\, dx^\sigma = g_{\sigma\mu}\zeta^{\,\mu}\, dx^\sigma$ is the dual 1-form of the vector field $\zeta$. The boundary term thus vanishes,   
\begin{align}
Q_\Sigma\big(\d\zeta; \b A^{\bs u} \big) = -  2 \int_{\Sigma} \d_\nu\,\zeta_\sigma\, {\sf T}^\nu\w dx^\sigma.
\end{align}
It then turns out that in the normal case,  the metric formulation has  no dual Noether charges of  diffeomorphisms.
\medskip

From the above and the results of section \ref{4D gravity}, we immediately deduce that there are no Lorentz charges associated to $L_{\text{{\tiny LQG}}}^{\bs u}$ since $\big(\bs \theta_\Sigma^{\text{{\tiny LQG}}}\big)^{\bs u}$ is $\SO$-basic, but  the latter induces the dressed Noether $\t\G$-charges
\begin{align}
Q_\Sigma^{\text{{\tiny LQG}}}\big(\alpha; \b A^{\bs u} \big) &= Q_\Sigma^{\text{{\tiny EC+$\Lambda$}}}\big(\alpha; \b A^{\bs u} \big) + \tfrac{1}{\upgamma} Q_\Sigma^{\text{{\tiny Holst}}}\big(\alpha; \b A^{\bs u} \big), \notag \\[1mm]
											&=  \int_{\d\Sigma} \sqrt{|\det(\bs g)|} \ \ \alpha \bs g\-  \bullet dx \w dx^T + \tfrac{1}{\upgamma}\Tr\big(   \alpha \cdot dx \w dx^T \! \cdot g \big) \notag\\
												& \qquad - 2 \int_\Sigma  \sqrt{|\det(\bs g)|} \ \ \alpha  \bs g\- \bullet {\sf T} \w dx^T  + \tfrac{1}{\upgamma}\Tr\big( \alpha \cdot {\sf T} \w dx^T\!  \cdot g \big), \\
											&= 	 \int_{\d\Sigma} \sqrt{|\det(\bs g)|} \ \ \alpha \bs g\-  \bullet dx \w dx^T + \tfrac{1}{\upgamma}\Tr\big(   \alpha \cdot dx \w dx^T\!  \cdot g \big)_{\ |N}, \notag\\
			&=  \int_{\d\Sigma} \sqrt{|\det(\bs g)|} \ \epsilon_{\,\mu\nu\alpha\beta}\ \alpha^{\,\mu\nu}\, dx^{\,\alpha} \w dx^{\,\beta}\ +\  \tfrac{1}{\upgamma} \, \alpha^{\,\mu\nu}\, dx^{\,\mu} \w dx^{\nu}_{\ \ \ |N}.  
\end{align}
As usual the dressed presymplectic 2-form is s.t. $\iota_{\alpha^v} \big(\bs \Theta_\Sigma^{\text{{\tiny LQG}}}\big)^{\bs u} = - \bs d Q_\Sigma^{\text{{\tiny LQG}}}\big(\alpha; \b A^{\bs u} \big)$ and induces a Poisson bracket via \eqref{dressed-Poisson-bracket-2nd-kind},  so that the Poisson algebra of charges is isomorphic to Lie$\t\G$.

In case $\alpha=\d\zeta$ for $\zeta$ generating an active diffeomorphism, 
\begin{align}
Q_\Sigma^{\text{{\tiny LQG}}}\big(\d\zeta; \b A^{\bs u} \big) &= Q_\Sigma^{\text{{\tiny EC+$\Lambda$}}}\big(\d\zeta; \b A^{\bs u} \big) + \tfrac{1}{\upgamma} Q_\Sigma^{\text{{\tiny Holst}}}\big(\d\zeta; \b A^{\bs u} \big), \notag \\[1mm]
											&=  \int_{\d\Sigma} \sqrt{|\det(\bs g)|} \ \ \d\zeta \bs g\-  \bullet dx \w dx^T + \tfrac{1}{\upgamma}\Tr\big(   \d\zeta \cdot dx \w dx^T\!  \cdot g \big) \notag\\
												& \qquad - 2 \int_\Sigma  \sqrt{|\det(\bs g)|} \ \ \d\zeta  \bs g\- \bullet {\sf T} \w dx^T  + \tfrac{1}{\upgamma}\Tr\big( \d\zeta \cdot {\sf T} \w dx^T \! \cdot g \big), \\
&= 	 \int_{\d\Sigma} \sqrt{|\det(\bs g)|} \ \ \d\zeta \bs g\-  \bullet dx \w dx^T + \tfrac{1}{\upgamma}\ d\zeta^*_{\ \ |N}, \notag\\
			&=  \int_{\d\Sigma} \sqrt{|\det(\bs g)|} \ \epsilon_{\,\mu\nu\alpha\beta}\ \nabla^{[\mu}\zeta^{\nu]}\, dx^{\,\alpha} \w dx^{\,\beta}_{\ \ \ |N}.  
\end{align}
That is, in the normal case $Q_\Sigma^{\text{{\tiny LQG}}}\big(\d\zeta; \b A^{\bs u} \big) = Q_\Sigma^{\text{{\tiny EC+$\Lambda$}}}\big(\d\zeta; \b A^{\bs u} \big)$ reduces to the Komar integral and there are no ``dual"  charges of diffeomorphisms coming from the Holst term. This converges with some of the recent results of \cite{Oliveri-Speziale2020-II}. The Poisson algebra of these charges  is isomorphic to Lie$\Diff(\M)\simeq \Gamma(T\M)$. 
\clearpage

{
\normalsize 
 \bibliography{Biblio2}
}

\end{document}